\documentclass[aps,rmp,reprint,normalem,longbibliography,nofootinbib]{revtex4-2}
\usepackage{bm}
\usepackage{bm,dcolumn,amsmath,graphicx,url,textcase}
\usepackage{epsfig}
\usepackage{epstopdf}
\usepackage{euscript}
\usepackage{amsfonts}
\usepackage{float}
\usepackage{braket}
\DeclareMathAlphabet{\mathpzc}{OT1}{pzc}{m}{it}
\usepackage{threeparttable}   
\usepackage{longtable}                  

\usepackage{textpos}
\usepackage{amssymb,mathrsfs,amsthm}
\usepackage{booktabs}
\usepackage{array}
\usepackage{tabu}
\usepackage{dcolumn}
\usepackage{rotating}
\usepackage{ulem}
\usepackage{soul}

\usepackage{upgreek}
\usepackage{textcomp}
\usepackage{xcolor}
\usepackage{datetime2}
\usepackage{feynmp-auto}  
\usepackage{tikz}  
\usepackage[T1]{fontenc}
\usepackage{titlesec}
\usepackage{nicefrac}
\usepackage{color}
\usepackage{cancel}
\usepackage{appendix}
\usepackage{makecell}
\usepackage[colorlinks=true,linkcolor=blue,citecolor=blue,urlcolor=blue,bookmarksopen,breaklinks,linktoc=page]{hyperref}
\usepackage{orcidlink}

\renewcommand{\theparagraph}{\alph{paragraph}.}

\titleformat{\paragraph}[runin]{\normalfont\normalsize\itshape}{\theparagraph}{0.4em}{}[{\mbox{}\rule[0.5ex]{1.2em}{0.8pt}}]

\DeclareMathAlphabet{\mathpzc}{OT1}{pzc}{m}{it}

\def\sV{{\ensuremath{\EuScript V}}}
\def\sN{{\ensuremath{\EuScript N}}}
\def\mr{\mathrm}
\def\bs{\boldsymbol}
\def\mc{\mathcal}

\newcommand{\LC}[1]{ \textcolor{orange}{LC: #1}}

\newcommand{\YS}[1]{\textcolor{cyan}{YS: #1}}

\usepackage[]{ezedits}
\defineEdit{DK}{\color[rgb]{1.0,0.0,0.0}}{\color[rgb]{1.0,0.0,0.0}}
\defineEdit{WJ}{\color[rgb]{0.0,1.0,0.0}}{\color[rgb]{0.0,1.0,0.0}}

\newcommand{\appropto}{\mathrel{\vcenter{
  \offinterlineskip\halign{\hfil$##$\cr
    \propto\cr\noalign{\kern2pt}\sim\cr\noalign{\kern-2pt}}}}}
\renewcommand{\v}[1]{\boldsymbol{#1}}		

\def\HIM{Helmholtz Institute Mainz, 55099 Mainz, Germany}
\def\GSI{GSI Helmholtzzentrum für Schwerionenforschung GmbH, 64291 Darmstadt, Germany}
\def\JGU{Johannes Gutenberg University, Mainz 55128, Germany}
\def\SHU{International Center of Quantum Artificial Intelligence for Science and Technology (QuArtist)\\ and
Department of Physics, Shanghai University, 200444 Shanghai, China}
\def\FMUV{Faculty of Mathematics, University of Vienna, Oskar-Morgenstern-Platz 1, 1090 Vienna, Austria}
\def\GPGUV{Gravitational Physics Group, University of Vienna, W\"{a}hringer Stra{\ss}e 17, 1090 Vienna, Austria}
\def\LUSTC{CAS Key Laboratory of Microscale Magnetic Resonance and School of Physical Sciences, University of Science and Technology of China, Hefei, Anhui 230026, China}
\def\CUSTC{CAS Center for Excellence in Quantum Information and Quantum Physics, University of Science and Technology of China, Hefei, Anhui 230026, China}
\def\USTC{University of Science and Technology of China, Hefei, Anhui 230026, China}
\def\UNSW{School of Physics, University of New South Wales, Sydney, New South Wales 2052, Australia}
\def\CSUEB{Department of Physics, California State University -- East Bay, Hayward, California 94542-3084, USA}

\def\PNPI{Petersburg Nuclear Physics Institute of NRC ``Kurchatov Institute'', Gatchina 188300, Russia}
\def\ETU{St.Petersburg Electrotechnical University LETI, Prof. Popov Str. 5, 197376 St. Petersburg, Russia}
\def\USyd{School of Physics, The University of Sydney, Sydney, New South Wales 2006, Australia}
\def\UC{Department of Physics, University of California at Berkeley, Berkeley, California 94720-7300, USA}
\begin{document}

\title{Spin-dependent exotic interactions}

\date{\today}

\author{Lei Cong}
\email{congllzu@gmail.com}
\thanks{The author contribute equally to this paper.}
\affiliation{\HIM}
\affiliation{\GSI}
\affiliation{\JGU}
\affiliation{\SHU}

\author{Wei Ji}
\email{wei.ji.physics@gmail.com}
\thanks{The author contribute equally to this paper.}
\affiliation{\HIM}
\affiliation{\GSI}
\affiliation{\JGU}

\author{Pavel Fadeev}
\affiliation{\JGU}

\author{Filip Ficek}
\affiliation{\FMUV}
\affiliation{\GPGUV}

\author{Min Jiang}
\affiliation{\LUSTC}
\affiliation{\CUSTC}

\author{Victor V.~Flambaum \orcidlink{0000-0001-8643-7374}}
\affiliation{\UNSW}

\author{Haosen Guan}
\affiliation{\USTC}

\author{Derek~F.~Jackson Kimball \orcidlink{0000-0003-2479-6034}}
\affiliation{\CSUEB}

\author{Mikhail G.~Kozlov \orcidlink{0000-0002-7751-6553}}
\affiliation{\PNPI}
\affiliation{\ETU}

\author{Yevgeny V.~Stadnik \orcidlink{0000-0002-3544-5160}}
\affiliation{\USyd}

\author{Dmitry Budker \orcidlink{0000-0002-7356-4814}}
\affiliation{\HIM}
\affiliation{\GSI}
\affiliation{\JGU}
\affiliation{\UC}

\begin{abstract}

Novel interactions beyond the four known fundamental forces in nature (electromagnetic, gravitational, strong and weak
interactions),
may arise due to ``new physics'' beyond the standard model, 
manifesting as a ``fifth force''.
This review is focused on spin-dependent fifth forces mediated by exotic bosons such as spin-0 axions and axionlike particles and spin-1 $Z'$ bosons, dark photons, or paraphotons. Many of these exotic bosons are candidates to explain the nature of dark matter and dark energy, and their interactions may violate fundamental symmetries. 
Spin-dependent interactions between fermions mediated by the exchange of exotic bosons have been 
investigated in a variety of experiments, particularly at the low-energy frontier. 
Experimental methods and tools 
used to search for exotic spin-dependent interactions,
such as atomic comagnetometers, torsion balances, nitrogen-vacancy spin sensors, and precision atomic and molecular spectroscopy, 
are described.
A complete set of interaction potentials, 
derived based on quantum field theory with minimal assumptions and characterized
in terms of reduced coupling constants,
are presented. 
A comprehensive summary
of existing 
experimental and observational constraints on exotic
spin-dependent 
interactions is given, illustrating the current research landscape and promising directions of further research.
\end{abstract}

\maketitle

\begin{sloppypar}
\tableofcontents
\end{sloppypar}
\newpage

\newpage


\section{Introduction}

\subsection{Background}\label{Sec:Intro-bg}
The contemporary standard model (SM) of particle physics \cite{barr_particle_2016,griffiths_introduction_2020,particle_data_group_review_2022} has been remarkably successful. It describes the interactions among fundamental particles, with a notable exception of gravity at high energies, and has made many predictions that have later been proven to be correct at high levels of precision.
Within the framework of the SM,
four fundamental interactions or forces are recognized: the strong 
force, weak 
force, electromagnetism, and gravity. Furthermore, electromagnetic and weak forces appear disparate at low energies but merge into a single force at high energies.

While there is currently no direct evidence for any fundamental interactions or forces beyond the established four,\footnote{Note that the discovery of the Higgs boson \cite{aad_observation_2012,chatrchyan_observation_2012}
indirectly suggests the presence of ``fifth force'' mediated by the exchange of Higgs bosons \cite{salam_higgs_2022}. Because of its relatively large mass ($\sim$125\,GeV/$c^2$) the Higgs-mediated force acts only over short distances, and has not yet been directly observed. 
However, the strength of this force is proportional to the product of the masses of the interacting fermions and this force gives a noticeable contribution to the energy of mesons containing heavy quarks; see, for example, \citet{flambaum_two_2011}.
} there is in principle no argument that would rule out such a possibility. Instead, it is possible to constrain the strengths of such hypothetical forces using precise experimental measurements \cite{fischbach_reanalysis_1986,fischbach_six_1992}. 
Fifth forces are generally classified into two broad categories: those that depend on the spins of at least one of the interacting particles (spin-dependent forces) \cite{dobrescu_spin-dependent_2006,fadeev_revisiting_2019} and those that are independent of spins \cite{adelberger_tests_2003}. 
While both types of forces may arise within the same theoretical frameworks, in the current review we focus on spin-dependent forces.

\subsection{Motivation for searching for exotic spin-dependent interactions }\label{Sec:Intro-motiv}



\subsubsection{Role of intrinsic spin in gravity}\label{Sec:Intro-motiv-spin-gravity}

The idea of intrinsic spin, which is a particular kind of angular momentum, comes from relativistic quantum mechanics or quantum field theory. 
However, in contrast to orbital angular momentum, it cannot be understood as arising from the physical rotation of an object. 
In the early days of the development of quantum mechanics, 
soon after the discovery of the Pauli spin matrix formalism,
researchers tried to understand how to incorporate the concept of spin into the framework of general relativity, which describes the force of gravity, attempting to extend our understanding of gravity to the quantum level.

One way to incorporate spin into general relativity is through the concept of spacetime torsion \cite{de_sabbata_spin_1994}.
Torsion quantifies the twisting of a coordinate system as it is transported along a curve. 
In Einsteinian general relativity, mass (energy) generates curvature of spacetime, but the torsion is zero. 
However, in an extension of general relativity proposed by \citet{cartan_generalization_1922}, torsion is introduced as an additional aspect of spacetime: torsion is caused by the spin of particles and can also influence particle dynamics
\cite{hehl_general_1976}. 
This extension, known as Einstein-Cartan theory, is still a subject of ongoing research.

Consequently, one theoretical impetus encouraging experimental searches for exotic spin-dependent interactions is speculation concerning the possibility of a torsion-related spin-gravity coupling manifesting as a gravitational dipole moment (GDM) of elementary fermions (\citealp{morgan_direct_1962}; \citealp{kobzarev_gravitational_1963}; \citealp{leitner_parity_1964}; \citealp{dass_test_1976}; \citealp{peres_test_1978}), or a gravitomagnetic force \cite{mashhoon_gravitomagnetic_2021}. 
If a spin-gravity coupling exists, then particles with a nonzero GDM would experience a spin-dependent force in the presence of a gravitational field.

Gravitomagnetism is a particularly interesting case to consider, as it provides a clear testable example of the most straightforward way to extend general relativity to incorporate intrinsic spin (\citealp{silenko_semiclassical_2005}; \citealp{adler_gravitomagnetism_2012}; \citealp{mashhoon_spin_2013}).
General relativity predicts that massive bodies drag spacetime around with them as they rotate, which consequently leads to precession of macroscopic gyroscopes. This is the Lense-Thirring effect \cite{lense_uber_1918} 
measured, for example, by the Gravity Probe B mission \cite{everitt_gravity_2011}. 
It remains an experimentally open question as to whether intrinsic spins undergo general relativistic precession as do macroscopic rotating objects, and there have been recent proposals to develop experimental probes to answer this question \cite{fadeev_gravity_2021}.\footnote{Note also the related work of \citet{trukhanova_search_2023}, who investigated the effects of torsion on the electromagnetic field, specifically the ``electric'' and ``magnetic'' components of the torsion field. 
They derived modified Maxwell's equations that describe how the electromagnetic field behaves in the presence of torsion, and studied propagation of electromagnetic waves under the influence of a uniform, homogeneous torsion field. 
They showed that in the presence of a nonzero torsion field, such a wave exhibits a Faraday effect, in which the polarization of the wave is rotated, and used astrophysical data to set a bound on the possible coupling/interaction strength of cosmic axial torsion fields.} 

When considering the possibility of the existence of gravitational torsion and how it could be detected using spin-based sensors, a subtle issue arises. Namely how does one differentiate torsion from new spin-dependent forces that are mediated by exotic bosons that have no particular connection to gravity? 

On the one hand, effects that manifest at a level consistent with the strength of the gravitational force and with universal coupling proportional to mass would suggest a connection. Another characteristic that would strongly hint that a newly found spin-dependent interaction was connected to torsion would be if it satisfied the equivalence principle. This would mean that accelerating frames would be indistinguishable from gravitational fields in terms of the spin-precession effects they induce \cite{singh_einstein_1997}.
On the other hand, in many models of torsion where the coupling strength is a free parameter, the distinction may be purely semantic. An example of this is the appearance of torsion in scalar-vector-tensor extensions of general relativity \cite{heisenberg_systematic_2019,aoki_ghostfree_2019}. Therefore, measurement of a coupling between spins and mass could be equally well described by particular models of gravitational torsion as well as an exotic spin-0 field with both scalar and pseudoscalar interactions which would generate a monopole-dipole coupling.

\subsubsection{Ultralight bosons as dark matter}\label{Sec:Intro-motiv-DM}

Cosmological and astrophysical observations indicate that the majority of matter is invisible and non-luminous ``dark matter'' -- estimated to constitute over 80\% of the matter mass fraction in the Universe (\citealp{spergel_dark_2015}; \citealp{bertone_new_2018}).
To understand the microscopic nature of this dark matter is a major objective of cosmology, astrophysics and particle physics, as it could provide insight into the early Universe and uncover new physical laws (\citealp{bertone_particle_2005}; \citealp{feng_dark_2010}; \citealp{chadha-day_axion_2022}). 
The evidence for the existence of dark matter is based entirely on its gravitational effects, which can be observed on a galactic scale and beyond, such as in measurements of galactic rotation curves \cite{rubin_rotation_1970},
observations of the velocity dispersion of galactic clusters \cite{zwicky_masses_1937}
and colliding galactic clusters \cite{clowe_direct_2006},
the formation of large-scale structure in the Universe (\citealp{springel_simulations_2005}; \citealp{angulo_large-scale_2022}),
and studies of the cosmic microwave background radiation \cite{bennett_nine-year_2013,planck_collaboration_planck_2016,hinshaw_nine-year_2013}.
Ultimately, to understand the microscopic nature of dark matter, it is crucial to try to measure its non-gravitational interactions with particles and fields of the SM. 

A potential explanation for dark matter is that it consists of ultralight bosons, with masses less than about one eV.
If such bosons indeed exist, they could interact with spins and thus mediate spin-dependent interactions as described in the present review.
The hypothesis that dark matter is made of ultralight bosons is among the most compelling motivations to search for new spin-dependent interactions, as discussed, for example, by \citet{jackson_kimball_search_2023}.

Like the vast majority of galaxies observed to date, it is believed that the stars of the Milky Way are embedded within and rotate through a spherical dark matter halo (\citealp{navarro_structure_1996-1}; \citealp{jiao_detection_2023}).
In addition to searching for new forces mediated by such ultralight bosons, if the ultralight bosons constitute the dark matter, one can attempt to directly detect the interaction of the bosonic dark matter with quantum sensors.
This is the strategy of ``haloscope'' experiments such as the long-running Axion Dark Matter eXperiment, ADMX 
\cite{admx_collaboration_search_2018,admx_collaboration_extended_2020}.

Thanks to the progress made in superconducting microwave cavities, magnetic resonance techniques, atomic interferometry, magnetometry, and atomic clocks, a range of experimental ideas have been proposed to use quantum sensors based on these technologies to look for bosonic dark matter candidates with masses between $10^{-22}$ and $10^{-3}$ eV \cite{semertzidis_axion_2022,adams_axion_2022,antypas_new_2022}.
Additionally, ways to investigate ultra-heavy, composite dark matter objects with both astrophysical and terrestrial measurements have been developed (\citealp{pustelny_global_2013}; \citealp{derevianko_hunting_2014}; \citealp{roberts_search_2017}; \citealp{afach_search_2021,afach_what_2023}). 

Such dark matter haloscope experiments and their connections to searches for boson-mediated spin-dependent interactions that are the focus of this review are briefly summarized in Sec.\,\ref{METH2.DM}. 
The ``dark matter connection'' provides a strong motivation to search for exotic spin-dependent interactions. 

\subsubsection{The strong-\texorpdfstring{$CP$}{CP} problem}\label{Sec:Intro-motiv-spin-0}

One of the first and most influential proposals for an exotic boson that couples to spin was the axion (\citealp{weinberg_new_1978}; \citealp{wilczek_problem_1978}; \citealp{kim_weak-interaction_1979}; \citealp{zhitnitsky_possible_1980}; \citealp{shifman_can_1980}; \citealp{dine_simple_1981}).
The axion is a consequence of a proposed solution to the strong-$CP$ problem in quantum chromodynamics (QCD) \cite{peccei_constraints_1977, peccei_mathrmcp_1977}, where $CP$ is the combined charge-conjugation ($C$) and parity ($P$) symmetry. 
The strong-$CP$ problem, reviewed by \citet{peccei_strong_2008}, is a so-called ``fine-tuning'' problem of why the observable $CP$-violating phase in the QCD Lagrangian, $\theta_{\rm{QCD}}$, is extremely small, 
presently constrained to be smaller than $10^{-10}\,$rad, 
as determined through measurements of 
the $CP$-violating permanent electric dipole moment (EDM) of the neutron \cite{abel_measurement_2020} and the EDM of the mercury atom \cite{graner_reduced_2016}.
Naively, one would expect $\theta_{\rm{QCD}}$ to be much larger, namely $\mathcal{O}(1)$. 
Notably, axions are also a leading dark matter candidate \cite{preskill_cosmology_1983,abbott_cosmological_1983,dine_not-so-harmless_1983}, see also \citet{jackson_kimball_search_2023}.  

To solve the strong-$CP$ problem, \citet{peccei_constraints_1977,peccei_mathrmcp_1977} proposed introducing a new global chiral $U(1)$ symmetry, subsequently referred to as the Peccei-Quinn (PQ) symmetry. In this scenario, the $CP$-violating phase $\theta_{\rm{QCD}}$ is not a constant but instead evolves dynamically and tends to a value close to zero due to the spontaneous breaking of the PQ symmetry \cite{weinberg_new_1978,wilczek_problem_1978}. 
In this model, the $CP$-violating phase in the QCD Lagrangian is coupled to the dynamical axion field $a$ such that $\theta_{\rm{QCD}} \rightarrow \theta_{\text{eff}} = \theta_{\rm{QCD}} + {a}/{f_a}$, where $f_a$ is the axion decay constant ($f_a$ is proportional to the energy scale of the PQ symmetry breaking). The quantum of this field is a spin-0 particle known as the axion. Due to instanton\footnote{Instantons refer to classical solutions of the QCD field equations with nontrivial topologies.} effects \cite{vainshtein_abc_1982,coleman_uses_1979}, 
the axion field acquires an effective potential of the form $V_\textrm{eff} \propto - \cos(\theta_\textrm{eff})$, for which the energy (as well as the amount of $CP$ violation) is minimized at $\theta_\textrm{eff} = 0$.\footnote{Initially, the axion model was called the Peccei-Quinn-Weinberg-Wilczek (PQWW) model \cite{peccei_constraints_1977,peccei_mathrmcp_1977,weinberg_new_1978,wilczek_problem_1978}. However, this model predicted unobserved consequences for existing particle properties, so it was quickly ruled out by experiments. Subsequently, it was proposed to increase the energy scale of the PQ symmetry breaking in the model characterized by a larger $f_a$. 
This made the axion much lighter and is the essence of the Dine-Fischler-Srednicki-Zhitnitsky (DFSZ) axion model 
(\citealp{zhitnitsky_possible_1980}; \citealp{dine_simple_1981}),
a so-called ``invisible axion'' model. 
Another invisible axion model is the Kim-Shifman-Vainshtein-Zakharov (KSVZ) model (\citealp{kim_weak-interaction_1979}; \citealp{shifman_can_1980}). 
Note that the KSVZ model assumes that the axion couples at tree level only to a super-heavy, as yet unobserved, quark, while the SM fermions do not couple to the axion at tree level.}

The axion typically couples to the axial-vector current of fermions through a derivative interaction of the form $ \partial_{\mu} \left( {a}/{f_a} \right) \bar{\Psi} \gamma^{\mu} \gamma^5 \Psi $, where $ \bar{\Psi} $ is the Dirac adjoint of the fermion field $\Psi$. 
This coupling to the axial-vector current indicates that the axion momentum interacts with the spin of the fermion. 
On macroscopic scales, the axion field can be interpreted as mediating monopole-dipole and dipole-dipole interactions as discussed by \citet{moody_new_1984}; note also early work by \citet{anselm_possible_1982} discussing a long-range dipole-dipole interaction mediated by ``arions'' -- particles closely related to axions. 
\citet{moody_new_1984} noted that a spin-0 field $\phi$ can couple to fermions in two ways: through a scalar vertex or through a pseudoscalar vertex. 
In the nonrelativistic limit, a fermion coupling to a scalar vertex behaves like a monopole at leading order, while a fermion coupling to a pseudoscalar vertex behaves like a dipole. 
This is related to the fact that in the center-of-mass (CM) frame of the particles, there are only two vectors that can be used to form a scalar or pseudoscalar quantity: spin and momentum. 
If the vertex does not involve the fermion spin at leading order in the nonrelativistic approximation, it is a monopole coupling, but if it does involve the spin, it depends on the dot (scalar) product of the spin and momentum, which is a $P$-odd (parity-odd) pseudoscalar term. 
Therefore, it is the pseudoscalar coupling of particles such as axions or axionlike particles that is responsible for generating new dipole interactions. 

The insights of \citet{moody_new_1984} sparked considerable interest in the search for exotic spin-dependent interactions; see, for example, discussions of theoretical developments in Sec.\,\ref{Sec:formalism} and experimental research in Secs.\,\ref{subsec_g_Pg_S} and \ref{subsec_g_Pg_P}.

\subsubsection{Quantum theories of gravity and axionlike particles (ALPs)}
\label{Sec:Intro-motiv-quantum-gravity}

Physicists have long sought to develop a unified theory of the four fundamental forces of nature: gravity, electromagnetism, the weak force, and the strong force. 
The aim of such a unified theory is to create a single theoretical framework that can explain all interactions in the universe. 
However, combining gravity with the other forces, which are described by quantum field theory in the SM, has thus far been unsuccessful due to the fact that the theory of general relativity breaks down at small distances (high energies).

String theory is a leading hypothesis for unifying gravity with the SM of particle physics (\citealp{zwiebach_first_2004}; \citealp{green_superstring_2012-1,green_superstring_2012}).
It proposes that the matter in the universe is composed of ``strings'' that vibrate in additional dimensions with very small (compared to everyday) 
length scales. 
 
Axionlike particles (ALPs), spin-0 bosons with properties similar to those of axions \cite{jaeckel_low-energy_2010}, have been suggested to exist in string theory as excitations of quantum fields that extend into the additional spacetime dimensions beyond the four we are familiar with \cite{bailin_kaluza-klein_1987,svrcek_axions_2006}. 
\citet{arvanitaki_string_2010}
proposed that, due to the complexity of the extra-dimensional manifolds of string theory, there should be many ultralight ALPs, possibly spanning each decade of mass down to the present-day Hubble scale of $10^{-33}$~eV, referred to as an ``axiverse.''

An important feature of ALPs, not only those from string theory but those arising generically in beyond-SM physics scenarios, is that the relationship between the ALP mass and the ALP decay constant $f_a$, and subsequently the ALP's interaction strength with SM particles and fields, is not as strongly constrained by theory as in the case of QCD axions \cite{marsh_axions_2017}. 
This opens a broader theoretically motivated parameter space to search for evidence of exotic spin-dependent interactions.

\subsubsection{The matter-antimatter asymmetry of the universe}
\label{Sec:Intro-motiv-matter-antimatter}

One of the most important problems in theoretical physics is to identify the cause of the matter-antimatter asymmetry of the Universe 
(\citealp{dine_origin_2003}; \citealp{canetti_matter_2012}).
There is strong evidence that there are no large concentrations of antimatter at any scale in the present universe
\cite{cohen_matter-antimatter_1998}.
\citet{sakharov_violation_1967} 
proposed that the baryon density may not be a result of some initial condition, but could in fact be explained by particle physics interactions in the early universe. Theory suggests that in order to generate the observed baryon density in the universe (and the lack of anti-baryons), some baryon-number- and lepton-number-violating interactions and, particularly, new sources of $CP$-violation are necessary \cite{shaposhnikov_baryon_1987}.
The presence of exotic bosons could potentially provide a solution to this problem of baryogenesis.

A theoretical model to explain baryogenesis using axions was developed by \citet{co_axiogenesis_2020}, and a closely related ALP-based theory of baryogenesis was discussed by \citet{co_lepto-axiogenesis_2021} and \citet{co_predictions_2021}.
In these axiogenesis and ALP-cogenesis models, explicit symmetry breaking in the early universe leads to a time-dependent $\theta_{\text{eff}}$ (i.e., $d\theta_{\text{eff}}/dt \neq 0$, implying a ``rotation'' of the axion or ALP fields). A non-zero $d\theta_{\text{eff}}/dt$ at the weak scale satisfies the out-of-equilibrium and $CP$-violation conditions for baryogenesis, and creates the baryon-antibaryon asymmetry via QCD and electroweak effects in the case of axions, or via couplings with photons, nucleons, and/or electrons in the ALP case.
This motivates searches for axions and ALPs, including probing their existence through spin-dependent forces discussed here.

\subsubsection{The hierarchy problem}
\label{Sec:Intro-motiv-hierarchy-problem}

One of the most perplexing issues in theoretical physics is the hierarchy problem: why is gravity so much weaker than the other forces? The core of this conundrum is why the observed Higgs mass ($m_h \approx 125$~GeV) is so much less than the Planck mass ($M_\textrm{Pl} \sim 10^{19}$~GeV), as one would anticipate that quantum corrections would make the effective Higgs mass closer to the Planck scale \cite{hamada_bare_2013}.
Attempts to address the hierarchy problem include, for instance, supersymmetric models \cite{dimopoulos_softly_1981}
and models involving large (sub-mm) extra dimensions \cite{arkanihamed_hierarchy_1998,arkani-hamed_phenomenology_1999,randall_large_1999}.

\citet{graham_cosmological_2015} suggest that the hierarchy problem can be solved by a dynamic relaxation of the Higgs-boson mass from the Planck scale to the electroweak scale in the early universe. 
This process is powered by inflation and a coupling of the Higgs boson to a spin-0 particle, known as the relaxion. The relaxion could be either the QCD axion or an ALP.

Inflation in the early universe causes the relaxion field to evolve in time, and the coupling between the relaxion and the Higgs field generates a periodic potential for the relaxion once the Higgs' vacuum expectation value becomes nonzero. 
This periodic potential creates large enough barriers that the time evolution of the relaxion halts and the effective mass of the Higgs boson settles at its observed value. 
The electroweak-symmetry-breaking scale is a special point in the evolution of the Higgs boson mass, which explains why the Higgs mass eventually settles at the observed value, relatively close to the electroweak scale and far from the Planck scale.

\subsubsection{Dark energy}
\label{Sec:Intro-motiv-DE}

Cosmological observations such as the redshift-distance measurements of Type 1A supernovae (\citealp{riess_observational_1998}; \citealp{perlmutter_measurements_1999})
and measurements of the cosmic microwave background anisotropy \cite{planck_collaboration_planck_2020}
indicate that the Universe recently entered a phase of accelerated expansion.
This accelerated expansion is thought to be caused by an entity with negative pressure that permeates the Universe --- colloquially referred to as ``dark energy''. 
The simplest (though perhaps not the most natural) explanation of the observed accelerated expansion is the presence of a nonzero cosmological constant parameter $\Lambda$ in the field equations of general relativity, which has the equation of state $w = p / \rho = -1$, where $p$ and $\rho$ are the pressure and energy density, respectively, associated with the $\Lambda$ term.
Another possible candidate to explain the dark energy is quintessence \cite{ratra_cosmological_1988}: 
an ultralight spin-0 field. 
The pressure and energy density associated with a spin-0 field $\phi$ are given by $p = \dot{\phi}^2/2 - V(\phi)$ and $\rho = \dot{\phi}^2/2 + V(\phi)$, respectively. 
Here $\dot{\phi}^2/2$ is the kinetic energy term, while $V(\phi)$ denotes the potential energy. 
For a sufficiently light scalar with mass $m_\phi \sim 10^{-33}~\textrm{eV}$ that is comparable to the present-day Hubble parameter, the scalar field evolves slowly with $\dot{\phi}^2/2 \ll V(\phi)$, and so the equation of state for the scalar field satisfies $w = p / \rho \approx -1$, which is consistent with observations. 
Such quintessence fields may have spin-dependent interactions with matter.

\citet{arkani-hamed_ghost_2004} proposed a modification of gravity that suggests dark energy is a ``ghost'' condensate, a scalar field $\phi$ with a constant velocity that permeates the Universe.
The ghost condensate spontaneously violates Lorentz invariance, which means that there is a preferred frame where the spatial distribution of $\phi$ is isotropic. 
This is similar to how the cosmic microwave background radiation or any other cosmological fluid violates Lorentz invariance. 
What sets the ghost condensate apart is that, unlike other cosmological fluids, it does not become diluted as the universe expands.
Thus the ghost condensate behaves similarly to a cosmological constant. 
The interaction between the ghost condensate and matter leads to apparent violations of Lorentz symmetry and the emergence of new long-range spin-dependent interactions. 
\citet{flambaum_scalar-tensor_2009} further suggested that if the quintessence field has pseudoscalar couplings to matter and is interpreted as a spin-0 component of gravity, there would be a coupling between spin and gravity (much like that described in Sec.~\ref{Sec:Intro-motiv-spin-gravity}). This implies that fermions would have gravitational dipole moments leading to spin precession frequencies in the Earth's gravitational field of $\sim 10$~nHz. 
It should be noted that most other theories proposing that cosmic acceleration is due to quintessence require some level of fine-tuning, such as invoking a nonzero cosmological constant. 
These theories often need some sort of screening mechanism to avoid constraints from gravity tests conducted in astrophysics and in laboratories \cite{martin_quintessence_2008}.

\subsubsection{Other mysteries suggesting new bosons}
\label{Sec:Intro-motiv-other-mysteries}

The theoretical ideas leading to the proposed existence of the axion as discussed in Sec.\,\ref{Sec:Intro-motiv-spin-0} suggested new approaches to resolving other mysteries in particle physics.

As an example, the origin of the mass spectrum of quarks and leptons, as well as quark mixing in the electroweak interactions, is still an unsolved problem in particle physics. 
The SM considers the quark masses and mixing angles to be arbitrary constants that are not connected to any other parameters in the theory. It has been proposed by \citet{wilczek_axions_1982} that a more efficient explanation of quark and lepton masses could be achieved if they were associated with some spontaneously broken symmetry, possibly closely related to the PQ symmetry. By the same mechanism generating the axions, new bosons known as ``familons'' would appear \cite{wilczek_axions_1982,gelmini_does_1983} and are another type of ALP that could mediate spin-dependent interactions.

Another case is the question of the nonzero neutrino masses \cite{gonzalez-garcia_neutrino_2003}.
A possible explanation for the nonzero mass of neutrinos is that they are Majorana particles and their mass term violates lepton-number conservation \cite{chikashige_are_1981, gelmini_left-handed_1981}. This hypothesis also explains why the neutrino masses are much smaller than those of the charged leptons. It is conceivable that lepton number symmetry is broken, leading to the emergence of massive scalar (majorons) or vector bosons \cite{witten_neutrino_1980}
that could mediate spin-dependent interactions.

As noted in Sec.\,\ref{Sec:Intro-motiv-quantum-gravity}, attempts to quantize gravity such as string theory often predict the existence of new spin-0 and spin-1 bosons. 
New spin-1 bosons are generally associated with new $U(1)'$ gauge symmetries (in addition to that associated with electromagnetism), and depending on whether this $U(1)'$ gauge symmetry is preserved or broken, there can appear bosons that are either massless or massive, respectively. 
Exotic massless spin-1 bosons are usually called paraphotons $\gamma'$ (\citealp{holdom_two_1986}; \citealp{appelquist_nonexotic_2003}; \citealp{dobrescu_massless_2005}).\footnote{Note, however, that there are cases in the literature where the paraphoton terminology is used to describe massive vector bosons \cite{okun_limits_1982}.}
Exotic spin-1 bosons that are coupled to axial-vector currents as well as vector currents are often called $Z^\prime$ bosons \cite{langacker_physics_2009}, in analogy to the $Z$ boson of the SM. 
Other types of massive spin-1 bosons are referred to as dark or hidden photons. 
Spin-1 bosons can either have direct couplings to fermions, or they can be noninteracting (sterile), but still generate detectable effects via ``kinetic mixing'' with real photons  -- a process analogous to neutrino oscillations.
Spin-1 bosons that couple to SM fermions only through their kinetic mixing with a ``real'' electromagnetic field (\citealp{holdom_two_1986}; \citealp{pospelov_bosonic_2008}; \citealp{jaeckel_force_2013})
are often referred to as hidden photons.\footnote{Note that the terminology of \emph{dark photons}, \emph{hidden photons}, and $Z^\prime$ \emph{bosons} occasionally have contradictory and overlapping definitions in the literature.  All these spin-1 bosons with nonzero masses are proposed as possible constituents of dark matter (\citealp{davoudiasl_dark_2012}; \citealp{bento_classes_2024}).}

\subsection{Foundations of the exotic fifth force theory }\label{Sec:Intro-Th}

Fundamental interactions are mediated (carried) by intermediate bosons specific to the type of interaction. The strong force is carried by gluons, the weak force is carried by the $Z^0$ and $W^\pm$ bosons, and electromagnetism is carried by photons. It is anticipated that gravitational interactions are mediated by massless spin-2 particles, gravitons, which describe gravity well at low energies (specifically, energies much smaller than the Planck scale $\Lambda_\textrm{Pl} \approx 10^{19}$\,GeV).
The basics of the quantum theory of gravity at low energies are described in a paper by \citet{feynman_quantum_1963}; one-loop quantum corrections to Newtonian gravity are discussed, for example, by \citet{kirilin_quantum_2002}.
In the early 20th century, the concept of bosons interacting with fermions was established, with photons as massless spin-1 particles mediating electromagnetic interactions \cite{peskin_introduction_2019}. 
\citet{yukawa_interaction_1935} expanded on this concept to include massive spin-0 bosons (pions), giving rise to the Yukawa potential, also known as the Yukawa description of the strong nuclear force. 
This laid the foundation for understanding the role of bosons in mediating fundamental forces. As theoretical advancements continued, in the 1960s, a low-energy quantum theory of gravity emerged \cite{feynman_quantum_1963}. 

\citet{moody_new_1984} introduced a general framework of hypothetical bosons and explored three potentials resulting from the exchange of spin-0 bosons with various interactions with fermions, known as monopole-monopole ($V_1|_{ss}$ in this review, see Sec.\,\ref{Subsec:Fformalism}), monopole-dipole ($V_{9+10}|_{ps}$) and dipole-dipole ($V_3|_{pp}$) potentials. \citet{dobrescu_spin-dependent_2006} later expanded this framework to encompass 16 potentials, employing a different approach based on symmetries. 
Note that, while the symmetry-based approach determines the overall spin/tensor structure of potentials, it cannot uniquely predict the power-law dependence on the interparticle distance/separation. \citet{dobrescu_spin-dependent_2006} also discussed the classification of the potentials in terms of coupling constants. 

A more recent study by \citet{fadeev_revisiting_2019} revisited the previous potentials, providing a unified framework for studying the effects of these bosons and their interactions, regardless of the specific theories or models that predict their existence. 
Rather than sorting them into the 16 groups by their mathematical spin-momentum structure \cite{dobrescu_spin-dependent_2006}, \citet{fadeev_revisiting_2019} chose to sort the potentials by their types of physical couplings. 
This classification is more intuitive from a physicist’s point of view since one is ultimately interested in the physical coupling constants of a particular model. 

\begin{figure*} [!htbp]
\begin{center}
\includegraphics[width=1\textwidth]{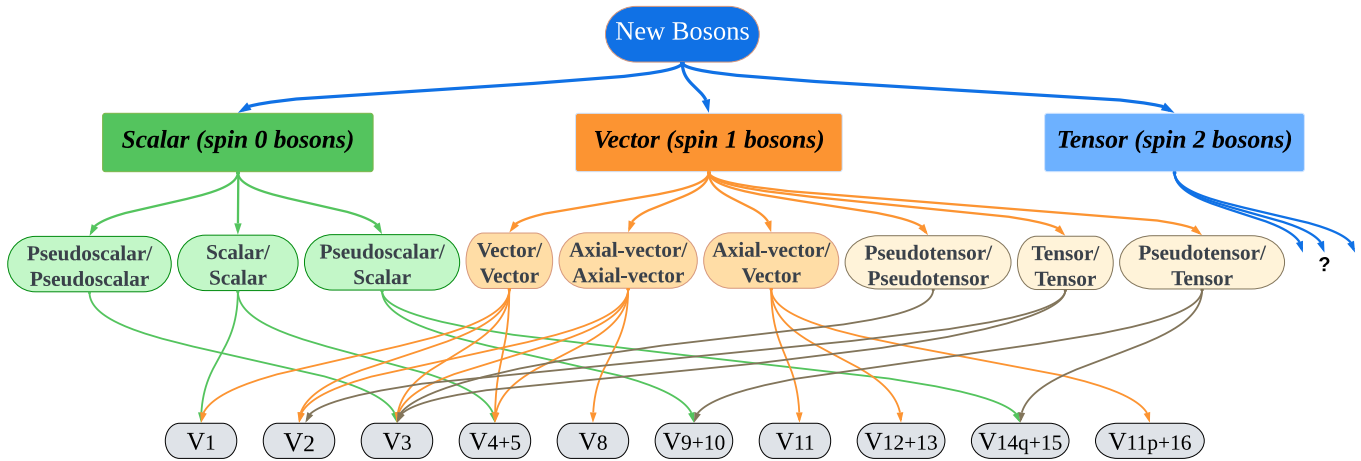}
\end{center}
\caption{(color online). 
New bosons carrying exotic interactions between fermions. 
The green blocks in the left (the branch with scalar spin-0 bosons) were studied by \citet{moody_new_1984}. 
The potentials $V_i$ in the bottom row are related to $\sV_i$ originally proposed by \citet{dobrescu_spin-dependent_2006} and widely studied by many experiments, see Sec.\,\ref{Sec:formalism} and \ref{Sec:limits_main}. 
This latest theoretical framework provides an overview of the classification of boson potentials. } 
\label{New-bosons}
\end{figure*}

The latest classification framework, as depicted in Fig.\,\ref{New-bosons}, links to the potentials $\sV_i$ described by \citet{dobrescu_spin-dependent_2006}. 
Three possible bosons, i.e., a massive vector spin-1 boson, scalar spin-0 boson, and massless vector spin-1 boson are further studied in three different combinations each, leading to nine different combinations of coupling constants, namely $g_Vg_V$, $g_Ag_A$, $g_Ag_V$,  $g_pg_p$, $g_sg_s$, $g_pg_s$, $g_{\tilde{T}} g_{\tilde{T}}$, $g_Tg_T$ and $g_T g_{\tilde{T}}$. 
Each of them maps to a combination of potential forms proposed in \citet{dobrescu_spin-dependent_2006}. 

Based on this framework, in the results section (Sec.\,\ref{Sec:limits_main}), we discuss which potential terms $\sV_i$ provide the most stringent bounds for a particular combination of couplings. 
Generally speaking, the most stringent bounds come from the lowest-order terms like $V_2$ or $V_3$.  
Higher-order terms like $V_{4+5}$ and $V_8$ generally give less stringent bounds, but are still useful. 
These higher-order terms probe different spin/velocity structures; in the event of a detection of a spin-dependent force, higher-order terms could be used to confirm which types of couplings the exchanged boson has. 
In addition, it is possible for a single experiment or set of measurements to be sensitive to a combination of $V_i$ potential terms \cite{fadeev_pseudovector_2022}. 

\subsection{Experimental searches for a fifth force}
\label{Sec:Intro-Exp}

There are two search philosophies: one is to make a precise comparison of experiment and theory, for instance, in the values of transition frequencies. The other is to look for interactions that are forbidden by some symmetries, for instance, parity (P) and/or time-reversal (T) invariance, or interactions that follow a specific, unusual, dependence of the interaction on particle distances, velocities and spins. The tell-tale symmetry violation or signature are powerful tools to identify the sought-after interaction
from the usual standard-model interactions. For such searches, it is more essential to have high sensitivity rather than high accuracy of the measurements (e.g., precise calibration of the apparatus). To reveal the specific signature of the force with respect to symmetries and parameter dependences, 
one benefits from the optimized design of the force sources and sensors, including proper reversals or modulations, with a representative example being the dedicated spin sources discussed in Sec.\,\ref{METH1.SOUR.} and \ref{METH1.SENS.}.
Both types of approaches have been highly productive in the search for exotic spin-dependent interactions, as discussed in this review, Sec.\,\ref{Sec:EXP.METH.1} and \ref{EXP.METH.2}.

\subsection{Discrete-symmetry-violating and -respecting interactions }

A general aspect of exotic long-range spin-dependent interactions to consider is whether they respect discrete symmetries such as charge-conjugation $C$, parity $P$, and time-reversal $T$. 
The weak interaction, for example, violates $P$ and also the combined symmetry $CP$.
On the other hand, the electromagnetic interaction respects $C$, $P$, and $T$.

The advantage of searching for discrete-symmetry-violating interactions is that background effects from the SM are strongly suppressed, as the short-range weak interaction is the only known symmetry-violating interaction in the SM.
On the other hand, searches for symmetry-respecting interactions offer the opportunity to directly constrain particular unknown coupling constants that cannot be directly constrained otherwise.
The idea is as follows. 
All of the long-range exotic potentials discussed here involve two coupling constants: one from the vertex representing the local interaction at the source and one from the vertex representing the local interaction at the sensor.
Symmetry-violating effects arise due to the involvement of two vertices with different symmetry characteristics. 
An example from the SM is the $P$-violating weak interaction that involves a vector and axial-vector coupling.
Thus an experimental measurement constraining a new symmetry-violating interaction necessarily only limits the product of two \emph{different} coupling constants, $g^Xg^Y$.
Such an experiment by itself does not constrain the individual coupling constants $g^X$ and $g^Y$ since one of them could be, for example, zero, and then the other is completely unconstrained by the measurement.
On the other hand, new symmetry-respecting interactions can be generated by local interactions of the same character and involve the square of a coupling constant $g^Xg^X$.
An experimental measurement constraining a new symmetry-respecting interaction provides a direct limit on a particular coupling constant $g^X$.

A similar point can be made for experiments measuring cross-species interactions as compared to same-species interactions. 
An experiment searching for an exotic interaction between electrons can constrain the square of the electron coupling constant, $g^eg^e$, and thus directly constrain $g^e$ itself, whereas an experiment searching for an exotic interaction between neutrons and electrons only constrains a product of coupling constants $g^eg^n$. Since either $g^n$ or $g^e$ could be zero, the experiment provides no direct constraint on the individual coupling constants.

Notably, the above argument can be usefully reversed, as discussed, for example, by \citet{raffelt_limits_2012}.
\citet{raffelt_limits_2012} took constraints on a scalar axion coupling $g_s$ from symmetry-respecting long-range force measurements and combined these with astrophysical constraints on the axion pseudoscalar coupling $g_p$ to constrain the product of coupling constants $g_sg_p$.
While many experiments have searched for long-range $P$- and $T$-violating monopole-dipole interactions proportional to $g_sg_p$, the constraints obtained by \citet{raffelt_limits_2012} remain among the most stringent. 

In this review, we present combined astrophysical-laboratory bounds for comparison with direct bounds from experiments; see figures in Sec.\,\ref{Sec:limits_main} and for more details, App.\,\ref{appendix_comp.}.

\subsection{Goals, scope and structure of this review }\label{Sec:Intro-GSS}

Our goal, after the general introduction to the topic, is to provide a concise description of the theoretical approaches to help the reader navigate through different frameworks and conventions. 
Tracing the development of the theoretical approaches, we present the latest unified approach and a set of spin-dependent interaction potentials. 
We then proceed to critically evaluate the significant body of recent experimental results, presenting the combined up-to-date\footnote{While there is no sharp cut-off, the data presented in this review are mostly those published through early 2024.} constraints on exotic interactions. 
The unified theoretical description enables a systematic comparison of results obtained in different experiments. 
Finally, we identify promising directions for future experimental work and discuss the significance and impact of a possible discovery.   

To limit the scope of the project, we focus on exotic spin-dependent interactions, only discussing the spin-independent ones when necessary for logical completeness and to compare the results for couplings that produce both types of interactions. 
Comprehensive reviews of spin-independent interactions can be found elsewhere (\citealp{adelberger_torsion_2009}; \citealp{antoniadis_short-range_2011}; \citealp{heil_spin_2013}; \citealp{salumbides_bounds_2014}; \citealp{lemos_submillimeter_2021}).

The most relevant earlier review is that of \citet{safronova_search_2018}, which examines spin-dependent exotic interactions in Section VII of that publication. 
However, there has been an ``explosion'' of both theoretical and experimental results in the years that have passed since that review. 
Our current review is significantly broader, more complete, up-to-date, and systematic.

The paper is organized as follows. In Sec.\,\ref{Sec:formalism}, we review the theoretical frameworks for spin-dependent exotic interactions. 
In Sec.\,\ref{Sec:EXP.METH.1}, we focus on dedicated source-sensor experiments, while in Sec.\,\ref{EXP.METH.2}, we discuss complementary experiments and observations for detecting exotic spin-dependent interactions. 
In Sec.\,\ref{Sec:limits_main}, we categorize the current best or most recent limits on spin-dependent potentials and coupling constants, based on the latest theoretical framework. 
Sec.\,\ref{Sec:Sum-Out} provides a summary and outlook.


\section{Theoretical frameworks}
\label{Sec:formalism}

\subsection{Early developments}
\label{Subsec:MWformalism}

During the development of quantum field theory in the early 20th century, it was realized that electromagnetic interactions could be described in terms of the exchange of a massless spin-1 mediator known as the photon \cite{peskin_introduction_2019}. 
Later, Yukawa generalized this concept to massive spin-0 bosons, the exchange of which gives rise to the Yukawa potential $V = -g^2 \exp(-M r)/ (4 \pi r)$, where $g$ is the dimensionless coupling constant and $M$ is the mass of the exchanged boson \cite{yukawa_interaction_1935}. 
In this case, the exchanged boson could in principle be either elementary or composite; in both cases, the exchanged boson can be treated as a single spin-0 degree of freedom on the length (or equivalently energy) scales involved. 
In the SM, for example, the strong nuclear force is mediated between nucleons by the exchange of pions and other mesons.

A low-energy quantum theory of gravity, in which gravitational interactions are mediated by the exchange of a massless spin-2 graviton, was developed in the 1960s 
(\citealp{feynman_quantum_1963}; \citealp{penrose_zero_1965}; \citealp{dewitt_quantum_1967}).

Since then, a number of studies of potentials induced by the exchange of exotic bosons have been undertaken. 
\citet{bouchiat_limit_1975} investigated the \textit{P},\textit{T}-odd potential arising from the exchange of a spin-0 boson, which interacts with one fermion via a scalar-type vertex and with another fermion via a pseudoscalar-type vertex. 
\citet{anselm_possible_1982} investigated the dipole-dipole potential arising from the exchange of a spin-0 boson that couples to fermions via a pseudoscalar-type vertex. 
\citet{moody_new_1984} explored the potentials arising from the exchange of a spin-0 boson that possesses both scalar and pseudoscalar-type interactions with fermions. 
In this case, three qualitatively different combinations of vertex types are possible, namely scalar-scalar, scalar-pseudoscalar and pseudoscalar-pseudoscalar. 
These lead, respectively, to monopole-monopole, monopole-dipole and dipole-dipole potentials/forces. 
The first of these is spin-independent at leading order and corresponds to the potential term $\mathcal{V}_1$ in the notation presented later on in Sec.\,\ref{Subsec:DMformalism}, see Eq.\,(\ref{DM_eqV1}). 
The latter two are already spin-dependent at leading order and correspond to the potential terms $\mathcal{V}_{9+10}$ and $\mathcal{V}_{3}$, respectively, see Eqs.\,(\ref{DM_eqV910}) and (\ref{DM_eq3}).

The above ideas \cite{bouchiat_limit_1975,anselm_possible_1982,moody_new_1984}, among others, inspired a generation of new experimental searches, including searches for spin-independent exotic forces [see, e.g., \citet{long_current_2003}; \citet{adelberger_tests_2003}; \citet{kapner_tests_2007}; \citet{antoniadis_short-range_2011}; \citet{costantino_exotic_2020}; \citet{lemos_submillimeter_2021}], as well as spin-dependent forces [see, e.g., \citet{wineland_search_1991}; \citet{venema_search_1992}; \citet{youdin_limits_1996} and \citet{ voronin_neutron_2009}; \citet{hoedl_improved_2011}; \citet{tullney_constraints_2013} and \citet{stadnik_nuclear_2015,terrano_short-range_2015}; \citet{crescini_improved_2017,crescini_quax-gpgs_2017,crescini_search_2022}; \citet{luo_proposed_2020}; \citet{xiong_searching_2021}]. 
In this review, we discuss the latest updates in theory and experiments.


\subsection{Approximations and limitations}
\label{Sec:th-AP}

In this section, we outline the limitations and approximations that we adopt in this review.\footnote{These approximation pertain to theory and interpretation of experiment; the experiments may have sensitivity to new physics going beyond these approximations, such as, for example, exotic high-spin bosons.} 
The frameworks in \citet{moody_new_1984}; \citet{dobrescu_spin-dependent_2006}; \citet{fadeev_revisiting_2019} provide 
an overview of the classification of single-boson-mediated potentials and set the stage for the investigation of the implications of these exotic interactions. 
In this review, we consider interactions between fundamental spin-1/2 fermions via the exchange of a single spin-0 or spin-1 boson. 
For example, we do not systematically discuss spin-2 boson exchange and many-boson-exchange processes. 
Let us consider in turn all the assumptions in this scenario and the limitations they may cause.

\subsubsection*{1. The spin of the intermediate bosons} 
We consider only spin-0 and spin-1 mediators. This brings up several questions.
\begin{itemize}

\item Why do we consider only these?  Up to now, we do not have a consistent ultraviolet (UV)-complete theory for higher-spin bosons. 
This does not mean that they do not exist, but this is beyond our scope here.

\item Can higher-spin propagators be considered?  
An effective low-energy theory can be used to analyze interactions mediated by spin-2 bosons. One can write down the spin-2 propagator to consider the exchange of the spin-2 bosons in the first order of perturbation theory. 
The interactions of spin-2 bosons with matter can in principle take on a different form than those for spin-0 or spin-1 bosons, but that does not necessarily mean that the structures of the resulting potentials will be fundamentally different. 
For example, the exchange of a massless spin-0 boson with scalar-type interactions to matter produces the same $1/r$ form of the non-relativistic potential as the exchange of a spin-2 graviton between two masses (\citealp{feynman_quantum_1963}; \citealp{penrose_zero_1965}; \citealp{dewitt_quantum_1967}), 
even though the structure of their underlying couplings to matter is different in the two cases (scalar vs. tensor). 
Likewise, the exchange of a spin-1 photon also leads to a non-relativistic potential with a $1/r$ form.\footnote{Unlike the attractive non-relativistic $1/r$ potential mediated by spin-0 or spin-2 boson exchange, the Coulomb potential mediated by spin-1 boson exchange may be either attractive or repulsive depending on the relative signs of the electric charges involved.}

\item Will such propagators bring new structures of potentials? The potentials we consider here present all structures that are linear in the spin and momentum of fermions with spin $s=1/2$. 
For example, if we consider short-range (contact) interactions, then one can obtain all possible spin-dependent potential structures from combinations of $\gamma$-matrices of the interacting fermions without any reference at all to the intermediate boson \cite{khriplovich_parity_1991}. 
However, depending on the spin of the boson, the potentials can appear in particular combinations with different sets of coefficients (see Sec.\,\ref{Subsec:Fformalism}). 
\end{itemize}

\subsubsection*{2. One-boson versus two-boson or two-fermion exchange} 

Based on empirical evidence, the interaction constant associated with new forces is likely to be small, in which case two-boson exchange is strongly suppressed compared to single-boson exchange since the former process contains an extra power of a small interaction constant.\footnote{There may be processes involving the exchange of an exotic boson and a SM boson (e.g., a photon). 
Such mixed-boson-exchange processes contain another small interaction constant (e.g., $\alpha$), and so are still suppressed compared to the single-exotic-boson exchange process.} 
It may be necessary to consider such processes when single-boson exchange is suppressed by some selection rules.
However, two-fermion-exchange processes, such as neutrino-pair-exchange forces, may be important, as single-fermion exchange does not produce an inter-particle potential; see, e.g., \citet{stadnik_probing_2018}; \citet{ghosh_probing_2020};
\citet{bolton_probing_2020};
\citet{costantino_neutrino_2020}; \citet{xu_short-range_2022}; \citet{dzuba_long-range_2022}; \citet{dzuba_erratum_2022}; \citet{munro-laylim_effects_2023}. 
In addition, the exchange of a pair of bosons with spin-dependent couplings to matter can generate a higher-order spin-independent contribution to the interaction energy between the two fermions, potentially allowing one to more stringently constrain spin-dependent couplings by reanalyzing data from existing experiments that search for spin-independent forces 
(\citealp{fischbach_constraints_1999}; \citealp{aldaihan_calculations_2017}).

Two-boson (and two-fermion) exchange-induced potentials have different dependences on the interparticle distance compared to single-boson potentials; see, for example, \citet{bauer_fifth_2024}. 
In addition, in the case of interactions between composite systems, such as nuclei, the pair of exchanged bosons can interact with different nucleons within the same nucleus. 
This leads to an increase in the possible types of potential structures that are not necessarily limited to the forms of the potentials $V_1$ -- $V_{16}$ and related potentials containing additional integral powers of fermion momenta discussed in Secs.\,\ref{Subsec:DMformalism} and \ref{Subsec:Fformalism} below. 

It is sometimes asked: Why cannot potentials between two fermions be mediated by the exchange of a single fermion mediator?
Interaction vertices with an odd number of fermion lines are forbidden by rotational symmetry, among other things. 
The vertex described by a corresponding Lagrangian term must be a Lorentz scalar, but three half-integer angular momenta couple to half-integer total angular momentum, which is not rotationally invariant. 
Therefore, Lagrangian terms must contain an even number of fermion fields and  
there are no potentials associated with single fermion exchange.

\subsubsection*{3. Distinction between macroscopic bodies and quantum particles with spin} 
 
For a spin-polarized macroscopic body (for instance, a permanent magnet), spins can be oriented along some internal axis, which is not necessarily the same as the axis around which the body rotates. For an elementary particle, all intrinsic vectors (magnetic dipole, electric dipole, and anapole moments) are oriented along the particle spin. This is also true for a composite quantum particle such as a nucleus, where the overall expectation value of the internal spins (i.e., the integral of the nucleon spin density) can only be oriented collinear with the total spin of the particle.\footnote{If we do not take the overall expectation value of nucleon spins  and consider nuclear spin density, the spins  may be oriented in different directions. For example,  in a spinless nucleus,  the spins of the nucleons are directed perpendicular to the nuclear surface along the radius. Such a spin hedgehog can be formed by a $P$,$T$-violating interaction that polarizes the nucleon spins along the radius, corresponding to the $\bf s \cdot r$ correlation \cite{flambaum_spin_1994}.}

\subsubsection*{4. Higher-spin fermions and composite systems} 
 
In nature, we have so far only observed elementary spin-1/2 fermions.  
In principle, elementary spin-3/2 fermions may arise in models beyond the SM, such as supersymmetric models, which goes beyond the scope of the present review. 
Higher-spin fermions can appear, however, as composite objects, such as nuclei and atoms. 

In higher-spin composite systems, part of the spin may be due to the orbital angular momenta of the constituent particles. 
The spin-dependent interactions that we consider here couple to the intrinsic spins of fundamental or elementary fermions (and to their orbital angular momenta via relativistic corrections).
To determine the size of the effect of a spin-dependent interaction, we can apply the usual rules of angular momentum addition to calculate the projection of the intrinsic spins along the total spin of the system. 
This discussion is also applicable for composite systems with a total spin of $1/2$, where an elementary-fermion spin can be directed opposite to other angular momenta\footnote{Strictly speaking, one has to integrate the interaction of the intermediate boson with composite particle spin density.}. 

When constructing interaction potentials based on symmetry consideration, one can consider structures with higher powers of a fermion spin $\v s= \v \sigma/2$. However, such structures can be reduced using recursive application of the Pauli-matrix identity $\sigma_i\sigma_j=\delta_{ij}I_2 + i\varepsilon_{ijk}\sigma_k$, where $I_2$ is the $2 \times 2$ identity matrix. 
Therefore, any potential term can be rewritten with at most one spin operator associated with a given fermion. 

When we consider an interaction, we assume that the interaction is between elementary particles. 
Interactions between composite particles, such as nuclei, can have different and more complicated structures involving higher-order tensors and higher powers of total spin. 
In this case, the interaction should still be expressed in terms of interactions between elementary particles, as discussed above. 
Otherwise, we will have an infinite number of free parameters, which are specific to individual experiments and which cannot generally be compared between different experiments.

\subsubsection*{5. Interactions between bosons}
In the case of exotic-boson-mediated interactions, there is no restriction for the interacting particles to be fermions. 
Exotic bosons may also interact with the standard-model gauge bosons, such as photons and gluons. 
For example, the scalar coupling of a spin-0 boson $\phi$ to the electromagnetic field tensor $F$ of the form $\phi F_{\mu\nu}F^{\mu\nu}$ couples $\phi$ to the Coulomb energy of an atom or nucleus, since $F_{\mu\nu}F^{\mu\nu}=2(|\v B|^2-|\v E|^2) \approx -2|\v E|^2$ for a nonrelativistic atom or nucleus, and in turn generates a Yukawa-type potential between two unpolarised bodies consisting of atoms or nuclei. 
In the case of a nucleus, the Coulomb binding energy scale is $\approx 0.7\,\text{MeV} \times Z(Z-1)/A^{1/3}$, which for an atomic number of $Z \gg 1$, greatly exceeds the contribution from the rest-mass-energies of atomic electrons, which is equal to $Z m_e$; see, for example, \citet{leefer_search_2016} for more details. 
Another example is boson-nucleon couplings that give rise to spin-independent or spin-dependent potentials between nucleons. 
In the present review, we treat boson-nucleon interactions as effective low-energy operators. 
However, at the fundamental level, the boson-nucleon couplings receive contributions from boson-quark, boson-photon and boson-gluon interactions. 

\subsection{Symmetry-based formalism (Dobrescu-Mocioiu framework)}
\label{Subsec:DMformalism}

Basing on symmetry arguments, \citet{dobrescu_spin-dependent_2006} investigated the possible spin-dependent forces between macroscopic objects, which arise from the underlying exchange of a boson between elementary spin-$1/2$ fermions. 
They showed that rotationally-invariant interactions between two spin-$1/2$ fermions (each with its own independent linear momentum) mediated by the exchange of new bosons can generate 16 independent potentials when considering single-boson exchange processes (corresponding to classical tree-level processes). 
These 16 potentials contain at most one power of each fermion spin.\footnote{In composite systems with \textit{total} spin greater than $1/2$, such as nuclei and atoms, higher powers of spin may appear in potentials.}

Eight of these potentials are invariant under a parity transformation and can be written in the nonrelativistic limit (small fermion velocity and low momentum transfer) as:

\begin{equation}\label{DM_eqV1}
{\sV}_{1} =\frac{1}{r}\, y(r) \, ,
\end{equation}

\begin{equation}
{\sV}_{2} =\frac{1}{r} \;{\v \sigma_X}\cdot {\v \sigma_Y^{\,\prime}} \, 
y(r) \, ,
\end{equation}

\begin{align}\label{DM_eq3}
{\sV}_{3} &= \frac{1}{ m^2\,r^3} \left[ 
{\v\sigma_X}\cdot {\v\sigma_Y^{\,\prime}} \left(1 - r\frac{d}{dr} \right) - \right. \nonumber\\
&\left.  3 \left( \v\sigma_X\cdot \hat{\v{r}}  \right)\,
\left({\v\sigma_Y^{\,\prime}}\cdot \hat{\v{r}}  \right) 
\left(1 - r\frac{d}{dr} + \frac{1}{3}r^2\frac{d^2}{dr^2} \right) \right] y(r) \, ,
\end{align}
\begin{equation}\label{DM_eqV45}
{\sV}_{4,5} =-\frac{1}{2 m\,r^2}
\left(\v\sigma_X \pm \v\sigma_Y^{\,\prime} \right) 
\cdot \left(\v v\times \hat{\v{r}}  \right)  
\left(1 - r\frac{d}{dr}\right) y(r) \, ,
\end{equation}
\begin{align}\label{DM_eqV67}
{\sV}_{6,7} &= -\frac{1}{ 2 m\,r^2}
\left[\left(\v\sigma_X\cdot{\v v}\right)\,
\left( \v\sigma_Y^{\,\prime} \cdot \hat{\v{r}}  \right) \pm \right. \nonumber\\
&\left.
 \left(\v\sigma_X\cdot  \hat{\v{r}}  \right)\, 
\left( \v\sigma_Y^{\,\prime} \cdot{\v v} \right) \right] 
\left(1 - r\frac{d}{dr}\right) y(r) \, ,
\end{align}
\begin{equation}\label{DM_eqV8}
{\sV}_8 = \frac{1}{r} \left(\v\sigma_X\cdot\v v\right)
\left(\v\sigma_Y^{\,\prime}\cdot\v v\right) \, y(r)  \, . \;\;
\end{equation}

The other eight potentials change sign under a parity transformation:
\begin{equation}\label{DM_eqV910}
{\sV}_{9,10} =-\frac{1}{2 m \, r^2}\, 
 \left(\v\sigma_X\pm\v\sigma_Y^{\,\prime} \right) \cdot \hat{\v{r}}  \,
\left(1 - r\frac{d}{dr}\right) y(r) \, ,  
\end{equation}
\begin{equation}
{\sV}_{11} =-\frac{1}{ m \, r^2} \, 
\left(\v\sigma_X\times \v\sigma_Y^{\,\prime}\right)\cdot \hat{\v{r}}  \,
\left(1 - r\frac{d}{dr}\right) y(r) \, ,
\end{equation}
\begin{equation}
{\sV}_{12,13} =
\frac{1}{2 r}\, \left(\v\sigma_X\pm \v\sigma_Y^{\,\prime}\right) \cdot {\v v} 
\; y(r) \, ,
\end{equation}
\begin{equation}
{\sV}_{14} =\frac{1}{r} \, 
\left(\v\sigma_X\times \v\sigma_Y^{\,\prime}\right) \cdot {\v v} 
\; y(r) \, ,
\end{equation}
\begin{widetext}
\begin{equation}
{\sV}_{15} =-\frac{3}{2 m^2 \, r^3} 
\left\{ \rule{0mm}{5mm}\left[ 
\v\sigma_X\cdot \left(\v v\times \hat{\v{r}}  \right) \right]
\, \left(\v\sigma_Y^{\,\prime} \cdot \hat{\v{r}}  \right)
+ \left(\v\sigma_X \cdot \hat{\v{r}}  \right) \, 
\left[ \v\sigma_Y^{\,\prime} \cdot \left(\v v\times \hat{\v{r}} 
\right) \right]\right\}  \times \left(1 - r\frac{d}{dr} + \frac{1}{3}r^2\frac{d^2}{dr^2}\right) y(r) \, ,
\end{equation}
\begin{equation}
\label{DM_eqV16}
{\sV}_{16} =   -\frac{1}{2m \, r^2} 
\left\{ \rule{0mm}{5mm}
\left[ \v\sigma_X \cdot \left(\v v\times \hat{\v{r}}  \right) \right]
\, \left(\v\sigma_Y^{\,\prime} \cdot \v v \right)
+ \left(\v\sigma_X \cdot \v{v} \right) \, 
\left[ \v\sigma_Y^{\,\prime} \cdot \left(\v v\times \hat{\v{r}} 
\right) \right]\right\} \left(1 - r\frac{d}{dr}\right) y(r) \, . \\
\end{equation}
\end{widetext}
Here $\v\sigma_X$ and $\v\sigma_Y^{\,\prime}$ [corresponding to $\v\sigma$ and $\v\sigma^{\,\prime}$ in \citet{dobrescu_spin-dependent_2006}, respectively] are vectors of Pauli matrices of the spins of the two fermions $X$ and $Y$,
$\v r$ is the position vector of the fermion with spin $\v{\sigma}_X$ and mass $m$ relative to the fermion with spin $\v\sigma_Y^{\,\prime}$, $\hat{\v{r}} $ is the corresponding unit vector, and $r$ is the length of $\v r$. 
In the case of single-boson-exchange forces within a Lorentz-invariant quantum field theory, $y(r)$ has the simple form: $y(r)=e^{-Mr}/{(4\pi)}$ where $M$ is the mass of the exchanged boson. 
$\v v$ is 
the relative velocity of the two objects in the case of interactions between macroscopic objects. 
Here the natural units $\hbar=c=1$ are used.

Of these 16 interactions, one is spin-independent, six involve the spin of one of the particles, and the remaining nine involve both particle spins. 
Ten of these 16 possible interactions depend on the relative momenta of the particles. 
Given that \citet{dobrescu_spin-dependent_2006} are ultimately interested in the potentials between slowly moving macroscopic objects, they start in the nonrelativistic approximation, abandoning explicit Lorentz invariance of the theory.

\citet{dobrescu_spin-dependent_2006} showed that their potentials could also be obtained by explicitly considering single-boson-exchange processes involving nine types of coupling combinations grouped into three sets, namely vector and pseudovector interactions of a spin-1 field, tensor and pseudotensor-type interactions of a spin-1 or spin-2 field, and scalar and pseudoscalar couplings of a spin-0 field, and sorting according to the scalar (rotational) invariants made up of the spins and momenta of the two fermions involved. 
We restrict our attention to three types of bosons in this review --- a massive spin-1 boson $Z'$, a massless spin-1 boson $A'$, and a spin-0 boson $\phi$ (which can be either massive or massless). 

Each boson has its own set of local Lorentz-invariant interactions with the standard-model fermions $\psi$:
\begin{equation} 
 \label{Lagrangian_vector}
\mathcal{L}_{Z'} = Z'_{\mu} \sum_\psi \bar{\psi}\gamma^{\mu}\left(g_V^\psi + \gamma_{5}  g_A^\psi  \right) \psi  \, ,
\end{equation}
\begin{equation} 
 \label{Lagrangian_tensor}
\mathcal{L}_{A'} = \frac{v_{h}}{\Lambda^{2}}P_{\mu\nu} \sum_\psi  \bar{\psi}\sigma^{\mu\nu} \left[ \mathrm{Re} (C_{\psi}) + i \gamma_{5} \mathrm{Im} (C_{\psi})  \right] \psi   \, ,
\end{equation}
\begin{equation} 
 \label{Lagrangian_scalar}
\mathcal{L}_{\phi} = \phi \sum_\psi \bar{\psi}  \left( g_s^\psi + i \gamma_{5} g_p^\psi \right) \psi  \, .
\end{equation}
Here $\psi$ denotes the fermion field (for instance, $\psi = e$ for an electron, $\psi = N$ for a nucleon, with $p$ denoting a proton and $n$ denoting a neutron), $P_{\mu\nu} = \partial_{\mu} A'_{\nu} - \partial_{\nu} A'_{\mu}$ is the field-strength tensor of the massless paraphoton field $A'_{\mu}$, $\sigma^{\mu\nu} = \frac{i}{2} [\gamma^\mu , \gamma^\nu]$, and $\gamma^\mu$, $\gamma_5 = i \gamma^0 \gamma^1 \gamma^2 \gamma^3$ are Dirac matrices. 
The dimensionless interaction constants $g_s^\psi$, $g_p^\psi$, $g_V^\psi$, $g_A^\psi$, $\mathrm{Re} (C_{\psi})$, $\mathrm{Im} (C_{\psi})$ parameterize the scalar, pseudoscalar, vector, pseudovector, tensor and pseudotensor interaction strengths, respectively. 
The Higgs vacuum expectation value is denoted as $v_h \approx 246\,\textrm{GeV}$ \cite{erler_electroweak_2004}, and $\Lambda$ is the ultraviolet energy cutoff scale for the Lagrangian in Eq.\,\eqref{Lagrangian_tensor}. 
Note that the derivative form 
of the pseudoscalar interaction in Eq.\,\eqref{Lagrangian_scalar} is also commonly used in the literature (\citealp{pospelov_bosonic_2008}; \citealp{stadnik_axion-induced_2014}; \citealp{abel_search_2017})
and gives rise to the same nonrelativistic potential, see \citet{fadeev_revisiting_2019} for details.\footnote{
We note that this is a specific result that does not necessarily hold in other cases, such as boson exchange between relativistic fermions, boson exchange in the presence of a background of such bosons (in-medium processes), or multi-boson exchange (quantum processes involving loops). 
The reason is that the derivative and non-derivative forms of the pseudoscalar interaction are only equivalent up to an electromagnetic term of the form $\phi F_{\mu \nu} \tilde{F}^{\mu \nu}$. 
See \citet{arza_electron_2023} and \citet{bauer_fifth_2024} 
for more details. 
} 

Note that the two terms in the paraphoton interaction in Eq.\,\eqref{Lagrangian_tensor} are similar to the interactions of an ordinary photon with the fermion anomalous magnetic moment and EDM which are produced by radiative corrections (or internal structure of composite particles in the case of anomalous magnetic moments of nucleons) \cite{berestetskii_quantum_1982,khriplovich_parity_1991}. 
In principle, such terms may also be added to Eq.\,\eqref{Lagrangian_vector} for $Z'$. The difference with the photon and paraphoton cases is that $Z'$ is a massive particle. 
This leads to an additional term in the numerator of the $Z'$ propagator ($q_{\mu} q_{\nu} / M^2$) and in the radial dependence of the nonrelativistic potential in Eq.\,\eqref{pseudotensor-pseudotensor_potential} produced by the fermion pseudotensor/pseudotensor interaction [such a $1/M^2$ term also appears in the axial-vector/axial-vector interaction, see Eq.\,(\ref{pseudovector-pseudovector_potential}) below]. 
This EDM-EDM type of interaction does not violate any symmetries, making it hard to separate its effects from other interactions, and may be assumed to be small, since even a first-order effect in an EDM interaction is already small and has not been observed yet. 

Building upon the three potentials ($\mathcal{V}_1$, $\mathcal{V}_{3}$ and $\mathcal{V}_{9+10}$) introduced by \citet{moody_new_1984}, \citet{dobrescu_spin-dependent_2006} demonstrated the existence of numerous new types of spin-dependent macroscopic forces that warranted experimental exploration. This was undertaken, for example, by \citet{heckel_new_2006}; \citet{heckel_preferred-frame_2008}; \citet{adelberger_torsion_2009}; \citet{ledbetter_constraints_2013}; \citet{hunter_using_2013}; \citet{yan_new_2013}; \citet{kim_experimental_2018}; \citet{ji_new_2018} and others. 

It would be beneficial to understand the limitations of the work conducted by \citet{dobrescu_spin-dependent_2006}.

\citet{dobrescu_spin-dependent_2006} were interested in interactions between (spin-polarized) macroscopic bodies. Therefore, they used non-relativistic approximation, considered only the long-range part of the potentials and assumed that all internal motions inside these bodies were averaged out, so that only the classical velocity $\v{v}$ of the whole body was left.
On the other hand, for microscopic particles, the relative velocity can no longer be described by a classical vector, but must be described by a quantum vector operator.
Authors such as \citet{ficek_constraints_2017}, \citet{dzuba_probing_2017}, \citet{stadnik_improved_2018} and \citet{fadeev_revisiting_2019} emphasized the importance of deriving potentials in the ``position representation'' for atomic-scale calculations, employing quantum treatment, i.e., with the momentum described as an operator. See more details in Sec.\,\ref{Subsec:Fformalism}.

Furthermore, the forms of these 16 single-boson-exchange potentials $\sV_{1-16}$ presented in \citet{dobrescu_spin-dependent_2006} by themselves form a complete set if we restrict ourselves to single-boson-exchange processes. However, $\sV_{1-16}$ are insufficient to describe all potentials induced by the simultaneous exchange of multiple bosons, which generate quantum (loop-level) forces instead of the classical (tree-level) forces that we focus on in this review. 
For instance, scalar-pair exchange leads to a modified Yukawa-type potential that has a $1/r^3$ power-law dependence on short length scales, but a $1/r^{5/2}$ dependence on long length scales; see, for example, \citet{ferrer_higgs-_1999}. 
Such scale-noninvariant power-law dependences cannot be matched onto the scale-invariant form of any of the 16 single-boson-exchange potentials within the Dobrescu-Mocioiu formalism.

\subsection{Recent developments}
\label{Subsec:Fformalism}

Using the more traditional particle-exchange formalism discussed in Sec.\,\ref{Subsec:MWformalism},
\citet{fadeev_revisiting_2019} systematically categorised 
a larger set of interactions. 
The potentials presented by \citet{fadeev_revisiting_2019} are relevant for both macroscopic-scale and (sub)atomic-scale experiments when searching for spin-dependent forces. 
For instance, they are useful for calculating new-force contributions to energy shifts in the spectra of atoms and nuclei (\citealp{ficek_constraints_2017}; \citealp{ficek_constraints_2018}).
Hence, they are more generic than the Dobrescu-Mocioiu potentials presented in Eqs.\,\eqref{DM_eqV1} -- \eqref{DM_eqV16} in Sec.\,\ref{Subsec:DMformalism}.

The form of these potentials is also helpful when comparing constraints on the various boson-fermion interactions coming from fifth-force searches with (indirect) constraints from other types of signatures, such as 
constraints coming from astrophysics or from a combination of laboratory and astrophysical bounds 
(see Sec.\,\ref{Sec:limits_main}), because we use the physical interaction constants $g_s$, $g_p$, $g_V$, $g_A$, parametrizing the scalar, pseudoscalar, vector and pseudovector interactions, rather than the coefficients $f_i$ from \citet{dobrescu_spin-dependent_2006}, see more in App.\,\ref{appendixA}. 

\begin{figure}[!htbp]
\centering
\includegraphics[width=0.27\textwidth]{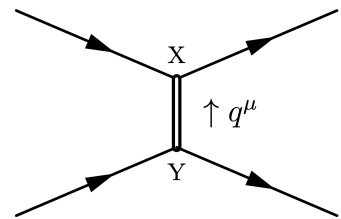}
\caption{Elastic scattering of two fermions with masses $m_X$ and $m_Y$ and spins $\v{s}_X$ and $\v{s}_Y^{\,\prime}$, respectively, mediated by the exchange of a boson of mass $M$ with four-momentum $q^{\mu} = (q^0,\v{q})$ that is transferred from fermion $Y$ to fermion $X$. 
} 
\label{fig:ES}
\end{figure}

We present these potentials below (in the natural units $\hbar=c=1$), 
grouping together interactions with similar features and identifying the dominant terms in the case of\\ 
(1) axial-vector/vector, axial-vector/axial-vector, vector/vector, \\
(2) tensor/tensor, pseudotensor/tensor, and pseudotensor/pseudotensor interactions, \\
(3) pseudoscalar/scalar, pseudoscalar/pseudoscalar, scalar/scalar, \\
generated by the exchange of a single 
massive spin-1 boson of mass $M$, massless spin-1 
boson, and massive or massless spin-0 boson, respectively, between fermions $X$ and $Y$ (see Fig.\,\ref{fig:ES}). 

\begin{widetext}
\begin{align}
\label{pseudovector-vector_potential}
\centering
V_{AV}(\v{r}) &= \underbrace{g^X_A g^Y_V \v{\sigma}_X \cdot \left\{ \frac{ \v{p}_X}{m_X} - \frac{ \v{p}_Y}{m_Y} , \frac{e^{-M r}}{ 8 \pi r} \right\} }_{V_{12+13}} +  \underbrace{g^X_V g^Y_A \v\sigma_Y^{\,\prime} \cdot \left\{ \frac{ \v{p}_Y}{m_Y} - \frac{ \v{p}_X}{m_X} , \frac{e^{-M r}}{ 8 \pi r} \right\} }_{V_{12-13}}  \nonumber \\
&\underbrace{ -\frac{g^X_A g^Y_V}{2}  \left(\v{\sigma}_X \times \v\sigma_Y^{\,\prime} \right) \cdot \hat{\v{r}}  \left( \frac{1}{r^2} + \frac{M}{r} \right)  \frac{e^{-M r}}{4 \pi m_Y} }_{V_{11}|_{AV}} \, +\underbrace{ \frac{g^X_Vg^Y_A }{2}  \left(\v\sigma_Y^{\,\prime} \times \v{\sigma }_X \right) \cdot \hat{\v{r}}  \left( \frac{1}{r^2} + \frac{M}{r} \right)  \frac{e^{-M r}}{4 \pi m_X} }_{V_{11}|_{VA}} \nonumber \\
& \underbrace{ - \frac{g_A^X g_V^Y}{4}  \left\{ \boldsymbol{\sigma}_X \cdot \boldsymbol{p}_X , \left\{ \v\sigma_Y^{\,\prime} \cdot \left[ \left( \frac{\v{p}_X}{m_X} - \frac{\v{p}_Y}{m_Y} \right) \times \hat{\v{r}} \right] , \left( \frac{1}{r^2} + \frac{M}{r} \right) \frac{e^{-M r}}{16 \pi m_X m_Y} \right\} \right\}}_{V_{11p+16}|_{AV} \sim \v{p}^2\sV_{11}+\sV_{16}} \nonumber \\
&  \underbrace{ - \frac{g_V^X g_A^Y}{4}  \left\{ \v\sigma_Y^{\,\prime} \cdot \boldsymbol{p}_Y , \left\{ \boldsymbol{\sigma}_X \cdot \left[ \left( \frac{\v{p}_X}{m_X} - \frac{\v{p}_Y}{m_Y} \right) \times \hat{\v{r}} \right] , \left( \frac{1}{r^2} + \frac{M}{r} \right) \frac{e^{-M r}}{16 \pi m_X m_Y} \right\} \right\}}_{V_{11p-16}|_{AV} \sim \v{p}^2\sV_{11}-\sV_{16}} \,,
\end{align}
\begin{align}
\label{pseudovector-pseudovector_potential}
\centering
&V_{AA}(\v{r}) = \notag\\
&\underbrace{- g^X_A g^Y_A  \v{\sigma}_X \cdot \v\sigma_Y^{\,\prime} \frac{e^{-M r}}{4 \pi r} }_{V_2} \notag \\
&\underbrace{ - \frac{g_A^X g_A^Y m_X m_Y}{ M^2} \left[ \v{\sigma}_X \cdot \v\sigma_Y^{\,\prime} \left[ \frac{1}{r^3} + \frac{M}{r^2} + \frac{4 \pi}{3} \delta(\v{r}) \right]  -  \left( \v{\sigma}_X \cdot \hat{\v{r}} \right) \left( \v\sigma_Y^{\,\prime} \cdot \hat{\v{r}} \right)  \left( \frac{3}{r^3} + \frac{3M}{r^2} + \frac{M^2}{r} \right)  \right] \frac{e^{-M r}}{4 \pi m_X m_Y} }_{V_3|_{AA}} 
\notag \\
&\underbrace{ - \frac{g_A^X g_A^Y}{4} \left\{ \boldsymbol{\sigma}_X \cdot \left( \frac{\boldsymbol{p}_Y}{m_Y^2} \times \hat{\boldsymbol{r}} \right), \left( \frac{1}{r^2} + \frac{M}{r} \right) \frac{e^{-Mr}}{8 \pi} \right\} }_{V_{4+5}|_{AA}} 
+ \underbrace{  \frac{g_A^X g_A^Y}{4} \left\{  \v\sigma_Y^{\,\prime} \cdot \left( \frac{\boldsymbol{p}_X}{m_X^2} \times \hat{\boldsymbol{r}} \right) , \left( \frac{1}{r^2} + \frac{M}{r} \right) \frac{e^{-Mr}}{8 \pi} \right\} }_{V_{4-5}|_{AA}}  \, \nonumber\\
& + \underbrace{g_A^X g_A^Y  \left[ \left\{ \frac{\boldsymbol{\sigma}_X \cdot \boldsymbol{p}_X}{m_X} , \left\{ \frac{\v\sigma_Y^{\,\prime} \cdot \boldsymbol{p}_Y}{m_Y} , \frac{e^{-M r}}{16 \pi r} \right\} \right\} - \frac{1}{2}\left\{ \frac{\boldsymbol{\sigma}_X \cdot \boldsymbol{p}_Y}{m_Y} , \left\{ \frac{\v\sigma_Y^{\,\prime} \cdot \boldsymbol{p}_Y}{m_Y} , \frac{e^{-M r}}{16 \pi r} \right\} \right\} -\frac{1}{2}\left\{ \frac{\boldsymbol{\sigma}_X \cdot \boldsymbol{p}_X}{m_X} , \left\{ \frac{\v\sigma_Y^{\,\prime} \cdot \boldsymbol{p}_X}{m_X} , \frac{e^{-M r}}{16 \pi r} \right\} \right\} \right] }_{V_8}\,,
\end{align}
\begin{align}
\label{vector-vector_potential}
&V_{VV}(\v{r})= \notag\\
& \underbrace{g^X_V g^Y_V \frac{e^{-M r}}{4 \pi r} }_{V_1|_{VV}}
+  \underbrace{\frac{g^X_V g^Y_V }{4}  \left[ \v{\sigma}_X \cdot \v\sigma_Y^{\,\prime} \left[ \frac{1}{r^3} + \frac{M}{r^2} + \frac{M^2}{r} - \frac{8 \pi}{3} \delta(\v{r}) \right]  -  \left( \v{\sigma}_X \cdot \hat{\v{r}} \right) \left( \v\sigma_Y^{\,\prime}\cdot \hat{\v{r}} \right)  \left[ \frac{3}{r^3} + \frac{3M}{r^2} + \frac{M^2}{r} \right]  \right] \frac{e^{-M r}}{4 \pi m_X m_Y} }_{(V_2 + V_3)|_{VV}} \notag \\
 & + \underbrace{ \frac{g_V^X g_V^Y}{4} \left\{ \boldsymbol{\sigma}_X \cdot \left( \frac{\boldsymbol{p}_X}{m_X^2} \times \hat{\boldsymbol{r}} \right) 
 - 2 \boldsymbol{\sigma}_X \cdot \left( \frac{\boldsymbol{p}_Y}{m_X m_Y} \times \hat{\boldsymbol{r}} \right)
, \left( \frac{1}{r^2} + \frac{M}{r} \right) \frac{e^{-Mr}}{8 \pi} \right\} }_{V_{4+5}|_{VV}}\notag\\
  & \underbrace{- \frac{g_V^X g_V^Y}{4} \left\{
  \v\sigma_Y^{\,\prime} \cdot \left( \frac{\boldsymbol{p}_Y}{m_Y^2} \times \hat{\boldsymbol{r}} \right) 
 - 2 \v\sigma_Y^{\,\prime} \cdot \left( \frac{\boldsymbol{p}_X}{m_X m_Y} \times \hat{\boldsymbol{r}} \right)
 , \left( \frac{1}{r^2} + \frac{M}{r} \right) \frac{e^{-Mr}}{8 \pi} \right\} }_{V_{4-5}|_{VV}} \, ,
\end{align}
and
\begin{equation}
\label{tensor-tensor_potential}
V_{TT}(\v{r}) =  \underbrace{\frac{4 v_h^2 \mathrm{Re}(C_X) \mathrm{Re}(C_Y) m_X m_Y}{\Lambda^4 } 
 \left[  \v{\sigma}_X \cdot \v\sigma_Y^{\,\prime} \left[ \frac{1}{r^3} - \frac{8 \pi}{3} \delta(\v{r}) \right]  - \left( \v{\sigma}_X \cdot \hat{\v{r}} \right) \left( \v\sigma_Y^{\,\prime}\cdot \hat{\v{r}} \right)  \frac{3}{r^3}  \right] \frac{1}{4 \pi m_X m_Y}  }_{(V_2 + V_3)|_{TT}} \, ,
\end{equation}

{\small
\begin{align}
\label{pseudotensor-tensor_potential}
&V_{\tilde{T}T}(\v{r}) = \notag\\
 &\underbrace{ \frac{4 v_h^2 \mathrm{Im}(C_X) \mathrm{Re}(C_Y) m_X m_Y }{ \Lambda^4 }  \left[ \left(\v{\sigma}_X \times \v\sigma_Y^{\,\prime} \right) \cdot \left\{ \frac{\v{p}_X}{m_X} - \frac{\v{p}_Y}{m_Y}  , ~ 
 \frac{1}{r^3} + \frac{4 \pi}{3} \delta(\v{r}) \right\}+  \left\{   \left( \frac{\v{p}_X}{m_X} - \frac{\v{p}_Y}{m_Y} \right)_i  , ~ \frac{3 \left(\v{\sigma}_X \cdot \hat{\v{r}} \right) \left(\v\sigma_Y^{\,\prime} \times \hat{\v{r}} \right)_i }{r^3} \right\} \right] \frac{1}{8 \pi m_X m_Y} }_{  V_{14q+15}|_{\tilde{T}T}  } \notag\\
 + &\underbrace{  \frac{4 v_h^2 \mathrm{Im}(C_Y) \mathrm{Re}(C_X) m_X m_Y }{ \Lambda^4 } \left[ \left(\v\sigma_Y^{\,\prime} \times \v{\sigma}_X \right) \cdot \left\{ \frac{\v{p}_Y}{m_Y} - \frac{\v{p}_X}{m_X}  , ~ 
 \frac{1}{r^3} + \frac{4 \pi}{3} \delta(\v{r}) \right\}+  \left\{   \left( \frac{\v{p}_Y}{m_Y} - \frac{\v{p}_X}{m_X} \right)_i  , ~ \frac{3 \left(\v\sigma_Y^{\,\prime} \cdot \hat{\v{r}} \right) \left(\v{\sigma}_X \times \hat{\v{r}} \right)_i }{r^3} \right\} \right] \frac{1}{8 \pi m_X m_Y} }_{ V_{14q-15}|_{\tilde{T}T}   }\notag \\
&\underbrace{-\frac{2 v_h^2 \mathrm{Im}(C_X) \mathrm{Re}(C_Y) m_X m_Y }{\Lambda^4} \frac{\v{\sigma}_X \cdot [\v{\nabla} \delta(\v{r})] }{ m_X m_Y^2 }  }_{ V_{9+10}|_{\tilde{T}T}  } +\underbrace{\frac{2 v_h^2 \mathrm{Im}(C_Y) \mathrm{Re}(C_X) m_X m_Y }{\Lambda^4} \frac{\v\sigma_Y^{\,\prime} \cdot [\v{\nabla} \delta(\v{r})] }{ m_Ym_X^2}  }_{ V_{9-10}|_{\tilde{T}T}  } \, , 
\end{align}
}
\begin{equation}
\label{pseudotensor-pseudotensor_potential}
V_{\tilde{T}\tilde{T}}(\v{r}) =  \underbrace{ \frac{4 v_h^2 \mathrm{Im}(C_X) \mathrm{Im}(C_Y) m_X m_Y}{\Lambda^4 } 
 \left[  \v{\sigma}_X \cdot \v\sigma_Y^{\,\prime}\left[ \frac{1}{r^3} + \frac{4 \pi}{3} \delta(\v{r}) \right]  -  \left( \v{\sigma}_X \cdot \hat{\v{r}} \right) \left( \v\sigma_Y^{\,\prime}\cdot \hat{\v{r}} \right)   \frac{3}{r^3}  \right]  \frac{1}{4 \pi m_X m_Y}  }_{V_3|_{\tilde{T}\tilde{T}}} \, . 
\end{equation}
and
\begin{align}
\label{scalar-pseudoscalar_potential}
&V_{ps}(\v{r})= \notag\\
&\underbrace{ - g^X_p g^Y_s \v{\sigma}_X \cdot \hat{\v{r}} \left( \frac{1}{r^2} + \frac{M}{r} \right) \frac{e^{-M r}}{8 \pi m_X} }_{V_{9+10}|_{ps}}   
+ \underbrace{g^Y_p g^X_s  \v\sigma_Y^{\,\prime} \cdot \hat{\v{r}} \left( \frac{1}{r^2} + \frac{M}{r} \right) \frac{e^{-M r}}{8 \pi m_Y} }_{V_{9-10}|_{ps}} \notag \\
& + \underbrace{\frac{g^X_p g^Y_s}{8} \left[ \left\{ \left( \v{\sigma}_X \times \v\sigma_Y^{\,\prime} \right) \cdot \frac{\v{p}_Y}{m_Y} , \frac{1}{r^3} + \frac{M}{r^2} + \frac{4 \pi}{3} \delta(\v{r})  \right\} -\left\{ \left( \v{\sigma}_X \cdot \hat{\v{r}} \right) \v\sigma_Y^{\,\prime} \cdot \left( \frac{\v{p}_Y}{m_Y} \times \hat{\v{r}} \right) , \frac{3}{r^3} + \frac{3M}{r^2} + \frac{M^2}{r}  \right\}\right]\frac{e^{-M r}}{8 \pi m_X m_Y}  }_{V_{14q+15}|_{ps}\sim \v{q}^2\sV_{14}+\sV_{15}}  \notag\\
&+ \underbrace{\frac{g^Y_p g^X_s}{8} \left[\left\{ \left( \v\sigma_Y^{\,\prime} \times \v{\sigma}_X \right) \cdot \frac{\v{p}_X}{m_X} , \frac{1}{r^3} + \frac{M}{r^2} + \frac{4 \pi}{3} \delta(\v{r}) \right\} -\left\{ \left( \v\sigma_Y^{\,\prime} \cdot \hat{\v{r}} \right) \v{\sigma}_X \cdot \left( \frac{\v{p}_X}{m_X} \times \hat{\v{r}} \right) , \frac{3}{r^3} + \frac{3M}{r^2} + \frac{M^2}{r} \right\} \right]\frac{e^{-M r}}{8 \pi m_X m_Y} }_{V_{14q-15}|_{ps}\sim\v{q}^2\sV_{14}-\sV_{15}} \, ,
\end{align}
\begin{equation}
\label{pseudoscalar-pseudoscalar_potential}
V_{pp}(\v{r}) = \underbrace{ -  \frac{g^X_p g^Y_p}{4}  \left[  \v{\sigma}_X \cdot \v\sigma_Y^{\,\prime}\left[ \frac{1}{r^3} + \frac{M}{r^2} + \frac{4 \pi}{3} \delta(\v{r}) \right]
- \left( \v{\sigma}_X \cdot \hat{\v{r}} \right) \left( \v\sigma_Y^{\,\prime}\cdot \hat{\v{r}} \right)  \left[ \frac{3}{r^3} + \frac{3M}{r^2} + \frac{M^2}{r} \right]   \right] \frac{e^{-M r}}{4 \pi m_X m_Y} }_{V_3|_{pp}} \, ,
\end{equation}
\begin{align}
\label{scalar-scalar_potential}
V_{ss}(\v{r}) = \underbrace{  -  g^X_s g^Y_s \frac{e^{-M r}}{ 4 \pi r} }_{V_1|_{ss}} 
+ \underbrace{ \frac{g_s^X g_s^Y}{4} \left\{ \boldsymbol{\sigma}_X \cdot \left( \frac{\boldsymbol{p}_X}{m_X^2} \times \hat{\boldsymbol{r}} \right), \left( \frac{1}{r^2} + \frac{M}{r} \right) \frac{e^{-Mr}}{8 \pi} \right\} }_{V_{4+5}|_{ss}} - 
\underbrace{ \frac{g_s^X g_s^Y}{4} \left\{ \v\sigma_Y^{\,\prime} \cdot \left( \frac{\boldsymbol{p}_Y}{m_Y^2} \times \hat{\boldsymbol{r}} \right) , \left( \frac{1}{r^2} + \frac{M}{r} \right) \frac{e^{-Mr}}{8 \pi} \right\} }_{V_{4-5}|_{ss}}\, , 
\end{align}
\end{widetext}
Here $\boldsymbol{\sigma}_{X}$ and $\v\sigma_Y^{\,\prime}$ are vectors of Pauli matrices of the spins $\boldsymbol{s}_i=\hbar {\boldsymbol{\sigma}_i}/2$ of the two fermions, 
$r$ is the distance between particles $X$ and $Y$, $\hat{\boldsymbol{r}}$ is the unit position vector directed from particle $Y$ to particle $X$, 
$M$ is the mass of the new boson, 
while $m_X$ and $m_Y$ and $\boldsymbol{p}_{X}$ and $\boldsymbol{p}_{Y}$ denote the masses and momenta of fermions $X$ and $Y$, respectively. 
$\left\{\Box,\Box\right\}$ denotes an anticommutator. 
In the two-body centre-of-mass reference frame, $\boldsymbol{p}_X= \frac{m_Xm_Y}{m_X+m_Y} (\boldsymbol{v}_X-\boldsymbol{v}_Y)$ and $\boldsymbol{p}_Y=\frac{m_X m_Y }{m_X+m_Y} (\boldsymbol{v}_Y-\boldsymbol{v}_X)$. 
Note that when the classical regime is relevant, the momentum should be treated as a classical variable. 
In such cases, the relative velocity, given by $\boldsymbol{v}=\boldsymbol{v}_X-\boldsymbol{v}_Y$, can be utilized. However, in the quantum treatment (\citealp{ficek_constraints_2017}; \citealp{ficek_constraints_2018}), the momentum is an operator and we substitute $\boldsymbol{p}_X=-i\hbar\boldsymbol{\nabla}_X$ and $\boldsymbol{p}_Y=-i\hbar\boldsymbol{\nabla}_Y$. 
In Eq.\,(\ref{pseudotensor-tensor_potential}), summation over the spatial index $i = 1,2,3$ is implied. 

In Sec.\,\ref{Sec:limits_main}, we delve further into the forms of the potential terms in the centre-of-mass frame and without the employment of anticommutators. 
The obtained equations are then in a common form used for macroscopic-scale experiments. 
Details of these calculations are provided in Appendix\,\ref{App:appendix_f_gg}. 

In this review, we introduce equations that, while originally derived from the expressions in \citet{fadeev_revisiting_2019}, have been modified slightly, mainly to address two points. 

First, we showcase that the potentials are written in a complete form, as opposed to an abbreviated form that lacks symmetry under the permutation of particle indices  $X \leftrightarrow Y$, especially for the potentials in Eqs.\,\eqref{pseudovector-vector_potential}, \eqref{scalar-pseudoscalar_potential}, and \eqref{pseudotensor-tensor_potential}. 
We reformulated the terms previously labelled as $\mathcal{V}_{9,10}$ to the more conventionally used form $V_{9\pm10}$. 
We use the notation $V_i$ to represent potential terms containing coupling constants and $\mathcal{V}_i$ to represent the corresponding spin-structure parts without coupling constants, the latter having been presented in the previous Sec.\,\ref{Subsec:DMformalism}. 

Second, we have incorporated several higher-order terms which, 
although they tend to play a sub-dominant role phenomenologically
as shown in the figures in Sec.\,\ref{Sec:limits_main}, have been under examination in previous experiments. 
These terms include $V_{4\pm5}|_{AA, VV, ss}$, $V_{8}$ and $V_{14q\pm15}|_{ps, \tilde{T}T}$. 
For $V_{4\pm5}$ terms, we have added them in $V_{VV}$, $V_{ss}$ and $V_{AA}$. Previously, \citet{fadeev_revisiting_2019} considered only  phenomenologically more important terms, i.e. the spin-independent $V_1$ terms in $V_{ss}$ [Eq.\,\eqref{scalar-scalar_potential}], $V_1$ and $V_2+V_3$ term in $V_{VV}$ [Eq.\,\eqref{vector-vector_potential}], and $V_2$ and $V_3$ term in $V_{AA}$ [Eq.\,\eqref{pseudovector-pseudovector_potential}]. We include the $V_{4\pm5}$ terms here in light of our focus on spin-dependent interactions in this review. 
For $V_8$ term, we added it in Eq.\,\eqref{pseudovector-pseudovector_potential}. 
For  $V_{14q\pm15}$ term, we added it in the pseudoscalar-scalar potential in Eq.\eqref{scalar-pseudoscalar_potential} and \eqref{pseudotensor-tensor_potential}.
See Sec.\,\ref{Subsec:Pairs-pots.} below for further discussion of paired terms like $V_{4\pm5}$ and $V_{14q\pm15}$.


\subsection{Pairs of potentials}
\label{Subsec:Pairs-pots.}

In this work, we present systematic potentials as formulated in Eqs.\,\eqref{pseudovector-vector_potential} -- \eqref{pseudotensor-pseudotensor_potential}, which are posited for subsequent use by researchers in this field. 
The annotations accompanying these equations elucidate the correlation between our potentials and the 16 potentials introduced by \citet{dobrescu_spin-dependent_2006}, see Sec.\,\ref{Subsec:DMformalism}. 

For many of the single potentials, such as $V_1$, $V_8$, the connection between our potentials and those of \citet{dobrescu_spin-dependent_2006} is generally straightforward, since, for example, the forms of $V_1$ and $\mathcal{V}_1$ match up to a normalisation factor. 
Regarding the potentials presented in pairs, they can be categorized into two groups: 
(1) A potential like $V_{4\pm5}$ ($V_{9\pm10}$) is a simple addition or subtraction of $\sV_4$ and $\sV_5$ ($\sV_9$ and $\sV_{10}$). 
Their paired presentation can be traced back to the choice by \citet{dobrescu_spin-dependent_2006} to represent their 16 potentials in a symmetric form and (anti)symmetric form with respect to exchange of the particle spins, which is further explained below. 
(2) We have introduced new notations, such as $V_{11p \pm 16}$ and $V_{14q \pm 15}$.
The terms $\mathcal{V}_{11}$ and $\mathcal{V}_{16}$ ($\mathcal{V}_{14}$ and $\mathcal{V}_{15}$) appear paired as a linear combination in $V_{11p \pm 16}$ ($V_{14q \pm 15}$), because they arise at the same order in the expansion in terms of the fermion momenta when the extra factor of $\v{p}^2$ ($\v{q}^2$) is counted together with $\mathcal{V}_{11}$ ($\mathcal{V}_{14}$). 
Here $\v{p}^2$ is a quadratic function of the fermion momenta $\v{p}_X$ and $\v{p}_Y$, while $\v{q}$ is the momentum transfer between the fermions (see Fig.\,\ref{fig:ES}), both understood to be taken in momentum space.

Among the 16 potentials presented by \citet{dobrescu_spin-dependent_2006}, those denoted with two indices, such as $\sV_{4,5}$, encompass linear combinations of unit vectors of spin operators for the two fermions, $\v{\sigma}_X$ and $\v{\sigma}_Y^{\,\prime}$, i.e., $\v{\sigma}_X \pm \v{\sigma}_Y^{\,\prime}$. 
This effectively results in two distinct potentials. Conversely, potentials identified by a single index, like $\sV_2$ and $\sV_8$, do not incorporate such linear combinations of fermion spins. In experimental setups employing, for example, only one polarized test mass, researchers can use $V_{4\pm5}$, a linear combination of $\sV_{4}$ and $\sV_{5}$. 
This principle also applies to $V_{9\pm10}$.

Meanwhile, we note that the overall structures of the $V_{4\pm5}$ potentials for the scalar/scalar, vector/vector and axial-vector/axial-vector cases are somewhat different [see Eqs.\,\eqref{pseudovector-pseudovector_potential} and \eqref{vector-vector_potential} and \eqref{scalar-scalar_potential}], even though they are the same for macroscopic bodies (see App.\,\ref{App:appendix_f_gg}).
Specifically, $V_{4\pm5}|_{ss}$ and the temporal part of $V_{4\pm5}|_{VV}$ 
(i.e., the part associated with the $\gamma^0$ vertex factors)
share the same structure. 
However, in the case of $V_{4\pm5}|_{VV}$, there is an additional spatial part 
associated with the $\gamma^i$ vertex factors,
which introduces extra terms that mix the spins and momenta of particles $X$ and $Y$ (a phenomenon not present in the scalar-scalar case). 
Similarly, $V_{4\pm5}|_{AA}$ and the spatial part of $V_{4\pm5}|_{VV}$ share the same form even though the overall structures of $V_{4\pm5}|_{AA}$ and $V_{4\pm5}|_{VV}$ are different. 
Although $V_{4\pm5}|_{AA}$, $V_{4\pm5}|_{VV}$ and $V_{4\pm5}|_{ss}$ are different, they have the same underlying structure in terms of $\sV_{4,5}$ [see Eq.\,\eqref{DM_eqV45}] defined by \citet{dobrescu_spin-dependent_2006}, as evident in Eqs.\,\eqref{gaga_V45}, \eqref{gvgv_V45new} and \eqref{gsgs_V45}, which are presented in the two-body centre-of-mass frame. 
We refer the reader to App.\,\ref{App:appendix_f_gg} where we demonstrate that these potentials have the same form as that described by \citet{fadeev_neue_2018} and \citet{dobrescu_spin-dependent_2006}. 

In the $V_{AV}$ potential, the $\v{p}^2\sV_{11}$ term arises at the same order in the expansion of the relativistic potential in terms of the fermion momenta as the velocity-dependent term $\sV_{16}$. 
In the centre-of-momentum frame, $\v{p}^2$ simplifies to $-\v{P}^2$, with $\v{P} = (\v{p}_{X,i}+\v{p}_{X,f})/2$, where $\v{p}_{X,i}$ and $\v{p}_{X,f}$ denote the momentum of fermion $X$ as it enters or exits the interaction vertex, respectively. Therefore, both of these terms should be inseparably included in searches for velocity-dependent spin-spin forces.

This is particularly important for the interpretation of searches involving identical fermion species, where the $\sV_{16}$ term vanishes, but the $\v{p}^2\sV_{11}$ term survives and gives the sole contribution in that case. Depending on whether we keep the $\v{p}^2\sV_{11}$ term or drop it in the analysis (as done, for example, by \citet{leslie_prospects_2014}; \citet{hunter_using_2014}; \citet{chu_search_2016}; \citet{ji_new_2018}; \citet{chu_search_2020}; \citet{xiao_exotic_2024}), we can arrive at completely different conclusions about whether or not there is an effect for identical fermions.

In the $V_{ps}$ potential, $V_{14q+15}$ is a combination of $\v{q}^2 \sV_{14}$ and $\sV_{15}$, see \citet{fadeev_neue_2018}. 
We combine them since they arise at the same order in the expansion of the potential. 
While the $\mathcal{V}_{14}$ structure proper does not arise in the single-boson-exchange processes considered by \citet{dobrescu_spin-dependent_2006,fadeev_neue_2018,fadeev_revisiting_2019}, a term with the same spin structure but containing two additional spatial derivatives (or equivalently an extra factor of $\v{q}^2$ in momentum space, 
where $\v{q}$ is the change in momentum between the final and initial state of a fermion)\footnote{On a related note, the $V_{9\pm10}|_{\tilde{T}T}$ terms in the tensor-pseudotensor potential in Eq.\,\eqref{pseudotensor-tensor_potential} have a similar $\sV_{9,10}$ origin as the $V_{9+10}|_{ps}$ terms in the pseudoscalar-scalar potential, except the former involve two extra spatial derivatives that correspond to an extra factor of $\v{q}^2$ in momentum space.}  
referred to as $V_{14a}$ in \citet{fadeev_neue_2018} [the term labelled as $\sV_{14}$ in \citet{fadeev_revisiting_2019}] appears at the same order as $\sV_{15}$: we combine this term with $\sV_{15}$ and refer to the resulting potential term as $V_{14q\pm15}$. 

This is important for experiments and proposals, such as those of \citet{leslie_prospects_2014}; \citet{hunter_using_2014}; \citet{chu_search_2016, chu_experimental_2020,chu_proposal_2022}; \citet{ji_new_2018}; \citet{xiao_exotic_2024}; \citet{huang_new_2024}, which studied $V_{14}$ (which does not exist in its simplest $1/r$ variant) and $V_{15}$ separately. 
Their sensitivity estimates/limits still apply for $V_{14q\pm15}$ with a possible $\mathcal{O}(1)$ correction factor due to the geometry of the experiment. Note, for experiments involving identical fermion species, the $\sV_{15}$ term vanishes, but the $\v{q}^2\sV_{14}$ term survives and gives the sole contribution.
For future experimental work, it is important to study the combined term $V_{14q+15}$.

Note that the potential terms $\sV_{6,7}$, defined in Eq.\,(\ref{DM_eqV67}), do not arise for any of the single-boson-exchange scenarios at the order of calculations that we consider. These terms are described by \citet{dobrescu_spin-dependent_2006, fadeev_neue_2018}. Neither $\sV_6$ nor $\sV_7$ identically vanish in the case of identical fermions.



\section{Dedicated source-sensor experiments}
\label{Sec:EXP.METH.1}

\subsection{General remarks} 
\label{Sec:source-sensor_general_remarks}

The basic concept of searches for spin-dependent exotic interactions can be understood by considering a fermion at one vertex as the source of a force-carrying intermediate boson,
and a fermion at the other vertex as the sensor, which is affected by the interaction and whose response is used as the signal for detection. For macroscopic systems, one needs to sum (integrate) the effects of all the source fermions and average the effect over the sensor. 

The 
approach is somewhat
different for atomic-scale 
measurements as compared to
macroscopic-scale experiments. 
On the atomic scale, the source and sensor are typically coupled together 
essentially constituting a single composite
system
to be probed by experiment,
for instance, the electrons and nucleus in the same atom. 
In this case, it is not necessary to define which particle is the sensor and which is the source. For macroscopic- or mesoscopic-scale experiments, the sensor and source are separated entities allowing individual design and optimization of each.

In terms of the way to detect or to set limits on an exotic interaction, there are two broad types of experiments: (1) direct searches for an excess of expected signal above a background; and (2) precise comparisons of theoretical predictions and experimental results that may reveal possible discrepancies. 

Experiments of the first type tend to be dedicated source-sensor searches (discussed in this Section), while those of the second type are often byproducts of precision measurements in molecular, atomic and subatomic studies (primarily discussed in Sec.\,\ref{EXP.METH.2}), which include precision measurements of spectroscopic or fundamental parameters, such as, for example, the muon $g-2$ experiments (see Sec.\,\ref{METH2.g-factor}) or precision spectroscopy of antiprotonic He (see Sec.\,\ref{METH2.PM.EA}).

In this Section, we begin with a discussion of pseudomagnetic fields arising from the exotic interactions and then give examples of source-sensor experiments. Then we discuss
(1) sources of the exotic field, which encompass spin-polarized and unpolarized materials; and
(2) sensors for detecting the exotic field, including vapor cells containing alkali atoms and/or noble-gas atoms, solid-state sensors utilizing spins from defects (e.g., color centers in diamond), as well as resonators using torsional and cantilever modes.

The physical manifestations of an ordinary magnetic field coupling to spins and the coupling of exotic fields to spins are analogous: both lead to spin-dependent energy shifts. 
However, there are crucial differences that can be exploited to distinguish exotic interactions from the well-known magnetic interactions. 
For example, exotic couplings are not expected to be proportional to magnetic moments: this opens the possibility of comagnetometry to use the distinct couplings to distinguish magnetic and nonmagnetic effects \cite{lamoreaux_electric_1989}.
Additionally, exotic fields may have different discrete symmetry properties as compared to magnetic fields.
Crucially, exotic fields generally do not interact in the same way as magnetic fields with various classes of magnetic shielding materials, for example, superconductors \cite{jackson_kimball_magnetic_2016}.
A closely related distinction is that exotic fields with spin-dependent couplings generally do not couple to orbital angular momentum, whereas magnetic fields couple to the orbital angular momentum of charged particles such as electrons.
Yet another difference is that, unlike magnetic fields, exotic fields may have nonzero divergence \cite{grabowska_detecting_2018}.

Even considering the above points,
in source-sensor experiments, the exotic interaction 
is often modeled
as an ``effective magnetic field'' acting on the spins of the sensor.
For example, consider the velocity-dependent interaction of an electron with a nucleon (discussed in detail in Sec.\,\ref{subsec_g_Ag_V}) described by
\begin{equation}
\begin{split}
V_{12+13}=g_A^eg_{V}^N \frac{\hbar}{4\pi}(\boldsymbol{\sigma}_e \cdot \boldsymbol{v})\left(\frac{e^{-r/\lambda}}{r}\right) \,, 
\end{split}
\label{eq.v1213example}
\end{equation}
which 
creates a spin-dependent energy shift equivalent to
an effective magnetic field 
acting on the sensor spin:
\begin{equation}
\begin{split}
\boldsymbol{b}=\frac{1}{\mu}\frac{g^e_Ag^N_V\hbar}{4\pi}\boldsymbol{v}\left(\frac{e^{-r/\lambda}}{r}\right)\, , 
\end{split}
\label{eq:beff}
\end{equation}
where $\mu$ is the magnetic moment of the atom.

After integrating this effective magnetic field over the entire mass source, we get the total effective magnetic field measured by the sensor. 
Note that we use $\boldsymbol{b}$ to represent the pseudomagnetic field from the fifth force and use $\v B$ to represent the normal magnetic field.

In a typical fifth-force experiment, if the fifth force is not detected, 
the experiment sets an upper bound on the strength of the interaction, which depends on the combined statistical and systematic uncertainties, characterized by the parameter $\sigma_B$. 
Correspondingly, the sensitivity scales as $g^e_Ag^N_V \propto g \mu r \sigma_B / (v e^{-r/\lambda})$. 
This indicates that, for $r \gtrsim \lambda$, achieving better experimental sensitivity to this term (i.e, better coupling-constant resolution) requires a shorter distance $r$ (for a fixed number of source particles), 
higher relative velocity $v$ and lower magnetic noise/systematics $\sigma_B$.
Since the apparent strength of the exotic interaction can be converted to an effective magnetic field via multiplying by $\mu^{-1}$, a smaller magnetic moment of the particles comprising the sensor is preferable in order to enhance the apparent strength of the effective magnetic field. The energy shift due to the exotic field does not depend on the magnetic moment; however, the advantage is that the sensitivity to an actual magnetic field is suppressed, which can reduce noise and/or systematics.

\subsection{Sources }\label{METH1.SOUR.}

Since the contributions to an exotic potential from the fermions in the source are additive and can be linearly superposed, the overall potential can be evaluated as an integral over the source volume:
\begin{equation}
    V_\textrm{eff}=\int_V D_i n(\v{r}) {y^i(\v{r})}  d\v{r} \, , 
\label{eq:int_source}
\end{equation} 
where $D_i$ represents the coefficients describing the exotic coupling,
such as $g_A^eg_{V}^N\hbar/4\pi$ for the case of Eq.\,\eqref{eq.v1213example}; 
$n(\v{r})$ is the fermion number density in the source; $y^i(\v{r})$ is the functional form for this type of coupling, which normally includes a Yukawa-type scaling factor $e^{-r/\lambda}$ and an additional polynomial factor, for instance, $\boldsymbol{\sigma}_e\cdot\boldsymbol{v} e^{-r/\lambda}/r$ for the case of Eq.\,\eqref{eq.v1213example}. 
It is always useful to position the source 
closer
to the sensor 
than $\lambda$
to take advantage of the larger value of $y^i(\v{r})$. A discussion of the scale of the sources and sensors can be found in Sec.\,\ref{METH1.SOUR.US} and Sec.\,\ref{METH1.SENS.MS}, respectively.

To increase the effect of an exotic interaction, one generally seeks to increase the fermion density rather than the size of the source in order to be able to access shorter interaction ranges. 
For an unpolarized source, $n_{n,p,e}(\v{r})= n_A  (\v{r}) {A}\chi_{n,p,e}$ with 
$n_A$ being the density of the atoms, $A$ the atomic number,
and $\chi_{n,p}=N_{n,p}/N_N$ the fraction of neutron or proton number in the nuclei. 
In light atoms, we have $\chi_p \approx \chi_n \approx 0.5$, whereas in heavy atoms, we have $\chi_p \approx 0.4$ and $\chi_n \approx 0.6$.  
In electrically neutral materials, the electron density is the same as the proton density. 
An example of the use of neutron and proton fractions can be found around Eq.\,\eqref{ANZ} in Sec.\,\ref{subsec_g_Ag_V}.

For polarized samples, for instance, polarized atoms, the polarized neutron (or proton or electron) number density can be defined according to
\begin{equation}
n^{s}_{n,p,e}(\v{r})=2 n_A (\v{r}) F{ \eta_{n,p,e}} P_{n,p,e}\,,
\label{eq.ns.frac}
\end{equation}
where $\eta_{n,p,e}$ is the fraction of total spin polarization contributed by the particle of interest, $F$ is the total angular momentum of the atom, and $\bar{\bm{s}}_{n,p,e}=\bm{F}\eta_{n,p,e}$ is the expectation value of the spin of neutrons (or protons or electrons) along the total angular momentum of the atom. 
The factor of 2 in Eq.\,\eqref{eq.ns.frac} appears because the spins of the neutron, proton and electron are 1/2.
$P_{n,p,e}$ is the spin polarization of the atoms, which equals to zero if the atoms are not polarized and equals to one if the atoms are fully polarized. Thus to increase $n^s(\v{r})$, one seeks atoms with high atomic density, high fraction of spin polarization $\eta_{n,p,e}$, and a high polarization $P_{n,p,e}$.
 Note that until now, there has been no uniformity in the literature regarding the definition of the spin fractions.\footnote{There are several other places where one finds inconsistencies in numerical factors. For instance, one such place is where $\bm{\sigma}$ is used in place of $\bm{s}$.} For example, \citet{stadnik_nuclear_2015} and \citet{brown_nuclear_2017}) used
$\bar{\bm{s}}$ and \citet{almasi_new_2020} normalized the nucleon spin by $\bar{\bm{s}}/\bm{s}$. In this review, $\eta$ is normalized with respect to the total atomic angular momentum according to $\eta=\bar{\bm{s}}/\bm{F}$, as done by \citet{jackson_kimball_nuclear_2015}. We recommend using this convention in future work. 
Specific examples of the use of the spin fractions are discussed in Sec.\,\ref{METH1.SOUR.NSS} and in Tab.\,\ref{tabel_source}.

In this section, we discuss unpolarized sources in Sec.\,\ref{METH1.SOUR.US}, and polarized 
nucleon
and electron sources in Secs.\,\ref{METH1.SOUR.NSS} and \ref{METH1.SOUR.ESS}, respectively.

\begin{table*}
\centering
\caption{Source materials used in experiments searching for exotic forces. $\eta_{n,p,e}$ are the fractions of spins in the total angular momentum, and $\eta'_e$ is the electron-spin fraction in magnetic moments. For unpolarized sources we use the nucleon density, whereas for polarized sources we use the atomic spin density $n_A$ for polarized atoms and $n_e$ for polarized electrons in solid materials (for ferromagnetic materials, the magnetization is typically $\sim 1$\,T). }
\renewcommand{\arraystretch}{1.8} 
\begin{tabular}{p{7cm}<{\centering} p{ 6cm}<{\centering}  p{4cm}<{\centering} p{0.1cm}<{\centering}} 
\hline
\hline
\textbf{Material} & \textbf{Major Property} & \textbf{Fermion Density} & \\
\hline
\textbf{Unpolarized}& \textbf{Mass Density $\rho_m$(g\,cm$^{-3}$)} & \textbf{Nucleon Density $n_N$(cm$^{-3}$\,)$^a$} &\\
Water (salt)  \cite{chu_laboratory_2013} & 1.0 & $0.6\times 10^{24}$ &\\
{SiO$_2$} (\citealp{rong_searching_2018}; \citealp{jiao_experimental_2021}) & 2.2 & $1.3 \times 10^{24}$ &\\
MACOR (ceramic) \cite{chu_laboratory_2013}&2.5  & $1.5\times 10^{24}$&\\ 
Zirconia  \cite{bulatowicz_laboratory_2013} & 5.9 & 3.6$\times 10^{24}$ & \\
BGO 
(\citealp{tullney_constraints_2013}; \citealp{chu_search_2016}) & 7.1 & $4.3\times 10^{24}$ & \\
Copper (\citealp{ni_search_1999}; \citealp{terrano_short-range_2015}) &9.0 & $5.4\times 10^{24}$&\\ 
Lead (Pb) (\citealp{lee_improved_2018}; \citealp{liang_new_2022}; \citealp{crescini_search_2022}) & 11.3 & $6.8\times 10^{24}$ &\\
$^{238}$U \cite{smith_short-range_1999} & 18.4 & $11\times10^{24}$&\\
Tungsten (W) \cite{wei_constraints_2022,arvanitaki_resonantly_2014}& 19.3 & $11.5\times10^{24}$&\\
\hline
\textbf{Polarized}& \textbf{Fraction of Spin } & \textbf{Atom/Spin Density  (cm$^{-3}$)$^b$} &\\
$^3$He \cite{vasilakis_limits_2009,glenday_limits_2008}& $\eta_n=0.87$, $\eta_p=-0.027$ \cite{friar_neutron_1990} & $n_A\approx 4.8\times 10^{19}$ \cite{vasilakis_limits_2009} \\
$^{87}$Rb \cite{wang_limits_2022}& $\eta_e=0.06$, $\eta_p=0.31$ low polarisation limit \cite{jackson_kimball_nuclear_2015}   & $n_A \approx3.8\times 10^{14}$ \cite{wang_limits_2022} 
& \\
SmCo$_5$ \cite{heckel_preferred-frame_2008,ji_searching_2017} & $\eta'_e=0.51$ \cite{heckel_preferred-frame_2008}& $n_e\approx 4.2\times 10^{22}$ @ 0.96\,T \cite{heckel_preferred-frame_2008}& \\
Alnico 5 \cite{heckel_preferred-frame_2008}   & $\eta'_e=0.95$\cite{heckel_preferred-frame_2008} & $n_e\approx 7.8\times 10^{22}$ @0.96\,T \cite{heckel_preferred-frame_2008}&\\
Iron (\citealp{heckel_limits_2013}; \citealp{ji_searching_2017}; \citealp{almasi_new_2020}) &$\eta'_e=0.96$ \cite{ji_searching_2017}&$n_e\approx 8.2\times 10^{22}$ @ 1\,T \cite{ji_searching_2017}&\\
Dy-Fe alloys \cite{ritter_experimental_1990,chui_experimental_1993} 
& --- & $n_e\approx 1.5\times 10^{22}$ \cite{chui_experimental_1993}& \\
DyIG \cite{leslie_prospects_2014} & --- & $n_e\approx 8\times 10^{20}$ \cite{leslie_prospects_2014}&\\
pentacene \cite{rong_constraints_2018}& --- &  $n_e\approx1.62\times 10^{18}$ \cite{rong_constraints_2018}&\\
\hline
\hline
\hline
\end{tabular}
\label{tabel_source}
\begin{tablenotes}
\item \textcolor{blue}{a} Calculated with $n_N=\rho_m/m_N$ where $\rho_m$ is the mass density of the material and $m_N=1.67\times 10^{-27}$kg is the average mass of neutron and proton mass.
\item \textcolor{blue}{b} Calculated with the total spin number divided by the volume, using the parameters found in the references.
\end{tablenotes}
\end{table*}

\subsubsection{Unpolarized sources}\label{METH1.SOUR.US}

As discussed previously, to obtain a higher nucleon number density, one would prefer a material with a mass density as high as possible. 
However, in practice, one should also consider electromagnetic interactions of the source with the sensor. 
For instance, one seeks to avoid ferromagnetic impurities in the material of an unpolarized source and to avoid using conductors to diminish induction effects. 

In terms of reducing magnetic impurities and conductivity, the best choices of materials are heavy insulators with low magnetic susceptibility. 
A typical material is bismuth germanate insulator (Bi$_4$Ge$_3$O$_{12}$ or BGO), which is a scintillator material that features the high atomic number of bismuth ($Z=83$) and a high mass density of $7.13 \,\textrm{g}/\rm{cm}^3$, an ultralow magnetic susceptibility of
$\chi_\textrm{mag}\approx-19\rm\,ppm$ \cite{tullney_constraints_2013}, 
and magnetic leakage negligible at the present levels of sensitivity. 
Various experiments searching for exotic forces in the sub-centimeter range 
(\citealp{tullney_constraints_2013}; \citealp{leslie_prospects_2014}; \citealp{chu_search_2016}; \citealp{su_search_2021}; \citealp{chu_proposal_2022}; \citealp{feng_search_2022}; \citealp{xiao_femtotesla_2023})
use this crystal.

If sufficient magnetic shielding can be applied or if electromagnetic effects can be sufficiently well controlled, then one would not necessarily need to worry about whether or not the source material is an insulator. 
In such cases, many high-density metals can be considered as well. 
For example, copper, tungsten, lead, and bismuth are good choices, considering their commercial availability and chemical and radiation safety. 
Several fifth-force search experiments use tungsten \cite{wei_constraints_2022}, lead (\citealp{youdin_limits_1996}; \citealp{leefer_search_2016}; \citealp{lee_improved_2018}), copper \cite{terrano_short-range_2015,ni_search_1999}, and even well-shielded magnets (\citealp{ji_searching_2017}, \citeyear{ji_new_2018}) as the mass source. 
Depleted uranium is also used as a source in some fifth-force searches; e.g., \cite{smith_short-range_1999}. Earth can also be used as an unpolarised source if one is searching for long-range forces with characteristic length scales $\lambda \gtrsim$ Earth's radius (\citealp{venema_search_1992}; \citealp{jackson_kimball_constraints_2017}; \citealp{zhang_search_2023}).
It has an overall nucleon number of about $3.6\times 10^{51}$, significantly higher than that of the lab-based sources. Similarly, the Sun and Moon can also be used as unpolarized sources \cite{heckel_preferred-frame_2008,leefer_search_2016,wu_new_2023}.

Water solution with paramagnetic salts such as MnCl$_2$ has also been used \cite{chu_laboratory_2013}. Such paramagnetic salt solutions can compensate the diamagnetism of water when the salt mass ratio is carefully chosen. Silicon and silicon oxide (quartz) SiO$_2$, ceramic or glass are also used due to their commercial availability, chemical stability and ease of manufacture \cite{chu_proposal_2022,chu_laboratory_2013,liang_new_2022}. 
The amorphous silica typically have a mass density of $\approx 2.2$\,g/cm$^3$, 
which is less than half that of BGO crystals and non-magnetic salts. 

The mass density and nucleon density of typical unpolarized sources used in experiments are presented in Tab.\,\ref{tabel_source}. 

\subsubsection{Neutron and proton spin sources}
\label{METH1.SOUR.NSS}

The free neutron is naturally thought of as a neutron spin source. 
However, since free neutrons are produced in accelerators or reactors and their lifetime is only about 15 minutes \cite{wietfeldt_colloquium_2011}, 
it is difficult to accumulate a large number of free neutrons in a possible source. 
On the other hand, free neutrons have good energy resolution and can propagate deep inside materials, and so they are widely used as sensors (\citealp{piegsa_limits_2012}; \citealp{yan_new_2013}; \citealp{parnell_search_2020}).
We discuss experiments with free neutrons in Sec.\,\ref{METH1.SENS.PBS} and Sec.\,\ref{subsec_g_Pg_S_n-N}. 

Nuclear-polarized materials, especially noble-gas atoms, are widely used as neutron sources because of their hyperpolarization and typically high fraction of spin-polarization from neutrons specifically. 
The valence nucleon (within the single-particle picture) in these nuclei is a neutron and the spin-polarization fraction due to neutrons is high. 
At the same time, the polarization fraction due to protons may be nonzero due to configuration mixing that involves core and valence nucleons. 

The typical methods of hyperpolarizing noble gases include:
\begin{itemize}
\item{Spin-Exchange Optical Pumping (SEOP): This method involves transferring angular momentum from optically pumped alkali metal atoms (such as Rb) to noble gas nuclei (such as Xe or $^3$He) via collisions \cite{walker_spin-exchange_1997}.}
\item{Metastability Exchange Optical Pumping (MEOP): This technique is primarily used for $^3$He. It involves exciting $^3$He atoms to a metastable state using an radio frequency discharge, followed by optical pumping with circularly polarized light to achieve hyperpolarization \cite{batz_fundamentals_2011}.
}
\item{Dynamic Nuclear Polarization (DNP): In this method, unpaired electron spins in a paramagnetic agent are polarized at low temperatures and high magnetic fields. The polarization is then transferred to the noble gas nuclei through microwave irradiation \cite{hooper_dynamic_2020}.
}
\end{itemize}
The most widely used noble-gas atoms are $^3$He, $^{21}$Ne and $^{129}$Xe. 
One particular noble-gas source used to search for an exotic dipole-dipole interaction involves $^3$He at a pressure of 12\,atm enclosed in a cylindrical cell with 4.3\,cm inner diameter and 12.8\,cm height, 
with a polarization of $^{3}$He of about $15\%$, corresponding to approximately $9\times 10^{21}$ polarized nuclei as the source \cite{vasilakis_limits_2009}. Polarized noble gases are also widely used as sensors rather than as sources due to their high magnetic sensitivity as discussed in Sec.\,\ref{METH1.SENS.AM}. 

Alkali atoms, such as K and Rb, are used as proton (and neutron) sources. A detailed procedure for deducing the spin fraction of alkali atoms is discussed by \citet{jackson_kimball_nuclear_2015}.
To determine the polarization fraction, one needs to consider the nuclear spin polarization in the atoms as well as the underlying contribution of protons and neutrons to the polarization in the nuclei. 
We discuss the polarization of alkali atoms in Sec.\,\ref{METH1.SENS.}.  
An example of a fifth force search that involves only protons is that of \citet{ramsey_tensor_1979} who studied the effect of the interaction between two protons in molecular hydrogen on the energy spectrum of the molecule. The source and sensor in this experiment were protons within one and the same hydrogen molecule.
The nucleon spin fraction of typical atoms used in exotic spin-dependent interaction research is presented in Tab.\,\ref{tabel_source}. 

\citet{jackson_kimball_nuclear_2015}, \citet{stadnik_nuclear_2015} and \citet{brown_nuclear_2017} discuss other atoms used in spin-dependent physics research.

\subsubsection{Electron spin sources}\label{METH1.SOUR.ESS}

It is desirable to have electron-spin sources with large spin densities that would at the same time
produce minimal magnetic fields at the sensor. 
Solid-state materials, especially ferromagnets and ferrimagnets, are commonly used as spin sources, because they typically have the highest electron-spin densities. 
However, to effectively use these materials as spin sources in precision searches for new physics, one needs to 
find a way to minimize
magnetic leakage. 

The dominant contributions to magnetism of a material sample can in general arise from both spin and orbital angular momenta of electrons. 
We denote the net magnetism due to orbital motion of electrons as $\v{M}_\textrm{o}$ and the net magnetism due to electron spins as $\v{M}_\textrm{s}$. An ideal electron spin source should have the smallest possible net total magnetism $\v{M}=\v{M}_\textrm{s}+\v{M}_\textrm{o}$, while having the largest possible net spin magnetism $\v{M}_\textrm{s}$. 
The net electron spin number in the spin source is estimated to be $\v{N}_\textrm{s} \approx \v{M}_\textrm{s}/\mu_\textrm{B} \approx \v{M}\eta'_{e}/\mu_B$. We use $\eta'_e$ to represent the ratio of the magnetic moment of the spin of interest to the atomic magnetic moment, whereas $\eta_e$ represents the ratio of spin to the total angular momentum of the atom, see Tab.\,\ref{tabel_source}.

The spin contributions of electrons on specific lattice sites can be deduced from their $g$-factors. 
The magnetism of $5d$ transition metals is typically dominated by the spin magnetic moments, due to quenching of orbital magnetism in the lattice field. 
For instance, the spin contribution in pure iron is $96\%$ \cite{ji_searching_2017}, while the spin contribution in Alnico is $95\%$ \cite{heckel_limits_2013}. 
Materials that are compounds of different elements including $5d$ and $4f$ metals, such as Sm and Dy, need to be carefully considered. 
For instance, in $\rm SmCo_5$, the fraction of spin magnetization in Co is around 80\% \cite{heckel_preferred-frame_2008}, while the spin magnetization of Sm is approximately opposite to that of its orbital magnetization; and in DyIG, the magnetism of Fe is assumed to be almost entirely from its spin, while the Dy $4f$ electron orbital momenta are protected by the outer electron shells in the atom,
and $\eta'_{e}\approx 73\%$ \cite{leslie_prospects_2014}.
The effective $\eta'_e$ can be calculated by integrating over all the atoms in these materials, as shown in Tab.\,\ref{tabel_source}.

A Dy-Fe alloy, Dy$_{6}$Fe$_{23}$, was proposed by \citet{ritter_experimental_1987}
and subsequently used in torsion-pendulum-based fifth-force search experiments (\citealp{ritter_experimental_1990}; \citealp{pan_experimental_1992}; \citealp{chui_experimental_1993}).
Dy-Fe is a ferrimagnetic material synthesized by melting stoichiometric quantities of metallic iron and metallic dysprosium. 
The compensation temperature, i.e., the temperature when $\v{M}_\textrm{o} \approx - \v{M}_\textrm{s}$, of such a material is around 250\,K to 289\,K,
which is close to room temperature. 
Other alloys of such Dy-Fe material, such as $\rm DyFe_3$, $\rm Dy_2Fe_{17}$ and $\rm Dy Fe_2$ are also possible low-magnetism, high-spin-density sources, 
but these require further characterization. A general Dy-Fe alloy has about 0.4-0.6 polarized electrons per atom \cite{ritter_experimental_1990}. 
A shortcoming of these materials is that they readily oxidize. 
An alternative chemically stable ferrimagnetic oxide material dysprosium iron garnet DyIG 
(Dy$_3$Fe$_5$O$_{12}$), which has about 0.06 polarized electrons per atom, has been proposed \cite{leslie_prospects_2014}. 
The compensation temperature of DyIG is around 225\,K. In contrast to the materials that have large net magnetisation, these materials show negligible net magnetisation, thus the net polarized-electron number can be calculated by multiplying the number of atoms and the polarized-electron fraction per atom. 
Other paramagnetic materials, such as Gd$_2$SiO$_5$ (GSO), which are used as sensors, are discussed in Sec.\,\ref{METH1.SENS.}.  

Ferrimagnetic materials are relatively difficult to synthesize and have lower electron spin densities than those of ferromagnetic materials. 
The shortcoming of ferromagnetic materials is that their magnetic leakage is large. 
To overcome this problem, other materials with different spin-contribution ratios can be used \cite{heckel_preferred-frame_2008}. 
The idea is similar to that of the compensation regime in ferrimagnetic materials, namely that the total magnetization is compensated, while the spin magnetization and orbital magnetization are not. 
The E\"{o}t-Wash-group setup used alternating stacks of magnets made from Alnico and $\rm SmCo_5$ as the source which has a high spin density and low magnetic leakage, see Fig.\,\ref{EW_setup}. 
The magnetically hard $\rm SmCo_5$ component is cut from commercially obtained material, which is initially magnetized to about 1\,T, whereas the Alnico component is magnetized upon assembly because it is a soft magnetic material \cite{heckel_preferred-frame_2008}. 
Following a similar idea, a soft magneto-magnetic material shielded/hard magnet structure, specifically iron shielded SmCo$_5$ (ISSC), was proposed by \citet{ji_searching_2017} and subsequently utilized by \citet{ji_new_2018,ji_constraints_2023} and \citet{almasi_new_2020} in searches for exotic spin-dependent interactions. 
Its advantage is that it can be manufactured individually, it has negligible magnetic leakage, and one can freely choose a combination of different sets of such sources for specific purposes. 

For long-range interactions beyond the laboratory scale, \citet{hunter_using_2013} proposed an approach that uses Earth as a polarized spin source to investigate spin-spin interactions. 
The Earth is a large-scale source of polarized electron spins because electrons in
paramagnetic minerals 
are polarized by
Earth's geomagnetic field. 
Geophysical models show that there are $\sim 10^{49}$ unpaired electron spins in Earth, with an excess of $\sim 10^{42}$ electron spins polarized antiparallel to Earth's magnetic field.

\citet{hunter_using_2013} use the known strength of the geomagnetic field and a model of Earth’s composition (called the pyrolite compositional model), as well as temperature and other parameters to create a map of the electron-spin polarization everywhere within the Earth and thus are able to reliably use
polarized geoelectrons as a spin source in searches for long-range exotic spin-dependent interactions. 
Compared to typical laboratory spin sources, the total number of polarized electrons is a factor of $\sim 10^{17}$ greater.
However, laboratory sources are usually located 
$\lesssim$
a meter from the detection apparatus, while polarized geoelectrons are $\sim$ a few thousand kilometers away, meaning that the ratio of characteristic distances is $\sim 10^7$. 
Thus, for spin-spin potentials (e.g., $V_{2}$ and $V_{11}$) that fall off as $1/r^n$ at large distances with $n = 1,2$, in which case the suppression of the sensitivity due to the increased distance is either 7 or 14 orders of magnitude, respectively, 
employing geoelectrons as the spin polarized source
can result in substantially improved sensitivity  for sufficiently long-range interactions;
on the other hand, for potentials (e.g., the dipole-dipole potential $V_3$) which fall off as $1/r^3$ or faster at large distances, the suppression of the sensitivity due to the increased distance is $\sim$ 21 orders of magnitude, and so there is no net advantage in using geoelectrons as a spin-polarized source in that case. See more in Sec.\,\ref{Sec:limits_main}.

To search for fifth forces at the sub-micrometer length scale, low-density organic materials have been proposed such as polymeric organic molecules \cite{jiao_searching_2020,chen_optomechanical_2020}. \citet{rong_constraints_2018} used a single crystal of p-terphenyl doped with pentacene-d14, 0.05 mol\%  as electron spin source; the spin density of the sample was estimated to be $1.6\times10^{18}\rm{cm}^{-3}$. 
  
Electron-spin densities of different materials can be found in Tab.\,\ref{tabel_source}. 


\subsection{Sensors    }
\label{METH1.SENS.}


To search for spin-dependent forces manifesting in low order in the interaction (i.e., neglecting quadratic and higher-order effects), at least one part of the experiment, either the source or sensor, should have net spin (i.e., unpaired or polarized spins).  
If the sensor does not have net spin, for instance, mechanical sensors composed of unpolarized solids, then one can search for spin-dependent forces by using a spin-polarized source. 

In Sec.\,\ref{METH1.SENS.MS}, we discuss mechanical sensors that are composed of solid objects. 
These sensors are sensitive to forces (linear accelerations) and torques (angular accelerations). 
In Sec.\,\ref{METH1.SENS.AM}, we introduce atomic-spin-based sensors, which are typically in the gaseous state, that are sensitive to magnetic fields or effective (pseudo)magnetic fields. 

Generally, the sensors are read out optically. 
For instance, the angular position of mechanical torsion pendulums is read out via deflection of reflected light, while atomic magnetometers are read out using Faraday rotation, among other optical methods. 
If the sensor is spin based, the read-out scheme can also be magnetic, taking advantage of the magnetic moments of the fermions. 
A response of the spins to an external magnetic field generates a magnetic signal, which can be detected with another magnetometer. 
In this sense, spins work as transducers of exotic forces. 
In some cases, the transducer will generate a larger magnetic signal than the original exotic field, in which case the transducer essentially acts as an amplifier. 
We discuss the concept of spin amplification in Sec.\,\ref{METH1.SENS.AM.SMSA}. Electron-spin transducers are described in Sec.\,\ref{METH1.SENS.SS}. We present a summary of sensors used in fifth-force searches, including their amplification factors, probe methods and sensitivities in Table\,\ref{tabel_sensor}. 
We also introduce particle-beam sensors in Sec.\,\ref{METH1.SENS.PBS}.
See also the recent review of spin-based sensors for tests of fundamental physics by \citet{jackson_kimball_probing_2023}.


\begin{table*}
\centering
\caption{Sensors used in experiments searching for exotic forces. NMR = nuclear magnetic resonance, SC = self-compensation regime, FID = free induction decay, GGG = gadolinium gallium garnet; SQUID = Superconducting Quantum Interference Device. 
} 
\renewcommand{\arraystretch}{1.8} 
\begin{tabular}{c|c|c} 
\hline
\hline
\textbf{Material} & \textbf{Probe} & \textbf{Sensitivity to exotic field}  \\
\hline
 \textbf{Nuclear Spin}&\textbf{} & \textbf{}  \\
 $^{199}$Hg-$^{201}$Hg (Clock Comparison) \cite{venema_search_1992} &  Light &  290\,nHz \\
  $^{3}$He-$^{129}$Xe (Clock Comparison-Maser) \cite{glenday_limits_2008} &  Rb &  6.1\,nHz ($^3$He)$^{a}$ \\
  $^{3}$He-$^{129}$Xe (Clock Comparison-FID) \cite{tullney_constraints_2013} &  SQUID &  7.1\,nHz \,$^{b}$ \\
  $^{129}$Xe-$^{131}$Xe (Clock Comparison-FID) \cite{zhang_search_2023} &  Rb & 65\,nHz \\
  $^{3}$He (SC) \cite{vasilakis_limits_2009} &  K & $0.75 \,\rm{fT}/\sqrt{\rm{Hz}}$ \\
 $^{129}$Xe (NMR) \cite{su_search_2021}  & $^{87}$Rb  &  $18\,\rm{fT}/\sqrt{\rm{Hz}}$  \\ 
 $^{21}$Ne (SC) \cite{wei_constraints_2022} & Rb  & $1.5\,\rm{fT}/\sqrt{\rm{Hz}}$ \\
\hline
\textbf{Electron Spin}& \textbf{} & \textbf{} \\
TbF$_3$ \cite{ni_search_1999}& SQUID & $11\,\rm{fT/\sqrt{\rm{Hz}}}$\,$^{c}$ 
\\
GGG \cite{chu_search_2015} &  SQUID & $5\,\rm{aT}/\sqrt{\textrm{Hz}}$ (proposal)  \\
GSO \cite{crescini_search_2022} & SQUID &$53\,\rm{aT}/\sqrt{\textrm{Hz}}$ \\
Permalloy (\citealp{vorobyov_new_1988}) & SQUID &--  \\
Nitrogen Vacancy \cite{huang_new_2024}  & Fluorescence & $2\,\rm{nT}/\sqrt{\textrm{Hz}}$ \\
\hline
\hline
\end{tabular}
\begin{tablenotes}
\item $^a$ Have a sensitivity of about $10\,\rm{\mu Hz}/\sqrt{\rm{Hz}}$ at around 400\,$\mu$Hz frequency.
\item $^b$ The sensitivity of the SQUID magnetometer is $10\,\rm{fT}/\sqrt{\rm{Hz}}$.
\item $^c$ Deduced from $5\times 10^{-6}\phi_0/\sqrt{\rm{Hz}}$ \cite{ni_search_1999}, and $\phi_0=2.07 \times 10^{-15}\rm{Wb}$ is the magnetic flux quantum.
\end{tablenotes}
\label{tabel_sensor}
\end{table*}


\subsubsection{Mechanical sensors}\label{METH1.SENS.MS}

A mechanical resonator can be suspended in different ways. 
These include mechanical suspension, 
such as in Cavendish-style torsional balances suspended by a thin wire where the torsional degree of freedom is used; or cantilevers fixed at one end, where translational modes at the free end of the cantilever are commonly used. 
In recent years, various kinds of levitation methods have also been used \cite{gonzalez-ballestero_levitodynamics_2021},
such as optical levitation and diamagnetic levitation. 
Such sensors enable new methods of detection. 

The thermal-noise-limited force sensitivity of a mechanical sensor at angular frequency $\omega$ is \cite{bachtold_mesoscopic_2022}:
\begin{equation}
\delta F(\omega)=\sqrt{\frac{4k_B T M_0\omega_0 }{Q}}\,,
\end{equation} 
where $k_B$ is the Boltzmann constant, $T$ is the temperature, $M_0$ is the mass of the oscillator, $\omega_0$ is the resonance frequency, and $Q$ is the quality factor. 
The sensitivity of the sensor can be increased by conducting experiments in a cryogenic vacuum, i.e., by reducing $T$ and increasing $Q$. 

Note that, generally, smaller sensors have the advantage of smaller mass and thus higher force sensitivity or torque sensitivity, while larger sensors generally have better acceleration sensitivity [$\delta a(\omega)=\delta F(\omega)/M_o$] and angular acceleration sensitivity \cite{bachtold_mesoscopic_2022}. 
Given that the detection of fifth forces, encompassing both spin-dependent and spin-independent varieties, necessitates the measurement of the aggregate effect on mechanical sensor masses or spins, the resultant effect is normalized by the number of fermions of a specific type within the detector. Consequently, when assessing (angular) acceleration, larger sensors generally confer a superior sensitivity to exotic forces (assuming all other variables remain constant).
However, smaller sensors have the advantage of being able to probe shorter-range forces, taking advantage of the larger Yukawa-type scaling factor [see Eq.\,\eqref{eq:int_source}] that can effectively enhance the short-range sensitivity. However, one needs to be careful about the specifics of the experiment. For example, there is no enhancement in the size of the effect when going to shorter ranges with proportionally smaller sensors (and sources), unless the potential is sufficiently highly singular. Assuming that the separation of the source and the sensor is larger than their sizes, for a potential $V$ with a power-law scaling $\propto 1/r^n$, the size of the signal scales as $\sim R_\textrm{source}^3 R_\textrm{sensor}^3 / r^n$. 
Assuming that the source and sensor both have a characteristic linear dimensional size of $\sim R$, the signal scales as $\sim R^6/r^n$, and so the induced force scales as $\propto R^6/r^{n+1}$. 
We see that we can only win with smaller sensors (and smaller sources) if $n \ge 5$.
However, it is notable that potentials derived from single-boson-exchange mediated by massive bosons involve Yukawa-type scaling factor $\propto e^{-Mr}$, and thus are naturally sufficiently singular so that there is indeed an advantage to matching the size of the source and sensor to the length scale of the interaction.
Another example where we can win by using smaller systems on the (sub)atomic scale is in the case of neutrino-pair-exchange forces in atoms and nuclei, with the potential scaling as $V \propto 1/r^5$ \cite{stadnik_probing_2018}. 

\paragraph{Torsion pendulum}
\label{Sec:torsion_pendula_sensors}

The torsion balance for the purpose of detecting small forces was independently developed by Charles Coulomb and John Michell in the 1770s-1780s \cite{jungnickel_cavendish_1996,gillies_torsion_1993}.
The basic structure of a torsion balance consists of a rod horizontally suspended from a soft thin fiber. 
An external force applied asymmetrically to one end of the bar generates a torque that twists the system until the torque from the fiber balances the external torque. 
The external torque and force can be deduced by measuring the rotation angle. 
The torsion balance was among the first apparatuses that could measure feeble forces and enabled many breakthrough discoveries in physics. 
Coulomb used this device to measure the electrostatic force between charges and later established Coulomb's law, while Henry Cavendish conducted an experiment in the 1790s to measure the gravitational force between two masses in the laboratory and precisely determined the value of the universal gravitational constant \cite{cavendish_experiments_1798}.  

Another famous torsion balance is the E\"{o}tv\"{o}s balance, which consists of two masses at opposite ends of a rod that is hung from a thin fiber,
enabling detection of acceleration in two spatial dimensions. 
Lor\'{a}nd E\"{o}tv\"{o}s conducted an experiment around the year 1885 to precisely measure the difference between inertial mass and gravitational mass, which was later reinterpreted as a test of the weak equivalence principle (WEP) with an unprecedented precision for its time (\citealp{eotvos_mathematische_1883}, \citeyear{eotvos_beitrage_1922}; \citealp{will_confrontation_2014}).
To mitigate the effects of low-frequency noise in the torsion balance, measurements can be performed around the torsional resonance frequency, in which case this is referred to as a resonant torsion pendulum.
In practice, the terms ``torsion balance'' and ``(static) torsion pendulum'' are used interchangeably in the literature \cite{gillies_torsion_1993}.
Torsion pendulums are still among the most sensitive sensors of forces (accelerations) and torques. 
The smallest gravitational force measured in the laboratory to date was between two millimeter-radius gold spheres, which was measured using a miniature version of a torsion pendulum \cite{westphal_measurement_2021}. 

We take the E\"{o}t-Wash (a clever play on words) setup as an example to explain the principles behind torsion-pendulum experiments. 
The setup of \citet{adelberger_new_1987} was initially designed to search for spin-independent fifth forces and to test the weak equivalence principle WEP, following the earlier landmark tests of the WEP using similar apparatuses by the Princeton \cite{roll_equivalence_1964} and Moscow \cite{braginskii_verification_1972} groups.  
The idea of searching for spin-dependent forces with torsion pendulums was proposed in \cite{newman_proceedings_1982}, and was later implemented using spin-polarized Dy$_6$Fe$_{23}$ masses (\citealp{ritter_experimental_1987}, \citeyear{ritter_experimental_1990}).
Subsequent experiments conducted by the E\"{o}t-Wash group involved a similar concept \cite{heckel_preferred-frame_2008,hoedl_improved_2011,terrano_short-range_2015}. 

\begin{figure} [!htbp]
\begin{center}
\includegraphics[width=0.42\textwidth]{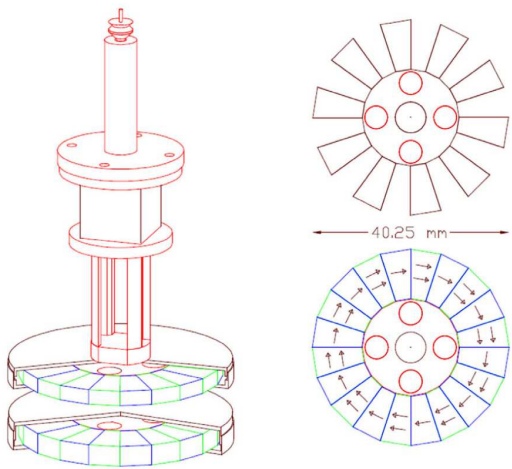}
\end{center}
\caption{(color online). 
The spin-polarized torsion pendulum experimental setup from \citet{terrano_short-range_2015}. 
\textbf{Left}: The torsion pendulum (top) and the spin source (bottom). The torsion pendulum is suspended by a thin fiber, and a mirror cube is placed in the middle to detect the rotation angle. Laser light is reflected by the mirror and further detected by a position-sensitive sensor (not shown). Two sets of spin sources are covered by $\mu$-metal shielding (cut away to shown the internal sources). 
\textbf{Right}: The unpolarized source (top) and spin-polarized source (bottom). The unpolarized source is composed of gapped poles of copper to produce mass densities that alternate in the rotating frame, while the spin-polarized source consists of poles composed of Alnico (high density) and SmCo$_5$ (low density), which have different electron spin densities, with the arrows indicating the spin direction and density. 
The red circles represents the calibration cylinders.} 
\label{EW_setup}
\end{figure}

The spin-pendulum setup of the E\"{o}t-Wash group is shown in Fig.\,\ref{EW_setup}. 
As shown in the left panel, the E\"{o}t-Wash spin pendulum is suspended with a tungsten fiber with a diameter of around 30\,$\mu$m, and it can freely rotate in the horizontal plane. 
There are two sets of test objects, both covered with a $\mu$ metal sheath to protect them from the effects of electromagnetic interactions. 
Because one of the polarized objects is attached to the pendulum system and is directly probed, this set can be considered as the sensor, while the other polarized object is taken as the source. 
The structures of two types of test objects are shown in Fig.\,\ref{EW_setup} (right), with one test object being spin-polarized and the other unpolarized. 
For a spin-polarized source, an alternating configuration of 10 pairs of different spin-density materials composed of Alnico and $\rm SmCo_5$ is used, allowing one to tune the signal frequency associated with an exotic spin-dependent force to 10 times that of the rotational frequency, which can help mitigate the effects of low-frequency noise. 
The spin-unpolarized source had a similar segmented structure. 
Details of these spin sources can be found in Sec.\,\ref{METH1.SOUR.}. 

A typical sensitivity of a torsion pendulum to torques is $\sim \rm 10\,aN \cdot m$ after one day of integration \cite{terrano_short-range_2015}.
The torque on the pendulum was determined by performing a harmonic analysis of the twist angle $\theta$ as a function of the attractor angle $\phi=\omega t$, using a data analysis procedure similar to that described by \citet{kapner_tests_2007}. 

The E\"{o}t–Wash team reported results on tests of the WEP \cite{adelberger_new_1987}, tests of non-Newtonian gravity at short distances \cite{kapner_tests_2007,hoyle_submillimeter_2004}, as well as searches for spin-dependent fifth forces associated with the potential terms $V_{4+5}$, $V_{9+10}$ and $V_{12+13}$ 
(\citealp{heckel_new_2006}, \citeyear{heckel_preferred-frame_2008}, \citeyear{heckel_limits_2013}; \citealp{adelberger_torsion_2009}; \citealp{terrano_short-range_2015}); 
see Sec.\,\ref{Sec:limits_main} for further details. 
 
\paragraph{Microscale mechanical oscillators}\label{METH1.SENS.MS.MMO}

To search for exotic forces on the microscale, one may use micro- and nano-fabricated sensors \cite{bachtold_mesoscopic_2022}. 
They can be used in cantilever mode \cite{chiaverini_new_2003} or in torsional mode \cite{long_upper_2003}, depending on the geometry of the sensor. 

The typical force sensitivity of microscale cantilever sensors is $\sim 100 \,\rm aN/\sqrt{Hz}$, \cite{chiaverini_new_2003}, while nanoscale resonators have achieved force sensitivities well below the $\rm aN/\sqrt{Hz}$ level in a cryogenic system \cite{moser_ultrasensitive_2013}. 

Microcantilever experiments to test gravity at short distances were conducted by \citet{chiaverini_new_2003}, where 0.355\,$\mu$m thick, 50\,$\mu$m wide and 250\,$\mu$m long single-crystal silicon oscillators were used as force sensors while a driven oscillating source mass consisting of alternating 100\,$\mu$m wide gold and silica bars was used to generate an alternating gravitational-type force.
In the torsional resonator experiment of \citet{long_upper_2003}, the lead source mass was modulated at high frequency by a piezoactuator. 
In addition, the cantilever itself does not depend on polarized spins; materials that have high net spin densities, such as ferromagnetic materials, can be used in the source \cite{ding_constraints_2020} or the sensor \cite{ren_search_2021} to search for spin-dependent interactions. 

\paragraph{Levitated sensors}

Levitated particles are sensitive to forces and torques: sensors based on levitated spherical particles have achieved force sensitivities of $\rm 1.6\,aN/\sqrt{Hz}$ \cite{ranjit_zeptonewton_2016}, while levitated anisotropic particles can have torque sensitivities exceeding $\rm 10^{-27}\,N\cdot m/\sqrt{Hz}$ \cite{gonzalez-ballestero_levitodynamics_2021}. These sensors also showed potential for high acceleration sensitivity, for example $\sim \rm 10^{-10}\,g/\sqrt{Hz}$ \cite{timberlake_acceleration_2019}.
Due to their exceptional sensitivity, levitated sensors have been suggested and used for investigating short-range forces (\citealp{geraci_short-range_2010}; \citealp{chen_ultrasensitive_2021},\citealp{yin_experiments_2022}).


Because these types of optically levitated sensors do not have net spin, they are not normally used to search for spin-dependent forces; 
however, if a spin-polarized source is placed nearby, then it is possible to search for the monopole-dipole term $V_{9+10}$. 
An emerging type of levitated spin-based sensor is the levitated ferromagnetic sensor \cite{jackson_kimball_precessing_2016}, which is predicted to have an unprecedented sensitivity to spin-dependent forces due to the high spin density in the ferromagnetic material and its strongly correlated electron spins. 
The remarkable sensitivity of levitated ferromagnetic sensors is a result of the rapid averaging of quantum noise \cite{vinante_surpassing_2021}.
This allows the levitated ferromagnetic torque sensor to ameliorate the detrimental effects of spin relaxation and spin-projection noise that are present in gas-phase atomic magnetometers.
Levitated ferromagnetic sensors have been proposed to search for spin-dependent fifth forces \cite{fadeev_ferromagnetic_2021} and dark matter \cite{ahrens_levitated_2024}, 
and potentially even test gravitational frame-dragging with intrinsic spins \cite{fadeev_gravity_2021}.


\subsubsection{Atomic magnetometers}
\label{METH1.SENS.AM}


Atomic magnetometers use atomic vapors as the sensing medium. 
The atoms can be paramagnetic, typically ground-state alkali atoms or metastable noble gases with unpaired electron spins. An alternative is ground-state diamagnetic atoms with non-zero nuclear spin. In some cases, a combined electron- and nuclear-spin medium is most beneficial. In this section,
we discuss the most commonly used alkali atom magnetometers and nuclear-spin sensors, such as mercury and noble-gas magnetometers. 

\paragraph{Alkali-metal optically pumped magnetometers}\label{METH1.SENS.AM.AOPM}
In a typical setup, alkali atoms are enclosed in a glass or a fused silica cell and are pumped and probed with light. 
A classic review of optical pumping was given by \citet{happer_optical_1972}, 
and a review of the nonlinear magneto-optical effects that are at the heart of modern atomic magnetometers was given by \citet{budker_resonant_2002}.
Atomic spins precess under the influence of an external magnetic field and thus generate an oscillating magnetization that can be detected optically by Faraday rotation or by optical absorption \cite{dupont-roc_detection_1969}. 
Various types of alkali magnetometers work in different regimes to optimize specific parameters such as sensitivity, dynamic range, or bandwidth. 
For a review and explanation of the principles of atomic magnetometers, see (\citealp{budker_optical_2007}; \citealp{budker_optical_2013}).

The fundamental sensitivity of an atomic magnetometer is limited by the spin-projection noise, which can be characterised as
 \cite{budker_sensing_2023}:
\begin{equation}
    \delta B\approx \frac{\hbar}{g_F\mu_\textrm{B} \sqrt{2F}}\left(\frac{\Gamma}{N t}\right)^{1/2} \, ,
    \label{eq:sensitivity}
\end{equation}
where $g_F$ is Land\'{e} factor, $\mu_B$ is the Bohr magneton,  $F$ is the total angular momentum of the atom, $\Gamma$ is the spin-relaxation rate, $N$ is the number of the spins used in the measurement, and $t$ is the measurement time. 
To measure the magnetic field with better sensitivity, larger atomic numbers, longer coherence time, and longer integration time are desired. 
To increase the number of atoms, one can increase the saturated vapor pressure above an alkali-metal sample in the liquid or solid state by heating the vapor cell non-magnetically, for instance with hot flowing air, or using an AC electric heater operating at frequencies far removed from those of interest for the measurement. 
Furthermore, a larger vapor cell can be used to increase the volume and thereby increase the total number of atoms $N$ participating in the measurement. 
Efforts are also directed towards decreasing the spin-relaxation rate (increasing the spin precession coherence time $T_2$). 
Typical ways to increase coherence time include coating the walls of the vapor cell with anti-relaxation materials, adding a buffer gas to slow down diffusion to the cell walls, and using larger cells \cite{walker_spin-exchange_1997}. 

\begin{figure} [!htbp]
\begin{center}
\includegraphics[width=0.54\textwidth]{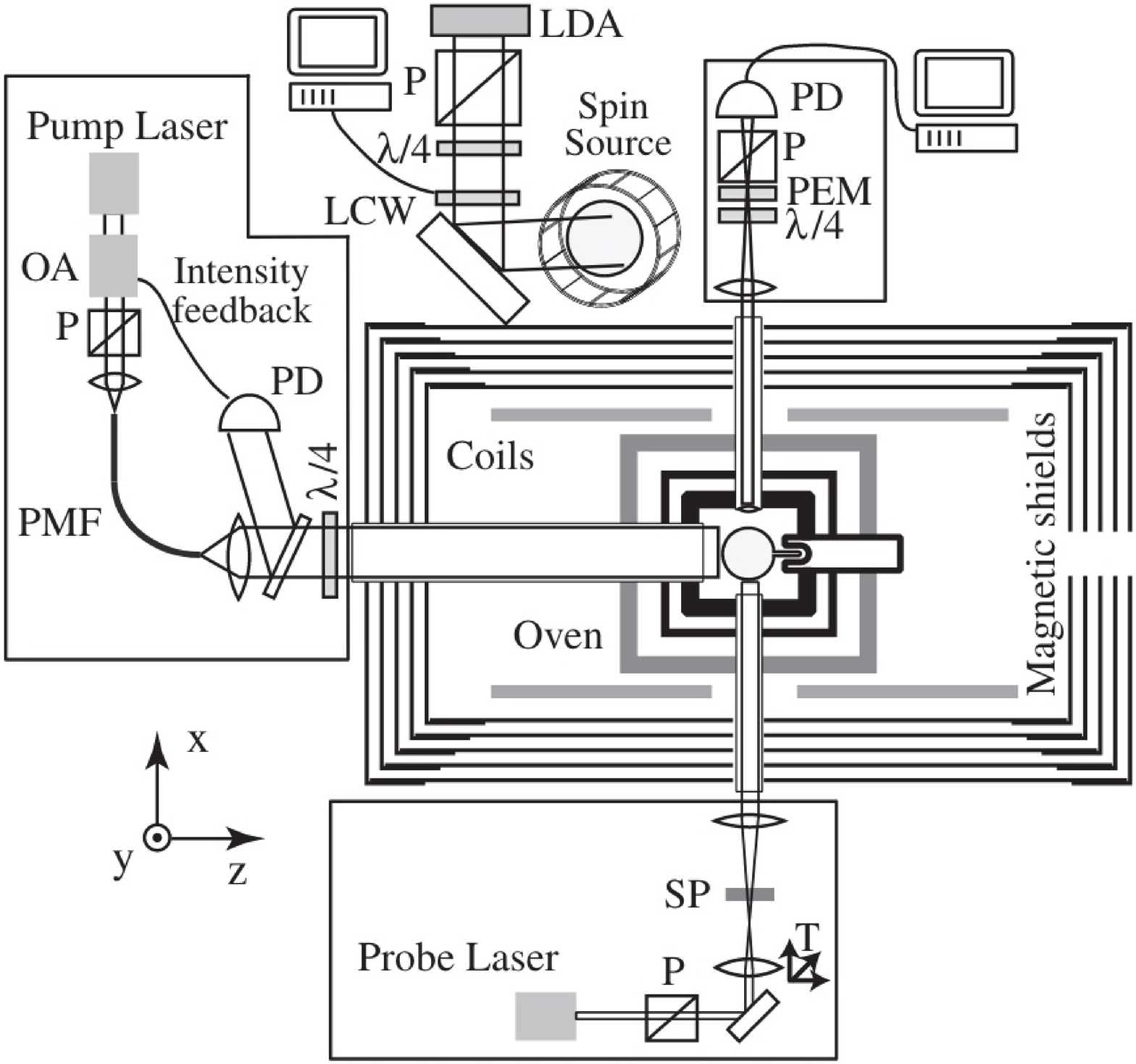}
\end{center}
\caption{ The atomic comagnetometer experimental setup from \citet{vasilakis_limits_2009} used in a search for an exotic spin-dependent interaction. OA: optical amplifer, PMF: Polarization maintaining fiber, P: polarizaer, LDA: laser diode array, LCW: liquid crystal wave plate, PD: photodiode, SP: stress plate to control polarization of the probe beam, T: Translation stage to shift the probe beam, PEM: photoelastic modulator, $\lambda/4$: quarter-wave plate.}\label{fig.comagnetometer} 
\end{figure}

A typical optically pumped atomic magnetometer,
as shown in Fig.\,\ref{fig.comagnetometer}, uses a pump light beam and a probe light beam, and the vapor cell is enclosed within a multi-layer $\mu$-metal shielding to ensure the cell is in near-zero-field conditions and also to guarantee low residual magnetic noise. 
Coils are positioned inside the magnetic shielding to generate a well-controlled magnetic field. 
In the case illustrated in Fig.\,\ref{fig.comagnetometer},
pump laser light is circularly polarized, while the probe laser light is linearly polarized. 
There are many variations on optical pumping and probing schemes, including single-beam arrangements \cite{budker_nonlinear_1998}
and different choices of light polarizations optimized for different measurement goals \cite{ben-kish_dead-zone-free_2010},
as well as many different modulation schemes \cite{budker_nonlinear_2002,gawlik_nonlinear_2006}.

In this Review, we pay special attention to alkali magnetometers operating in the spin-exchange relaxation-free (SERF) regime, which are widely used in spin-dependent fifth-force searches. 
The SERF regime was discovered in 1973 (\citealp{happer_spin-exchange_1973}; \citealp{happer_effect_1977}) 
and first demonstrated in a magnetometer by the Princeton group in 2002 \cite{allred_high-sensitivity_2002}. 
In the low-density case where the Larmor precession rate is faster than the spin-exchange rate [see, for example, \citet{budker_sensitive_2000}],
when collisions transfer alkali atoms between ground-state hyperfine levels, the precessing atomic spins dephase since the ground state Land\'e factors $g_F$ have opposite signs.
In contrast, in the high-alkali-density SERF regime, the spin-exchange rate is faster than the Larmor precession rate: 
the atoms are collisionally transferred back-and-forth between ground-state hyperfine levels many times during a single Larmor period. Thus the atoms collectively precess with a time-averaged spin and magnetic moment (non-zero due to the different statistical weights of the two ground-state Zeeman manifolds). In this way, the SERF regime avoids dephasing due to spin-exchange collisions, allowing operation at much higher atomic vapor densities (and hence larger $N$) without an overly large spin-relaxation rate $\Gamma$.

The requirements of the SERF regime are a relatively high alkali atom density
such that the spin-exchange rate is fast, and a sufficiently small magnetic field such that the Larmor frequency is small.

Alkali atoms have valence electrons and nuclei with unpaired spins. 
The hyperfine interaction and the spin-temperature distribution should be considered to deduce the fraction of atomic spin due to the electron spin $\eta_e$ and nuclear spin $\eta_N$. 
In the nuclei, finding the fraction of neutron spin $\eta_n$ and proton spin $\eta_p$
requires input from nuclear theory \cite{jackson_kimball_nuclear_2015,brown_nuclear_2017}, although reasonable estimates are obtained for valence nucleons based on the single-particle approximation \cite{schmidt_uber_1937}.

Data from experiments with thermal-vapor alkali spin sensors, such as alkali spin-exchange relaxation rates \cite{jackson_kimball_constraints_2010} and frequency-shift noise \cite{xiao_exotic_2024}, were used to search for exotic forces down to the atomic scale. \citet{kim_experimental_2019} and \citet{chu_experimental_2020} used BGO and DyIG sources, respectively, in conjunction with atomic magnetometers to search for spin-dependent force at a similar range. Atomic magnetometers in conjunction with SmCo$_5$ and iron-based electron spin sources were used to search for 10\,cm- and longer-range forces \cite{ji_constraints_2023}. Experiments using alkali comagnetometers are discussed in Sec.\,\ref{Sec:limits_main}.

\paragraph{Nuclear spin magnetometers}

Some atomic magnetometers used in precision searches for exotic spin-dependent interactions employ nuclear-spin-polarized diamagnetic atoms as sensors, owing to the paired nature of their electron spins, which results in cancellation of the electronic magnetic moment and exclusive manifestation of the nuclear magnetic moment. Consequently, the atomic magnetic moment is significantly reduced, often by several orders of magnitude, in comparison to magnetometers based on paramagnetic atoms such as alkalis.
A small magnetic moment means better energy resolution for Zeeman-like interactions experiencing the same magnetic field noise; for further details, see the discussion around Eq.\,\eqref{eq:beff}. 
Vapor cells containing noble gas or mercury are often used because of the achievable high vapor density and long spin-coherence time. 
Commonly used noble-gas atoms include the spin-$1/2$ atoms $\rm ^{3}$He and $\rm ^{129}$Xe and the spin-$3/2$ atoms $\rm ^{21}$Ne and $\rm ^{131}$Xe, with multiple species sometimes used together to 
measure, compensate, and subtract noise from
fluctuations in the magnetic field, i.e., operating in the comagnetometer mode, 
see Sec.\,\ref{METH1.SENS.AM.AC} for details.

Noble-gas atoms can be polarized using spin-polarized alkali atoms via spin-exchange collisions \cite{walker_spin-exchange_1997}. 
The readout can be performed \textit{in situ} using alkali atoms \cite{vasilakis_limits_2009,wei_constraints_2022} or it can be done using external sensors such as superconducting quantum interference devices (SQUIDs) \cite{tullney_constraints_2013}. The \textit{in situ} readout method takes advantage of the Fermi contact interaction between the alkali electron and the noble gas nuclei. These sensors can work in the comagnetometer mode (see Sec.\,\ref{METH1.SENS.AM.AC}) or spin-amplifier mode (see Sec.\,\ref{METH1.SENS.AM.SMSA}). 

\paragraph{Atomic comagnetometers}\label{METH1.SENS.AM.AC}

Comagnetometers involving two (or more) 
magnetically sensitive atomic species
operating together are widely used in fundamental physics, including in searches for fifth forces \cite{lee_improved_2018,wei_constraints_2022}, dark matter \cite{wu_search_2019,lee_laboratory_2023}
and testing Lorentz invariance and $CPT$ violation \cite{brown_new_2010}. 

Comagnetometers can be configured to be insensitive to normal magnetic fields, while retaining sensitivity to exotic physics interactions that do not directly couple to the magnetic moment of the atoms. 
A comagnetometer can utilise two atomic species, for instance, two species of noble gases, such as $^{3}$ He and $^{129}$ Xe, or two isotopes of the same species, such as $^{129}$Xe and $^{131}$Xe. 
A comagnetometer can also use two atomic levels in a single atom, such as the two hyperfine levels of $^{87}$Rb \cite{wang_single-species_2020}
and $^{133}$Cs \cite{bevington_dual-frequency_2020},
or even molecular states with different total nuclear angular momentum in an isotopologue of acetonitrile \cite{wu_nuclear-spin_2018}, which was used for ultralight dark matter search \cite{wu_search_2019}.
The spin-polarized torsion pendulum system described in Secs.\,\ref{METH1.SOUR.ESS} and \ref{Sec:torsion_pendula_sensors} can also generally be thought of as comagnetometers because the spin and orbital magnetic moments cancel each other, which renders the spin pendulum insensitive to normal magnetic fields whilst being sensitive to exotic pseudo-magnetic fields.

In this section, we focus on atomic comagnetometers. 
For earlier reviews of comagnetometers in the context of fundamental physics, see \citet{terrano_comagnetometer_2022}; \citet{huang_axion-like_2024}.
With regard to the basic operating principle, comagnetometers can be characterized according to the way the comparison is done. \citet{terrano_comagnetometer_2022} distinguish the ``clock-comparison'' and  the ``quantization-axis'' self-compensation (SC) method, where the magnetization axis of the compensating species automatically rotates to cancel an external field.

The clock-comparison method typically compares the precession frequencies of two atomic species; since the gyromagnetic ratios of these species are not the same, their precession frequencies are also different. 
By comparing the two precession frequencies, one can cancel out the sensitivity to the magnetic field, whilst retaining sensitivity to an exotic field. 
The general dynamics of the two species (which could even be different states in the same atom) can be described according to the relations $\omega_1= \gamma_1(B +  b_1)$ and $\omega_2=\gamma_2 (B+ b_2) $, 
neglecting any other sources of non-magnetic spin precession (which inevitably lead to a wide range of systematic errors), and where we assume the directions of the magnetic field vector and pseudo-magnetic field vector are aligned along the comagnetometer sensitive axis.
Here $\omega_1$, $\omega_2$ and $\gamma_1$, $\gamma_2$ are the precession frequencies and gyromagnetic ratios of the two species, respectively. 
The two species sense the same magnetic field $B$, but typically see different effective pseudo-magnetic fields due to the exotic spin-dependent interaction. 
The exotic potential $V$ will be sensed by the two species as effective magnetic fields $b_1 = V \, \eta_1 /\gamma_1 \hbar$, and $b_2 = V \, \eta_2 /\gamma_2 \hbar$, respectively (where $\eta_1$ and $\eta_2$ are the fractions of spin polarization for these two species).
The frequency ratio can be expressed as 
\begin{align}\label{Eq:fre-ratio}
    \frac{\omega_1}{\omega_2} &= \frac{\gamma_1}{\gamma_2} \frac{B + b_1}{B + b_2} \approx \frac{\gamma_1}{\gamma_2} \left( 1 + \frac{b_1}{B} - \frac{b_2}{B}  \right) \notag\\
    &\approx \frac{\gamma_1}{\gamma_2} \left[ 1 + V\left( \frac{\eta_1}{\hbar\omega_1} - \frac{\eta_2}{\hbar\omega_2} \right) \right]~.
\end{align}
Here, the approximate sign indicates that this is to leading order in $\eta_{1,2}V$; i.e., discarding second- and higher-order terms. 
From Eq.\,\eqref{Eq:fre-ratio}, we can extract the exotic spin-dependent potential $V$ by comparing two precession frequencies. 



The comagnetometer of \citet{venema_search_1992} employing both $\rm ^{199}Hg$ and $\rm ^{201}Hg$ atoms is of the clock-comparison type. 
The atoms are polarized with ultraviolet light [the physics of how nuclei are polarized in such a process is discussed by \citet{budker_atomic_2010} in Prob.\,3.20]. 
A typical atomic density of Hg is on the order of $10^{13} - 10^{14}\,\rm{cm}^{-3}$, similar to that of alkali atoms but lower than that of noble gas atoms.
In contrast to the indirect probing of the noble-gas spins in alkali-noble gas sensors, Hg atoms are directly probed with linearly polarized light via 
optical rotation.
\citet{venema_search_1992} used the Earth as the mass source of the long-range exotic force. In such experiments, since the source cannot be modulated, a typical approach involves modulating the sensitive axis of the sensor. This is done in a configuration where the Earth rotation axis is parallel to the sensitive axis of the comagnetometer, minimizing the uncertainties due to tilt errors caused by gyroscopic effects associated with the Earth rotation. The magnetic field direction is periodically inverted to modulate 
the relative sign of the magnetic and non-magnetic torques causing precession.
Other experiments that used Earth as the source employed a $^{85}$Rb-$^{87}$Rb dual-isotope alkali comagnetometer \cite{jackson_kimball_constraints_2017} or a $^{129}$Xe-$^{131}$Xe dual-isotope noble-gas comagnetometer \cite{zhang_search_2023} to search for long-range forces. \citet{wu_nuclear-spin_2018} proposed using the $^{1}$H-$^{13}$C nuclei in liquid-state acetonitrile that can work as a comagnetometer for long-range forces, and \citet{wang_single-species_2020} proposed using dual hyperfine levels of $^{87}$Rb. 

Some experiments searching for local Lorentz invariance, or testing the $CPT$ symmetry, also employ modulation of the comagnetometer sensitive direction. Such experiments include $^{129}$Xe-$^3$He comagnetometery \cite{allmendinger_new_2014}, $^{199}$Hg-$^{133}$Cs comagnetometery \cite{peck_limits_2012}, and  K-$^3$He and K-Rb-$^{21}$Ne comagnetometers with self-compensation  \cite{brown_new_2010,smiciklas_new_2011}.  In some cases, the data from such experiments can be re-analyzed and re-interpreted as a search for long-range exotic interactions sources by the Earth, the Sun, or the Moon. For instance, \citet{hunter_using_2013,hunter_using_2014} reanalyzed the data from \citet{peck_limits_2012} and \citet{heckel_preferred-frame_2008} assuming geoelectrons as the source of exotic fields, and \citet{wu_new_2023} reanalyzed the data from \cite{allmendinger_new_2014} taking the solar and lunar masses as sources.

Other experiments probing shorter-scale distances include those of \citet{bulatowicz_laboratory_2013} who used $^{129}$Xe-$^{131}$Xe comagnetometers with a SQUID probe and zirconia rod source to search for mm-scale forces, \citet{tullney_constraints_2013} who used a $^{3}$He-$^{129}$Xe co-magenetomter and BGO sources, and  \citet{feng_search_2022} who used $^{129}$Xe-$^{131}$Xe-Rb comagnetometer and BGO sources to search in a similar force range.

A self-compensating SERF comagnetometer was first demonstrated by \citet{kornack_dynamics_2002}. 
The configuration of a typical alkali-noble-gas comagnetometer setup is shown in Fig.\,\ref{fig.comagnetometer}.
It operates on the principle that for sufficiently slow changes in the magnetic field, the noble-gas spins follow the changing magnetic field direction. 
If the noble-gas magnetization and applied magnetic field are tuned so that they approximately cancel one another, the alkali atoms experience a net zero field environment, which is automatically undisturbed if the magnetic field adiabatically changes direction since the noble-gas magnetization follows the field and compensates it.
To leading order, the response of the probe alkali atoms (specifically, the polarization along the $x$-axis) is given by \cite{vasilakis_limits_2009}: 
\begin{equation}
    P_x^e=\frac{P_z^e\gamma_e}{R_{tot}}\left(b_y^N-b_y^e+\frac{\Omega_y}{\gamma_N}\right)\,,
    \label{eq.SC}
\end{equation}
where $P_x^e$ and $P_z^e$ are the polarizations of alkali along x- and z- axis, while $b_y^N$ and $b_y^e$ are the pseudomagnetic field from the exotic field,
along the y-axis that are sensed by the alkali and noble gas nuclei, respectively, $R_{tot}$ is the total relaxation rate, $\Omega_y$ is the frequency of the apparatus rotation along the $y$ axis (the magnetometer is also essentially an atomic gyroscope), $\gamma_e$ and $\gamma_N$ are the gyromagnetic ratio of alkali and noble gas atoms respectively.
Previous comagnetometers mainly suppressed low-frequency magnetic noise.
A novel atomic comagnetometer \cite{qin_new_2024}, based on coupling between alkali-metal and noble-gas spins, extends this suppression to higher frequencies, up to 100\,Hz.

Among the fifth-force experiments that employed self-compensating comagnetometers are that of \citet{vasilakis_limits_2009} who used polarized $^{3}$He as the source and placed constraints on dipole-dipole interactions between nucleon spins; the experiment of \citet{lee_improved_2018} who used lead sources to place constraints on monopole-dipole interactions between nucleon spin and nucleon mass; the experiment of \citet{almasi_new_2020} who used a an iron shielded SmCo-magnet source and placed constraints on dipole-dipole interactions between neutrons and electrons; and the work of \citet{wei_constraints_2022} who utilized a tungsten-ring source and placed constraints on velocity-dependent interactions between unpolarized nucleons and neutron/proton spins. 



\paragraph{Noble-gas spin masers and amplifiers}\label{METH1.SENS.AM.SMSA}


In addition to the comagnetometer regime previously discussed, polarized nuclei can operate in various other regimes tailored for specific objectives. For example, resonance feedback from coils and resonance circuits is utilized in the maser regime, as detailed by \citet{glenday_limits_2008}. In the nuclear magnetic resonance (NMR)-amplifier mode, the effective magnetic field of an exotic interaction can be resonantly amplified at the NMR frequency. This category of sensors includes those employing external pickup-coil readout methods \cite{budker_proposal_2014,arvanitaki_resonantly_2014}, as well as those utilizing an alkali \textit{in-situ} readout method, also referred to as a ``spin-based amplifier'' by \citet{jiang_search_2021,su_search_2021}.
The idea of the spin-based amplifier is that a feeble field acting upon a highly polarized dense spin sample with a long relaxation time rotates the magnetization of the sample. This rotation is detected with a magnetometer that senses the field of the polarized sample that is much larger than the field that caused the rotation. This way, the real or effective magnetic is amplified. 


Generally, electron spins can also serve as transducers of exotic fields. For instance, ferromagnetic materials can be used to augment the coupling of electron spins with exotic fields, as demonstrated by \citet{chui_experimental_1993} and \citet{gramolin_search_2021}. These applications are further discussed in the section on solid-state electron spin sensors (Sec.\,\ref{METH1.SENS.SS}).

A nuclear spin-based sensor, with spins initially oriented along a leading magnetic field, develops transverse magnetization when subjected to a resonant oscillating magnetic field, as described, for example, by \citet{arvanitaki_resonantly_2014}.
The magnetization of the nuclei can be read out with external or \textit{in-situ} magnetometers.

\textit{In-situ} readout of the alkali-noble gas magnetometers offers the advantage of an additional enhancement factor due to the Fermi-contact interaction between noble gases and alkali atoms. Amplification factors for magnetic fields and pseudo-magnetic fields in a $^{129}$Xe-Rb system have been reported in the range of 100-5400 \cite{su_review_2022}. 
This amplification scheme is generic and can be applied to a wide range of noble gas isotopes, which could further improve signal amplification by factors of $10^4$ with $^{21}$Ne and $10^6$ with $^3$He \cite{jiang_search_2021}. While both spin-based amplifiers and self-compensating comagnetometers (\citealp{vasilakis_limits_2009}; \citealp{lee_improved_2018}; \citealp{almasi_new_2020}; \citealp{xu_constraining_2023})
utilize overlapping spin ensembles (e.g., $^{129}$Xe-$^{87}$Rb), the former have a wider range of operational bias fields. In the K-Rb-$^{21}$Ne system, \citet{wei_dark_2023} identified a hybrid spin resonance distinct from the self-compensating regime. Under the conditions of hybrid spin resonance, the nuclear and electron precession frequencies coincide, leading to an increased linewidth of the noble gas, enabling dark matter search over a comparably much wider mass range \cite{wei_dark_2023}. Alkali-noble gas spin amplifiers using different types of species are useful in searches for axion-like dark matter (\citealp{jiang_search_2021}, \citeyear{jiang_floquet_2022}; \citealp{wei_dark_2023}; \citealp{huang_axion-like_2024}),
as well as exotic spin-dependent forces \cite{su_search_2021,wang_search_2023-1,wang_search_2023}.

Such type of spin amplifiers use a scheme in which nuclear spins and the detector spatially overlap in the same vapor cell, providing two significant advantages: nuclear spins can be directly hyperpolarized to achieve a polarization of 0.1–0.3 by spin-exchange optical pumping; nuclear spin signals can be enhanced due to a large Fermi contact enhancement factor, detected \textit{in-situ} with an atomic magnetometer.


Other 
magnetic resonance experiments for exotic physics searches include the CASPEr experiment that uses hyperpolarized liquid-state Xe \cite{garcon_cosmic_2017}
and the ARIADNE experiment that uses polarized $^3$He \cite{arvanitaki_resonantly_2014} 
for short-range spin-dependent forces from both spin-polarized and spin-unpolarized sources. 


\subsubsection{Solid-state electron spin sensors}
\label{METH1.SENS.SS}

\paragraph{Ferromagnetic material}
Generally, the 
magnetism
of a ferromagnet is dominated by unpaired electron spins, and thus ferromagnets can be 
used as sources and sensors based on polarized
electron spins
for dark-matter (\citealp{gramolin_search_2021}; \citealp{bloch_scalar_2023})
and exotic spin-dependent force searches \cite{fadeev_ferromagnetic_2021}.


The core of an electromagnet consists of a magnetically soft material with oriented magnetic domains producing magnetization that reinforces the applied field, thereby enhancing the applied magnetic field. 
A magnet made of a hard magnetic material, if it can freely move in space with small mechanical energy dissipation, can also be treated as an amplifier and can be used as a high-sensitivity magnetometer 
(\citealp{jackson_kimball_precessing_2016}; \citealp{fadeev_ferromagnetic_2021}; \citealp{vinante_surpassing_2021}; \citealp{ahrens_levitated_2024}).
In their pioneering experiments, \citet{vorobyov_new_1988} used ferromagnetic permalloy material as a transducer and a SQUID to read out its magnetization change. 
\citet{chui_experimental_1993} utilized paramagnetic terbium fluoride (TbF$_3$) as a transducer material and Dy$_6$Fe$_{23}$ as a spin-polarized source to search for electron-spin-dependent forces. 
\citet{quax_collaboration_axion_2020} (QUaerere AXion, or: QUest for AXions) used ferromagnetic yttrium iron garnet (YIG) as an amplifier in their axion dark matter search, but in their case by searching for the axion coupling to the electron spins in YIG. They also used Gd$_2$SiO$_5$, which has a magnetic susceptibility of $\chi \approx 0.7$ as a transducer for exotic interaction research; see Sec.,\ref{gpgs_e-N}. 
The experiment of \citet{gramolin_search_2021} used a ferromagnetic material to enhance the currents generated by ALP dark matter via the ALP-photon coupling.
In addition, paramagnetic materials, for instance Gd$_{3}$Ga$_{5}$O$_{12}$, which has a susceptibility of $\chi \approx 0.53$ at 3.1\,K, are also proposed as transducers of exotic spin dependent forces even without amplification \cite{chu_search_2015}. In these experiments, the signals/magnetization from the transducer materials are measured with SQUID magnetometers.

\paragraph{Sensing with nitrogen-vacancy (NV) centers in diamond}\label{METH1.SENS.SS.NV}

The NV center in the diamond-crystalline lattice is a point defect with unique optical and magnetic properties, enabling the detection of small magnetic fields with high sensitivity and on short distance scales \cite{degen_quantum_2017,barry_sensitivity_2020}. 
NV-diamond-based sensors have emerged in recent years as a promising platform to probe exotic interactions at the micrometer scale 
(\citealp{rong_searching_2018}; \citealp{rong_constraints_2018}; \citealp{chen_optomechanical_2020}; \citealp{rong_observation_2020}; \citealp{jiao_experimental_2021}; \citealp{chen_ultrasensitive_2021}; \citealp{liang_new_2022}; \citealp{chu_proposal_2022}; \citealp{guo_searching_2024}; \citealp{huang_new_2024}).
NV-center magnetometry, and especially single-NV-center magnetometry, offers high spatial resolution, down to atomic scale, exceeding that of atomic vapor-cell magnetometry. 
Single-NV-center magnetometers have magnetic-field sensitivity down to $\sim 0.5\,\rm nT/\sqrt{Hz}$ \cite{zhao_sub-nanotesla_2023}. Although not as sensitive as atomic magnetometers (whose sensitivities reach sub-fT$\rm/\sqrt{Hz}$ for macroscopic fields), diamond sensors are useful for probing exotic forces at much shorter length scales.

\begin{figure} [!htbp]
\begin{center}
\includegraphics[width=0.3\textwidth]{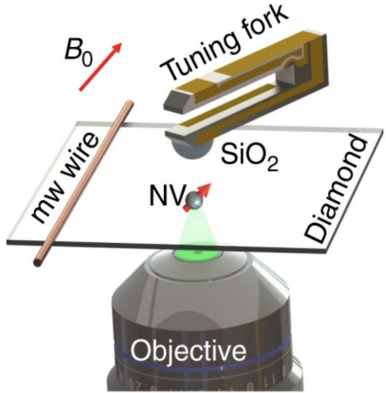}
\end{center}
\caption{NV-diamond magnetometer setup used by \citet{rong_searching_2018}. 
From top to bottom: 
The source is a $\rm SiO_2$ hemisphere, which is driven to oscillate vertically by a tuning fork, thereby modulating the exotic force. 
The sensor is an NV magnetometer on a diamond rectangular plate. 
The NV center is pumped with a green laser and emits red fluorescent light, which is collected via the same objective that delivers the pump light. 
A wire is used to deliver microwaves to the NV center driving a transition between two Zeemann levels. 
The direction of the source movement can be changed to search for different types of exotic interactions 
(\citealp{rong_searching_2018}, \citeyear{rong_observation_2020}; \citealp{jiao_experimental_2021}; \citealp{liang_new_2022}).
} \label{nvexperimentsetup}
\end{figure}
One typical experiment using an NV-based sensor \cite{rong_searching_2018} is shown in Fig.\,\ref{nvexperimentsetup}. 
A $\rm SiO_2$ hemisphere is utilized as an unpolarized mass source (see Sec.\,\ref{METH1.SOUR.}), which is driven to oscillate in the vertical direction, thereby tuning the relative velocity and the distance from the source to the sensor. 
As a result, an otherwise static exotic interaction can be modulated into a sinusoidal signal in the kHz frequency range.
A spin-echo protocol \cite{rong_searching_2018} or continuous-wave protocol \cite{liang_new_2022} can be used to detect the oscillating pseudo-magnetic field. 

In another experiment \cite{rong_constraints_2018}, a laser-pumped single crystal of $p$-terphenyl-doped pentacene-$d_{14}$ is used as a source of polarized electron spins. 
The NV sensor experiences an effective magnetic field induced by polarized electrons, resulting from the coupling between the two systems, including the magnetic dipole-dipole interaction and an exotic axial-vector dipole-dipole interaction, 
with the magnetic dipole-dipole interaction given by 
\begin{equation}
H=-\frac{\mu_0\gamma_e\gamma_e\hbar^2}{(16\pi r^3)}[3(\boldsymbol{\sigma}_e \cdot \hat{\boldsymbol{r}})(\boldsymbol{\sigma}_e^{\,\prime} \cdot \hat{\boldsymbol{r}})-\boldsymbol{\sigma}_e \cdot \boldsymbol{\sigma}_e^{\,\prime}]\,,
\end{equation}
where $\mu_0$ is the magnetic constant (permeability of free space). The authors selected a single-NV center in their sample and measured the overall coupling strength to the polarized-electron source using a pulse sequence that measures the phase shift between NV sublevels that results from spin-spin interactions. 
This phase shift is then converted to the population of the NV-center $\ket{m_s=0}$ state and measured. 
The experiment was repeated for three single-NV centers at different locations with respect to the source. 
Based on a fitting of their data to the phase shift caused by the effective magnetic field 
\begin{align}
&b_\textrm{eff}(r,\theta)=\nonumber\\&-\frac{\mu_0 \gamma_e \hbar}{(8\pi r^3)} \times [3{\cos}^2(\theta)-1] + {g_A^eg_A^e} \frac{e^{-\frac{r}{\lambda}}} {({2\pi \hbar \gamma_e r})}\,,
\end{align}
\citet{rong_constraints_2018} obtained an upper limit on the exotic axial-vector dipole-dipole interaction constant product $g_A^eg_A^e$, which was treated as a fitting parameter. 

An outlook towards future NV-based experiments is given by \citet{chu_proposal_2022} and \citet{du_single-molecule_2024}. For further discussion, see Sec.\,\ref{Sec:limits_main}.




\subsubsection{Particle beams}
\label{METH1.SENS.PBS}

Beyond the spin-based sensors mentioned previously,
an extensive array of other spin-based sensors are employed or suggested for the exploration of exotic spin-dependent interactions. These include particle beam sensors
(\citealp{piegsa_limits_2012}; \citealp{yan_new_2013}; \citealp{schulthess_new_2022}; \citealp{snow_searches_2022}).
Certain particle beam sensors, such as muon and neutron beams, show sensitivity to the interactions of exotic fields. Moreover, some beams are tailored for specific objectives, for instance, neutron beams are deployed in experiments aimed at discovering a permanent neutron electric dipole moment (EDM) \cite{abel_measurement_2020}. 


Having no electric charge and other advantageous properties such as their small polarizability, neutrons are an attractive system for fundamental experiments. In fact, 
cold and ultracold neutrons have been used as probes for various exotic interactions as discussed in comprehensive reviews by \citet{sponar_tests_2021} and \citet{snow_searches_2022}.
Polarized beams of slow neutrons with energies below $0.025$\,eV propagating through liquid $^4$He have been used to search for potential new interactions over mesoscopic distances.
For example, experiments have established an upper limit on the strength of possible long-range $P$-odd interactions between neutrons and both nucleons and electrons, derived from data on the search for parity violation in neutron spin-rotation within liquid $^4$He \cite{yan_new_2013}. 
Experiments involving ultracold neutrons with energies below $10^{-7}$\,eV (and $^{199}$Hg atoms as a comagnetometer) have been used to probe axionlike dark matter \cite{abel_search_2017} and search for a permanent EDM of the neutron \cite{abel_measurement_2020}. 
Applying a strong electric field allowed placing new limits on the coupling of axion dark matter to gluons, achieved using a Ramsey-type apparatus designed for cold neutrons \cite{schulthess_new_2022}.
Other current and proposed neutron-based experiments could open new avenues for the exploration of exotic interactions 
(\citealp{nesvizhevsky_quantum_2002}, \citeyear{nesvizhevsky_neutron_2008}; \citealp{baesler_constraint_2007}; \citealp{pokotilovski_neutron_2011}; \citealp{piegsa_proposed_2012}; \citealp{cronenberg_gravity_2015}; \citealp{haddock_search_2018}; \citealp{fujiie_development_2024}). 
A summary of existing neutron beam experiments that search for spin-dependent exotic interactions is given in Sec.\,\ref{subsec_g_Pg_S_n-N}. 

\section{Complementary experiments and observations to detect exotic spin-dependent interactions}\label{EXP.METH.2}
\subsection{General remarks }

In addition to the source-sensor experiments discussed in Sec.\,\ref{Sec:EXP.METH.1}, various other experiments, not necessarily initially dedicated to searching for fifth forces, can be used to search for such forces, as we discuss here. 
Among such precision measurements at subatomic, atomic, and molecular scales are experiments related to high-precision spectroscopy of various exotic atoms, parity violation, searches for EDMs, isotope shifts ad so on. 

In analyzing such experiments, it should be recognized that, while the range of an interaction is given by the inverse mass of the intermediate boson responsible for the interaction, the effects of the interaction can be observable on much larger scales. 
An example of such a ``cascade of scales'' is the spin-dependent atomic parity violation effect due to the nuclear anapole moment (Sec.\,\ref{METH2.PV}). 
Indeed, weak interactions are mediated by $W$ and $Z$ bosons, the masses of which (80 and 91\,GeV/$c^2$, respectively) correspond to an interaction range of $\sim 10^{-16}$\,cm, which is much smaller than the nuclear size (between 10$^{-13}$ and 10$^{-12}$\,cm). However, the 
exchange of mesons withing the nucleus ``spreads'' the influence of parity nonconservation (PNC) interactions over the entire nucleus, generating a parity-violating anapole moment. 
The presence of the nuclear anapole moment in turn affects the atomic electrons and can be (and indeed was) detected in atomic experiments.

\subsection{Precision measurements}
\label{METH2.PM}

\subsubsection{Trapped ions     }
\label{METH2.PM.TI}

Trapped ions have been extensively studied for their applications in precision measurements \cite{cairncross_precision_2017,blaum_high-accuracy_2006}, quantum computing (\citealp{cirac_quantum_1995}; \citealp{blatt_entangled_2008}; \citealp{harty_high-fidelity_2014}; \citealp{bruzewicz_trapped-ion_2019})
and quantum simulation 
(\citealp{blatt_quantum_2012}; \citealp{lv_quantum_2018}; \citealp{cong_selective_2020}; \citealp{monroe_programmable_2021}).
Precision measurements using trapped ions refer to the use of ion traps and precise ion control to make accurate measurements of various physical quantities, with applications in tests of fundamental physics 
(\citealp{wineland_search_1991}; \citealp{kotler_constraints_2015}; \citealp{cairncross_precision_2017}),
atomic clocks 
(\citealp{rosenband_frequency_2008}; \citealp{ludlow_optical_2015}; \citealp{kozlov_highly_2018}; \citealp{safronova_search_2019}; \citealp{burt_demonstration_2021}),
and more. 

\citet{wineland_search_1991} searched for an exotic dipole-monopole interaction by examining the hyperfine resonances of approximately 5000 trapped and cooled $^9$Be$^+$ ions. The nucleons in the Earth served as the source, while the ions acted as a spin sensor. An applied magnetic field was switched between being directed either parallel or antiparallel to the direction of gravity, and any contribution of the exotic interaction would be revealed by a difference in the resonance frequency of the transition between the $^2 S_{1/2} \ket{F=1,M=0}$ and $^2 S_{1/2} \ket{F=1,M=-1}$ states of the $^9$Be$^+$ ions. Additionally, \citet{wineland_search_1991} studied the dipole-dipole interaction with the same ions, using electron spins in an electromagnet as the dipole source. 

Experiments involving ion traps can also detect small deviations from the predictions of established physical theories, potentially leading to discovery of new physics beyond the SM. \citet{kotler_measurement_2014} confined two ions of $^{88}$Sr$^+$ at a separation of one micrometer using a linear radio-frequency Paul trap. The ions evolved under the magnetic dipole-dipole interaction and reached a final state that encoded information about the strength of the interaction, which could be determined by interrogating the ions. This technique enabled precise measurement of the magnetic dipole-dipole interaction between the ions.
\citet{kotler_constraints_2015} compared their experimental result \cite{kotler_measurement_2014} on the strength of the interaction to a theoretical calculation: the small uncertainty was translated into new constraints on the strength of an axial-vector- or pseudoscalar-mediated interaction between electrons. 

The work of \citet{wineland_search_1991} and \citet{kotler_constraints_2015} emphasized trapped and cooled ions as a useful platform for exotic interaction searches, benefiting from the controllability of trapped ions and the relatively straightforward analysis of systematic errors.


\subsubsection{Atomic spectroscopy }\label{METH2.PM.AHS}

Spectroscopic experiments are arguably the most accurate experiments in physics. 
The relative inaccuracy of a routine spectroscopic experiment is about $10^{-8}$, or better, while the best atomic clocks have a fractional inaccuracy on the order of $10^{-19}$. 
This accuracy corresponds to potentially high sensitivity to exotic interactions, provided that there is a theoretical standard-model prediction for the observed transition of a comparable accuracy. 
At present, this is possible only for a limited number of simple systems (\citealp{karshenboim_hyperfine_2011,karshenboim_constraints_2010,karshenboim_precision_2010}; \citealp{ficek_constraints_2017}; \citealp{ficek_constraints_2018}) which consist of only a few particles and ideally lack hadronic particles.
(As discussed in Sec.\,\ref{Sec:Intro-Exp}, this applies to searches that rely on theory-experiment comparisons; high experimental precision alone may be sufficient for searches that do not strongly rely on highly accurate theory.) 

Even in the case of hydrogen, the accuracy of the theory is limited by the fact that the proton is not an elementary particle, but rather is a hadronic state consisting of quarks bound together via their interactions with gluons. 
One way to bypass this limitation is to look at the difference of the effect for different quantum numbers (\citealp{karshenboim_precision_2010}; \citealp{fadeev_pseudovector_2022}). 
For example, the hyperfine splitting $\Delta_\mathrm{hf}(n)$ of the $s$-states scales with the principal quantum number $n$ as $\propto 1/n^3$. Therefore, one can look, for example, for the difference $\Delta_{21}=8\Delta_\mathrm{hf}(2s)-\Delta_\mathrm{hf}(1s)$. Exotic spin-spin interactions, which decrease with the distance as $1/r$ rather than $1/r^3$, scale differently with $n$ and contribute to the non-zero $\Delta_{21}$.\footnote{There are also residual contributions to $\Delta_{21}$ from nuclear structure and other standard-model effects \cite{karshenboim_hyperfine_2002,karshenboim_hyperfine_2002-1,yerokhin_electron_2008}.}

\subsubsection{Spectroscopy of exotic atoms }\label{METH2.PM.EA}

Fine structure in simple atoms is less sensitive to the nuclear size and can be calculated more accurately. 
This allows direct comparison of the theoretical prediction with the experiment. 
The most accurate calculations of the fine structure are possible for systems such as positronium and muonium, which are devoid of on-shell hadronic particles.

Exotic atoms incorporate particles other than the common protons, neutrons, and electrons. 
The simplest examples include positronium, which is a bound state of a positron and an electron, or muonium -- a bound state of an antimuon and electron. 
Their simplicity and the fact that they contain unusual components make them an ideal laboratory for fundamental physics. 
Comparison of theoretical calculations and experimental measurements for various transitions in such systems has led to constraints on interactions between the constituents of positronium (\citealp{frugiuele_current_2019}; \citealp{fadeev_pseudovector_2022})
and muonium (\citealp{frugiuele_current_2019}, \citeyear{frugiuele_muonic_2022}; \citealp{ohayon_precision_2022}; \citealp{fadeev_pseudovector_2022}).

Simple atoms and molecules may be used to test theories with extra dimensions which predict highly singular gravitational forces on short length scales; see, for example, \citet{arkanihamed_hierarchy_1998} and \citet{arkani-hamed_phenomenology_1999}. Limits on the size and number of extra dimensions have been obtained by \citet{salumbides_constraints_2015} and \citet{dzuba_effects_2022}.

A more complicated exotic atom is antiprotonic helium. 
It is an atomic system consisting of an electron and an antiproton orbiting around a helium nucleus \cite{yamazaki_antiprotonic_2002}. 
Such atoms are usually produced by slowing down antiprotons in a helium environment, leading to the replacement of one of the electrons in a He atom. 
Of the resulting atoms, a small fraction have antiprotons in orbits with quantum numbers $(n,l)=(38,37)$, giving metastable states with lifetimes on the order of microseconds. 
These states were used for measurements of various antiproton properties, such as its magnetic moment (\citealp{pask_antiproton_2009}; \citealp{nagahama_sixfold_2017}; \citealp{smorra_parts-per-billion_2017})
and mass \cite{hori_buffer-gas_2016}. 
As such, antiprotonic helium plays an important role in testing $CPT$ invariance \cite{hayano_antiprotonic_2007}. 

Antiprotonic-helium spectroscopy has yielded the first constraints on semileptonic spin-dependent interactions between matter and antimatter. \citet{ficek_constraints_2018} compared the experimental measurements of transition energies between two states within the hfs of the antiprotonic helium $(n,l)=(37,35)$ state \cite{pask_antiproton_2009} with quantum electrodynamics (QED) based predictions \cite{korobov_fine_2001}. 
This led to constraints on several types of exotic interactions between an electron and an antiproton, including higher-order velocity-dependent interactions. 

At present, there is good agreement between theory and experiment for the fine structure of normal and exotic atoms, which can be used to constrain exotic spin-dependent interactions at distances on the order of a Bohr radius (and greater). 



\subsubsection{Nuclear magnetic resonance    }\label{METH2.PM.NMR}

Similarly to the case of precision atomic and molecular spectroscopy, high-precision NMR measurements can also be sensitive to spin-dependent exotic forces. One can illustrate this with the measurement of spin-spin coupling in the isotopologues of the hydrogen dimer, e.g., HD \cite{ledbetter_constraints_2013}.

The conventional dipole-dipole interactions between nuclear spins are averaged in liquid and gas samples as a result of rapid (on the time scale given by the strength of the interaction) tumbling of molecules. However, spin-spin interactions are still present (\citealp{hahn_chemical_1951}; \citealp{gutowsky_nuclear_1953})
due to second-order hyperfine interaction \cite{ramsey_interactions_1952}: the interaction with a nuclear spin perturbs the electronic wavefunction of the molecule, and this perturbation is ``carried'' by the electrons to the second nucleus, which ``feels'' this perturbation, once again, due to the hyperfine interactions. This interaction of the form $J\,\bm{I}_1\cdot\bm{I}_2$ (where \textit{J} is the interaction strength) and is often referred to as scalar or \textit{J}-coupling. For simple molecules such as hydrogen, \textit{J}-couplings can be accurately calculated and a comparison between the experimental and theoretical result may reveal the presence of an exotic force.

Although the early analysis \cite{ledbetter_constraints_2013} of the existing data and calculations produced upper limits on the strength of exotic interactions, the experimental work of \citet{neronov_nmr_1975}; \citet{neronov_determination_2014}; \citet{ garbacz_indirect_2016}; \citet{neronov_determination_2018} actually claimed a significant disagreement with theory and a possible discovery of a pseodoscalar boson (an axion with a mass in the vicinity of 1\,keV). However, subsequent theoretical work \cite{puchalski_nuclear_2018} considered nonadiabatic corrections ignored in the earlier theory and brought the experiment and theory into agreement with each other.

\subsection{Atomic and molecular parity-violation experiments    }
\label{METH2.PV}

\subsubsection{Atomic parity violation} 
\label{METH2.PV.APV}
\citet{safronova_search_2018} reviewed spin-dependent interactions based on an electron-nucleus interaction resulting in atomic parity violation (APV). 
In the nonrelativistic limit, the parity-violating term in the Hamiltonian is proportional to $\v{s} \cdot \v {p}$, where $\v{s}$ is the electron or nuclear spin and $\v{p}$ is the electron momentum. 
The APV effect is due to the exchange of a $Z$ boson that was predicted by the Weinberg-Salam electroweak theory. Observation of APV by \citet{barkov_observation_1978} played an important role in establishing the SM.   
This is an excellent example showing that precision measurements conducted at low energies can provide an alternative way of exploring high-energy physics, apart from the fruitful but expensive accelerator-based experiments and astrophysical observations. 

Weak neutral currents in the standard model arise from the exchange of a heavy $Z$-bozon. 
On the atomic scale, it corresponds to a short-range interaction. 
This leads to a strong $Z$-dependence of parity violation (PV) effects. 
For neutral atoms, the dominant nuclear-spin-independent PV effects scale with the nuclear charge $Z$ faster than $\propto Z^3$ \cite{bouchiat_i_1974,bouchiat_parity_1975}. 
Because of this scaling, most APV experiments have been performed on heavy atoms. 
At present, APV has been observed in Bi \cite{barkov_observation_1978}, Pb \cite{meekhof_high-precision_1993}, Tl \cite{edwards_precise_1995,vetter_precise_1995}, Cs \cite{wood_measurement_1997}, and Yb \cite{tsigutkin_observation_2009}. 
All observed effects are in agreement with the predictions of the SM. 
The most accurate measurement \cite{wood_measurement_1997} and theoretical calculations 
(\citealp{dzuba_relativistic_1985}, \citeyear{dzuba_summation_1989}, \citeyear{dzuba_summation_1989-2}, \citeyear{dzuba_summation_1989-1}, \citeyear{dzuba_high-precision_2002}, \citeyear{dzuba_long-range_2022}; \citealp{blundell_high-accuracy_1990}; \citealp{flambaum_radiative_2005}; \citealp{porsev_precision_2009}; \citealp{sahoo_new_2021}, \citeyear{sahoo_reply_2022})
have been made for Cs, where agreement is at the sub 1\% level. 

APV experiments in Cs were pioneered by \citet{bouchiat_i_1974}, see also \citet{bouchiat_atomic_2012}. 
At the time of the later definitive work of \citet{wood_measurement_1997}, the sensitivity of this experiment to new physics, such as, for example, a second heavier $Z'$ boson, was comparable to that of collider-based searches at the Tevatron. 

Any exotic parity-violating electron-nuclear interactions would contribute to APV. 
So far, these studies, carried out mostly in heavy atoms, have not revealed significant deviations from the predictions of the SM. 
Therefore, APV experiments can be used to derive constraints on exotic PV electron-nuclear potentials, see, for example, \citet{davoudiasl_dark_2012}; \citet{roberts_limiting_2014,roberts_parity-violating_2014}; \citet{dzuba_probing_2017}. 
Often these experiments have been used to place limits on the mass of a second heavy $Z'$ boson assuming that it has the same coupling constants as the standard model $Z$ boson. 
An explicit limit on the coupling constants of a possible light $Z'$ boson was placed by \citet{antypas_isotopic_2019}. 
In this work, the PV effect was measured for four isotopes of ytterbium and the scaling of the PV amplitude with the number of neutrons was found to be in agreement with the predicted dependence of the weak nuclear charge $Q_\textrm{W}$ on the number of neutrons $N$.  
Earlier limits on a low-mass $Z'$ boson were obtained by \citet{dzuba_probing_2017}, which combined new atomic calculations with experimental data from previous APV experiments. 

There were also several attempts to observe PV effects in hydrogen
(\citealp{hinds_parity_1977}; \citealp{dunford_parity_1978}; \citealp{fortson_atomic_1984})
where the near degeneracy of opposite-parity levels strongly enhances parity-violating mixing. 
$Z$-boson exchange leads to an effectively contact electron-nuclear interaction. 
Atomic matrix elements of such an interaction scale approximately as $Z^3$ (somewhat more rapidly than $\propto Z^3$ due to relativistic effects), where $Z$ is the nuclear charge. 
Because of this, PV effects in atomic hydrogen are much smaller than those in heavy atoms and the experimental sensitivity $30-40$ years ago was insufficient to observe such effects. 
For light bosons, the nuclear-spin-independent PV effects scale as the first power of $Z$ because the analogue of the weak charge of the nucleus linearly grows with $Z$ and there is practically no additional $Z$ dependence from short-distance effects within the atom. 
Due to this weaker scaling with $Z$, hydrogen can be appreciably sensitive to PV interactions with a range comparable to the Bohr radius or larger (i.e., boson masses below $\sim 1$\,keV). 
Recently, interest in APV experiments using hydrogen has revived (\citealp{rasor_laser-based_2020}; \citealp{li_feasibility_2023}).
Taking into account significantly increased experimental accuracy, these experiments may be competitive in studying particular types of exotic PV interactions. 

\begin{figure}
\centering
\includegraphics[width=0.3\textwidth]{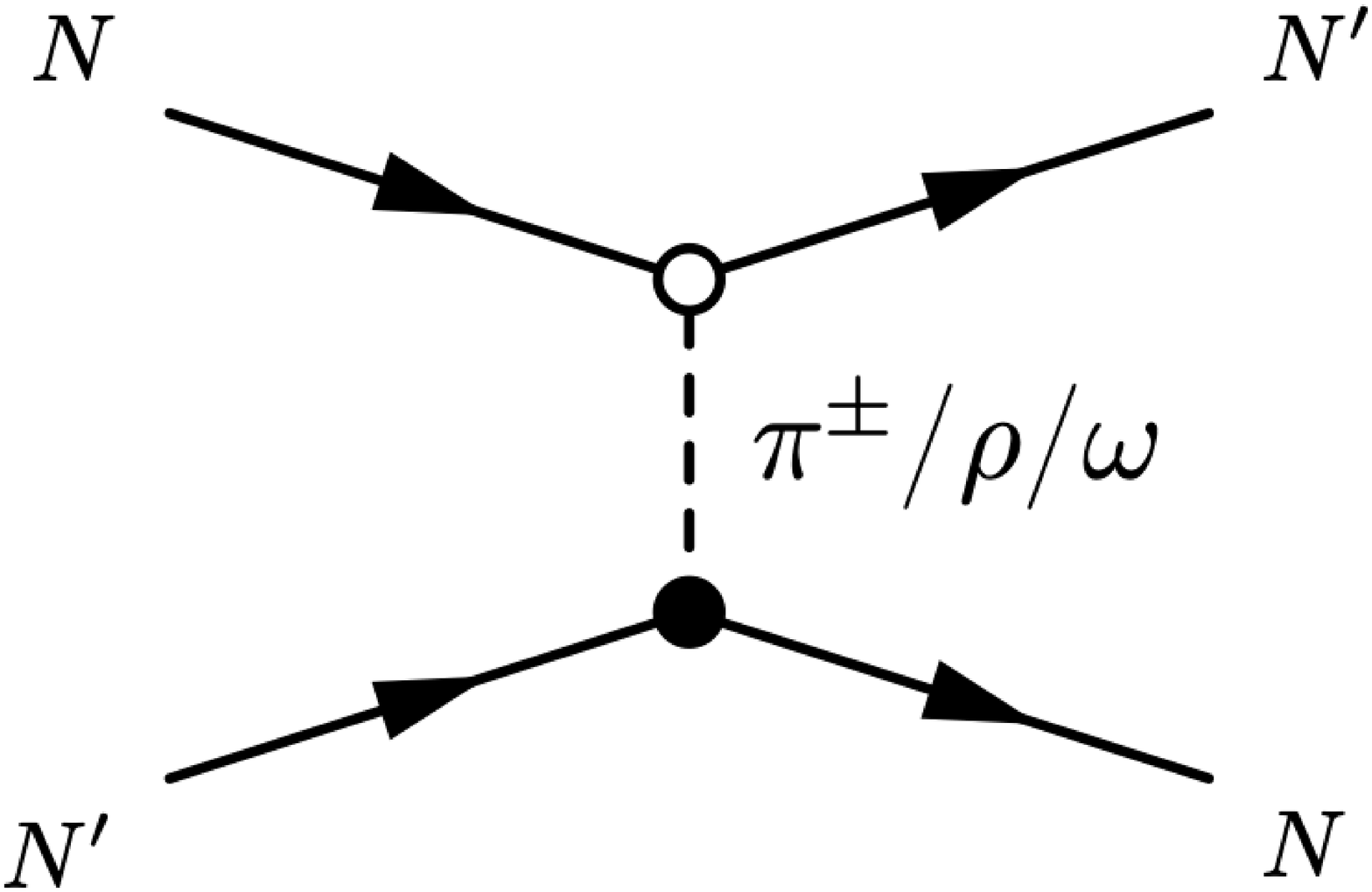}

\includegraphics[width=0.3\textwidth]{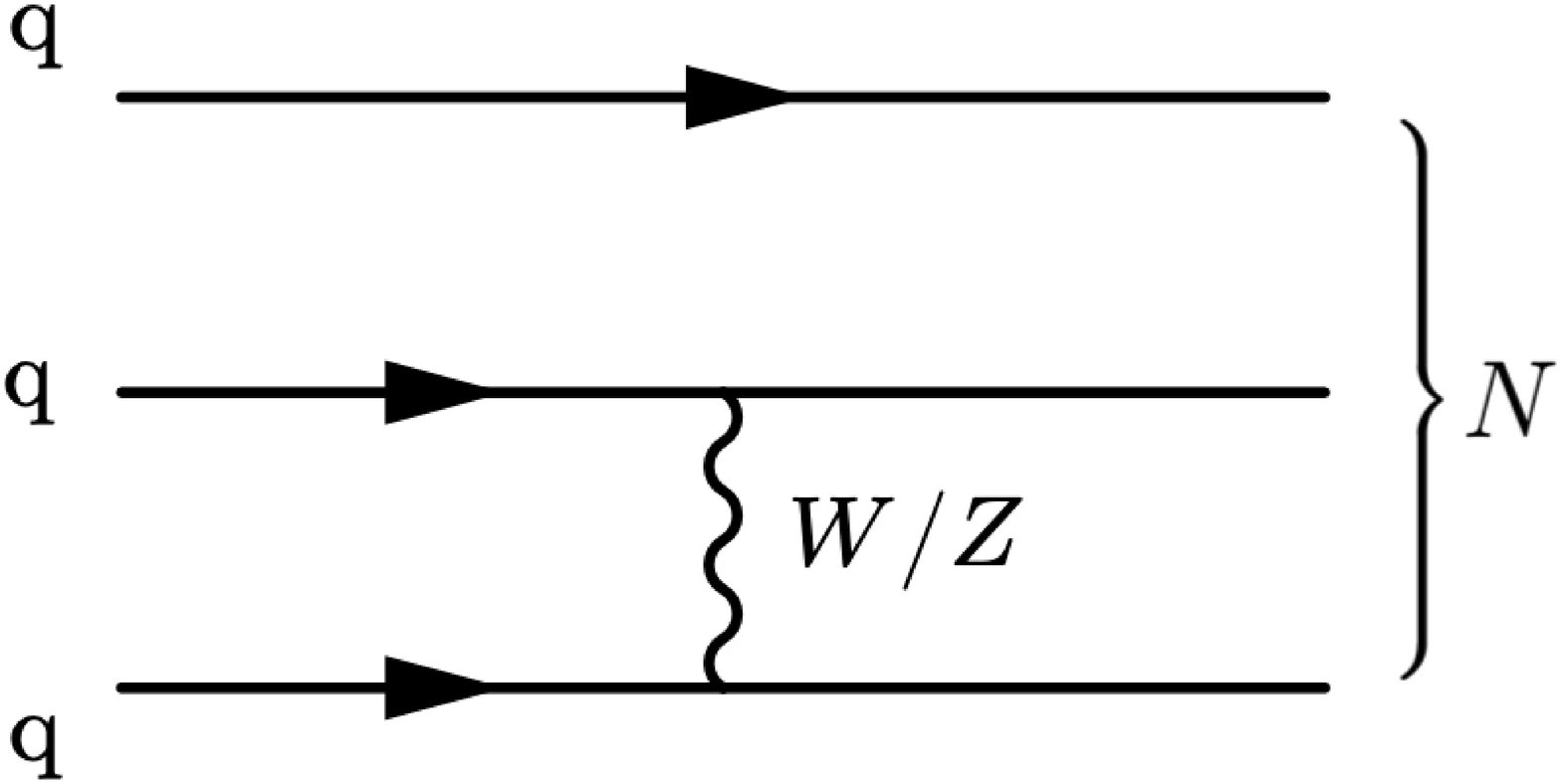}
\caption{\textbf{Top:} $P$-violating and $T$-conserving interactions between nucleons ($N$) are mediated by the exchange of mesons, such as $\pi^\pm$, $\rho$ and $\omega$ mesons. 
In the diagram, one vertex (denoted by a white circle) is parity violating and is induced by the exchange of $Z$ or $W$ bosons between constituent quarks, while the other vertex (denoted by a black circle) is parity conserving. 
\textbf{Bottom:} Representative quark-level process that induces the parity-violating meson-nucleon vertices via the exchange of a $Z$ or $W$ boson between two quarks within a single nucleon. 
}\label{fig:PVandPC}
\end{figure}

The work of \citet{wood_measurement_1997} on Cs APV also reported the discovery of the nuclear anapole moment \cite{zeldovich_electromagnetic_1958} which had been predicted to dominate nuclear-spin-dependent APV in heavy atoms and molecules \cite{flambaum_p-odd_1980,flambaum_nuclear_1984}.
The anapole moment is produced by parity-violating nuclear forces (see Fig.\,\ref{fig:PVandPC}) and generates APV in atoms via magnetic interaction with atomic electrons. 
Comparison of predicted and observed values of the Cs nuclear anapole moment allowed to obtain limits on exotic nuclear interactions (\citealp{roberts_limiting_2014,roberts_parity-violating_2014}; \citealp{dzuba_probing_2017}).
Note that the latter two references \cite{roberts_limiting_2014,roberts_parity-violating_2014}
discuss nuclear interactions with a background cosmic field that may be associated with the ``dark sector''. Such studies are also carried out with chiral molecules, see Sec.\,\ref{METH2.PV.MPV}.

The short-range electron-electron PV interactions in atoms have a weaker $Z$-scaling and are suppressed by the factor $\sim 1/(4Z)$ compared to the contribution of the electron-nucleus PV interaction because the electron-electron interaction does not profit  from a $Z$-fold increase of the valence-electron density inside the nucleus
\cite{sushkov_nonconservation_1978}.  
This is why all PV effects observed in atoms so far are caused by the PV electron-nuclear interaction or intranuclear interactions as in the case of the anapole. 
Exotic PV interactions do not necessarily have a strong scaling with $Z$ and so there is not necessarily significant suppression of $e$-$e$ interactions. 
Therefore, both $e$-$N$ and $e$-$e$ PV interactions can be studied in light atoms (and molecules) with close levels of opposite parity. 
For example, parity violation measurements may be used to search for axions-induced forces
(axion-induced effects in atoms, molecules and nuclei: parity non-conservation, anapole moments, electric dipole moments, and spin-gravity and spin-axion momentum couplings \cite{stadnik_axion-induced_2014}.
We note that parity violation has also been observed in $e$-$e$ scattering \cite{slac_e158_collaboration_observation_2004,moller_collaboration_moller_2014}.

Finally, we note that radiative corrections can lead to long-range parity-violating interactions (many orders of magnitude larger range than that for $Z$-boson exchange)  as discussed by \citet{dzuba_long-range_2022}.

\subsubsection{Molecular parity violation} \label{METH2.PV.MPV}

Prediction of the parity-violating weak neutral currents within the electroweak theory led to the emergence of a whole new field of atomic physics. However, the first dedicated experiment to search for the parity non-conservation in atomic-molecular physics was performed much earlier: \citet{bradley_absence_1962} found that mixing of levels of opposite parity in the  O$_2$ molecule  was less than $3 \times 10^{-8}$ [see also a review of early PV calculations and measurements by \citet{fortson_atomic_1984}]. 

Molecules, particularly diatomic ones like BaF and RaF radicals,
are considered promising systems for studying PNC effects 
\cite{sushkov_parity_1978-1,kozlov_topical_1995}.
These molecules have closely spaced energy levels of opposite parity, which can be mixed by electroweak interactions. Molecular PNC effects have not yet been directly observed, even though various experimental approaches have been proposed, such as Stark interference and optical rotation techniques. 

A variety of new approaches to molecular parity-violation experiments are being developed \cite{safronova_search_2018}. For example, \citet{demille_using_2008} introduced a Stark interference technique for measuring nuclear spin-dependent parity violation in diatomic molecules, applicable to a broad spectrum of atomic systems.
Most recently, \citet{blanchard_using_2023} propose to measure the effect induced by the $P$-violating, $T$-conserving interaction of the nuclear anapole moment with electrons in a new type of experiment aimed at probing the nuclear spin-dependent PNC effect using a non-chiral molecule like TlF. 



In chiral molecules, one can look for the difference in the transition frequencies of the enantiomers caused by the parity-violating electron-nuclear or electron-electron interactions.\footnote{For systems with well-defined parity, there are no energy shifts that are first order in a parity-violating interaction that mixes opposite-parity states. However, a molecule with a given chirality is in a superposition of opposite-parity states and first-order energy shifts are indeed possible. Observation of parity violation in molecules, chiral or otherwise, remains an unmet challenge to this day, see, for example, \citet{quack_parity_2005,schwerdtfeger_search_2010,safronova_search_2018,berger_parity_2019}.}
In the SM, such effects can be caused by $Z$-boson exchange and are suppressed because of the short range of the weak interaction. 
In the electroweak theory, the couplings of the photon and $Z$ boson are of the same order, but an effective dimensional coupling constant for the $Z$ boson, namely the Fermi constant, accounts for the contact nature of the interaction and is proportional to $1/M_Z^2$ \cite{khriplovich_parity_1991}. 
This suppression rapidly decreases with the nuclear charge $Z$. 
For example, the spin-independent energy shift scales as $Z^5$ \cite{gajzago_weak_1974,zeldovich_energy_1977}, or, more precisely, as $Z_1^3Z_2^2$, where $Z_1$ and $Z_2$ are the nuclear charges of the two heaviest nuclei in the molecule \cite{hegstrom_calculation_1980}. 
Note that this scaling requires two heavy atoms with similar nuclear charge number if main-group compounds are considered [the single-center theorem, see \citet{hegstrom_calculation_1980}]. 
On the other hand, the exotic long-range PNC potential terms $V_{12\pm13}$ decrease with the distance as $1/r$ for sufficiently small boson mediator masses. 
Their atomic amplitudes do not have such 
strong $Z$-scaling as the respective amplitudes of the conventional contact PNC interaction \cite{dzuba_probing_2017} (see also the discussion of atomic parity violation in Sec.\,\ref{METH2.PV.APV}). 
Similarly, we can expect weaker $Z$-scaling in chiral molecules. 
For example, the interaction of a chiral molecule with a $P$-odd cosmic field, discussed below, has two terms, which scale as $Z^4 \alpha^5$ and $Z^2 \alpha^3$, respectively \cite{gaul_parity-nonconserving_2020}.
In the case of light chiral molecules, the conventional $Z$-boson effects are too small 
to be observed in current experiments. 
Therefore, observation of an enantiomeric energy difference in light chiral molecules 
near the present level of sensitivity would be an indication of an exotic long-range force. 

The energies and transition frequencies of chiral molecules can become time-dependent in the presence of a $P$-odd cosmic field  
(\citealp{roberts_limiting_2014}, \citeyear{roberts_parity_2015}; \citealp{gaul_chiral_2020}, \citeyear{gaul_parity-nonconserving_2020}).
If such a field oscillates at the frequency $\omega=Mc^2/\hbar$, where $M$ is the mass of the constituent boson of the cosmic field, 
the enantiomeric energy splitting also acquires an oscillating component. 
Such effects have not specifically been sought experimentally. 
Nevertheless, some constraints on such $P$-odd cosmic fields have been derived \cite{gaul_chiral_2020} from the published results of a spectroscopic experiment on the chiral molecule CHBrClF \cite{daussy_limit_1999}. Similar work with atoms \cite{roberts_parity-violating_2014} is discussed in Sec.\,\ref{METH2.PV.APV}.

A theoretical discussion of $P$- and $CP$-violating effects in chiral molecules and projections for future experiment were recently given by \citet{baruch_constraining_2024-1}.   

We conclude that molecular PNC experiments present as yet unexplored opportunities to search for exotic interactions.


\subsection{EDM experiments }\label{METH2.EDM}


In the classical picture, an EDM $\boldsymbol{d}$ is a measure of the separation of positive and negative electric charges in a system. \citet{purcell_possibility_1950} proposed the idea that elementary particles, such as electrons, might possess a permanent EDM in addition to their magnetic dipole moment. 
They suggested searching for EDM of elementary particles as a test of $P$ violation. 
Later, it was realized that EDMs also imply time-reversal invariance violation \cite{landau_conservation_1957}. 
Thus, the existence of an EDM of an elementary particle violates both the $P$ and $T$ symmetries. 
A variety of $P,T$-violating processes within and beyond the SM can generate EDMs of elementary and composite particles, including leptons, nucleons, nuclei, atoms and molecules \cite{ginges_violations_2004,pospelov_electric_2005,engel_electric_2013}. 
Within the SM, EDMs of elementary particles are produced by high-order radiative corrections and are extremely small compared to foreseeable experimental sensitivities. 
Therefore, EDM measurements at present provide a clean probe of ``new physics''. 


Since an EDM is a vector quantity, it must be directed along the total angular momentum of a system, in accordance with the Wigner-Eckart theorem. 
The EDM of a composite system (nucleon, nucleus, atom) may be due to the $P,T$-violating forces between the constituent particles, which must be spin-dependent. 
Such an interaction is proportional to $\v{s} \cdot \v{v}$, where $\v {s}$ is the spin of the particle and $\v {v}$ is a polar vector. 
In the case of a short-range $P,T$-violating interaction (mediated by a heavy boson), the vector $\v{v}$ may be directed along the gradient of the nuclear or electron density, see, for example, \citet{ginges_violations_2004}. 
A similar interaction between nucleons produces a nuclear EDM.
 However, the electron shielding effect needs to be taken into consideration \cite{schiff_measurability_1963}. It turns out that the electric field of a nuclear EDM is completely screened by atomic electrons outside the nucleus. Inside a nucleus of finite size, this field survives and is known as the Schiff-moment field \cite{flambaum_nuclear_2002}.
This field polarizes the atom and creates an atomic EDM. 
Such EDMs have been experimentally constrained for Hg \cite{graner_reduced_2016}, Xe (\citealp{rosenberry_atomic_2001}; \citealp{allmendinger_measurement_2019}; \citealp{sachdeva_new_2019}),
Ra \cite{bishof_improved_2016} and Yb \cite{zheng_measurement_2022} atoms and the TlF molecule (\citealp{hinds_electric_1980}; \citealp{cho_search_1991}; \citealp{dzuba_electric_2002}).
These experiments give us information about $P,T$-odd interactions in the hadronic sector including limits on the QCD vacuum angle parameter $\theta$, the proton and neutron EDMs, quark EDMs, and quark chromo-EDMs (the quark chromo-EDM interacts with the gluon fields inside a nucleon, while the ordinary EDM interacts with the electric field). 
They also give us limits on the electron-nucleus $P,T$-odd interactions proportional to nuclear spin. 

Such an interaction between electrons and the nucleus proportional to the electron spin \cite{stadnik_improved_2018} or nuclear spin \cite{dzuba_new_2018} may be produced by axion exchange. 
Therefore, measurements of atomic EDMs provide us with information about axions (see Sec.\,\ref{subsec_g_Pg_S}). 

An atomic EDM $d_A$ may also be produced by the interaction of an electron EDM $d_e$ with the atomic electric field. 
This atomic EDM increases rapidly with nuclear charge $Z$, at a rate faster than $Z^3$ (\citealp{sandars_enhancement_1966}; \citealp{flambaum_enhancement_1976}).
The electron EDM enhancement factor $K=d_A/d_e$ in atoms with a valence electron with angular momentum $J$ may be estimated as  \cite{flambaum_enhancement_1976,ginges_violations_2004}:
\begin{equation} 
\label{da}
|K|=|d_A/d_e| \sim \frac{10 Z^3 \alpha^2 R(Z \alpha)}{J(J+1/2)(J+1)^2}\,,
\end{equation}
where the relativistic factor $R(Z \alpha)$ increases with the nuclear charge $Z$ and tends to unity as $Z\alpha \to 0$, with $\alpha$ being the electromagnetic fine-structure constant. 
As a result, the atomic EDM of a heavy atom may exceed the bare electron EDM by up to three orders of magnitude; e.g., in Cs, $K \approx +130$; in Tl, $K \approx -500$; in Fr, $K \approx +1000$. 
Note that for an $s_{1/2}$ electron, $K>0$, while for a $p_{1/2}$ electron, $K<0$, since electron spin in $p_{1/2}$ state is directed opposite to the total electron angular momentum $J$.  

In molecules, this enhancement can be several orders of magnitude larger due to mixing of closely spaced rotational levels of opposite parity by the interaction of $d_e$ with the intrinsic molecular electric field \cite{sushkov_parity_1978-1}.  
Note, however, that this picture of an enhanced molecular EDM in a rotational molecular state corresponds to the limit of a weak external electric field. 
In actual experiments, the electric field $\v{E}$ is sufficiently strong to polarize the molecule along the electric field, since the interaction $\v{D} \cdot \v{E}$ of the intrinsic dipole moment $\v{D}$ of a polar molecule with the external electric field $\v{E}$ is comparable to or even exceeds the small interval between opposite-parity rotational molecular levels. 
In this case, the electron EDM $d_e$ interacts with the so-called intrinsic effective field of a polar molecule, $\v{\epsilon}$, which is directed along the external electric field $\v{E}$ but may exceed it in size by several orders of magnitude [a similar molecular enhancement has been explored in measurements of the proton EDM in TlF molecular experiments, see, for example, \citet{hinds_electric_1980}].\footnote{Note that the effective field $\epsilon$ should not be confused with the real intrinsic electric field inside a polar molecule. For example, the effective field $\epsilon$ and atomic EDM $d_A$ vanish in the nonrelativistic limit and the electron EDM becomes unobservable, in accordance with the Schiff theorem  \cite{schiff_measurability_1963}.}

Indeed, $\epsilon$ may reach  $\sim 10^{11}$\,V/cm in heavy polar molecules such as ThO \cite{skripnikov_communication_2013,fleig_electron_2014,skripnikov_combined_2016}, 
since $\epsilon$ is proportional to $d_A$ in Eq.\,(\ref{da}) and rapidly increases with $Z$. 
The energy shift is $\delta= -d_e \epsilon \sim d_A  D/R^3$, where $R$ is the distance between the atoms in a molecule.
As a result, the most stringent limits on the electron EDM and short-range electron-spin-dependent $P,T$-violating interactions between electrons and nucleons were obtained in experiments with molecular ThO \cite{acme_collaboration_improved_2018} and HfF$^+$ \cite{roussy_improved_2023}.

The basic idea behind the exploration of the $P,T$-violating EDMs of atoms or molecules is to search for a shift in the energy of an atomic or molecular state that scales as an odd power of the applied electric field; linearly in the low-field limit.
This is in contrast with the $P,T$-even energy shifts, which do not change sign upon reversal of the electric field  $\boldsymbol{E}$. 

If one simultaneously applies both electric and magnetic fields in the experiment, in the weak-field limit, the energy shift experienced by the state of interest takes the form $\delta \varepsilon = - \boldsymbol{d} \cdot \boldsymbol{E} - \boldsymbol{\mu} \cdot \boldsymbol{B}$, where $\boldsymbol{\mu}$ is the magnetic moment of the state and $\boldsymbol{B}$ is the magnetic field experienced by the state. 
By first applying the electric and magnetic fields parallel to one another and then reversing the relative direction of the fields, one can elucidate the size of the EDM or place a constraint on it. 
The observed effect in this case manifestly breaks $P$ and $T$ invariance, since the correlation $\boldsymbol{E} \cdot \boldsymbol{B}$ is a $P,T$-odd quantity. 
This traditional approach has been adopted in numerous EDM experiments, including experiments using ultracold neutrons \cite{baker_improved_2006,abel_measurement_2020}, 
as well as atomic xenon \cite{allmendinger_measurement_2019,sachdeva_new_2019}, 
mercury \cite{graner_reduced_2016} 
and thallium \cite{regan_new_2002}. 
A more recent approach employing parity-doublet molecular states that function as an internal comagnetometer has been utilised in experiments using 
PbO \cite{eckel_search_2013},
ThO \cite{acme_collaboration_order_2014,acme_collaboration_improved_2018} and HfF$^+$ \cite{roussy_improved_2023}. 
The advantage of this alternative approach is that the direction of the internal electric field experienced by the molecular electrons can be reversed without necessarily reversing the direction of the applied electric field, eliminating a major source of systematic uncertainty. 

\citet{alarcon_electric_2022} provide a recent overview of experimental efforts to measure the EDMs of fundamental particles, along with a discussion of the importance of EDM measurements in probing physics beyond the SM. 

To conclude this section, we discuss a point emphasized by \citet{chupp_electric_2015} [see also \citet{fleig_model-independent_2018} and \citet{degenkolb_global_2024}]: it is possible in principle that more than one source of ``new physics'' contributes to an atomic or molecular EDM. 
For instance, in some models, an intrinsic electron EDM appears together with $P,T$-odd electron-nucleon forces, both of which contribute to the EDM of an atom or a molecule. 
To account for such a possibility, one needs to construct two- or higher-dimensional parameter-exclusion regions, which generally enlarges the allowed ranges for individual parameters. 
A combination of different experiments with different relative sensitivities to individual EDM sources can be used to disentangle the individual contributions. 
While such analyses will become essentially important once an EDM is discovered, it is our opinion that the minimalistic approach where limits on individual new-physics sources are derived on the assumption of a unique source are, in most cases, adequate until the time when new physics is discovered. 
We believe, this approach should be favored due to its simplicity and a straightforward way to compare the relative sensitivity of different experiments. 
The minimalistic approach is additionally justified since, within the same model, different mechanisms often give contributions of different order of magnitude. The electron-nucleus $T,P$-violating interaction (a tree-level effect) normally dominates over the electron EDM (loop-level).

\subsection{Particle \texorpdfstring{$g$}{g}-factors}
\label{METH2.g-factor}


The $g$-factor of a particle relates its magnetic dipole moment $\v{\mu}$ to its angular momentum $\v{J}$ according to: 
\begin{equation}
    \label{g-factor_definition}
    \v{\mu} = g \frac{e}{2m} \v{J} \, , 
\end{equation}
where $e$ is the elementary electric charge (equal to the magnitude of the charge of the electron) and $m$ is the particle or reference mass. 
The $g$-factor can be either positive or negative depending on the (anti)particle species; e.g., $g_e \approx -2$ for the electron, while $g_{e^+} \approx +2$ for the positron. 
Note that in atomic physics, the electron spin $g$-factor $g_s$ is often defined as the absolute value of $g_e$; i.e., $g_s = |g_e| = -g_e$. 
It is common to parameterize the magnetic moment of the electron in terms of the Bohr magneton $\mu_\textrm{B} = e \hbar / (2 m_e)$ and to parameterize the magnetic moments of nucleons in terms of the nuclear magneton $\mu_\textrm{N} = e \hbar / (2 m_p)$.

In the case of a body of mass $m$ and electric charge $q = e$ orbiting on a classical trajectory, 
the $g$-factor associated with the orbital motion of the body is $g = 1$. 
This $g$-factor value remains unchanged in the quantum theory of elementary spin-$1/2$ particles, including leptons and quarks, with respect to their orbital angular momentum. 
In the case of the intrinsic spin angular momentum of an elementary spin-$1/2$ particle, however, the $g$-factor value is $g = 2$ in the Dirac theory. 

Small deviations from the Dirac value of $g = 2$ arise due to quantum radiative (loop) corrections, parameterized by the anomalous magnetic moment $a = (g-2)/2$. 
The leading one-loop correction to the electron $g$-factor in QED was computed by Schwinger and found to be $a = \alpha / (2\pi)$ \cite{schwinger_quantum-electrodynamics_1948}. 
Corrections to the $g$-factor of the electron have been computed to the five-loop level in QED, as well as to lower orders in weak and hadronic processes \cite{aoyama_theory_2019}, 
and provide the basis for a stringent test of the SM via comparison with experiment \cite{fan_measurement_2023}. 
In the case of the muon, there is presently some disagreement between experiment and theoretical predictions for the $g$-factor of the muon, potentially hinting at possible new physics beyond the SM \cite{keshavarzi_muon_2022}.

\begin{figure}
\centering
\includegraphics[width=0.27\textwidth]{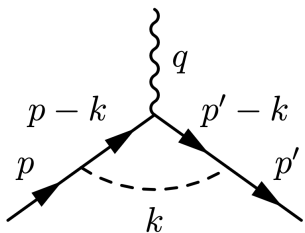}
\caption{Modification of a lepton's electromagnetic vertex by exchange of a `new' boson (dashed line), which can give rise to contributions to the magnetic and/or electric dipole moment(s) of the particle.
The wavy line represents a photon and the solid lines represent the lepton. 
The labels indicate the four-momenta of the particles. 
Figure modified from \citet{yan_constraining_2019}.
}
\label{yan2019-fig}
\end{figure}

The existence of new bosons with couplings to the SM according to Eqs.\,\eqref{Lagrangian_scalar} -- \eqref{Lagrangian_tensor} would modify the $g$-factors of fermions via quantum loop-level processes involving these virtual bosons (see Fig.\,\ref{yan2019-fig} for an example of a one-loop-level process). 
These loop processes are distinct from the 
tree-level forces that are the focus of the present review. 
By studying these loop-level processes, one can place constraints on interactions of exotic bosons. 
The $g$-factors of leptons (such as the electron and muon) provide sensitivity to much feebler couplings of new bosons than the $g$-factors of nucleons and nuclei, principally due to the elementary (i.e., non-composite)  
and non-hadronic nature of the former, which results in cleaner and more precise standard-model predictions. 
For discussions and calculations of the contributions of new bosons to lepton $g$-factors, see, e.g., \citet{karshenboim_constraints_2014}; \citet{chen_muon_2017}; \citet{yan_constraining_2019}.

The $g$-factor of a particle can be experimentally determined from measurements of the Larmor spin-precession frequency $\nu_\textrm{s}$ and cyclotron frequency $\nu_\textrm{c}$ of the particle, according to the relation: 
\begin{equation}
    \label{g-factor_measurement}
    \frac{g}{2} = \frac{\nu_\textrm{s}}{\nu_\textrm{c}} \, . 
\end{equation}
In the case of stable particles, measurements of $g$-factors can be performed using cold particles in Penning traps. 
Penning-trap experiments have been used to measure the $g$-factors of the electron (\citealp{hanneke_new_2008}; \citealp{fan_measurement_2023}),
proton \cite{schneider_double-trap_2017}
and antiproton \cite{smorra_parts-per-billion_2017}. 
In the case of antiprotons, measurements should be performed in ultra-high-vacuum conditions in order to minimise the probability of antiproton-proton annihilation via collision of the antiproton with residual gas in the experimental chamber. 
In the case of intrinsically unstable particles with short lifetimes, measurements of $g$-factors can be performed using relativistic particles in storage rings in order to increase their lifetime in the laboratory frame of reference. 
Storage-ring experiments have been used to measure the $g$-factors of the muon and antimuon \cite{muon_g-2_collaboration_final_2006,muon_g-2_collaboration_measurement_2021,muon_g2_collaboration_measurement_2023}. 

In Fig.\,\ref{V1-fig}, we present the constraints on exotic interactions extracted from the $g-2$ experiment-to-theory comparison, as obtained by \citet{delaunay_probing_2017} and \citet{frugiuele_current_2019}.

\subsection{Isotope shifts    }\label{METH2.IS}

New interactions conserving $P$ and $T$ symmetries can be searched for using the spectra of simple atomic systems (hydrogen, antiprotonic helium, muonium, positronium), where the accuracy of both theory and experiment is high, and so any unaccounted contribution would be noticeable, see Sec.\,\ref{METH2.PM.AHS}, \ref{METH2.PM.EA}, \ref{Sec:limits_main} and \citet{potvliege_deuterium_2023}. 

In heavy atoms, the best theoretical accuracy is about 0.1\% (which is much worse than in simple atomic systems), and so the comparison of experiment and theory generally leads to mild limits. 
An exception, however, involves isotope shifts. 
Consider, for example, two atomic transitions. 
If we plot an appropriately normalized frequency of one transition versus another transition, then to a high accuracy, the resulting plot will be a linear function. 
Such a graph is called a King plot \cite{king_isotope_1984}. 
A new interaction mediated, for instance, by a scalar particle can violate the linearity of a King plot. 
Thus, King plots may be used as a tool to search for new Yukawa-type interactions between electrons and the nucleus \cite{berengut_probing_2018}. 
The main challenge from the point of view of theory here is to accurately calculate the small corrections that produce nonlinearities of the King plot within the SM, see, for example, \citet{flambaum_isotope_2018}. 


A number of recent experiments have searched for King-plot nonlinearities in chains of stable zero-nuclear-spin isotopes, see, for example, \citet{rehbehn_sensitivity_2021}; \citet{figueroa_precision_2022,hur_evidence_2022,ono_observation_2022}. 

The analyses of King-plot nonlinearities have so far been limited to zero-nuclear-spin isotopes to avoid the complications of including the first- and second-order effects of hfs that would need to be evaluated to high accuracy in order to extract the isotope shifts for nonzero-nuclear-spin isotopes. 
Ongoing theoretical work \cite{ginges_notitle_2023} holds promise for theory to reach the required level of accuracy. 
This raises a question: could precision isotope-shift comparisons involving nonzero-spin isotopes be useful to search for exotic spin-dependent interactions? 
Corrections to King plots, after averaging over first-order hyperfine and exotic spin-dependent interactions, contain smaller second-order terms. 
It remains to be seen if the high accuracy in isotope-shift measurements could eventually compensate this smallness. 
\subsection{Dark matter experiments    }\label{METH2.DM}

Throughout this review we discuss experiments where a local source and a spin-based sensor are used to search for exotic spin-dependent interactions mediated by ``new'' bosons (Sec.\,\ref{Sec:EXP.METH.1}). 
Another possibility is that these new bosons could have been abundantly generated in the early universe and therefore constitute all or part of the dark matter responsible for gravitational effects observed throughout the universe as noted in Sec.\,\ref{Sec:Intro-motiv-DM}. 
In such cases, the presence of these new bosons could be directly detected through their spin-dependent interactions with electrons or nuclei; in other words, no ``source'' (other than the Big Bang that produced the relic background of dark matter) is required for the experiment.\footnote{In contrast to the virtual (off-shell) bosons mediating interactions between a source and a sensor, relic bosons of cosmological origin are ``on-shell.''} 
Detectors designed to measure the interaction of particles comprising the dark matter halo within which the Milky Way galaxy is presumably embedded are known as \emph{haloscopes}. Haloscope experiments using spin-based sensors are reviewed by \citet{graham_experimental_2015,graham_spin_2018}, \citet{safronova_search_2018}, \citet{jackson_kimball_search_2023}, \citet{flambaum_searches_2023}, and \citet{jackson_kimball_probing_2023}.
Here we only briefly outline key features of such searches and their relationship to the source/sensor-type experiments that are the focus of this review. For further information, see the aforementioned reviews and references therein.

At the outset, we emphasize that most reported constraints on exotic bosons and their interactions derived from haloscope experiments are based on the additional assumption that the exotic bosons make up the majority of dark matter.
The key point is that haloscope experiments are actually sensitive to a combination of the interaction strength and (the square root of) the fraction of the dark matter density composed of the bosonic field probed.
Thus reported constraints on the coupling constants from haloscope experiments are weakened if the exotic bosons do not in fact make up a significant fraction of dark matter.
Interpretation of limits from source/sensor-type experiments (Sec.\,\ref{Sec:EXP.METH.1}) explicitly do \emph{not} require this assumption. 
In this sense, constraints on exotic spin-dependent interactions from source/sensor-type experiments are independent of cosmological models and assumptions and, consequently, more robust. 

A crucial characteristic determining the physical manifestation of bosonic dark matter is the relationship between the boson mass $M$ and the local dark matter density $\rho_\textrm{DM}$. 
At the position of our solar system within the Milky Way galaxy, the estimated average dark matter density is $\rho_\textrm{DM} \approx 0.3-0.6~\textrm{GeV}/c^2$ 
(\citealp{strigari_reconstructing_2009}; \citealp{pato_systematic_2010}; \citealp{catena_novel_2010}; \citealp{salas_dark_2021}),
corresponding to an average mass density
of about one hydrogen atom per a few cubic centimeters. 
Note that if dark matter is made up mostly or entirely of a species of particle with mass $M \lesssim 10$\,eV/$c^2$, then in order to agree with observations it must be a boson. 
This is because for the inferred $\rho_\textrm{DM}$, the Fermi velocity of particles with $M \lesssim 10$\,eV/$c^2$ exceeds the galactic escape velocity. 
If such new bosons have negligible self-interactions and do not form composite structures, they can be treated as independent particles through which the Earth moves, leading to daily and annual modulations of their flux, as discussed, for example, by \citet{freese_colloquium_2013} and \citet{gramolin_spectral_2022}. 
Importantly, such ultralight bosons will on average have extremely high mode occupation number in the local vicinity of Earth. 
In this case, the phenomenology of ultralight bosonic dark matter (UBDM) is well-described by a field oscillating at approximately the Compton frequency, $\omega_c = Mc^2/\hbar$. 
More massive dark matter candidates [such as weakly interacting massive particles, WIMPs --- see reviews by \citet{aleksandrov_search_2021} and \citet{feng_dark_2010}] behave as incoherent individual particles and experiments commonly employ particle detectors that measure, e.g., energy deposition or ionization signals to find them. 
In contrast, UBDM acts as a coherent entity on the scale of the sensor (and potentially also on the scale of terrestrial networks of sensors). 
Consequently, precision measurement techniques, similar to those used in searches for macroscopic-range exotic interactions discussed in this review, can be employed to search for UBDM. 
For up-to-date constraints on UBDM from a wide variety of experiments from all over the world, see \citet{ohare_cajohareaxionlimits_2020-1}, 
where references describing the experiments and associated limits can also be found. See also the reviews by \citet{sikivie_invisible_2021}, \citet{adams_axion_2022}, and \citet{antypas_new_2022}.

Dark matter composed of axions or ALPs (Secs.\,\ref{Sec:Intro-motiv-spin-0} and \ref{Sec:Intro-motiv-quantum-gravity}), spin-0 particles that can possess scalar and pseudoscalar couplings to fermions \cite{moody_new_1984}, provide an illustrative example of the close connection between searches for exotic spin-dependent interactions and dark matter searches using spin-based sensors as haloscopes.
An axion (or ALP) field $a(\bs{r},t)$ is described, approximately, by
\begin{align}
    a(\bs{r},t) = a_0 \cos\left( \bs{k}\cdot\bs{r} - \omega_c t + \phi_0 \right) \, , 
    \label{eq:axion-field}
\end{align}
where $\bs{k} \approx M \bs{v} /\hbar$ represents the nonrelativistic axion wave vector (with $\bs{v}$ being the velocity of the field relative to the sensor), $\phi_0$ is a random phase offset, and $a_0$ is the average oscillation amplitude of the field. 
The average oscillation amplitude of the field can be estimated by assuming that the average energy of the axion field corresponds to the local dark matter energy density $\rho_\textrm{DM}$ \cite{jackson_kimball_search_2023}:
\begin{align}
    a_0^2 \approx \frac{2\hbar^2}{c^2}\frac{\rho_\textrm{DM}}{M^2} \, . 
    \label{eq:axion-field-amplitude}
\end{align}
The axion field exhibits a finite coherence time due to the random kinetic energy of the constituent axions. 
This leads to a broadening of the line shape to $\delta \omega/\omega \sim v^2/c^2 \sim 10^{-6}$ as discussed, for example, by \citet{krauss_calculations_1985} and \citet{gramolin_spectral_2022}.\footnote{The line is not only broadened, but also develops an asymmetric shape, as discussed, e.g., by \citet{derevianko_detecting_2018} and \citet{gramolin_spectral_2022}.} 
Additionally, there are stochastic fluctuations in the amplitude, phase, and $\bs{k}$ of the axion field, as pointed out by \citet{centers_stochastic_2021};  \citet{masia-roig_intensity_2023} and \citet{flambaum_fluctuations_2023}.
In the latter work, limits on dark matter were obtained from measured variances of energy-shift fluctuations (statistical errors). Axion signal may be separated using measurements of higher moments and correlators of energy-shift fluctuations.

The canonical QCD axion discussed in Sec.\,\ref{Sec:Intro-motiv-spin-0} naturally couples to the gluon field and produces an oscillating EDM along the nuclear spin orientation \cite{graham_new_2013,stadnik_axion-induced_2014}, 
leading to an interaction described by the Hamiltonian:
\begin{align}
\mc{H}_\textrm{EDM} =  g_d a(\bs{r},t) {\bs{E}}^\star \cdot \bs{\sigma}_N\,,
\label{eq:oscillating-EDM}
\end{align}
where $g_d$ is an associated coupling parameter\footnote{It should be noted that there can be multiple contributions to the coupling parameter $g_d$, which depend on the theoretical model assumed. \citet{graham_new_2013} and \citet{budker_proposal_2014} discuss contributions from oscillating nucleon EDMs, but there can also be significant contributions to $g_d$ from axion-induced $P,T$-violating intranuclear forces as described by \citet{stadnik_axion-induced_2014} and \citet{stadnik_manifestations_2017-1}. For the canonical QCD axion with the usual axion-gluon coupling, the contributions to $g_d$ from axion-mediated intranuclear interactions are dominant in most nuclei, as they arise at the tree level whereas oscillating nucleon EDMs are induced at the one-loop level.} and ${\bs{E}}^\star$ is the electric field acting on the spin \cite{budker_proposal_2014}. 
Axions and ALPs can also couple directly to electron ($e$) or nuclear ($N$) spins $\bs{\sigma}_{e,N}$ through a gradient interaction \cite{graham_new_2013,stadnik_axion-induced_2014},
described by the Hamiltonian
\begin{align}
\mc{H}_\textrm{grad} = g_{aNN} \bs{\nabla} a(\bs{r},t) \cdot \bs{\sigma}_{N} + g_{aee} \bs{\nabla} a(\bs{r},t) \cdot \bs{\sigma}_{e}\,,
\label{eq:gradient-interaction}
\end{align}
where $g_{aNN}$ represents the axion-nucleon coupling and $g_{aee}$ represents the axion-electron coupling.
The coupling constants in Eq.\,\eqref{eq:gradient-interaction} are related to the pseudoscalar coupling constants in Eq.\,\eqref{Lagrangian_scalar} according to the relations $g_{aNN} = g_p^N / (2m_N)$ and $g_{aee} = g_p^e / (2m_e)$. 
Here the connection to the topic of this review is evident: the coupling constants $g_{aNN}$ and $g_{aee}$ describing the spin-dependent dark matter interaction are closely related to the pseudoscalar coupling constants constrained by source/sensor experiments described in Secs.\,\ref{subsec_g_Pg_S} and \ref{subsec_g_Pg_P}. See, for example, \citet{jackson_kimball_search_2023} for further discussion of the relationship between these effects. 

An example of a haloscope experiment employing a spin-based sensor is the Cosmic Axion Spin Precession Experiment, CASPEr (\citealp{budker_proposal_2014}; \citealp{aybas_search_2021}).
The similarity in form between the interaction between spins and an axion dark matter field [Eqs.\,\eqref{eq:oscillating-EDM} and \eqref{eq:gradient-interaction}] and the Zeeman interaction of spins with an oscillating magnetic field suggests the use of magnetic resonance techniques in the search for axion dark matter. 
This is the approach taken by the CASPEr experiment, which uses NMR to detect $a(\bs{r},t)$ through its coupling to nuclear spins. Experiments that involve NMR measure the dynamics of nuclear spin in the presence of an applied bias field $\bs{B}_0$, which determines the Larmor frequency $\Omega_L = \gamma_N B_0$ ($\gamma_N$ is the nuclear gyromagnetic ratio). 
The hypothetical presence of the continuous oscillating field $a(\bs{r},t)$ [Eq.\,\eqref{eq:axion-field}] along with the aforementioned spin-dependent interactions make the situation analogous to continuous-wave (CW) NMR \cite{aybas_quantum_2021}. 
In the experiment of \citet{aybas_search_2021}, 
a fraction of the $^{207}$Pb nuclear spins in a ferroelectric\footnote{A ferroelectric crystal is used to provide a relatively large effective electric field ${\bs{E}}^\star$ to enhance the EDM signal.} 
PMN-PT crystal [chemical formula (PbMg$_{1/3}$Nd$_{2/3}$O$_3$)$_{2/3}$-(PbTiO$_3$)$_{1/3}$] are initially oriented along an applied magnetic field $\bs{B}_0$ due to their equilibrium thermal polarization. 
The applied magnetic field $\bs{B}_0$ is scanned, and if there was a detectable interaction between the nuclear spins and the axion dark matter, then when $\Omega_L \approx \omega_c$ a resonance would occur. 
Such an interaction would cause the spins to tip away from the direction of $\bs{B}_0$ and precess at $\Omega_L$, generating a time-dependent magnetization that can be measured using, e.g., induction through a pick-up loop. 
The experiment of \citet{aybas_search_2021}
found no evidence for dark matter axions with masses $162\,{\textrm{neV}}/c^2 \lesssim M \lesssim 166\,{\textrm{neV}}/c^2$, and set upper limits on the parameters $g_d$ and $g_{p}^{N}$ under the assumption that axions are dark matter with the nominal local density. Since the original CASPEr proposal, there have been several haloscope experiments using NMR and electron spin resonance (ESR) methods for the detection of UBDM fields (\citealp{crescini_operation_2018}; \citealp{quax_collaboration_axion_2020}; \citealp{jiang_floquet_2021}; \citealp{bloch_new_2022}).

These haloscope experiments using NMR and ESR 
typically are relatively narrow-band and require scanning the applied magnetic field to search for dark matter bosons of different masses, which correspond to UBDM fields oscillating at different Compton frequencies \cite{zhang_frequency-scanning_2024}.
An alternative approach particularly useful for relatively small Compton frequencies ($\lesssim 1$\,kHz) is to use a broad-band spin-based atomic magnetometer or comagnetometer \cite{terrano_comagnetometer_2021,wei_dark_2023}.
These experiments can simultaneously search for axion dark matter in a broad range of masses (as opposed to resonant-type searches that need to scan over specific and typically limited mass ranges).
There have been several such broadband searches carried out in recent years using a variety of spin-based sensors
(\citealp{abel_search_2017}, \citeyear{abel_search_2023}; \citealp{wu_search_2019}; \citealp{terrano_constraints_2019}; \citealp{garcon_constraints_2019}; \citealp{smorra_direct_2019}; \citealp{roussy_experimental_2021}; \citealp{jiang_search_2021}; \citealp{schulthess_new_2022}; \citealp{wei_dark_2023}; \citealp{bloch_constraints_2023}; \citealp{xu_constraining_2023}) 
and there are new proposals to expand the sensitivity of such searches \cite{fierlinger_ramsey_2024}.

The haloscope experiments referenced above rely on models that assume the UBDM behaves as a collection of non-interacting bosons described by the standard halo model \cite{evans_refinement_2019}.
These models neglect any small-scale structures in the dark matter halo beyond the natural stochastic fluctuations \cite{centers_stochastic_2021}.
Consequently, the sensors in these experiments are assumed to be continuously exposed to the UBDM field. 
However, it is possible that the energy density of the UBDM field is concentrated in larger composite structures that form, for example, due to self-interactions among the dark matter bosons (including gravity), or topological effects. 
In such scenarios, Earth would spend most of its time in regions of space with little or no dark matter. 
As a result, Earth would only occasionally and briefly encounter dark matter, resulting in infrequent and short-lived signals in dark matter detectors. 
In theory, a single sensor could detect such transient events. 
However, sensor networks offer more robust detection and verification of potential signals and reduced background of spurious signals. 

The Global Network of Optical Magnetometers for Exotic physics searches (GNOME) was the first such sensor array proposed to target such large composite dark matter structures (\citealp{pospelov_detecting_2013}; \citealp{pustelny_global_2013}; \citealp{afach_characterization_2018}, \citeyear{afach_search_2021}).
The GNOME experiment conducts simultaneous measurements of spin-dependent interactions of composite dark matter objects 
(\citealp{pospelov_detecting_2013}; \citealp{jackson_kimball_searching_2018}; \citealp{grabowska_detecting_2018})
using atomic magnetometers and comagnetometers in magnetically shielded environments at widely separated locations. The GNOME experiment looks for correlated signals across the network with particular spatio-temporal patterns, associated with different hypothetical exotic-physics scenarios \cite{afach_what_2023},
which help differentiate genuine signals from unrelated background noise. Similar experiments using atomic clock networks have searched for scalar dark matter 
(\citealp{derevianko_hunting_2014}; \citealp{wcislo_experimental_2016}, \citeyear{wcislo_new_2018}; \citealp{roberts_search_2017}, \citeyear{roberts_search_2020}).
There have also been related proposals by \citet{stadnik_enhanced_2016} to carry out a network-based clock-cavity comparison and by \citet{stadnik_searching_2015} and \citet{grote_novel_2019} to use a network of laser interferometers for dark matter searches.
Yet another application of sensor networks for dark matter detection is the idea of \citet{budker_axion_2020} to search for correlated signals from relativistic axions emitted from the decay of axion-quark ``nuggets'' \cite{zhitnitsky_nonbaryonic_2003}.
The use of astrophysical pulsar networks, which have much larger ``apertures'' than terrestrial networks, was discussed by \citet{stadnik_searching_2014}.


Note also that spin-based sensors can be used in the search for more massive, particle-like dark matter candidates with masses $M\gg 1$\,eV$/c^2$. 
For instance, \citet{fitzpatrick_effective_2013} describes spin-dependent interactions that could be used for the detection of WIMP dark matter. 
\citet{rajendran_method_2017} propose to use nitrogen-vacancy-center spin spectroscopy in diamonds to measure WIMP-induced damage tracks and perform directional detection of WIMPs. 
This would enable searches for WIMPs below the background limits set by the neutrino floor. Another example is the proposal of \citet{lyon_single_2024}
to use magnetic moment measurements for single-$^3$He-atom detection to find $^3$He atoms evaporated from a van der Waals liquid-helium film by energy transferred in WIMP collisions. 
Related ideas involving spin-dependent interactions of dark matter particles inducing observable collective effects in condensed matter systems are explored by \citet{trickle_effective_2022}. 
Note that \citet{fitzpatrick_effective_2013} and \citet{trickle_effective_2022} categorize boson-mediated spin-dependent interactions of WIMP dark matter in a manner similar to the categorization of boson-mediated spin-dependent potentials between fermions discussed in Sec.\,\ref{Sec:formalism} of this review, illustrating the close connection between the detection of wave-like UBDM and particle-like WIMPs.


In summary, the existence of dark matter is one of the most compelling reasons to search for the existence of exotic bosons, and there are a plethora of experiments testing the UBDM hypothesis, many employing spin-based sensors. There is a strong overlap and cross-fertilization between the underlying theories and the experimental techniques employed in searches for exotic spin-dependent interactions and spin-based dark matter haloscope searches.

\subsection{Astrophysics}\label{METH2.Astro.}

Astrophysical phenomena can provide powerful complementary probes of new interactions, 
either alone or in combination with very stringent laboratory limits on spin-independent interactions from gravitational experiments \cite{adelberger_improved_2013}. 
One should note that a light boson of mass $M$, which couples to ordinary matter (such as photons or fermions), can be copiously produced in the collisions or interactions of ordinary-matter particles in the hot environment of a stellar interior if $M \lesssim T$, where $T$ is the temperature of the stellar interior. 
The production and subsequent emission of light bosons from the interior of a star constitutes a local energy sink, which, if too draining, would drastically affect the properties or evolution of that star; in particular, stellar observations and predictions could differ noticeably. 
In the absence of discrepancies between stellar observations and predictions, limits can be placed on a variety of new interactions, see, for example, the reviews by \citet{raffelt_astrophysical_1990,raffelt_particle_1999}. 
A variety of stellar systems can be used, including ``active'' stars (such as the Sun) that ``burn'' fuel via nuclear fusion to generate an outward pressure that counteracts the inward gravitational pressure of the star and thereby maintain hydrostatic equilibrium, as well as ``dead'' stars (such as white dwarfs and neutron stars) which no longer consume fuel but rather maintain an outward pressure due to the degeneracy pressure of the dense electron or neutron content.



\begin{table*}
\caption{Summary of astrophysical bounds on interactions of light bosons with fermions. The stellar/astrophysical bounds are approximately independent of the boson mass for boson masses up to the stellar core temperature [e.g., $\sim 10$\,keV in \citet{raffelt_particle_1999} here]; at higher boson masses, these bounds degrade exponentially rapidly with the boson mass. Our preference with these types of astrophysical bounds is to generally quote only one significant figure to reflect the nature of the astrophysical uncertainties.
These referenced astrophysical bounds are presented in the main results figures in Sec.\,\ref{Sec:limits_main} either by themselves or in combination with stringent laboratory limits on spin-independent interactions that probe or are related to the $V_1$ term, see App.\,\ref{appendix_comp.} for more details.
} 
\renewcommand{\arraystretch}{1.8} 
\begin{tabular}{c|c|c|c|c} 
\hline
\hline
Referenced & Coupling constant & Constraint & Mass range & Reference \\
\hline
$\checkmark$&$g_A^e$ & $\lesssim M/(10^9~\textrm{GeV})$ & $M \lesssim 10 ~\textrm{MeV}$  &\citet{dror_dark_2017} \\
$\checkmark$&$g_A^N$ & $\lesssim M/(10^8~\textrm{GeV})$ & $M \lesssim 100 ~\textrm{MeV}$ &\citet{dror_dark_2017}\\
\hline
$\checkmark$&$g_V^e$ &  $\lesssim 0.9 \times 10^{-14}$  &$M \lesssim 10 ~\textrm{keV}$   &\citet{raffelt_particle_1999}\\
$\checkmark$&$g_V^N$ &  $\lesssim 3 \times 10^{-11}$    &$M \lesssim 10 ~\textrm{keV}$   &\citet{raffelt_particle_1999}\\
\hline
$\checkmark$&$g_p^e$ &$\lesssim 1.4 \times 10^{-13}$ &$M\lesssim 10 ~\textrm{keV}$ &\citet{bertolami_revisiting_2014}\\ 
            &$g_p^e$ &$\lesssim 1.5 \times 10^{-13}$ &$M\lesssim 10 ~\textrm{keV}$ &\citet{straniero_rgb_2020,capozzi_axion_2020}\\
$\checkmark$&$g_p^n$ &$\lesssim 3 \times 10^{-10}$   &$M \lesssim 10~\textrm{keV}$ &\citet{beznogov_constraints_2018}\\
$\checkmark$&$g_p^N$ &$\lesssim 10^{-9}$             &$M \lesssim 30~\textrm{MeV}$ &\citet{carenza_improved_2019}\\ 
            &$g_p^\mu$&$ \lesssim 10^{-8}$           &$M \lesssim 30~\textrm{MeV}$ &\citet{bollig_muons_2020}\\
\hline
            &$g_s^e$    &$\lesssim 1.3 \times 10^{-14}$   &$M \lesssim 10 ~\textrm{keV}$  &\citet{raffelt_particle_1999}\\
            &$g_s^N$    &$\lesssim 4 \times 10^{-11}$     &$M \lesssim 10 ~\textrm{keV}$  &\citet{raffelt_particle_1999}\\
            &$g_s^{\mu}$&$\lesssim 5 \times 10^{-9}$      &$M \lesssim 100 ~\textrm{keV}$ &\citet{caputo_muonic_2022}\\
$\checkmark$&$g_s^e$    &$\lesssim 1.1 \times 10^{-15}$   &$M \lesssim 10 ~\textrm{keV}$  &\citet{hardy_stellar_2017}\\
$\checkmark$&$g_s^N$    &$\lesssim 2 \times 10^{-12}$     &$M \lesssim 10 ~\textrm{keV}$  &\citet{hardy_stellar_2017}\\
\hline
\hline
\end{tabular}\label{tabel_Astrophysical-bounds}
\end{table*}

The couplings and mass of a boson generally depend on environmental conditions such as the temperature and matter density \cite{jaeckel_need_2007}.
Astrophysical bounds arise from phenomena that typically occur in conditions that are vastly different from those in the laboratory, including the system temperature, density, size, gravitational potential, and so on. 
This implies that astrophysical and laboratory bounds are complementary, arising from different sets of assumptions. 
For example, solar bounds can be evaded altogether if a boson has a sufficiently strong self-interaction \cite{jain_evading_2006} -- the so-called ``chameleon'' mechanism, while the high density of ordinary matter in stars can inhibit the production of bosons due to an increase in the boson mass inside the star \cite{derocco_exploring_2020}. 
If the boson of interest is a composite (rather than elementary) particle, then astrophysical bounds may be affected at the high temperatures (corresponding to high momenta) in stars, whereas laboratory experiments conducted at much lower temperatures would be unaffected. 
Additionally, laboratory conditions are often much cleaner and better controlled than in astrophysical environments. 
In particular, rarer astrophysical phenomena, such as supernovae explosions, may require additional assumptions that may not be valid, potentially invalidating particular types of astrophysical bounds; see, for example, \citet{bar_is_2020}. 
These potential loopholes in astrophysical constraints, coupled with the intrinsic value of controlled laboratory experiments and the large range of theoretical ideas which can generate exotic spin-dependent interactions, motivate the need to search for such interactions in laboratory experiments. 

In Tab.\,\ref{tabel_Astrophysical-bounds}, we present the current constraints on the axial-vector coupling constants $g_A$, vector coupling constants $g_V$, pseudoscalar coupling constants $g_p$ and scalar coupling constants $g_s$ derived from astrophysical observations. 

\section{Current limits on spin-dependent potentials and coupling constants}
\label{Sec:limits_main}

\subsection{Overview of the experimental search results}
\label{Sec:limits_main_overview}

For a given type of the interaction, two factors are crucial in the interpretation and impact of the research. 
The first is the composition of the field source and detector (i.e., which particles interact via the exotic force). 
The second is the spatial separation (distance) between the source and sensor.  
The distance is crucial as it determines the interaction ranges the experiment is most sensitive to, or equivalently, the range of masses of the exotic boson mediating the interaction. 


\textbf{Particle constituents}: In searches for exotic forces, many laboratory experiments focus on interactions between and among electrons and nucleons, the most common stable massive particles available. 
The relative strengths of the couplings of exotic bosons to protons and neutrons are model-dependent. 
The couplings may differ, for example, due to the different quark contents.
For example, in the KSVZ axion model \cite{kim_weak-interaction_1979,shifman_can_1980}, the coupling of the axion to the neutron is strongly suppressed compared to the axion's coupling to the proton \cite{di_cortona_qcd_2016}. 
On the other hand, if there is an (approximate) isotopic SU(2) invariance of the boson couplings to the nucleon sector (as occurs, for example, in the case of the coupling of the neutral pion to nucleons in the SM), then the size of the couplings to neutrons and protons can be (approximately) the same. 
If one considers the three common fermion species, namely the electron ($e$), proton ($p$), and neutron ($n$), then there are six possible pairs of fermions that can appear in single-boson-exchange potentials (see Sec.\,\ref{Sec:formalism}).
depending on the particle species involved in the interaction. 
These pairs are electrons and electrons ($e$-$e$), electrons and protons ($e$-$p$), electrons and neutrons ($e$-$n$), protons and protons ($p$-$p$), neutrons and neutrons ($n$-$n$), and neutrons and protons ($n$-$p$). 


In recent years, fifth-force searches have been extended to antimatter particles, such as $e^+$ and $\overline{p}$, in hitherto unexplored fermionic pairs, including electrons and positrons ($e$-$e^+$) (\citealp{kotler_constraints_2015}; \citealp{fadeev_pseudovector_2022})
and electrons and antiprotons ($e$-$\overline{p}$) (\citealp{ficek_constraints_2018}; \citealp{fadeev_pseudovector_2022}).
Additionally, investigations of exotic interactions involving other subatomic particles, including muons ($\mu$-$\mu$) \cite{yan_constraining_2019} and the antimuon (positively-charged muon) ($e$-${\mu}^+$) 
(\citealp{frugiuele_current_2019}; \citealp{fadeev_pseudovector_2022}),
have been undertaken. 

Furthermore, when the interaction involves fermions and both protons and neutrons (e.g., bound inside a nucleus or as part of a larger microscopic or macroscopic body), or the specific nucleon cannot be identified, we simply refer to the proton/neutron part by ``nucleon'' ($N$), with an implicit understanding that we refer to a weighted nucleon coupling constant, such as that in Eq.\,\eqref{ANZ}. 


\textbf{Distance}: Current experiments allow for proximity between a source and sensor ranging from astrophysical distances down to meters, millimeters, nanometers, and even subatomic scales. 
The longest-range experiments use the Sun, Earth or Moon (\citealp{wineland_search_1991}; \citealp{venema_search_1992}; \citealp{heckel_preferred-frame_2008}; \citealp{hunter_using_2013};  \citealp{leefer_search_2016};
\citealp{jackson_kimball_constraints_2017};
\citealp{zhang_search_2023}; \citealp{wu_new_2023};
\citealp{clayburn_using_2023})
as source masses. 

Mid-range experiments use laboratory-scale sources such as ferromagnets \cite{wineland_search_1991}, polarized helium-3 gases \cite{vasilakis_limits_2009}, and lead masses (\citealp{youdin_limits_1996}; \citealp{lee_improved_2018}).

Long- and mid-range experiments employ various sensors, including comagnetometers \cite{xiao_femtotesla_2023, wei_constraints_2022,kim_experimental_2018}, NV centers in diamond \cite{jiao_experimental_2021, liang_new_2022}, and cantilever-based instruments \cite{ding_constraints_2020,chiaverini_new_2003}.

The shortest-range experiments probe atomic or molecular structure (\citealp{ramsey_tensor_1979}; \citealp{jackson_kimball_constraints_2010}; \citealp{ledbetter_constraints_2013}; \citealp{ficek_constraints_2017}, \citeyear{ficek_constraints_2018}; \citealp{dzuba_probing_2017}; \citealp{stadnik_improved_2018}; \citealp{fadeev_pseudovector_2022}).

We follow the following rules for presenting results: First, the constraints delineated herein represent the strictest constraints on the couplings that have been attained through direct and complementary experimental methods (see Secs.\,\ref{Sec:EXP.METH.1} and \ref{EXP.METH.2}). 
Secondly, theoretical proposals are relegated to the subsections devoted to future perspectives at the end of their respective sections.
Thirdly, significant or high-quality experimental results reported on arXiv recently will be incorporated into the figures. 
Fourthly, for reference, we include spin-independent constraints derived from the $V_1$ potential term (see App.\,\ref{appendix_V1}), as well as astrophysical and combined astrophysical-laboratory constraints (see App.\,\ref{appendix_comp.}). 
Finally, in this section, we present constraints with the same confidence level (C.L.) as the original papers. 
While early experiments set constraints with 1$\sigma$, the trend in recent experiments is to use a C.L. of 95\%. 
In the interest of standardization, we suggest that future experiments use a 95\% C.L., if possible.

\subsection{Axial-vector/Vector interaction \texorpdfstring{$g_Ag_V$}{gAgV}} 
\label{subsec_g_Ag_V}

Experimental constraints on the exotic axial-vector/vector interaction constants $g_Ag_V$ can be derived from the $V_{12+13}$, $V_{11}$ and $V_{11p+16}$ potential terms, which we previously presented in Eq.\,\eqref{pseudovector-vector_potential}. 
Here, we present the potentials in SI units, reintroducing $\hbar$ and $c$, and we replace $M$ with $\frac{\hbar}{c \lambda}$ 
for clarity and consistency with the traditionally used potential forms. 

\begin{widetext}
\begin{equation}\label{V12+13-V_AV}
\begin{aligned}
V_{12+13}|_{AV}=  g^X_A g^Y_V \hbar\, \v{\sigma}_X \cdot \left\{ \frac{ \v{p}_X}{m_X} - \frac{ \v{p}_Y}{m_Y} , \frac{e^{-{r}/{\lambda}}}{ 8 \pi r} \right\} \Rightarrow g_A^Xg_V^Y\frac{\hbar}{4\pi}\boldsymbol{\sigma}_X\cdot\boldsymbol{v}\frac{e^{-{r}/{\lambda}}}{r} \, , 
\end{aligned}
\end{equation}

\begin{equation}
\label{V11-V_AV}
\begin{aligned}
V_{11}|_{AV}
&=-g_A^Xg_V^Y\frac{\hbar^2}{8\pi m_Y} \left( \boldsymbol{\sigma}_X\times\v\sigma_Y^{\,\prime} \right) \cdot \hat{\boldsymbol{r}}\left(\frac{1}{r^2}+\frac{1}{\lambda r}\right)e^{-{r}/{\lambda}}\,,
\end{aligned}
\end{equation}

\begin{equation}
\label{V11-16-V_AV}
\begin{aligned}
V_{11p+16}|_{AV} &= - \frac{g_A^X g_V^Y}{4} \frac{\hbar^2}{c^2} \left\{ \boldsymbol{\sigma}_X \cdot \boldsymbol{p}_X , \left\{ \v\sigma_Y^{\,\prime} \cdot \left[ \left( \frac{\v{p}_X}{m_X} - \frac{\v{p}_Y}{m_Y} \right) \times \hat{\v{r}} \right] , \left( \frac{1}{r^2} + \frac{M}{r} \right) \frac{e^{-{r}/{\lambda}}}{16 \pi m_X m_Y} \right\} \right\} \\
&\Rightarrow -g_A^X g_V^Y \frac{\hbar^2}{16 \pi (m_X+m_Y) c^2 } (\boldsymbol{\sigma}_X\cdot\boldsymbol{v}) [\v\sigma_Y^{\,\prime} \cdot (\boldsymbol{v}\times \hat{\boldsymbol{r}})] \left(\frac{1}{r^2}+\frac{1}{\lambda r}\right)e^{-{r}/{\lambda}} \, . 
\end{aligned}
\end{equation}
\end{widetext}

In this context, the expressions in the respective first parts of the equations for $V_{12+13}$ and $V_{11p+16}$ serve as a general form, suitable for both microscopic and macroscopic experiments. 
In contrast, the expressions presented in the second parts of the equations (after $\Rightarrow$) for $V_{12+13}$ and $V_{11p+16}$ conform to the more commonly used form for macroscopic experiments and 
are obtained from the respective first parts by expressing the fermion momenta in terms of the relative velocity $\v{v} = \v{v}_X -\v{v}_Y$ between the two fermions and by treating the velocity quantum operator as a classical variable (which removes the need for anticommutators, since classical position and velocity variables commute).
This enables a convenient comparison with existing experimental results. 
Other forms of the potentials can be found in Appendix~\ref{appendixA}. 
Especially, the relationship between $g_A^Xg_V^Y$ and $f_i^{XY}$, originating from \citet{dobrescu_spin-dependent_2006} and \citet{fadeev_neue_2018}, are listed in App.\,\ref{App:appendix_f_gg_table}.

Since $V_{11p+16}|_{AV}$ has not been studied earlier, we also present the traditional form $V_{16}$ below: 
\begin{widetext}
\begin{equation}
\label{V16-V_AV}
\begin{aligned}
V_{16} &=(-g_A^Xg_V^Y+g_V^Xg_A^Y)\frac{\hbar^2}{32\pi (m_X+m_Y)c^2} \left( [\boldsymbol{\sigma}_X \cdot (\boldsymbol{v}\times \hat{\boldsymbol{r}})] (\v\sigma_Y^{\,\prime}\cdot\boldsymbol{v})  +(\boldsymbol{\sigma}_X\cdot\boldsymbol{v}) [\v\sigma_Y^{\,\prime} \cdot (\boldsymbol{v}\times \hat{\boldsymbol{r}})]\right) \left(\frac{1}{r^2}+\frac{1}{\lambda r} \right) \, e^{-{r}/{\lambda}} \, ,
\end{aligned}
\end{equation}
\end{widetext}
which originates from the Dobrescu-Mocioiu formalism in Sec.\,\ref{Subsec:DMformalism}, as described by \citet{fadeev_neue_2018}.

Among these interactions, the odd-parity spin-velocity term $V_{12+13}$ has gained notable attention due to its unique properties. 
This interaction has the form of a Yukawa potential multiplied by the factor $\boldsymbol{\sigma}\cdot\boldsymbol{v}$. 
The idea of an exotic interaction being both spin- and velocity-dependent is intriguing because it may provide the only access to some particular form of coupling that might vanish in the static (velocity-independent) case. 
As an illustration, in the case where the velocity-dependent interaction might be enhanced due to relativistic effects, Lorentz invariance would probably require a temporal-type component to also be present if there is a spatial vector-like quantity like velocity involved. 
For example, in the theory of relativity, the momentum of a particle appears as the spatial components of the particle's 4-momentum, with the temporal component corresponding to the energy of the particle.
Recently, \citet{wu_new_2023} found that limits on exotic spin-dependent interactions induced by the Sun can be obtained using results from searches for Lorentz-invariance violation.%

Another profound aspect of $V_{12+13}$ is its parity-violating nature. 
The violation of parity symmetry in the weak interaction sector was first observed through the seminal experiment \cite{wu_experimental_1957}, which measured the directional anisotropy of emitted electrons in the beta decay of spin-polarized cobalt-60 nuclei. 
Parity-violating interactions mediated by the exchange of $W$ and $Z$ bosons within the standard model are generally restricted to short ranges.
In contrast, $V_{12+13}$ presents an opportunity to explore parity-symmetry-violating interactions on macroscopic scales. 
For example, the pioneering E\"{o}t-Wash experiments (\citealp{heckel_new_2006}, \citeyear{heckel_preferred-frame_2008}) were able to probe parity-violating interactions on astronomical scales, see Sec.\,\ref{METH1.SENS.MS} and subsequent subsections. 
These experiments not only established upper limits on boson-exchange forces related to $V_{12+13}$, 
but also on interactions that violate rotational invariance.
In addition, \citet{dzuba_probing_2017} probed new bosons associated with the $V_{12+13}$ term at the atomic scale using APV experiments. 

In molecules, the $V_{11}$ term can dominate over the $V_{12+13}$ term for internucleon interactions due to the slow velocities of the nuclei involved \cite{baruch_constraining_2024-1}.
 
The combined axial-vector/vector interaction generates the ${V}_{16}$ term that is parametrically suppressed compared to the ${V}_{12+13}$ and ${V}_{11}$ terms in Eqs.\,\eqref{V12+13-V_AV} and \eqref{V11-V_AV} by the respective factors $\mathcal{O}[\hbar v/(m_X + m_Y) r c^2]$ and $\mathcal{O}(v^2/c^2)$.
Searches for the $\mathcal{V}_{16}$ potential term have been performed and reported by \citet{hunter_using_2014,ji_new_2018,xiao_exotic_2024}. 
\citet{hunter_using_2014} investigated the interactions of spin-polarized geoelectrons with the spins of electrons, neutrons and protons in the laboratory. 
\citet{ji_new_2018} used SmCo$_{5}$ spin sources and a SERF comagnetometer and set constraints for interactions between electrons. \citet{xiao_exotic_2024} investigated the interactions between the electron spins of Rb atoms in a magnetometer in the presence of thermal motion and set constraints for interactions between electrons.
Note that all three of theses papers placed bounds on the $V_{16}$ term between two electrons. 
However, the $V_{16}$ potential term in Eq.\,\eqref{V16-V_AV} should vanish for identical fermions due to the cancellation of the combinations of the coupling constants in the first parentheses. To get a nonzero result for identical fermions, one should also include the term $p^2 V_{11}$ in the analysis. The $p^2 V_{11}$ term arises together with the $V_{16}$ term at the same order and, unlike the $V_{16}$ term, does not vanish for identical fermions.
For future experimental work (or, potentially, a re-analysis of existing data), it is important to study the $V_{11p+16}|_{AV}$, see more discussion in Sec.\,\ref{Subsec:Pairs-pots.}.

The exotic interactions described by the $V_{12+13}$, $V_{11}$ and $V_{11p+16}|_{AV}$ terms could be mediated by massive spin-1 bosons, including hypothetical Z$^{\prime}$ bosons \cite{dzuba_probing_2017}. 

Figure\,\ref{gvga-fig} shows the current laboratory constraints on an exotic axial-vector/vector interaction $g_A^Xg_V^Y$ between the studied particle pairs: (a) $e$-$e$, (b) $e$-$N$, $e$-$n$, and $e$-$p$ (c) $n$-$p$, $n$-$N$, $n$-$n$, $p$-$N$, where $A$ and $V$ denote the axial-vector and vector couplings, respectively, while $e$, $N$, $p$ and $n$ denote electron, nucleon, proton and neutron, respectively. 
The horizontal axes at the bottom of the figures show the range of interaction $\lambda$, which is inversely proportional to the mass $M$ of the boson (shown on the top axis) mediating the interaction, $\lambda=\hbar/(Mc)$. 
The vertical axes show the dimensionless product of coupling parameters $|g_Ag_V|$. 

Since the constraints are for various fermion pairs and can arise from different potential terms, we use different line colours to represent different fermion pairs and different line types to represent bounds from different potential terms (e.g., constraints from $V_{12+13}$ and $V_{11}$ are represented by solid and dotted lines, respectively.)
In addition, in each subplot, we use text with an arrow on the side to indicate the corresponding fermion pairs and the reference; as an example, ``Heckel 2008 (e-N)'' refers to the data from searches for exotic interaction between electrons and nucleons in the reference \citet{heckel_preferred-frame_2008} (all the other constraints for the $e$-$N$ interaction have the same color). 
The faint, gray lines represent combined bounds included for reference, where the bounds on $g_A$ come from astrophysical measurements, while the bounds on $g_V$ come from searches based on the spin-independent $V_1$ term. 
Details can be found in Sec.\,\ref{METH2.Astro.}, as well as Appendices\,\ref{appendix_V1} and \ref{appendix_comp.}.
All the other figures in Sec.\,\ref{Sec:limits_main} are presented in a similar manner.

We note that $g_V^N$ is often not precisely defined in the literature, including much of the primary literature we discuss in the current section. 
A natural definition of $g_V^N$ 
that is normalised with respect to the total number of nucleons reads: 

\begin{equation}
\label{ANZ}
g_V^N=(\sN g^n_V+Zg^p_V)/A\,,
\end{equation}
where $A=\sN+Z$ is the total number of nucleons, $\sN$ is the neutron number, and $Z$ is the proton number. 
This definition appears in \citet{dzuba_probing_2017} and is used implicitly in \citet{kim_experimental_2019}. 
If the neutron and proton coupling parameters are equal, then $g_V^N = g_V^n = g_V^p$. 
Other definitions are also used in the literature. 
For example, \citet{yan_new_2013} and \citet{yan_searching_2015} study interactions of neutrons with nucleons and electrons by measuring neutron spin rotation in $^4$He and use the unnormalised definition $g_V=2g_V^p+2g_V^n+2g_V^e$. 
\citet{su_search_2021} studied exotic interactions between polarized neutrons and unpolarized nucleons. 
The authors 
assumed that the coupling to unpolarized fermions is the same for neutrons and protons and is zero for electrons in the unpolarized mass.

Equation\,\eqref{ANZ} applies to both scalar and vector couplings, but only at the leading order of approximation when the scalar or vector coupling gives a spin-independent contribution to the vertex, in which case the spin-independent effect sums coherently over the nucleon content of the nucleus. On the other hand, if we consider the spin-dependent correction to a scalar or vector vertex, or if we consider a single pseudoscalar or axial-vector vertex which are spin-dependent at leading order, then Eq.\,\eqref{ANZ} generally does not apply due to pairing of nucleon spins in a nucleus, which means that there is no longer coherent addition over nucleon spins. As an example, consider the $V_{4+5}$ term for the case of nucleon-nucleon interactions, see Eqs.\,(\ref{gaga_V45}), (\ref{gvgv_V45new}) and (\ref{gsgs_V45}), where searches involve a spin-unpolarized source mass and a nuclear-spin-polarized probe (see Secs.\,\ref{subsec_g_Ag_A}, \ref{subsec_g_Vg_V} and \ref{subsec_g_sg_s} for further experimental details).  In this case, we can coherently sum over the nucleon-spin-independent contribution of the source mass similarly to Eq.\,(\ref{ANZ}); on the other hand, the nuclear-spin-dependent contribution associated with the probe is typically dominated by a single unpaired nucleon spin for most non-deformed nuclei.

\subsubsection{\texorpdfstring{$e$-$e$}{e-e}}
\label{Sec:A-V_e-e}

Figure\,\ref{gvga-fig}\,(a) shows the most stringent constraints on the exotic axial-vector/vector interaction parameters $g_A^e g_V^e$ between electrons. 

\citet{heckel_limits_2013} set constraints on interactions between the polarized electrons in their spin torsion pendulum and polarized electrons in stationary laboratory sources. 
The details of a torsion pendulum can be find in Sec.\,\ref{METH1.SENS.MS}. 
The stationary sources surrounding the pendulum in different configurations contribute to the different potential terms $V_{11}$, $V_{2}$ and $V_{3}$. 
\citet{heckel_limits_2013} set the best limit on exotic spin-spin interactions of electrons mediated by bosons with $\lambda>9.9\times 10^{-3}$\,m (masses up to 20\,\textmu eV) 
via the $V_{11}$ term, except at $\lambda\approx 10^6$\,m (masses around 100 feV, 1 feV = $10^{-15}$ eV) 
where the result from \citet{hunter_using_2013} via the $V_{11}$ term is comparably precise.

\citet{hunter_using_2013} examine interactions via the $V_{11}$ potential term between these spin-polarized geoelectrons and the spin-polarized electrons in the laboratory experiment of \cite{heckel_preferred-frame_2008}. 
By combining the pyrolite model for electron spins within Earth and experimental results, \citet{hunter_using_2013} set 
constraints on $g_A^e g_V^e$ in the range $10^{3}<\lambda<10^{11}$\,m.

\subsubsection{\texorpdfstring{$e$-$N$}{e-N}}
\label{Sec:A-V_e-N}

\begin{figure} [!htbp]
\begin{center}
\includegraphics[width=0.48\textwidth]{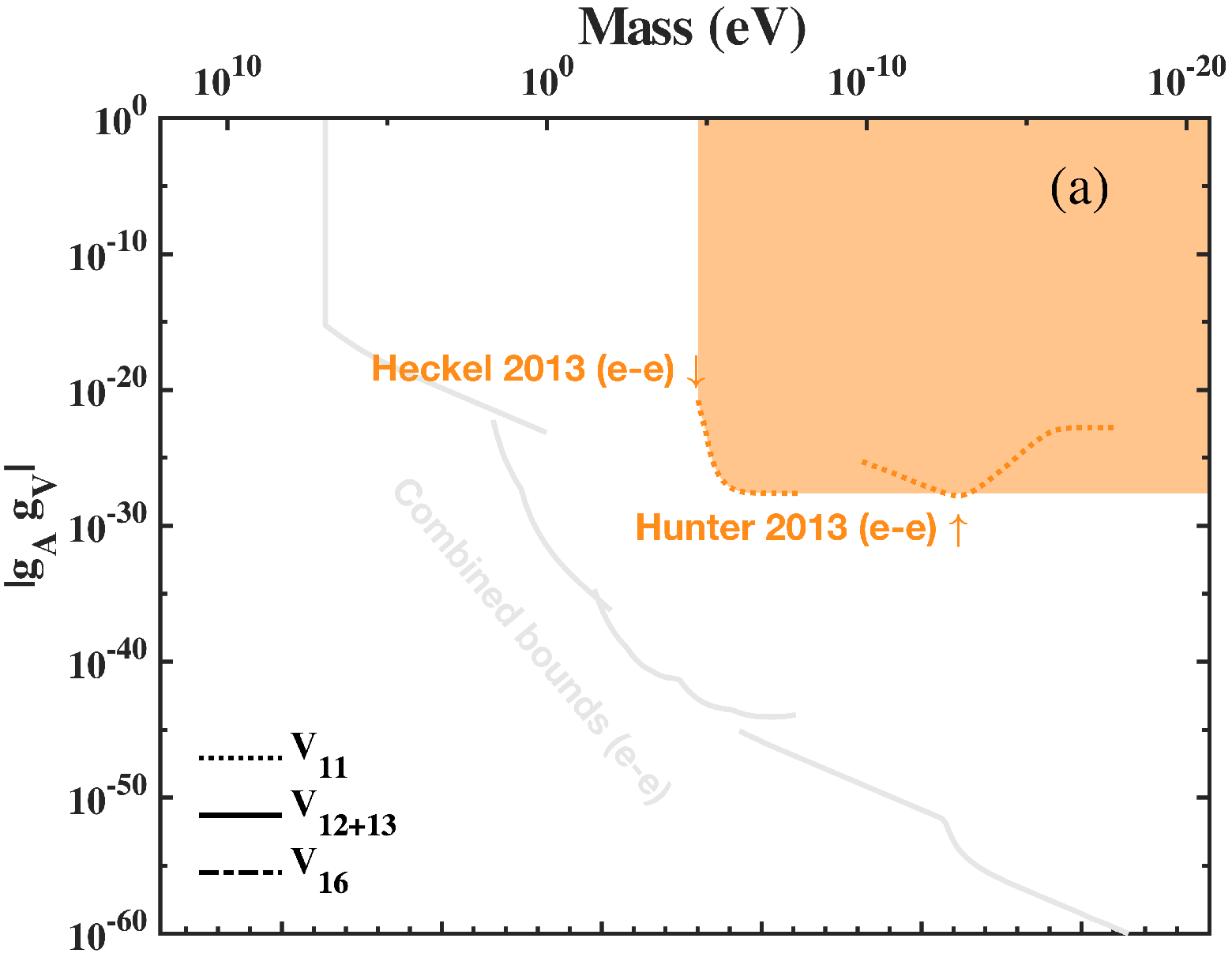}
\includegraphics[width=0.48\textwidth]{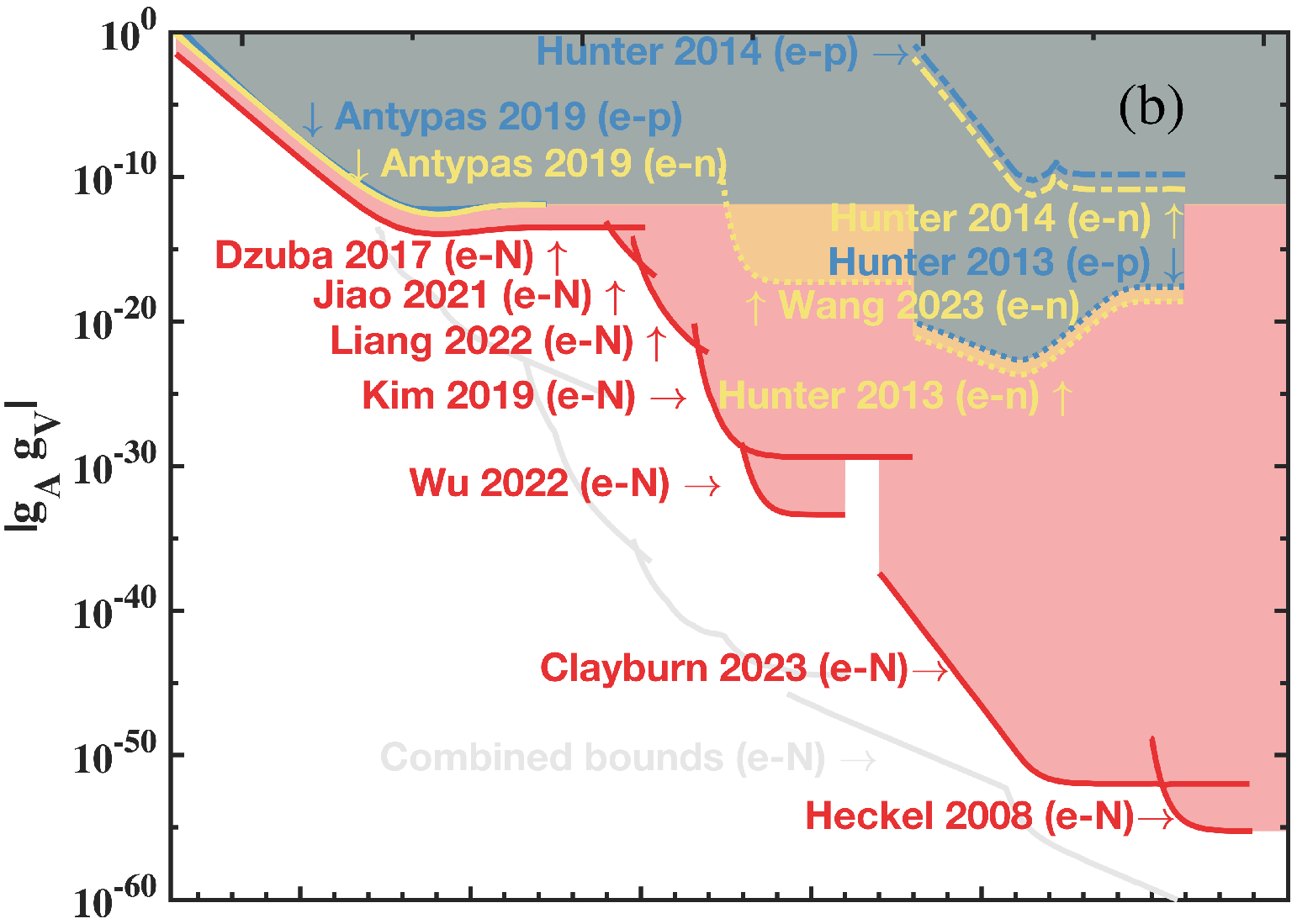}
\includegraphics[width=0.48\textwidth]{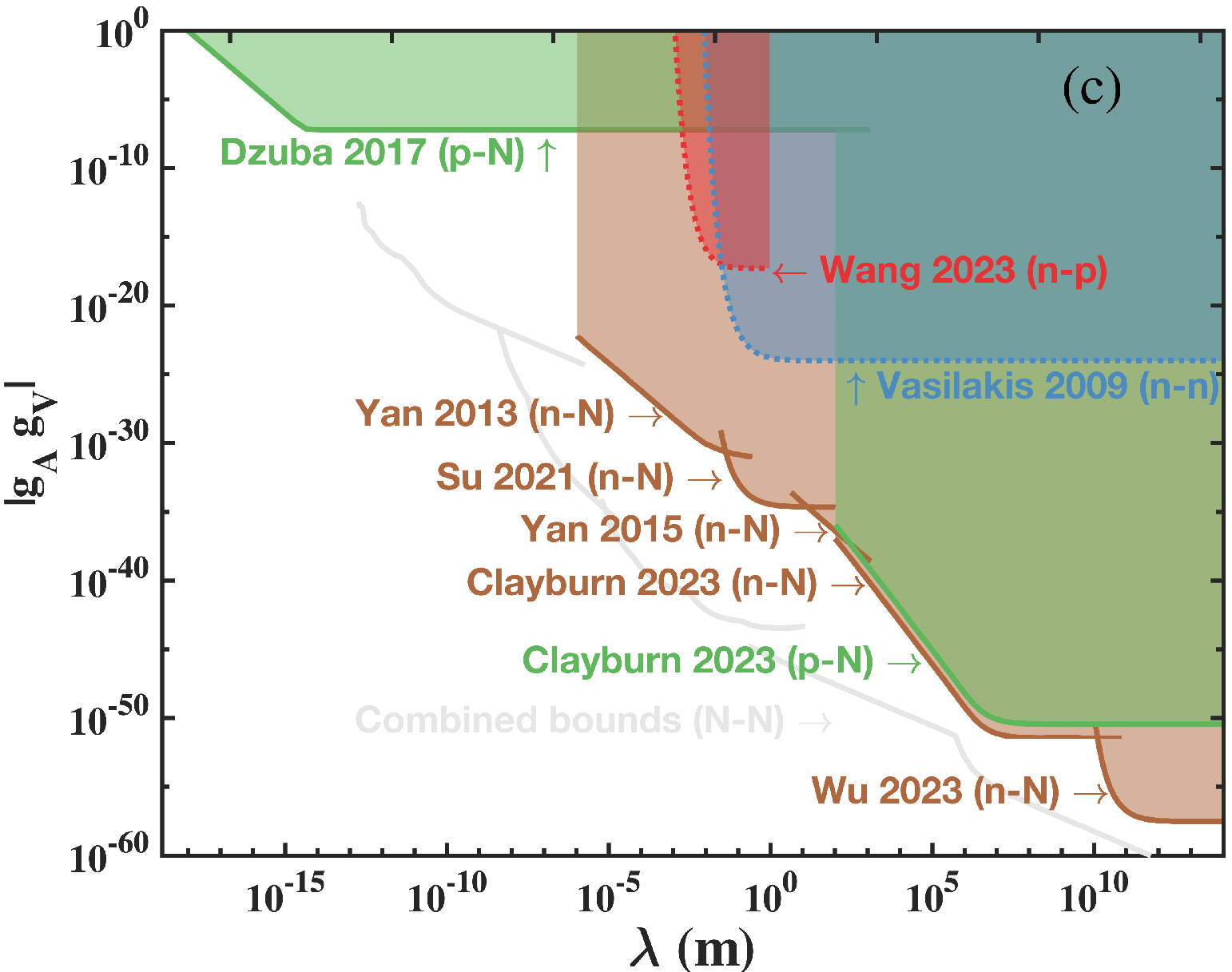}
\end{center}
\caption{
Constraints, depicted by coloured regions, on the coupling constant product $g_A g_V$ as a function of the interaction range $\lambda$ shown on the bottom x-axis. 
The top x-axis represents the new spin-1 boson mass $M$. 
Here, the subscripts $A$ and $V$ denote the axial-vector and vector couplings, respectively. 
Constraints shown for \textbf{(a)} $e$-$e$, \textbf{(b)} $e$-$N$, \textbf{(c)} $N$-$N$ couplings. 
Results include $V_{11}$ (dotted line), $V_{12+13}$ (solid line), and $V_{16}$ (dash-dotted line) 
terms. 
Combined bounds (faint, gray lines) are elaborated in Appendix\,\ref{appendix_comp.}. 
}
\label{gvga-fig}
\end{figure}

For the exotic axial-vector/vector interaction $g_A^e g_V^N$ between the particle pairs $e$-$N$, the most stringent constraint 
for the smallest boson masses
is set by the E\"{o}t–Wash group \cite{heckel_preferred-frame_2008}, as shown in Figure \ref{gvga-fig}\,(b). 
This was also the first dedicated experiment to set constraints on $g^e_Ag^N_V$ and is one of the most sensitive techniques 
at large values of $\lambda$
until now. 
The technique is based on a torsion pendulum containing $\approx 10^{23}$ spin-polarized electrons 
that constitute the sensor.
The experiment constrained 
exotic interactions between the polarized electrons in the pendulum and unpolarized matter in the Earth, Moon, or Sun. 
A detailed description of the experimental setup can be found in Sec.\,\ref{METH1.SENS.MS}. 
The experiment established limits on the 
coupling $|g^e_Ag^N_V| < 1.2 \times10^{-56}$ for $\lambda>1$ astronomical unit (AU). 

In the range $10^2\,\textrm{m}<\lambda<2.0\times10^{10}$ m, \citet{clayburn_using_2023} utilized the results from the E\"{o}t–Wash torsion-pendulum experiment \cite{heckel_preferred-frame_2008} 
and treated Earth as a rotating, unpolarized source \cite{hunter_using_2013, hunter_using_2014}
to set the most stringent constraints on $g_A^e g_V^N$. 
The constraint arises via the spin-velocity interaction term $V_{12+13}$. 
A discussion of their model is presented in Sec.\,\ref{Sec:A-V_e-e}. 
\citet{clayburn_using_2023} also established stringent bounds on the strength of the $V_{12+13}$ coupling between nucleons, as well as bounds on $V_{4+5}$ for the $e$-$N$ interaction and interactions between nucleons.

In the range of $\lambda$ around $\sim 0.1\,\textrm{m}$, the most stringent constraints on $g_A^eg_V^N$ are given by \citet{kim_experimental_2019} and \citet{wu_experimental_2022}. 
\citet{kim_experimental_2019} searched for the odd-parity spin- and velocity-dependent interaction term $V_{12+13}$ 
by using
an optically pumped alkali-metal atomic vapor
magnetometer
\cite{chu_search_2016}. 
Their approach probes exotic interactions between optically polarized electrons in an atomic vapor and unpolarized or polarized particles of a solid-state mass in the vicinity of the atomic vapor. \citet{wu_experimental_2022} uses an array of atomic magnetometers instead of a single one in order to increase the statistics and cancel the common-mode noise. They use rotationally modulated source masses while the previously used unpolarized BGO mass in \citet{kim_experimental_2019} was translated up and down. This allows for removing possible slow drifts which enabled 
 an improvement by a factor of up to 20,000 on some of the limits on $g^e_ Ag^N_V$.

In the range of $\lambda$ around $10^{-5}$\,m, the most stringent constraints are given by \citet{jiao_experimental_2021}  and \citet{liang_new_2022} using NV center in diamonds. 
\citet{jiao_experimental_2021} use a magnetometer that is based on a single NV center which is located close to the surface of a diamond. 
The source mass is a half-sphere lens made of fused silica. 
The NV spin detects an effective magnetic field created by the exotic spin-dependent interaction resulting from the moving nucleons. 
This experiment imposes constraints on the exotic spin- and velocity-dependent interaction within the force range of the order of $\sim 1-100$\,\textmu m. 
\citet{liang_new_2022} improved the magnetometer by using ensembles of NV centers in diamond yielding improved magnetic field sensitivity.
It is worth noting that \citet{ren_search_2021} employ a sensitive cantilever with a magnetic tip to directly measure the force between polarized electrons and unpolarized nucleons at this scale, and the obtained constraints approach the limits set with NV magnetometers.

Constraints at the shortest range are given by \citet{dzuba_probing_2017} by calculating PV effects in atoms and molecules (see Sec.\,\ref{METH2.PV}) which is induced by the exchange of vector bosons between electrons and nucleons. 
The vector-mediated atomic PV effect is calculated using the same relativistic many-body approaches as in earlier calculations of standard-model PV effects, but for the operator $V_{12+13}$ instead. 
The detailed calculation starts with the relativistic Hartree-Fock-Dirac method, including electron-core polarization corrections calculated in the random phase approximation framework, followed by the use of either the correlation potential method or the combination of the configuration interaction and many-body perturbation theory methods, depending on the species of atoms. 
The obtained results for Cs, Yb, and Tl were used to constrain the parity-violating axial-vector/vector electron-nucleon interaction, while the determination of the nuclear anapole moment of $^{133}$Cs was used to constrain the parity-violating axial-vector/vector proton-neutron interaction. 
Such interactions at the atomic and subatomic length scales allow one to effectively  
probe larger boson masses compared to experiments using spin-polarized torsion pendula, atoms and NV centers 
that probe interactions at longer ranges.

\subsubsection{\texorpdfstring{$e$-$n$}{e-n}}
\label{g_Vg_A_e-n}

In the range $10^3 \, \textrm{m} < \lambda < 10^{11}$\,m, the most stringent constraints on the exotic axial-vector/vector interaction $g_A^e g_V^n$ between the particle pairs $e$-$n$ were obtained by \citet{hunter_using_2013} by investigating the $V_{11}$ term interaction between geoelectrons and neutrons in the sensor, as shown in Figure~\ref{gvga-fig}\,(b). 
Note that for the $V_{11}$ term, it is possible to extract a bound on $g_A^X g_V^Y$ by simply multiplying the bound on $g_V^X g_A^Y $ by the fermion mass ratio $m_Y / m_X$, see Eq.\,\eqref{pseudovector-vector_potential}.

In their next work, \citet{hunter_using_2014} investigated the velocity-dependent spin-spin interaction term $V_{16}$ using their model of the electron spins within Earth discussed in \citet{hunter_using_2013}. 
To date, \citet{hunter_using_2014} remains the only the paper to place constraints on $g_A^eg_V^n$ via the $V_{16}$ term. 
Their latest work \citet{clayburn_using_2023} is based on these three related experiments (\citealp{venema_search_1992}; \citealp{heckel_preferred-frame_2008}; \citealp{peck_limits_2012}),
but set constraints via the spin-velocity interaction $V_{12+13}$ and $V_{4+5}$. 
However, since some experiments have already been updated, e.g. \citet{zhang_search_2023} measured the ratio of nuclear spin-precession frequencies between $^{129}$Xe and $^{131}$Xe, setting a more stringent constraint on the coupling energy between a nucleon spin and gravity than \citet{venema_search_1992}, it would be an opportune time to update the results of \citet{hunter_using_2013}; \citet{hunter_using_2014}; \citet{clayburn_using_2023}
in the near future.

For the interaction range $3\times 10^{-3}\,\textrm{m}<\lambda<10^3$\,m,  \citet{wang_search_2023} set the most stringent constraint by investigating the odd-parity spin-spin interaction term $V_{11}$ induced by the exchange of a hypothetical spin-1 $Z'$ boson. 
They developed a technique based on a quantum spin amplifier 
(see Sec.\,\ref{METH1.SENS.AM.SMSA}) 
to resonantly search for exotic interactions. 
In this experiment, the spin source uses polarized electron spins from $^{87}$Rb atoms, while the spin sensor uses polarized neutron spins from isotopically enriched polarized $^{129}$Xe gas. 

In the range $2.1\times10^{-19}\,\textrm{m} < \lambda < 3 \times10^{-3}$\,m, \citet{antypas_isotopic_2019} use isotopic-comparison data 
in APV experiments
to provide the most stringent individual constraints on the electron–neutron interaction 
via the $V_{12+13}$ term.
\citet{antypas_isotopic_2019} provide a measurement of the APV effect in four nuclear-spin-zero isotopes of Yb, namely $^{170}$Yb, $^{172}$Yb, $^{174}$Yb and $^{176}$Yb (see Secs.\,\ref{METH2.PV}). 
Such an isotopic variation in atomic parity violation confirms the predicted SM weak nuclear charge ($Q_\textrm{W}$) scaling with the number of neutrons, and additionally offers information about an additional $Z'$ boson. 
This is similar to the earlier result of \citet{dzuba_probing_2017}, which combined new atomic calculations with experimental data from previous APV experiments to place constraints on $g_A^e g_V^N$. 
In fact, here \citet{antypas_isotopic_2019} consider the PV effects arising from the exchange of a vector boson $Z'$ between the electrons and nucleons also in the form of the $V_{12+13}$ interaction like in \citet{dzuba_probing_2017}. 
The interaction effectively modifies the SM weak charge: $Q_\textrm{W} = Q_\textrm{W}^\textrm{SM} + Q_\textrm{W}^\textrm{exotic}$, with 
the former term being approximately proportional to the number of neutrons and the latter term being proportional to $g_A^eg_V^N$. 
\citet{antypas_isotopic_2019} use their data for the observed proton contribution to the PV amplitude to constrain $g_A^eg_V^p$. 
These bounds are then combined [through Eq.\,(10) in their paper] with constraints on the effective $g_A^e g_V^N$ coupling that come from the 
analysis of the Cs PV results in \citet{wood_measurement_1997} to give limits on the electron–neutron interaction $g_A^e g_V^n$. 

\subsubsection{\texorpdfstring{$e$-$p$}{e-p}}
As shown in Figure\,\ref{gvga-fig}\,(b) and discussed in Sec.\,\ref{g_Vg_A_e-n}, \citet{antypas_isotopic_2019} also provide the constraints on the electron-proton interaction via the potential term $V_{12+13}$ over the range $\lambda>3.9\times10^{-19}$\,m. 
In the range $10^3 \, \textrm{m} < \lambda < 10^{11}$\,m, the most stringent constraints on $g_A^e g_V^p$ were obtained by \citet{hunter_using_2013}. 
In addition, \citet{hunter_using_2014} also investigated the velocity-dependent spin-spin potential term $V_{16}$ 
and provided constraints on the electron-proton potential over the range $\lambda>1.1\times10^{-3}$\,m.

\subsubsection{\texorpdfstring{$n$-$N$}{n-N}}
Figure~\ref{gvga-fig}\,(c) shows the most stringent constraints on the exotic axial-vector/vector interaction $g_A^n g_V^N$ between the particle pair $n$-$N$.

At the astronomical distances $\lambda > 1.3\times 10^{10}$\,m, \citet{wu_new_2023} analysed existing data from laboratory measurements (\citealp{brown_new_2010}; \citealp{allmendinger_new_2014})
on Lorentz and $CPT$ violation to derive upper limits on exotic spin-dependent interactions of the form $V_{12+13}$ induced by the Sun (or Moon) as source masses and Earth's rotation as a modulation. 
\citet{brown_new_2010} and \citet{allmendinger_new_2014} originally searched for a sidereal signal induced by a Lorentz-Invariance- and $CPT$-violating background field.
\citet{wu_new_2023} explored the possibility of exotic spin-dependent interactions involving axial-vector/vector couplings $g_A g_V$, as well as axial-vector/axial-vector couplings $g_A g_A$ and pseudoscalar/scalar couplings $g_p g_s$, mediated by new particles. 

For the interaction range $10^{2}\,\textrm{m} < \lambda < 1.3 \times 10^{10}$\,m, \citet{clayburn_using_2023} utilized data from the 
$^{199}$Hg-$^{133}$Cs comagnetometer experiment of (\citealp{peck_limits_2012}; \citealp{hunter_using_2013})
and the Earth model in (\citealp{hunter_using_2013}; \citeyear{hunter_using_2014})
 to set the most stringent constraints on $g_A^ng_V^N$ via the $V_{12+13}$ term.

For the force range $6\,\textrm{m} < \lambda < 10^2$\,m, \citet{yan_searching_2015} used the best available $T_2$ measurement of polarized $^3$He gas atoms as the polarized sensor and Earth as an unpolarized source to present the most stringent experimental upper bound on $g_A^n g_V^N$. 
The physical effects of $V_{12+13}$ were considered to produce a velocity-dependent pseudomagnetic field, 
fluctuations in which would change the spin relaxation time of the $^3$He nuclear spins.
Thus it is possible to detect or constrain new physics by measuring the longitudinal spin relaxation time ($T_1$) or the transverse relaxation time ($T_2$) of polarized noble gases 
(\citealp{pokotilovski_limits_2010}; \citealp{petukhov_polarized_2010}; \citealp{fu_limits_2011}).
Since the noble gas is sealed in a still glass cell, $\langle \v v \rangle$ is zero, but $\langle v^2 \rangle$ is not. 
The nonzero $\langle v^2 \rangle$ produced by the velocity-dependent pseudo-magnetic field changes the spin relaxation times of polarized noble gases. 
Thus, although there is no bulk motion of either the polarized or unpolarized masses in the experiments, it is still possible to constrain the velocity-dependent interactions $V_{12+13}$.

For the interaction range $0.05\,\textrm{m} < \lambda < 6$\,m, the most stringent constraints \cite{su_search_2021} come from measurements using a spin-based amplifier (see Sec.\,\ref{METH1.SENS.AM.SMSA}) to search for effects due to pseudomagnetic fields associated with the interaction term $V_{12+13}$ between polarized neutrons and unpolarized nucleons. 
In their experiment, the polarized 
neutrons are mainly from $^{129}$Xe (where 73\% of the nuclear spin is due to neutron spins)
and the source of unpolarized nucleons is mainly from a high density BGO crystal.

For the force range $10^{-6}\,\textrm{m} < \lambda < 0.05$\,m, the most stringent constraints on $g_A^n g_V$ come from a cold neutron beam experiment \cite{yan_new_2013} (see Sec.\,\ref{METH1.SENS.PBS}.). 
\citet{yan_new_2013} derived an upper bound on the product of couplings $g_A^ng_V$ 
[the definition of $g_V$ is discussed below Eq.\,\eqref{ANZ}] 
in the mesoscopic range based on their prior experiments with neutron spin rotation in liquid $^4$He \cite{snow_upper_2011}. 
\citet{yan_new_2013} take advantage of the fact that the interaction potential proportional to $\boldsymbol{\sigma}\cdot\boldsymbol{v}$ violates parity and, as a result, when slow neutrons pass close to the surface 
of an unpolarized bulk material, the exotic interaction causes a rotation of the plane of polarization, which can be measured using the separated oscillating fields method \cite{ramsey_molecular_1950}. 

In addition, there are also two sets of constraints on the neutron-nucleon interaction derived from the odd-parity spin-spin interaction term $V_{11}$. 
\citet{vasilakis_limits_2009} presents limits on the neutron-neutron coupling $g_A^ng_V^n$ in the limit of a massless spin-1 particle to be $9.8\times 10^{-25}$ 
[represented in Fig.\,\ref{gvga-fig}(c)]. 
Further description of the experiment \citet{vasilakis_limits_2009} can be found in Sec.\,\ref{subsec_g_Pg_P}, as it gives the most stringent constraint on $g_p g_p$ for the lightest boson masses. 
For the force range $10^{-3}\,\textrm{m} < \lambda < 1$\,m, \citet{wang_search_2023} set the most stringent constraints on the odd-parity spin-spin interaction $V_{11}$. 
The authors consider valence protons within the Rb nuclei as 
the spin source. 
Further details about \citet{wang_search_2023} can be found in Sec.\,\ref{g_Vg_A_e-n}.

\subsubsection{\texorpdfstring{$p$-$N$}{p-N}}

As shown in Figure~\ref{gvga-fig}\,(c), \citet{dzuba_probing_2017}; \citet{clayburn_using_2023}
also set limits on proton-nucleon ($p$-$N$) interactions via the term $V_{12+13}$ mediated by a vector boson. For further details of these searches, see Sec.\,\ref{Sec:A-V_e-N}.

\subsubsection{Future perspectives}

There are many intriguing possibilities for the field of probing axial-vector/vector couplings in the future.

One of the promising directions involves the application of NV-center-based sensors (see Sec.\,\ref{METH1.SENS.SS.NV}) for micrometer-scale experiments, where the exotic interaction between the polarized electron spin and a moving nucleon source can be explored by measuring the 
induced magnetic field experienced by the electron spin quantum sensor \cite{jiao_experimental_2021,liang_new_2022}. 

In particular, \citet{chu_proposal_2022} have proposed a new method to probe spin-dependent interactions based on NV centers in diamond. The experimental setup involves a spin source or mass source 
attached to a mechanical oscillator above the NV sensor, similar to that in \citet{rong_searching_2018}. 
However, they employ multipulse quantum sensing protocols, including the XY8-N sequence \cite{smits_two-dimensional_2019,glenn_high-resolution_2018}, to improve the sensitivity. 
\citet{chu_proposal_2022} analyzed two exotic spin-velocity interactions ($V_{12+13}$ and $V_{4+5}$) and three velocity-dependent spin-spin interactions ($V_{6+7}$, $V_{14}$, and $V_{15}$), 
and in all cases predict better sensitivity than existing experimental results over certain microscopic ranges (and in some cases, over certain macroscopic ranges as well). 

In addition, \citet{chu_proposal_2022} also identified a new spin source based on hyperpolarized $^{13}$C nuclear spins in a thin diamond membrane. 
This material offers the advantages of high spin polarization efficiency and low stray magnetic fields due to the small magnetic moment of $^{13}$C and the particular membrane geometry used. 

Also noteworthy are 
a series of reinterpretations of existing older datasets.
For instance, \citet{clayburn_using_2023} utilized the torsion-pendulum experiment results of \cite{heckel_preferred-frame_2008} and the model of geoelectrons described in \citet{hunter_using_2013} and \citet{hunter_using_2014} to obtain new constraints on $V_{12+13}$ and $V_{4+5}$. 
\citet{wu_new_2023} reinterpreted the Lorentz and $CPT$ violation results in (\citealp{brown_new_2010}; \citealp{allmendinger_new_2014})
to place new bounds on the exotic spin-dependent interaction terms $V_{12+13}$, $V_{4+5}$ and $V_{9+10}$. 
Additionally, new atomic/molecular spectroscopy data could be used to set new and improved limits on exotic spin-dependent interactions. 
For example, \citet{antypas_isotopic_2019} studied vector-mediated PV effects in Cs, Yb and other elements, and set the most stringent constraints on atomic and subatomic length scales. 
Similarly, the ongoing experiments about Ra$^+$ ion \cite{nunez_portela_towards_2013} and Fr (\citealp{dzuba_calculation_1995}; \citealp{safronova_high-precision_2000}; \citealp{dzuba_calculations_2001}; \citealp{dzuba_calculation_2011-1}; \citealp{tandecki_offline_2014}) by the Manitoba-TRIUMF-Maryland team (\citealp{aubin_frpnc_2012}; \citealp{aubin_atomic_2013}),
as well as other future PV experiments with atoms (\citealp{williams_method_2013}; \citealp{leefer_towards_2014}) and molecules (\citealp{isaev_laser-cooled_2010}; \citealp{cahn_zeeman-tuned_2014}) are expected to prompt the exploration of new forces.


\subsection{Axial-vector/Axial-vector interaction \texorpdfstring{$g_Ag_A$}{gAgV}  }
\label{subsec_g_Ag_A}

Experimental constraints on the exotic axial-vector/axial-vector combination of parameters $g_A g_A$ can be derived from the potential terms $V_{2}$, $V_{3}$, $V_{4+5}$ and $V_{8}$ [see Eq.\,(\ref{pseudovector-pseudovector_potential})]: 
\begin{equation}
\label{gaga_V2}
\begin{aligned}
V_{2}=-g_A^Xg_A^Y \frac{\hbar c}{4\pi}\boldsymbol{\sigma}_X\cdot\v\sigma_Y^{\,\prime}\frac{1}{r}\,e^{-{r}/{\lambda}} \,.
\end{aligned}
\end{equation}

\begin{widetext}
\begin{equation}
\label{gaga_V3}
\begin{split}
V_{3}|_{AA}
= - \hbar c {g_A^X g_A^Y \lambda^2}  \left[ \v{\sigma}_X \cdot \v\sigma_Y^{\,\prime}\left[ \frac{1}{r^3} + \frac{1}{\lambda r^2} + \frac{4 \pi}{3} \delta(\v{r}) \right] -  \left( \v{\sigma}_X \cdot \hat{\v{r}} \right) \left( \v\sigma_Y^{\,\prime}\cdot \hat{\v{r}} \right)  \left( \frac{3}{r^3} + \frac{3}{\lambda r^2} + \frac{1}{\lambda^2 r} \right)  \right] \frac{e^{-r/\lambda}}{4 \pi}\,,
\end{split}
\end{equation}

\begin{equation}
\label{gaga_V45}
\begin{split}
V_{4+5}&|_{AA}= - \frac{g_A^X g_A^Y}{4} \frac{\hbar^2}{c} \left\{ \boldsymbol{\sigma}_X \cdot \left( \frac{\boldsymbol{p}_Y}{m_Y^2} \times \hat{\boldsymbol{r}} \right), \left( \frac{1}{r^2} + \frac{1}{\lambda r} \right) \frac{e^{-r/\lambda}}{8 \pi} \right\} 
\Rightarrow g_A^Xg_A^Y \frac{\hbar^2}{16\pi c}\frac{m_X}{m_Y(m_X+m_Y)}\boldsymbol{\sigma}_X \cdot(\boldsymbol{v} \times \hat{\boldsymbol{r}}) \left (\frac{1}{r^2}+\frac{1}{\lambda r} \right ) e^{-r/\lambda} \, , \\
\end{split}
\end{equation}

\begin{equation}\label{gaga_V8}
\begin{aligned}
V_{8}&= g_A^X g_A^Y \frac{\hbar}{c} \left[ \left\{ \frac{\boldsymbol{\sigma}_X \cdot \boldsymbol{p}_X}{m_X} , \left\{ \frac{\v\sigma_Y^{\,\prime} \cdot \boldsymbol{p}_Y}{m_Y} , \frac{e^{-r/\lambda}}{16 \pi r} \right\} \right\} - \frac{1}{2}\left\{ \frac{\boldsymbol{\sigma}_X \cdot \boldsymbol{p}_Y}{m_Y} , \left\{ \frac{\v\sigma_Y^{\,\prime} \cdot \boldsymbol{p}_Y}{m_Y} , \frac{e^{-r/\lambda}}{16 \pi r} \right\} \right\} -\frac{1}{2}\left\{ \frac{\boldsymbol{\sigma}_X \cdot \boldsymbol{p}_X}{m_X} , \left\{ \frac{\v\sigma_Y^{\,\prime} \cdot \boldsymbol{p}_X}{m_X} , \frac{e^{-r/\lambda}}{16 \pi r} \right\} \right\} \right] \\
&\Rightarrow -g_A^Xg_A^Y\frac{\hbar}{8\pi c}(\boldsymbol{\sigma}_X\cdot\boldsymbol{v})( \v\sigma_Y^{\,\prime}\cdot\boldsymbol{v})\frac{1}{r}e^{-{r}/{\lambda}} \, . \\ 
\end{aligned}
\end{equation}
\end{widetext}

Similarly to Sec.\,\ref{subsec_g_Ag_V}, for the $V_{4+5}$ [Eq.\,(\ref{gaga_V45})] and $V_8$ [Eq.\,(\ref{gaga_V8})] potential terms, the second equations conform to the usual form used for macroscopic-scale experiments. 
The $V_{4+5}$ potential corresponds to 
the potential term associated with the $f_{\bot}$ coefficient
in the notation of \citet{dobrescu_spin-dependent_2006}.

Spin-1 particles can be associated with dimension-four couplings of a light $Z^\prime$ boson with mass $m_{Z^\prime}$ 
to standard-model fermions, see Eq.\,\eqref{Lagrangian_vector}.\footnote{The dimension of an operator is determined by the number of fields appearing in that operator, with each boson field contributing a factor of 1 to the operator dimension, while each fermion field contributes a factor of 3/2 \cite{peskin_introduction_2019}.} 
Such couplings give rise to all of the potential terms $V_2$, $V_3|_{AA}$, $V_{4+5}|_{AA}$ and $V_8$. 
Additionally, a paraphoton $\gamma^\prime$ that couples to fermions through dimension-six operators 
[see Eq.\,\eqref{Lagrangian_tensor}]
also leads to a potential similar to $V_3|_{AA}$ 
up to an overall scaling factor with the boson mass, see Eq.\,\eqref{pseudotensor-pseudotensor_potential}.
Furthermore, spin-dependent forces exhibit particular sensitivity to axial couplings of unparticles to fermions, a topic explored in \citet{vasilakis_limits_2009}. 
Unparticles are peculiar entities that lack a definite mass and can arise in scale-invariant theories \cite{georgi_unparticle_2007}.
It may worth exploring these directions further in existing experiments. 

\begin{figure*} [!htbp]
\begin{center}
\includegraphics[width=0.88\textwidth]{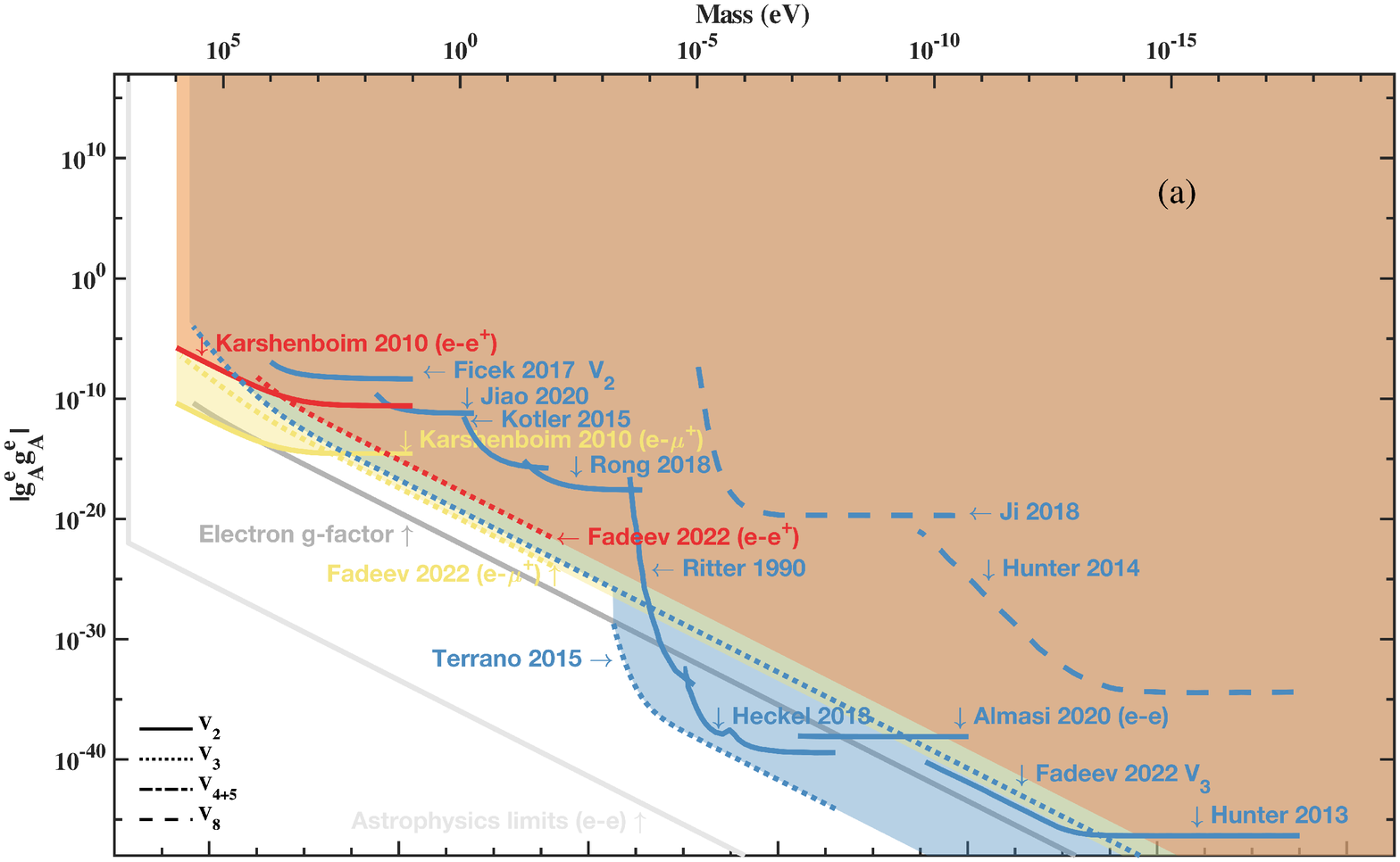}

\includegraphics[width=0.88\textwidth]{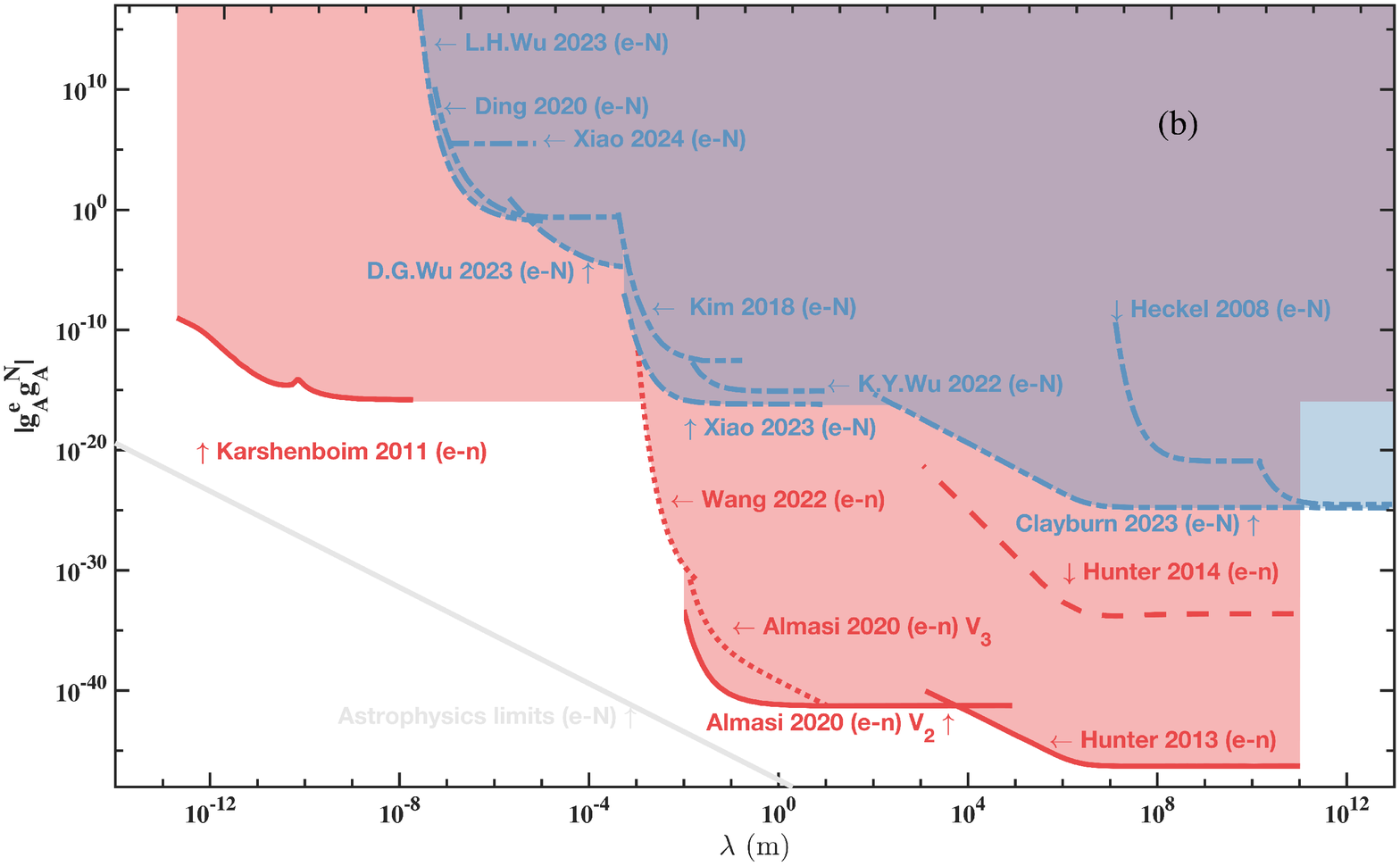}
\end{center}
\caption{
Constraints, depicted by coloured regions, on the coupling constant product $g_A g_A$ as a function of the interaction range $\lambda$ shown on the bottom x-axis. The top x-axis represents the new spin-1 boson mass $M$. The subscript $A$ refers to the axial-vector (pseudovector) coupling. 
Constraints are shown for \textbf{(a)} $e$-$e$, \textbf{(b)} $e$-$N$ couplings. Results are shown for $V_{2}$ (solid line), $V_{3}$ (dotted line), $V_{4+5}$ (dash-dotted line), and $V_8$ (dashed line) terms. 
Astrophysical bounds (faint, gray lines) are elaborated in Appendix~\ref{appendix_comp.}. 
Electron $g$-factor bounds (dark, gray line) are explained in Sec.\,\ref{METH2.g-factor}.
Besides the depicted bounds, \citet{ficek_constraints_2017} also placed constraints on the $V_{4+5}$ and $V_8$ terms, while \citet{xiao_exotic_2024} placed constraints on $V_8$, which are not shown for clarity.  
} 
\label{gaga-fig1}
\end{figure*}

\begin{figure*} [!htbp]
\begin{center}
\includegraphics[width=0.88\textwidth]{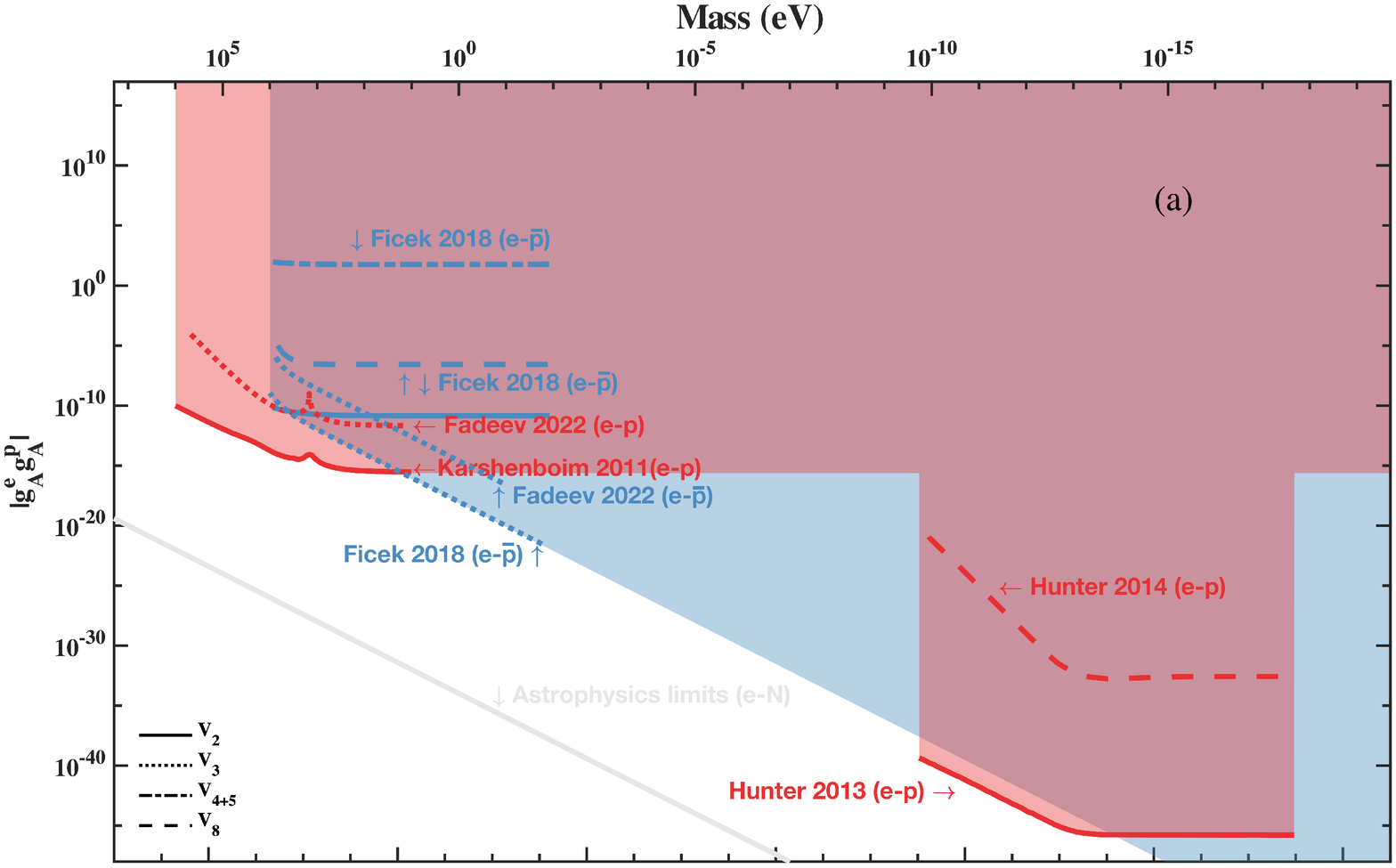}

\includegraphics[width=0.88\textwidth]{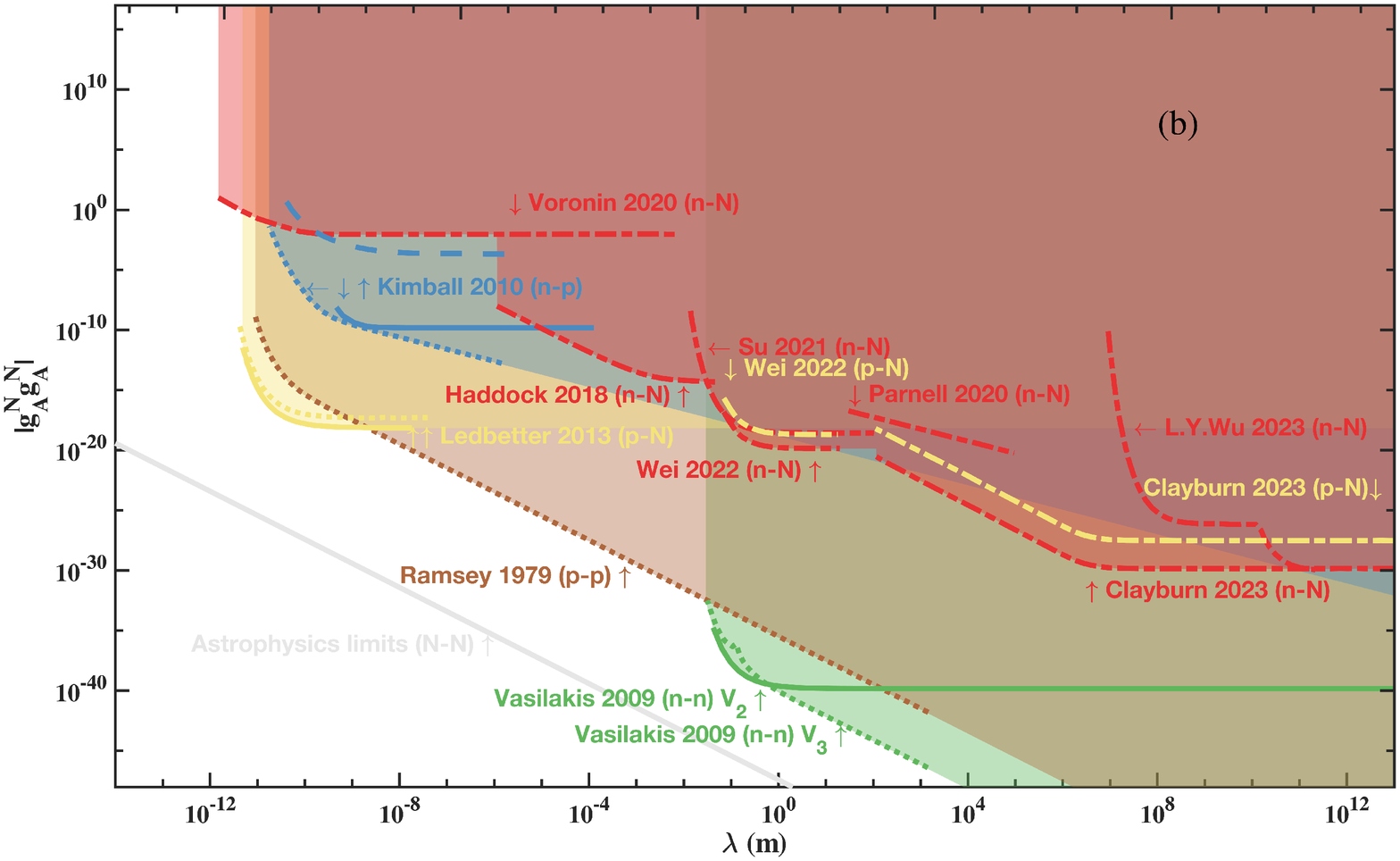}
\end{center}
\caption{
Constraints, depicted by coloured regions, on the coupling constant product $g_A g_A$ as a function of the interaction range $\lambda$ shown on the bottom x-axis. 
The top x-axis represents the new spin-1 boson mass $M$. 
The subscript $A$ refers to the axial-vector (pseudovector) coupling. 
Constraints are shown for \textbf{(a)} $e$-$p$, \textbf{(b)} $N$-$N$ couplings. 
Results are shown for $V_{2}$ (solid line), $V_{3}$ (dotted line), $V_{4+5}$ (dash-dotted line), and $V_8$ (dashed line) terms. 
Astrophysical bounds (faint, gray lines) are elaborated in Appendix\,\ref{appendix_comp.}.
}
\label{gaga-fig2}
\end{figure*}

In this section, Fig.\,\ref{gaga-fig1} shows the current laboratory constraints on 
$g_A^Xg_A^Y$ between the studied particle pairs (a) $e$-$e$, $e$-$e^+$ and $e$-$\mu^+$, (b) $e$-$N$ and $e$-$n$, and Fig.\,\ref{gaga-fig2} illustrates constraints for (a) $e$-$p$, and $e$-$\overline{p}$ (b) $p$-$N$, $p$-$p$, $n$-$p$, $n$-$N$ and $n$-$n$. Here $e^+$ and $\overline{p}$ stand for the positron (antielectron) and antiproton, respectively. 
According to the results presented in Figs.\,\ref{gaga-fig1} and \ref{gaga-fig2}, it is clear that $V_{2}$ and $V_{3}$ are the dominant leading terms.

While the term $V_{2}$ (as well as the terms $V_{4+5}$ and $V_8$) is frequently employed to explore possible axial-vector boson scenarios, the term $V_3$ is typically studied in the form $V_3|_{pp}$ in the context of pseudoscalar fields, see Eq.\,(\ref{gpgp_V3}). 
One of the primary reasons for this is the initial examination of the $V_{3}$ term in the context of spin-0 boson exchange by \citet{moody_new_1984}, which has persisted over time. 
\citet{anselm_possible_1982} considered the $V_3$ term in the context of massless pseudoscalar boson exchange a couple of years earlier. The boson there was a ``massless axion'' known as the ``arion''.
\citet{dobrescu_spin-dependent_2006} extended the study to the spin-1 case and demonstrated that the $V_3$ term can also appear in interactions mediated by axial-vector (and vector) bosons. 
Another historical reason for this focus is that, in many instances, the constraints on $g_A g_A$ derived from the $V_3$ term were not as stringent as those from the $V_2$ term if one used the formulae of \citet{dobrescu_spin-dependent_2006}. 
Later on, however, it was pointed out by \citet{malta_comparative_2016}
and \citet{fadeev_revisiting_2019} that the $V_3$ term in the axial-vector/axial-vector potential diverges in the limit of small boson masses as $\propto 1/M^2$ due to non-conservation of the axial-vector current, see Eq.\,\eqref{pseudovector-pseudovector_potential}. 
The enhancement of the $V_3$ term at small boson masses allows searches based on the $V_3$ term to surpass bounds from searches for the $V_2$ term in some cases (as well as generally surpassing the bounds derived from searches for the $V_{4+5}$ and $V_8$ terms), as can be seen in Figs.\,\ref{gaga-fig1} and \ref{gaga-fig2}.

The $V_{3}|_{AA}$ term in Eq.\,\eqref{gaga_V3} diverges as $\propto \lambda^2$ 
in the limit of a large $\lambda$ (or a small $M$), due to non-conservation of the axial-vector current. 
For comparison, we also present \citet{dobrescu_spin-dependent_2006}'s formula for the $V_{3}|_{AA}^{DM}$ (where DM is an abbreviation of Dobrescu and Mocioiu) term in the axial-vector/axial-vector potential below, 
which can be obtained by combining the formulae for $\sV_{3}$ and the coefficient $f_3$ in Eqs.\,(3.6) and (5.29) in their paper, respectively.
\begin{widetext}
\begin{equation}
\label{gaga_V3_initial}
\begin{split}
V_{3}|_{AA}^{DM}& = (g_A^X g_A^Y)_{DM} \frac{\hbar^3}{32\pi c}\frac{m_X^2+m_Y^2}{m_X^2 m_Y^2}\left[\boldsymbol{\sigma}_X\cdot\v\sigma_Y^{\,\prime} \left[\frac{1}{r^3} +\frac{1}{\lambda r^2} +\frac{4\pi}{3}\delta(r)\right]
-(\boldsymbol{\sigma}_X\cdot \hat{\boldsymbol{r}})(\v\sigma_Y^{\,\prime}\cdot \hat{\boldsymbol{r}})\left(\frac{3}{r^3}+\frac{3}{\lambda r^2}+\frac{1}{\lambda^2 r}\right)\right]e^{-{r}/{\lambda}} \, . 
\end{split}
\end{equation}
\end{widetext}
It is noteworthy that $V_{3}|_{AA}^{DM}$ exhibits qualitative differences compared to $V_{3}|_{AA}$, as discussed in \citet{fadeev_revisiting_2019} and further elucidated in \citet{fadeev_pseudovector_2022}. 
The relationship between their respective coefficients is given by
\begin{equation}
\label{gaga_V3_coefficients_relation}
{g_A^X g_A^Y} = - \frac{\hbar^2}{8 c^2}\frac{m_X^2+m_Y^2}{m_X^2 m_Y^2} \frac{1}{\lambda^2} {(g_A^X g_A^Y)_{DM}} \, . 
\end{equation}

The divergence with increasing $\lambda$ in $V_3|_{AA}$, Eq.\,\eqref{gaga_V3}, leads to greatly enhanced bounds on $g_A g_A$ at larger values of $\lambda$. 
For example, for $\lambda=1$\,m, 
the constraints on $g_A^e g_A^e$ are approximately 25 
orders of magnitude stronger than the bounds on $(g_A^Xg_A^Y)_{DM}$, 
since $\hbar^2/(2 c m_e \lambda)^2 \approx 4 \times 10^{-26}$ in this case.
This has a significant impact on the landscape of constraints arising from the $V_{3}|_{AA}$ term, as seen in Figs.\,\ref{gaga-fig1} and \ref{gaga-fig2}. 

The apparent divergence of limits on $g_A g_A$ can be circumvented by plotting constraints on the combination of parameters $g_A g_A \lambda^2$,
which should remain finite in a renormalisable theory in the limit of large $\lambda$  [since $g_A g_A/M^2$ should  remain finite as the mass of the mediating boson $M$ becomes very small, with $\lambda = \hbar/(Mc)$], as employed in \citet{fadeev_pseudovector_2022}. 

Regarding astrophysical bounds, \citet{dror_dark_2017} provide bounds on the electron and nucleon axial-vector couplings, $g_A^e \lesssim M/(10^9~\textrm{GeV})$ and $g_A^N \lesssim M/(10^8~\textrm{GeV})$, respectively. 
These bounds on $g_A$ likewise diverge at large values of $\lambda$ (corresponding to small boson masses). 
However, as mentioned above, one can instead present bounds on the combination $g_A g_A \lambda^2$, which is expected to remain finite in a renormalisable theory. 

As a consequence of the unusual behaviour of the $V_3|_{AA}$ term described above, the $V_2$ term does not necessarily give the most stringent limits on $g_Ag_A$ over the entire possible range of interactions. 
The critical force range occurs when $r \sim \lambda$. 
When $r \sim \lambda$, $V_2$ is comparable in magnitude to $V_3$, see Eqs.\,\eqref{gaga_V2} and \eqref{gaga_V3}. 
When $\lambda \gg r$, the extra $\lambda^2$ factor in $V_3$ generally leads to stronger limits than for $V_2$. 
Conversely, when $\lambda \ll r$, the $V_2$ term generally leads to stronger limits  than $V_3$. 
This behaviour can be readily observed in the plots of constraints on $g_A g_A$; for instance, the crossing of the lines corresponding to ``Almasi 2020 (e-n) $V_2$'' and ``Almasi 2020 (e-n) $V_3$'' in Fig.\,\ref{gaga-fig1}, and the crossing of the lines corresponding to ``Ficek 2018 (e-$\bar{p}$) $V_2$'' and ``Ficek 2018 (e-$\bar{p}$) $V_3$'' in Fig.\,\ref{gaga-fig2}. 
Experiments typically set constraints on both $V_2$ and $V_3$, see \citet{jackson_kimball_constraints_2010}; \citet{ledbetter_constraints_2013}; \citet{ficek_constraints_2018}; \citet{almasi_new_2020}; \citet{fadeev_pseudovector_2022}.

In this section, we mainly focus on a review of existing works related to the $V_{2}$ term, which was historically the most studied term in the content of axial-vector/axial-vector interactions. 
We also discuss the $V_{3}$ and $V_{4+5}$ terms, with further discussion of the related references presented in the subsequent Secs.\,\ref{subsec_g_Vg_V} and \ref{subsec_g_Pg_P}. 
Additionally, we discuss several results from the $V_{8}$ term here. 
However, since the $V_{8}$ term arises as a $\mathcal{O}(v^2)$ relativistic correction to the $V_{2}$ term in Eq.\,\eqref{pseudovector-pseudovector_potential}, it is generally much less important than the $V_2$ term in non-relativistic systems.

\subsubsection{\texorpdfstring{$e$-$e$, $e$-$e^+$ \& $e$-$\mu^+$}{e-e, e-e⁺ \& e-μ⁺}}
\label{subsec_g_Ag_A_ee}
In the range $\lambda> 10^{-2}$\,m, two of the experiments \cite{hunter_using_2013,heckel_limits_2013} searching for exotic axial-vector/vector interactions discussed in Sec.\,\ref{subsec_g_Ag_V} also place strong constraints on axial-vector/axial-vector interactions between electrons, $g_A^eg_A^e$, as shown in Figure\,\ref{gaga-fig1}\,(a). In addition, \citet{almasi_new_2020} also set constraints on $g_A^e g_A^e$; see Sec.\,\ref{sec:gAgA_en} for more details of their work.

In the range $9\times 10^{-4} \, \textrm{m} < \lambda < 10^{-2}$\,m, \citet{ritter_experimental_1990} performed an experiment with a spin-polarized torsion pendulum made from Dy$_6$Fe$_{23}$ (see Sec.\,\ref{METH1.SOUR.ESS}) to search for feeble forces associated with a possible exotic interaction between spin-polarized masses 
and set the most stringent limit on $g_A^e g_A^e$ until now. 
The constraints on the $V_2$ term for $g_Ag_A$ is based on the limit on axial photons summarized in Tab.\,VI of \citet{ritter_experimental_1990}. 
For further related discussion, see Sec.\,\ref{METH1.SOUR.ESS}. 


For the force range $10^{-5}\,\textrm{m} < \lambda < 9\times 10^{-4}$\,m, \citet{rong_constraints_2018} established the most stringent limit on the axial-vector-mediated interaction between electron spins, using single NV centers in diamond as sensors and a single crystal of $p$-terphenyl-doped pentacene-$d_{14}$ (see Sec.\,\ref{METH1.SOUR.ESS}), placed about 12 $\mu$m away from the sensor, as the source. More details of their experiments can be found in Sec.\,\ref{METH1.SENS.SS.NV} At $\lambda = 500\,\mu$m, their result is approximately a factor of 50 more stringent than the one set by \citet{kotler_constraints_2015} using trapped-ion measurements.



In the range $2.2\times 10^{-7}\,\textrm{m} < \lambda <10^{-5}$\,m, the most stringent constraint on the axial-vector/axial-vector interaction between electrons comes from the measurement of the dipole-dipole interaction between two trapped $^{88}$Sr$^+$ ions \cite{kotler_constraints_2015}, see Sec.\,\ref{METH2.PM.TI} for more details. 
In addition to this, using literature data pertaining to the ground-state hyperfine splitting interval in positronium, \citet{kotler_constraints_2015} set constraints on the electron-positron axial-vector/axial-vector and pseudoscalar/pseudoscalar interactions. 
The shift in the positronium hyperfine interval caused by an exotic spin-spin interaction was estimated by using first-order perturbation theory. 
The constraints were derived from the comparison of positronium hyperfine spectroscopy data 
(\citealp{mills_new_1975}; \citealp{mills_line-shape_1983}; \citealp{ritter_precision_1984}; \citealp{ishida_new_2014})
with QED predictions from theory \cite{eides_hard_2014}. 
Later on, the result on $g_A^e g_A^e$ has been surpassed by \citet{ficek_constraints_2017}; \citet{fadeev_pseudovector_2022} through investigation of the electronic structure of $^4$He. 
Also, see the later paragraph that discusses \citet{fadeev_pseudovector_2022}.
Similar methods based on spectroscopy of simple atomic systems have been utilised in \citet{karshenboim_hyperfine_2011,karshenboim_constraints_2010,karshenboim_precision_2010}, see the later paragraph and Sec.\,\ref{METH2.PM.AHS} for more details. 

In the range $3.5\times 10^{-9}\,\textrm{m} < \lambda < 2.2\times 10^{-7}$\,m, \citet{jiao_searching_2020} established the most stringent constraint for an axial-vector-mediated spin-dependent interaction between electrons by using a ``molecular ruler''. 
The molecular ruler is made up of two electron spins and a shape-persistent polymer chain; the distances between spins can be tailored by varying the length of the polymer chains connecting the spins. 
The coupling strength between the two electron spins, which are separated on the nanometer scale, is determined in double electron-electron resonance (DEER) experiments. 
The detected DEER signals contain information on the possible signal of the axial-vector exotic interaction on top of the magnetic dipole-dipole interaction. Therefore, by fitting the data from five samples of varying lengths, the coupling constant $g_A^e g_A^e$ can be determined or limits placed on it. 
It is worth mentioning that \citet{jiao_searching_2020} improved on the previous limit of \cite{luo_constraints_2017} obtained from scanning tunneling microscope and electron spin resonance (STM-ESR) experiments by an order of magnitude at $\lambda=200$\,nm. 
This research \cite{jiao_searching_2020} suggests using chemically synthesized molecules to search for exotic spin-dependent interactions at the nanometer scale as a promising approach. 

In the interaction range $2.0 \times 10^{-11}\,\textrm{m} < \lambda < 3.5 \times 10^{-9}$\,m, \citet{ficek_constraints_2017} established the most stringent constraints for dipole-dipole $V_2$ interactions between electrons at the atomic scale through investigation of the electronic structure of $^4$He. 
\citet{ficek_constraints_2017} compared the spectroscopic measurements of the helium 2$^3P_1$$-$2$^3P_2$ transition 
(\citealp{castillega_precise_2000}; \citealp{zelevinsky_precision_2005}; \citealp{borbely_separated_2009}; \citealp{feng_laser-spectroscopy_2015})
and the 2$^3S_1$$-$2$^3P_{0,1,2}$ transition \cite{pastor_absolute_2004} with the theoretical calculations of the corresponding energy structure \cite{pachucki_fine_2010}. 
The difference or level of agreement between experiments and theory limits the maximal possible energy contribution that may arise from exotic interactions. 
Thereby, \citet{ficek_constraints_2017} established constraints on possible exotic interactions that may be associated with the exchange of bosonic fields. 
\citet{ficek_constraints_2017} studied the $V_{2}$, $V_{3}$, $V_{4+5}$ and $V_{8}$ potential terms. 
In Fig.\,\ref{gaga-fig1}\,(a), we present the constraints arising from the $V_{2}$ 
term. 
The constraints arising from the 
$V_{4+5}$ and $V_{8}$ terms are all slightly milder than the $V_{2}$ constraint (not plotted for clarity), but these constraints are also important as they constitute the most stringent existing constraints for exotic axial-vector/axial-vector interactions between electrons. 
Is it worth mentioning that the bounds on the velocity-dependent terms $V_{4+5}$ and $V_8$ are competitive with the bound on the velocity-independent term $V_2$ in the case of helium, in contrast to typical macroscopic-scale experiments. This is because, in atoms, the typical speed of an orbiting electron is $v \sim \alpha c$, which is much larger than the relative speeds of constituent bodies in macroscopic experiments, see a related discussion of differences in speed by \citet{fadeev_revisiting_2019}. 
As atomic theory and experiments improve, He spectroscopy is expected to become an even more sensitive probe of exotic electron-electron interactions.

In the force range $\lambda > 2.0 \times 10^{-13}$\,m, \citet{karshenboim_constraints_2010} explored the axial-vector-mediated spin-dependent interactions by comparing theory and experimental results in atomic spectroscopy and established the most stringent constraints. 
He compared the experimental and predicted values for the hfs interval $\Delta_{\text{hfs}}(1s)$ of the ground state in muonium and positronium and placed bounds on $g_A g_A$ for the fermion combinations $e$-$\mu^+$ and $e$-$e^+$, respectively. 
\citet{karshenboim_constraints_2010} also studied hydrogen and other light hydrogenlike atoms (D, T and $^3{\rm{He}}^+$). 
\citet{karshenboim_precision_2010} used the specific difference $D_{21} = 8\Delta_{\text{hfs}}(2s) - \Delta_{\text{hfs}}(1s)$ \cite{karshenboim_hyperfine_2002-1}
to set better constraints in a larger force range compared to the constraints based on the ground state hfs interval alone. 
A combined result of constraints from the $1s$ hfs interval and $D_{21}$ can be found in his paper \cite{karshenboim_hyperfine_2011}.

Recently, experimental spectra of various simple atomic systems, including antiprotonic helium, muonium, positronium, helium, and hydrogen, have been reanalyzed and used to place constraints on exotic spin-dependent interactions between electrons and antiprotons ($e$-$\overline{p}$), electrons and antimuons ($e$-${\mu}^+$),
electrons and positrons ($e$-$e^+$), 
electrons and electrons ($e$-$e$), and electron and protons ($e$-${p}$) by \citet{fadeev_pseudovector_2022} following the theoretical framework in \citet{fadeev_revisiting_2019}.
These analyses were performed for the axial-vector/axial-vector potential $V_{AA}(\v{r})$, which receives contributions mainly from the $V_2|_{AA}$ and $V_3|_{AA}$ terms, as well as the pseudoscalar/pseudoscalar potential, for which the main contribution comes from the $V_3|_{pp}$ term, see Eq.\,\eqref{pseudoscalar-pseudoscalar_potential}. For clarity, we present the results from $V_3$. To explicitly see the effects of including both $V_2$ and $V_3$ terms, the readers are referred to the paper by \citet{fadeev_pseudovector_2022}.
In Figs.\,\ref{gaga-fig1}\,(a) and \ref{gaga-fig2}\,(a), we extracted and presented limits on $g_A g_A$, transformed from limits on $g_p g_p$, based on the $V_3|_{pp}$ term. The same approach is applied in a few other works mentioned below. Note that \citet{ficek_constraints_2017} also studied the $V_3$ term but without the $\delta$ term.

\citet{fadeev_pseudovector_2022}'s key contribution is the emphasis on the $V_3$ potential term, which is proportional to the inverse square of the mass $M$ of the mediating spin-1 boson, as discussed earlier in this Sec.\,\ref{subsec_g_Ag_A}. 
In addition, \citet{fadeev_pseudovector_2022} also emphasized that contact terms contribute to atomic spectroscopy. 
The contribution of short-range (contact) terms in atomic-scale experiments was discussed earlier in the paper \cite{fadeev_revisiting_2019}. 
Note that a similar contact term has been studied by \citet{karshenboim_hyperfine_2011}, with the form there applicable only for atomic $s$-wave states, whereas the form by \citet{fadeev_revisiting_2019} is more general and applicable to any atomic state. 
In some cases, the contact term gives the dominant or sole contribution to observable effects; for example, in the case of dipole-dipole interactions in the limit of a massless mediator, atomic $s$-wave states are only affected by the contact part of the interaction, while the contribution of the long-range part of the interaction vanishes.

In addition, the pseudoscalar-mediated spin-spin interaction term $V_{3}$ was used by \citet{terrano_short-range_2015} to place limits on the electron-electron coupling parameter $g_p^eg_p^e$, which has been transformed to $g_A^eg_A^e$ in our analysis. 
These limits were in the range $\lambda > 3\times 10^{-4}$\,m and are shown in Fig.\,\ref{gaga-fig1}\,(a). 
Further discussion of the experiment in \citet{terrano_short-range_2015} is given in Sec.\,\ref{subsec_g_Pg_P}. 

For the spin-spin velocity-dependent interaction term $V_8$, there are four sets of constraints on the electron-electron interaction parameter $g_A^e g_A^e$ given by 
\citet{hunter_using_2014}; \citet{ficek_constraints_2017}; \citet{ji_new_2018}; \citet{xiao_exotic_2024}.
\citet{hunter_using_2014} used spin-polarized geoelectrons to search for exotic interactions in the range $\lambda>10^3$\,m as discussed in Sec.\,\ref{subsec_g_Ag_V}. 
\citet{ji_new_2018} set the stringent limits on the $V_{8}$ term based on SmCo$_{5}$ spin sources and a SERF comagnetometer in the range $5 \times 10^{-2}\,\textrm{m}<\lambda<10^3$\,m. 
For the range $10^{-5}\,\textrm{m}<\lambda< \times 10^{-2}$\,m, \citet{xiao_exotic_2024} elucidated the role of particle thermal motion within their theoretical model and established constraints on interactions between electrons. For a shorter interaction range, \citet{ficek_constraints_2017} set the constraints. The latter two constraint lines are not plotted for the clarity of the figure. 

\subsubsection{\texorpdfstring{$e$-$N$}{e-N}}
In the range $\lambda>10^{7}$\,m, \citet{heckel_preferred-frame_2008} investigated the exotic spin- and velocity-dependent interaction term $V_{4+5}$ using their torsion pendulum experiments and  established the most stringent long-range limits on the coupling parameter product $g_A^e g_A^N$, as shown in Figure\,\ref{gaga-fig1}\,(b). 
Further experimental details are discussed in Sec.\,\ref{subsec_g_Ag_V}. 

For interaction ranges around $\lambda \sim 0.1$\,m, \citet{xiao_femtotesla_2023,wu_experimental_2022,kim_experimental_2018} conducted investigations on the $V_{4+5}$ term using atomic magnetometers (see Sec.\,\ref{METH1.SENS.AM}). 
\citet{kim_experimental_2018} used a SERF magnetometer to study the $V_{4+5}$ term and established the first limits on $g_A^e g_A^N$ in this interaction range, with further details discussed in Sec.\,\ref{subsec_g_sg_s}. 
\citet{wu_experimental_2022} improved the constraints for $\lambda>2\times 10^{-2}$\,m using an atomic-magnetometer array, see Sec.\,\ref{subsec_g_Ag_V} for more details. 
Recently, \citet{xiao_femtotesla_2023} further improved the constraints for $\lambda>7\times 10^{-4}$\,m by employing diffused optically polarized atoms, effectively reducing undesirable effects associated with optical pumping. 
Consequently, they established the most stringent limits on $g_A^eg_A^N$ in this range, with detailed discussions of the experiment provided in Sec.\,\ref{subsec_g_sg_s}.

For interaction ranges around $\lambda \sim 10^{-5}$\,m, \citet{ding_constraints_2020} searched for the exotic interaction term $V_{4+5}$ by measuring the force between a gold sphere and a microfabricated magnetic structure using a cantilever (see Sec.\,\ref{METH1.SENS.MS.MMO}). 
In the range around $10^{-4}$\,m, \citet{wu_improved_2023} employed an ensemble-NV-diamond magnetometer to investigate the $V_{4+5}$ interaction and established the most stringent limits on $g_A^e g_A^N$. 
For interaction ranges on the order of $10^{-7}$\,m, \citet{wu_spin-mechanical_2023} reported the use of an on-chip detector to search for hypothetical spin- and velocity-dependent interactions, by placing a micro-scale diamond with single NV spins above a micro-mechanical resonator. 
Detailed discussion of these experiments can be found in Sec.\,\ref{subsec_g_sg_s}.

\subsubsection{\texorpdfstring{$e$-$n$}{e-n}}\label{sec:gAgA_en}
In the range $\lambda> 10^{3}$\,m, \citet{hunter_using_2013} place the strongest constraints on the axial-vector/axial-vector interaction between electrons and neutrons, $g_A^e g_A^n$. 

In the interaction range $10^{-2}\,\textrm{m} < \lambda<10^{3}$\,m, \citet{almasi_new_2020} established the most stringent constraints on electron-nuclear spin-dependent forces, using a rotatable 
source made with a structure of
SmCo$_5$ 
covered by pure iron
[the same type as in \citeauthor{ji_searching_2017}, (\citeyear{ji_searching_2017}, \citeyear{ji_new_2018}), see Sec.\,\ref{METH1.SOUR.ESS} for more details.]  and a Rb-$^{21}$Ne comagnetometer \cite{smiciklas_new_2011}, see more details in Sec.\,\ref{METH1.SENS.AM}. 
To convert the measured value of anomalous fields to limits on spin-spin interactions, \citet{almasi_new_2020} express the energy shift due to the exotic potentials for neutrons, protons and electrons as $\eta_n V_n  + \eta_p V_p - V_e = \mu_{^{21}\textrm{Ne}} b_y^n - \mu_\textrm{B} b^e_y$, 
where $\mu_{^{21}\textrm{Ne}}$ denotes the magnetic moment of $^{21}$Ne, $\eta_n$ and $\eta_p$ 
denote the fractions of neutron and proton spin polarization in $^{21}$Ne, and $b_y^{n,e}$ are the anomalous magnetic fields that couple to the $^{21}$Ne nuclear spin and $^{87}$Rb electron spin, respectively, in the $y$ direction of the comagnetometer. 
\citet{almasi_new_2020} studied the exotic potential term $V_2$ ($V_3$) between $e$-$e$ and $e$-$n$, the results are shown in Fig.\,\ref{gaga-fig1}\,(a,\,b).
 

In the range $2 \times 10^{-13}\,\textrm{m} < \lambda < 10^{-2}$\,m, \citet{karshenboim_hyperfine_2011} set the most stringent constraints on spin-dependent interactions mediated by axial-vector light bosons. 
\citet{karshenboim_hyperfine_2011,karshenboim_constraints_2010,karshenboim_precision_2010} compared spectroscopic measurements of hyperfine structure intervals to quantum electrodynamics calculations (see Sec.\,\ref{METH2.PM.AHS}) for various atomic systems in order to derive constraints on axial-vector interactions. 
In contrast to his earlier work, \citet{karshenboim_hyperfine_2011} constructed a specific difference 
of the hyperfine structure intervals in the $1s$ and $2s$ states
that is only sensitive to the $V_2$ term, in order to disentangle the $V_2$ term from a contact term, which arises in the spin-spin interaction alongside the $V_2$ term. 
Previous constraints on spin-spin interactions were derived from data pertaining to the hyperfine structure interval of the $1s$ state in light hydrogenlike atoms \cite{karshenboim_constraints_2010,karshenboim_precision_2010}, where the accuracy was constrained either by the hyperfine structure internal measurements or by the uncertainty in the related contribution of nuclear effects. 
The theory of the 
specific difference of the $1s$ and $2s$ hyperfine intervals, $D_{21} = 8 E_\textrm{hfs}(2s) - E_\textrm{hfs}(1s)$, however, is practically insensitive to short-range contributions (including nuclear effects and associated uncertainties).
Due to the cancellation of the nuclear contributions, which have significant uncertainties, the theory of the difference becomes more accurate in the case when $\lambda$ is larger than atomic distances. 
In hydrogen, deuterium and the $^3$He$^+$ ion, this makes it possible to constrain the product of coupling constants, $g_A^e g_A^N$, for the spin-spin electron-nucleon interaction at a level below one part in $10^{16}$ in the sub-keV boson mass region ($\lambda > 2\times 10^{-10}$\,m). 
For super-keV boson masses, the constraints derived in \citet{karshenboim_hyperfine_2011} from data pertaining to the $D_{21}$ 
become weak and must be combined with the findings of \citet{karshenboim_constraints_2010}, which considered the $1s$ hyperfine structure interval. 
Although the constraints in \citet{karshenboim_constraints_2010} are milder in the keV mass range and at smaller masses, they are better suited for extension to higher masses such as the MeV range ($\lambda \approx 2\times 10^{-13}$\,m). 
As a result, the constraints on exotic spin-spin interactions at the shortest scales have been set by \citet{karshenboim_constraints_2010}. Note that \citet{karshenboim_hyperfine_2011} studied (in his App.\,C) a $V_3$-type short-range term, which contains a contact delta potential and grows like $\propto 1/M^2$ at small boson masses, similar to our $V_3|_{AA}$ term. 

Besides the $V_2$ term, there are a few other sets of constraints on exotic interactions between $e$-$n$ that have been derived from experiments via the $V_3|_{AA}$ and $V_8$ terms. 
In the case of the exotic dipole-dipole potential $V_3|_{AA}$, in the range $2 \times 10^{-2} \, \textrm{m} < \lambda < 10$\,m, \citet{almasi_new_2020} set the most stringent constraints on the coupling parameter product $g_A^eg_A^n$; 
in the interaction range $4 \times 10^{-4} \, \textrm{m} < \lambda < 2 \times 10^{-2}$\,m, \citet{wang_limits_2022} set the most stringent constraints on $g_A^e g_A^n$, which is described in more detail in Sec.\,\ref{subsec_g_Pg_P}. 
For the velocity-dependent exotic spin-spin potential term $V_8$, \citet{hunter_using_2014} used geoelectrons, which were discussed in Sec.\,\ref{subsec_g_Ag_V}, to set the only bounds thus far on $g_A^e g_A^n$.

\subsubsection{\texorpdfstring{$e$-$p$ \& $e$-$\overline{p}$}{e-p \& e-p̅}}
For the exotic axial-vector/axial-vector interaction between the fermion pair $e$-$p$, there are four sets of limits, as shown in Fig.\,\ref{gaga-fig2}(a). 
Two of these sets are for long interaction ranges and were derived by using geoelectrons to probe the potential terms $V_2$ \cite{hunter_using_2013} and $V_8$ \cite{hunter_using_2014} that we discuss in Sec.\,\ref{subsec_g_Ag_V}. 

The other two sets of constraints are for shorter interaction ranges and were obtained by comparing spectroscopic measurements of hyperfine splitting intervals to quantum electrodynamics calculations 
in atomic hydrogen
using the $V_2$ \cite{karshenboim_hyperfine_2011} 
and $V_3$ \cite{fadeev_pseudovector_2022} terms. 
For more details on Karshenboim's work and \citet{fadeev_pseudovector_2022}, see Sec.\,\ref{subsec_g_Ag_A_ee}. 
Note that improved constraints from hydrogen based on the analysis of the $D_{21}$ interval \cite{karshenboim_precision_2010} and the latest experimental results \cite{bullis_ramsey_2023} and theory \cite{yerokhin_electron_2008} are presented by \citet{cong_improved_2024}. 

For the exotic semi-leptonic e-$\overline{p}$ interaction, \citet{ficek_constraints_2018} compared theoretical calculations and spectroscopic measurements of hyperfine splitting intervals in antiprotonic helium ($^4$He$^+$$\bar{p}$) 
to set constraints on the $V_2$, $V_3$, $V_{4+5}$ and $V_8$ terms. 
They focused on antiprotonic helium with the antiproton in the $(n,l) = (37,35)$ state and the electron in the $(n,l)=(1,0)$ state. 
Here $n$ and $l$ denote the principal and orbital quantum numbers, respectively. 
The transition energies between the relevant hyperfine-structure states were measured experimentally by \citet{pask_antiproton_2009} and theoretically calculated by \citet{korobov_calculation_2008}. 
Comparison of these values gives $\Delta E = 0.3(1.3)$\,MHz (e.g. $\Delta E = 2.2$\,MHz at 90\%\,C.L.), which leads to the constraints presented in Fig.\,\ref{gaga-fig2}\,(a).

\subsubsection{\texorpdfstring{$N$-$N$}{N-N}}
\label{subsec_g_Ag_A:N-N}
In this section, we present limits on $g_A g_A$ for interaction between nucleons, including the combinations neutron-nucleon ($n$-$N$), proton-neutron ($p$-$N$), proton-proton ($p$-$p$), neutron-proton ($n$-$p$) and neutron-neutron ($n$-$n$), as shown from left to right in Fig.\,\ref{gaga-fig2}\,(b). 
Although many results are presented here, some of them, such as the results based on the $V_{3}$ term are discussed in detail in Sec.\,\ref{subsec_g_Pg_P}, while those pertaining to the $V_{4+5}$ term are discussed in Sec.\,\ref{subsec_g_sg_s}. 
Below, we introduce each limit individually.

\paragraph{\texorpdfstring{$p$-$N$ interactions}{p-N interactions}}
For the interaction range $6.5 \times 10^{-12}\,\textrm{m} < \lambda <  2.0 \times 10^{-8}$\,m (i.e., around the atomic scale), \citet{ledbetter_constraints_2013} derive constraints on the existence of the exotic spin-spin Yukawa potential term $V_2$ (as well as the dipole-dipole potential term $V_3$) and set the most stringent limits. 
The constraints are derived by comparing NMR measurements (\citealp{neronov_nmr_1975}; \citealp{beckett_temperature_1979}; \citealp{gorshkov_determination_1989}; \citealp{helgaker_nmr_2012})
with theoretical calculations 
(\citealp{oddershede_nuclear_1988}; \citealp{vahtras_indirect_1992}; \citealp{enevoldsen_correlated_1998}; \citealp{helgaker_nmr_2012})
of the $J$-coupling parameter in deuterated molecular hydrogen (HD), see Sec.\,\ref{METH2.PM.NMR} for more details. 
Note that for consistency with the per-nucleon normalisation in our Eq.\,\eqref{ANZ}, the constraints from \citet{ledbetter_constraints_2013}, which involved the combination $g_A^N = g_A^p + g_A^n$, are translated here to constraints on the combination $g_A^N = (g_A^p + g_A^n)/2$ in the case of deuteron.\footnote{We remark that the case of deuteron, which has two unpaired nucleon spins (one proton and one neutron), is special. Most non-deformed nuclei with non-zero spin have only one unpaired nucleon spin, in which case $g_A^N = g_A^{n,p}$ is already normalised correctly per-nucleon without the need for an extra factor of $1/2$.}
The constraints derived in this study outperform previous laboratory limits that were derived from studies of spin-exchange collisions between $^3$He and Na atoms \cite{jackson_kimball_constraints_2010}. 
\citet{ledbetter_constraints_2013} also suggest that spin systems with a small $J$-coupling may help to significantly improve these limits.
Chemically unbound systems, such as liquid $^3$He-H$_2$ mixtures \cite{hiza_liquid-vapor_1981}, can have nonzero $J$-coupling due to second-order hyperfine effects when the constituent species form van der Waals molecules. The $J$-coupling in these systems can be orders of magnitude smaller than that in HD. For relatively simple systems, the theoretical value of the $J$-coupling may be able to be calculated with an accuracy approaching that for HD. This combination of factors could significantly reduce the absolute uncertainty in the determination of anomalous energy shifts, and consequently improve sensitivity to exotic spin-dependent interactions.

Additionally, in the range $\lambda > 0.1$\,m, \citet{wei_constraints_2022} established a limit on $g_A^pg_A^N$ for the exotic potential term $V_{4+5}$ using a K-Rb-$^{21}$Ne comagnetometer, see Sec.\,\ref{subsec_g_sg_s} for more details. 

\paragraph{\texorpdfstring{$p$-$p$ interactions}{p-p interactions}}
For the range $\lambda >10^{-11}$\,m,
\citet{ramsey_tensor_1979} constrained the $V_3$ term based on spectroscopy experiments
experiments with the hydrogen molecule (H$_2$). 
By comparing the measurements of \cite{harrick_nuclear_1953} to theoretical calculations of the magnetic dipole-dipole interaction between the protons in H$_2$, \citet{ramsey_tensor_1979} set the only constraint on the exotic interaction between protons. 
Note that we have taken the limit presented in Fig.\,16 of Ref.\,\citet{safronova_search_2018} and convert this into a bound on $V_3|_{AA}$. 

\paragraph{\texorpdfstring{$n$-$p$ interactions}{n-p interactions}} In the interaction range $\lambda > 3.7 \times 10^{-10}$\,m, \citet{jackson_kimball_constraints_2010} compared experimental measurements (\citealp{soboll_spin_1972}; \citealp{borel_spin-exchange_2003})
and theoretical estimates (\citealp{walker_estimates_1989}; \citealp{tscherbul_collision-induced_2009})
of spin-exchange cross sections between Na and $^3$He atoms to constrain exotic spin-dependent forces. 
For the Na-$^3$He collisions, the interaction of colliding atoms is dominated by the spin-independent interatomic potential while spin exchange is dominated by a much smaller, spin-dependent potential arising from penetration of the Na valence electron wavefunction inside the $^3$He nucleus \cite{walker_spin-exchange_1997}.
One can calculate the contribution of exotic spin-dependent potentials generated by the exchange of new-force-mediating particles to the spin-exchange cross sections and thus constrain the range and coupling strengths of such hypothetical particles from the agreement between experiment and conventional theory. 
The work of \citet{jackson_kimball_constraints_2010} established limits on the previously unconstrained exotic spin-spin Yukawa-type $V_2$ potential term between neutrons and protons. 
In addition, the dipole-dipole potential term $V_3$ in the range $2.0 \times 10^{-11}\,\textrm{m} < \lambda < 2.0 \times 10^{-6}$\,m and the velocity- and spin-dependent interaction term $V_8$ in the range $\lambda > 4.9 \times 10^{-11}$\,m were also constrained.

\paragraph{\texorpdfstring{$n$-$N$ interactions}{n-N interactions}} In the range $\lambda>10^{-12}$\,m, \citet{haddock_search_2018}; \citet{voronin_constraint_2020}; \citet{parnell_search_2020}; \citet{su_search_2021}; \citet{wei_constraints_2022}; \citet{wu_new_2023}
collectively set the most stringent constraints via the potential term $V_{4+5}$, see Sec.\,\ref{subsec_g_Ag_V} for more details on \citet{su_search_2021} and Sec.\,\ref{subsec_g_sg_s} for details of the other experiments.

\paragraph{\texorpdfstring{$n$-$n$ interactions}{n-n interactions}}
For interaction ranges around $\lambda \sim 10$\,m, \citet{vasilakis_limits_2009} has set the most stringent constraints on the dipole-dipole interaction term $V_3|_{pp}$. 
In addition, this work also provided constraints on couplings to light vector bosons, including bounds on $g_A g_A$ via the $V_2$ term and bounds on $g_A g_V$ via the $V_{11}$ term. 
Further details about this experiment can be found in Sec.\,\ref{subsec_g_Pg_P}. 

\subsubsection{Future perspectives} 

There are many intriguing possibilities for probing exotic interaction mediated by new spin-1 axial-vector bosons in the future. 

For the $V_{2}$ term, \citet{chen_optomechanical_2020}; \citet{wang_proposal_2023}; \citet{guo_searching_2024}
propose novel experimental methods.
\citet{chen_optomechanical_2020} introduce a spin-resonator hybrid mechanical system, where an NV center is magnetically coupled to a cantilever resonator, and the probe absorption spectrum can be used to measure the exotic dipole-dipole interaction. 
\citet{wang_proposal_2023} propose a novel force sensor based on a cantilever to measure the exotic interactions in a periodic stripe magnetic source structure and a closed-loop magnetic structure integrated with the cantilever. 
The sensitivity to $g_A^e g_A^e$ in the interaction range $3 \times 10^{-8} \, \textrm{m} < \lambda < 10^{-3}$\,m is expected to be better than the sensitivities demonstrated in 
\citet{kotler_constraints_2015}; \citet{rong_constraints_2018}; \citet{jiao_searching_2020}.
\citet{guo_searching_2024} proposed to use a diamond-based vector magnetometer with ensembles of nitrogen-vacancy centers along different axes to search for exotic spin-spin interactions between electrons and nucleons. The constraints via $V_2$ can be improved in the interaction range 
$\sim 10^{-8} \, \textrm{m} < \lambda < 10^{-2}$\,m. 

Considering the significant impact of the $V_3$ term on the overall landscape of constraints realised in recent years, the NMR experiments proposed in \citet{arvanitaki_resonantly_2014}, which may establish unprecedented sensitivity to $g_A^e g_A^N$ and $g_A^N g_A^N$ for interaction ranges around $\lambda \sim 10^{-2}$\,m, appear particularly promising. 
More details about this proposed experiment can be found in Sec.\,\ref{gpgs_emerging}. 
In addition, \citet{guo_searching_2024} may also improve the constraints for $e$-$N$ via $V_3$
within the interaction range $ 10^{-8} \,\textrm{m} < \lambda < 10^{-4}$\,m.


\subsection{\texorpdfstring{Vector/Vector interaction $g_Vg_V$}{Vector/Vector interaction gVgV}}
\label{subsec_g_Vg_V}

Experimental constraints on the exotic vector/vector combination of parameters $g_V g_V$ come from the potential terms $V_1$, $V_{2}+V_{3}$ and $V_{4+5}$: 

\begin{equation}
\label{gVgV_V1}
V_1|_{VV} = g^X_V g^Y_V \frac{\hbar c}{4\pi}\frac{e^{-{r}/{\lambda}}}{r} \, , 
\end{equation}

\begin{widetext}
\begin{equation}
\label{gVgV_V23}
(V_2+V_3)|_{VV} = g_V^Xg_V^Y\frac{\hbar^3}{16\pi c m_Xm_Y}\left[ \v{\sigma}_X \cdot \v\sigma_Y^{\,\prime} \left[ \frac{1}{r^3} + \frac{1}{\lambda r^2} + \frac{1}{\lambda^2 r} - \frac{8 \pi}{3} \delta(\v{r}) \right]  -  \left( \v{\sigma}_X \cdot \hat{\v{r}} \right) \left( \v\sigma_Y^{\,\prime}\cdot \hat{\v{r}} \right)  \left( \frac{3}{r^3} + \frac{3}{\lambda r^2} + \frac{1}{\lambda^2 r} \right)  \right] e^{-{r}/{\lambda}} \, , 
\end{equation}
\begin{equation}
\label{gvgv_V45new}
\begin{aligned}
V_{4+5}|_{VV}&=\frac{g_V^X g_V^Y}{4} \frac{\hbar^2}{c} \left\{ \boldsymbol{\sigma}_X \cdot \left( \frac{\boldsymbol{p}_X}{m_X^2} \times \hat{\boldsymbol{r}} \right) 
 - 2 \boldsymbol{\sigma}_X \cdot \left( \frac{\boldsymbol{p}_Y}{m_X m_Y} \times \hat{\boldsymbol{r}} \right)
, \left( \frac{1}{r^2} + \frac{1}{\lambda r} \right) \frac{e^{-{r}/{\lambda}}}{8 \pi} \right\} \\
 &\Rightarrow g_V^X g_V^Y \frac{\hbar^2}{16\pi c}\frac{2m_X+m_Y}{m_X(m_X+m_Y)}\boldsymbol{\sigma}_X \cdot(\boldsymbol{v} \times \hat{\boldsymbol{r}}) \left(\frac{1}{r^2}+\frac{1}{\lambda r}\right) e^{-{r}/{\lambda}} \, , 
\end{aligned}
\end{equation}
\end{widetext}
with the second line of $V_{4+5}|_{VV}$ [Eq.\,\eqref{gvgv_V45new}] expressed in a commonly used form for macroscopic experiments.

We note that, similarly to the axial-vector/axial-vector case in Sec.\,\ref{subsec_g_Ag_A}, the potential associated with the vector/vector interaction also includes contributions from the $V_3$ and $V_{4+5}$ terms; however, there are key differences between the two cases. 
In the axial-vector case, the $V_3|_{AA}$ term diverges as $\propto \lambda^2$ as $\lambda$ increases -- there is no such divergence for the closely related $(V_2+V_3)|_{VV}$ term in the vector case. 
In the axial-vector case, $V_{4+5}|_{AA}$ only has spin-momentum correlations of the type $\boldsymbol{\sigma}_X \cdot \boldsymbol{p}_Y$ and $\v\sigma_Y^{\,\prime} \cdot \boldsymbol{p}_X$, whereas in the vector case, $V_{4+5}|_{VV}$ has additional correlations of the type $\boldsymbol{\sigma}_X \cdot \boldsymbol{p}_X$ and $\v\sigma_Y^{\,\prime} \cdot \boldsymbol{p}_Y$. 
See Sec.\,\ref{Subsec:Fformalism} for more and related discussion.

Prior analyses in the literature for the vector case have considered the sole term $V_3$ (defined in the equation below) instead of the combined term $V_2+V_3$: 
\begin{widetext}
\begin{equation}
\label{gVgV_V3}
\begin{aligned}
V_{3}|_{VV}=g_V^Xg_V^Y\frac{\hbar^3}{16\pi c m_Xm_Y}\left[ \v{\sigma}_X \cdot \v\sigma_Y^{\,\prime} \left[ \frac{1}{r^3} + \frac{1}{\lambda r^2}  + \frac{4 \pi}{3} \delta(\v{r}) \right]  -  \left( \v{\sigma}_X \cdot \hat{\v{r}} \right) \left( \v\sigma_Y^{\,\prime}\cdot \hat{\v{r}} \right)  \left( \frac{3}{r^3} + \frac{3}{\lambda r^2} + \frac{1}{\lambda^2 r} \right)  \right] e^{-{r}/{\lambda}} \, . 
\end{aligned}
\end{equation}
\end{widetext}
The difference between the $V_3$ limits reported in the literature and the (correct) $V_2+V_3$ limits that are appropriate for $g_Vg_V$ is relatively small on a logarithmic scale. Therefore, while future analyses should be done with the $V_2+V_3$ combination noted in Sec.\,\ref{Subsec:Fformalism} and in \citet{fadeev_revisiting_2019}, here in this review we choose to keep the $V_3$ bounds from earlier literature. 
For $Mr \ll 1$, when $\exp(-Mr) \approx 1$ and the constraints on $g_V g_V$ are generally at their strongest and are approximately independent of $M$, the typical relative difference between bounds on $V_{3}|_{VV}$ and $(V_2 + V_3)|_{VV}$ is $\mathcal{O}(Mr)^2 \ll 1$.
For $Mr \gg 1$, when the exponential character of the function $\exp(-Mr)$ dominates over the power-law terms, the relative difference in the shapes of exponential tails of limit curves for $V_{3}|_{VV}$ and $(V_2+V_3)|_{VV}$ is generally larger (but still small on the log-log scale).

\begin{figure} [!htbp]
\begin{center}
\includegraphics[width=0.48\textwidth]{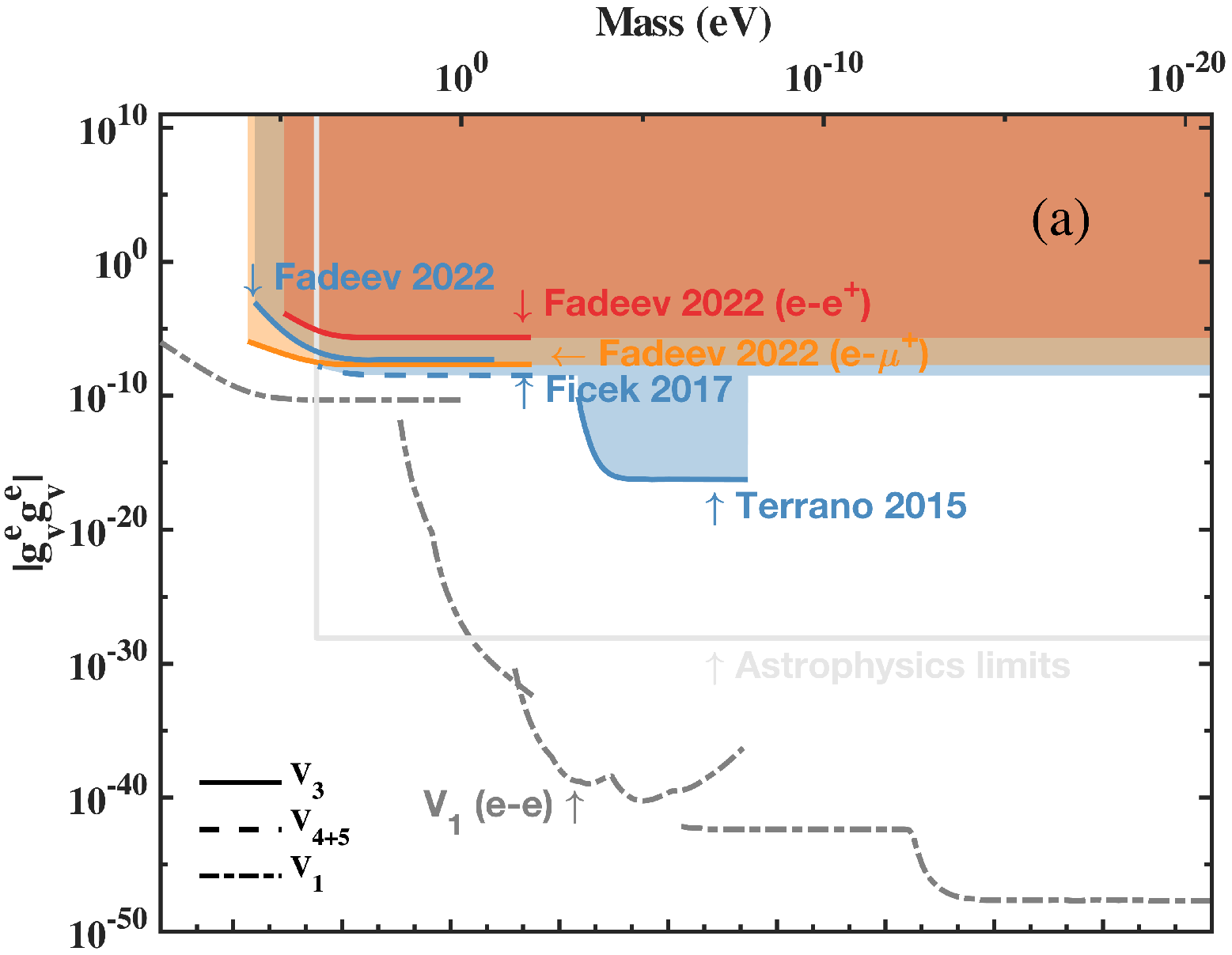}

\includegraphics[width=0.48\textwidth]{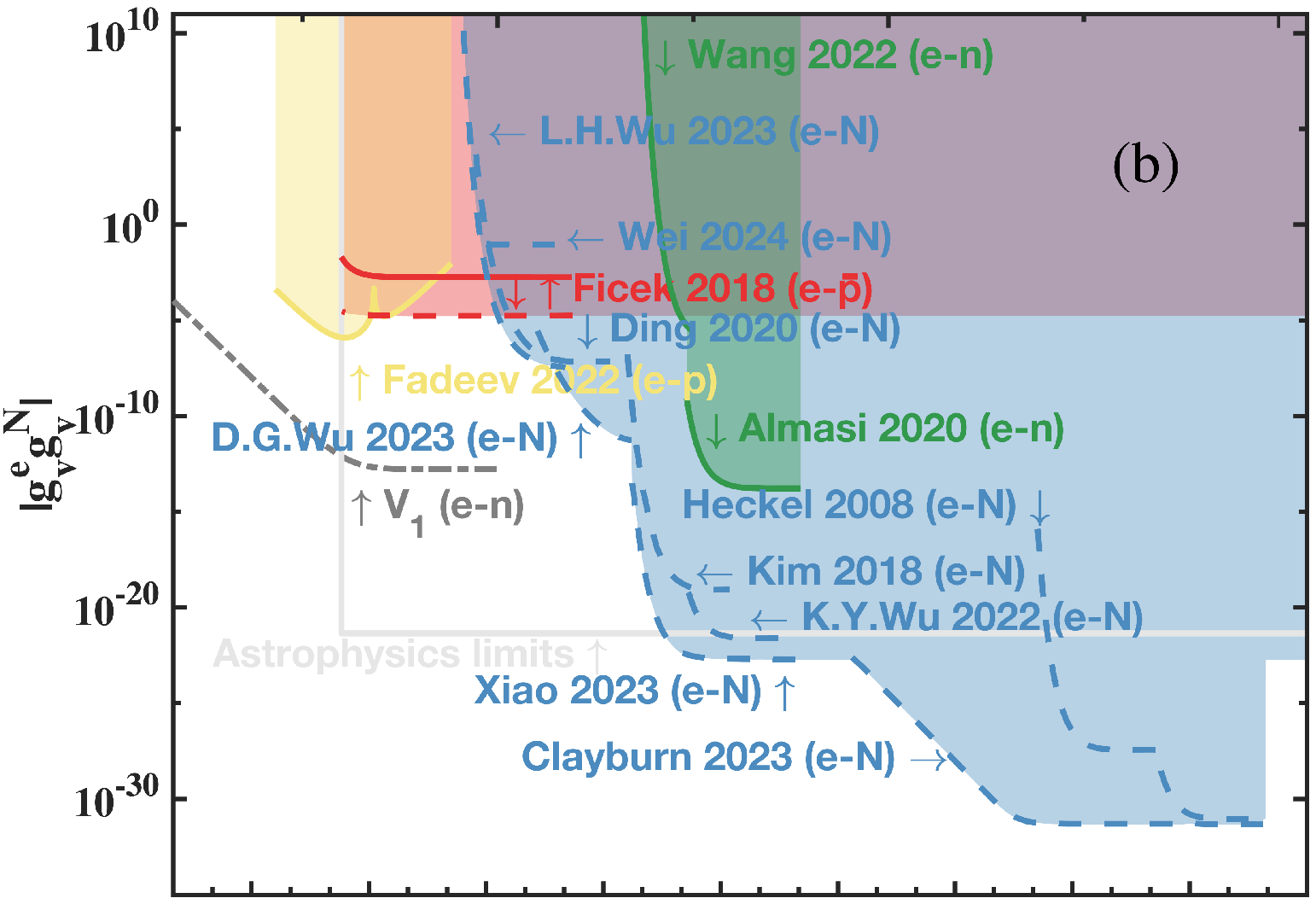}

\includegraphics[width=0.48\textwidth]{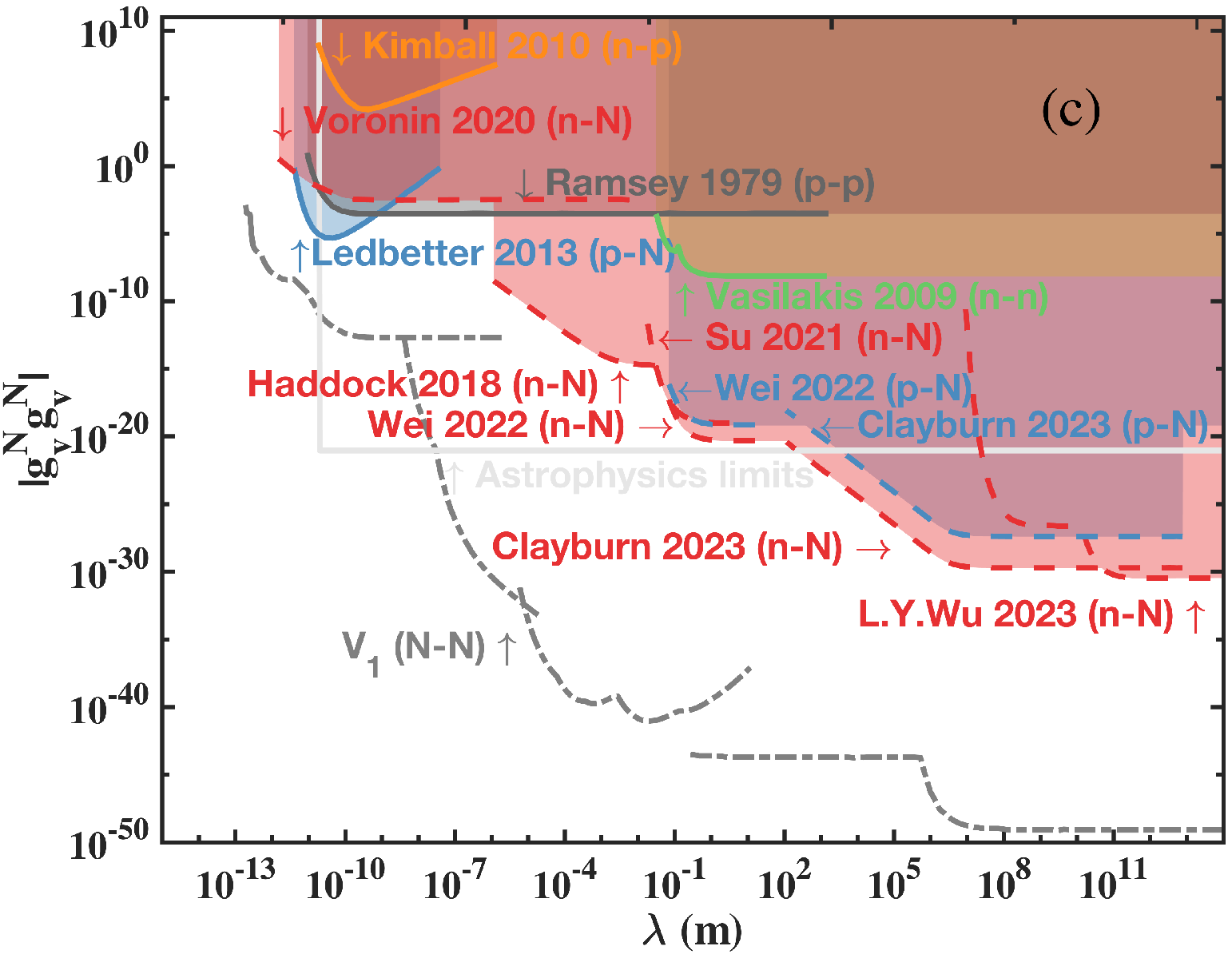}
\end{center}
\caption{Constraints, depicted by coloured regions, on the coupling constant product $g_A g_A$ as a function of the interaction range $\lambda$ shown on the bottom x-axis. 
The top x-axis represents the new spin-1 boson mass $M$.  
The subscript $V$ refers to the vector coupling. 
Constraints are shown for \textbf{(a)} $e$-$e$ (blue), $e$-${\mu}^+$ (green), $e$-$e^+$ (red); 
\textbf{(b)} $e$-$N$ (blue), $e$-$n$ (green), $e$-$p$ (yellow), $e$-$\overline{p}$ (red); 
\textbf{(c)} $n$-$N$, $p$-$N$, $p$-$p$, $n$-$p$ and $n$-$n$ (various colors). 
Results are shown for $V_3$ (solid line) and $V_{4+5}$ (dashed line) terms.
Additional bounds from the spin-independent $V_1$ term (dash-dotted, dark grey line) and astrophysics (solid, light grey line) are elaborated in Appendices\,\ref{appendix_V1} and \ref{appendix_comp.}.
} 
\label{gVgV-fig}
\end{figure}

Figure~\ref{gVgV-fig} shows the current laboratory constraints on $g_V^X g_V^Y$ between the studied particle pairs (a) $e$-$e$, $e$-$e^+$ and $e$-$\mu^+$, (b) $e$-$N$, $e$-$n$, $e$-$p$ and $e$-$\overline{p}$, and (c) $n$-$N$, $p$-$N$, $p$-$p$, $n$-$p$ and $n$-$n$. 
Constraints from the spin-independent $V_1$ term and astrophysical bounds are also presented for comparison, see Sec.\,\ref{METH2.Astro.} and Appendices\,\ref{appendix_V1} and \ref{appendix_comp.} for more details.

\subsubsection{\texorpdfstring{$e$-$e$, $e$-$e^+$ \& $e$-$\mu^+$}{e-e, e-e⁺ and e-μ⁺}}
As illustrated in Fig.\,\ref{gVgV-fig}\,(a), \citet{fadeev_pseudovector_2022} and \citet{terrano_short-range_2015} placed constraints on the spin-dependent $V_{3}$ interaction term for electrons. 
Both studies reported constraints on the pseudoscalar/pseudoscalar interaction term $V_3|_{pp}$; however, as $V_{3}|_{pp}$ has the same structure as the vector-vector interaction term $V_{3}|_{VV}$ in Eq.\,\eqref{gVgV_V3}, we have converted their constraints on $g_p^e g_p^e$ to constraints on $g_V^eg_V^e$. 
\citet{fadeev_pseudovector_2022} used an atomic spectroscopy approach, while \citet{terrano_short-range_2015} applied a torsion-pendulum method, see Secs.\,\ref{subsec_g_Ag_A} and \ref{subsec_g_Pg_P} for more details. 
Through investigations of the electronic structure of $^4$He, \citet{ficek_constraints_2017} 
establish constraints on the vector-vector interaction between electrons via $V_{4+5}$.

For the range $\lambda > 2\times 10^{-13}$\,m, \citet{fadeev_pseudovector_2022} searched for spin-dependent interactions with positronium and muonium spectroscopy and established the only constraints on $g_V^e g_V^{e^+}$ and $g_V^e g_V^{{\mu}^+}$ via the $V_{3}$ term. 
Further details regarding the work of \citet{fadeev_pseudovector_2022} can be found in Sec.\,\ref{subsec_g_Ag_A}. 

\subsubsection{\texorpdfstring{$e$-$N$, $e$-$n$, $e$-$p$ \& $e$-$\overline{p}$}{e-N, e-n, e-p and e-p̅}}

For the $g_A^eg_A^N$ coupling combination, limits were set by several studies, including those of \citet{heckel_preferred-frame_2008}; \citet{clayburn_using_2023} for $\lambda>5\times10^{2}$\,m, and 
\citet{kim_experimental_2018}; \citeauthor{wu_experimental_2022}, \citet{wu_experimental_2022}; \citet{wu_improved_2023,wu_spin-mechanical_2023}; \citet{ding_constraints_2020}; \citet{xiao_femtotesla_2023}
for the range $4 \times 10^{-8}\,\textrm{m} < \lambda < 5\times10^{2}$\,m. 
These studies were based on the exotic spin- and velocity-dependent interaction term $V_{4+5}$. 
Further details on the work of \citet{heckel_preferred-frame_2008,wu_experimental_2022}; \citet{clayburn_using_2023} can be found in Sec.\,\ref{subsec_g_Ag_V}, and that of \citet{kim_experimental_2018}; \citet{ding_constraints_2020}; \citet{xiao_femtotesla_2023}; \citet{wu_improved_2023,wu_spin-mechanical_2023} in Sec.\,\ref{subsec_g_sg_s}.

For the 
$g_V^e g_V^n$ coupling combination, in the range $10^{-3}\,\textrm{m} < \lambda < 10$\,m, current constraints come form \citet{almasi_new_2020} and \citet{wang_limits_2022}. 
These two experiments take advantage of magnetometer and spin-based amplifier techniques, respectively. 
More details about the work of \citet{almasi_new_2020} and \citet{wang_limits_2022} are given in Secs.\,\ref{subsec_g_Ag_A} and \ref{subsec_g_Pg_P}, respectively. 

For the $g_V^eg_V^p$ coupling-constant combination, we present the result obtained using the $V_3$ term of \citet{fadeev_pseudovector_2022}; an improved result is given by \citet{cong_improved_2024}.

For the $g_V^eg_V^{\overline{p}}$ coupling combination, constraints have been set on the $V_{3}$ term (\citealp{ficek_constraints_2018}; \citealp{fadeev_pseudovector_2022}) and $V_{4+5}$ term \cite{ficek_constraints_2018} for the exotic interaction between the e-$\overline{p}$ fermion pair by comparing theoretical calculations and spectroscopic measurements of the hyperfine structure in antiprotonic helium. 
Details of both works are described in Sec.\,\ref{subsec_g_Ag_A}.

\subsubsection{\texorpdfstring{$N$-$N$}{N-N}}
In this section, we present bounds on $g_V g_V$ for interactions between nucleons, including the combinations $n$-$N$, $p$-$N$, $p$-$p$, $n$-$p$, and $n$-$n$, 
as depicted in Fig.\,\ref{gVgV-fig}\,(c).

\paragraph{n-N interactions}
For the range $\lambda>10^{-12}$\,m, 
\citet{haddock_search_2018}; \citet{voronin_constraint_2020}; \citet{su_search_2021}; \citet{wei_constraints_2022}; 
\citet{wu_new_2023}; 
\citet{clayburn_using_2023}
together set the most stringent constraints on an exotic interaction between a neutron and a nucleon ($n$-$N$) via the $V_{4+5}$ potential term, 
see Sec.\,\ref{subsec_g_Ag_V} for more details on 
\citet{su_search_2021}; \citet{clayburn_using_2023}; \citet{wu_new_2023},
and Sec.\,\ref{subsec_g_sg_s} for more details on the work of 
\citet{haddock_search_2018}; \citet{voronin_constraint_2020}; \citet{wei_constraints_2022}.

\paragraph{p-N interactions} 
In the range $\lambda>5\times 10^2$\,m, \citet{clayburn_using_2023} 
provides the best limit on $g_V^p g_V^N$ for the exotic potential term $V_{4+5}$ by combining their Earth model \cite{hunter_using_2013,hunter_using_2014} with bounds extracted from \citet{peck_limits_2012}, which uses an Hg-Cs comagnetometer, and \citet{jackson_kimball_constraints_2017}, which use a dual-isotope Rb comagnetometer.

In the range $0.1\,\textrm{m} < \lambda < 5\times 10^2$\,m, 
\citet{wei_constraints_2022} established a limit on $g_V^p g_V^N$ for the exotic potential term $V_{4+5}$ based on a K-Rb-$^{21}$Ne comagnetometer, see Sec.\,\ref{subsec_g_sg_s} for more details. 

For the interaction range $4.9 \times 10^{-12}\,\textrm{m} < \lambda < 7.9 \times 10^{-8}$\,m (i.e., around the atomic scale), \citet{ledbetter_constraints_2013} established limits on $g_p^p g_p^N$ via the $V_3$ term by comparing NMR measurements with theoretical calculations of the coupling between the magnetic moments of the two nuclei in HD, see Sec.\,\ref{subsec_g_Ag_A} for more details about \citet{ledbetter_constraints_2013}. 
We have transformed the limits on $g_p^p g_p^N$ to limits on $g_V^p g_V^N$ here. 

\paragraph{p-p interactions} For the range $\lambda >10^{-11}$\,m,
\citet{ramsey_tensor_1979} set constraints on $g_V^p g_V^p$ for the potential term $V_{3}$ at the molecular scale based on molecular hydrogen (H$_2$) spectroscopy, see Sec.\,\ref{subsec_g_Ag_A}.

\paragraph{n-p interactions} 
For the range $\lambda > 2 \times 10^{-11}$\,m,
\citet{jackson_kimball_constraints_2010} compared experimental measurements and theoretical estimates of spin-exchange cross sections between Na and $^3$He atoms to set constraints on $g_V^n g_V^p$ via the potential term $V_{3}$, see Sec.\,\ref{subsec_g_Ag_A} for more details.

\paragraph{n-n interactions} 
For the range around $\lambda \sim 10$\,m, \citet{vasilakis_limits_2009} set constraints on the dipole-dipole interaction term $V_3$ between neutrons, see Sec.\,\ref{subsec_g_Pg_P} for more detail. 

\subsubsection{Future perspectives}
For the experiments aimed at probing exotic interactions mediated by new spin-1 vector bosons, there could be more specific research in the future. 
For historical reasons, we have presented limits on $V_3$ in Fig.\,\ref{gVgV-fig}. 
However, future analyses should be conducted using the correct combination $(V_2+V_3)|_{VV}$. 
For instance, the proposed ferromagnetic torque sensor experiment \cite{vinante_surpassing_2021}, which aims to improve upon \citet{terrano_short-range_2015}, could probe $g_V g_V$ via the $(V_2+V_3)|_{VV}$ term whilst simultaneously probing $g_p g_p$ via the $V_3$ term.
The same is true for the proposed NMR experiment \cite{arvanitaki_resonantly_2014}, which could probe $g_V^N g_V^N$ for interaction ranges around $\lambda \sim 10^{-2}$\,m, see Sec.\,\ref{subsec_g_Pg_S} for more detail. 

The study of $g_V g_V$ could also benefit from new experimental probes of the $V_{4+5}$ term. 
In the interaction range $10^{-8}\,\textrm{m} < \lambda < 10^{-3}$\,m, \citet{chu_proposal_2022} proposes to use an NV sensor to probe possible exotic interactions between electrons and nucleons, with the projected sensitivity expected to surpass that of \citet{wu_improved_2023} and \citet{wu_spin-mechanical_2023}. 
See Sec.\,\ref{subsec_g_Ag_V} for more details of the proposed experiments in \citet{chu_proposal_2022}.

\subsection{Limits for massless spin-1 bosons}
\label{subsec_massless}

In the preceding sections, we outlined constraints on massive spin-1 bosons. 
In this section, we focus on summarizing the constraints related to tensor-type interactions mediated by massless spin-1 bosons. The concept of a massless spin-1 boson (see Sec.\,\ref{Sec:Intro-motiv-other-mysteries} for more details) emerges from the introduction of a new U(1) gauge symmetry, as occurs, for example, for the massless photon in the SM. 
The interaction between a massless spin-1 boson and fermions would lead to new long-range forces. 

The potentials associated with the tensor/tensor interaction ($V_{TT}$), pseudotensor/tensor interaction ($V_{\tilde{T}T}$) and pseudotensor/pseudotensor interaction ($V_{\tilde{T}\tilde{T}}$) are presented in Eqs.\,\eqref{tensor-tensor_potential}\,--\,\eqref{pseudotensor-pseudotensor_potential} in Sec.\,\ref{Subsec:Fformalism}. 
We establish constraints on combinations of the dimensionless interaction constants $\mr{Re}(C_X)$ and $\mr{Im}(C_X)$. 
This is achieved by translating the current best constraints on the spin-0 and spin-1 interaction constants associated with the corresponding $V_i$ terms in Eqs.\,\eqref{pseudovector-vector_potential}\,--\,\eqref{scalar-scalar_potential} into constraints on the relevant (pseudo)tensor interaction constants appearing in Eqs.\,\eqref{tensor-tensor_potential}\,--\,\eqref{pseudotensor-pseudotensor_potential}.

Another interesting possibility arises in the case of a massless spin-1 boson with purely axial-vector interactions. In this case, the $V_3|_{AA}$ term, which is present in the $V_{AA}$ potential (see Eq.\,\eqref{pseudovector-pseudovector_potential}) for a massive spin-1 boson and diverges as $\propto 1/M^2$ in the limit of a small (but nonzero) boson mass, is absent, while the $V_2$ term and other terms remain, giving rise to a qualitatively different phenomenology in this case.

\subsubsection{\texorpdfstring{${\mathrm{Re}(C_X)\mathrm{Re}(C_Y)}/{\Lambda^4}$}{Re(CX)Re(CY)/Λ⁴}}

The potential $V_{TT}$ at leading order contains the term $V_2+V_3$, which is the same as the $V_2+V_3$ term in the potential $V_{VV}$ [Eq.\,(\ref{vector-vector_potential})] when taken in the massless limit $M \rightarrow 0$, but with different constant prefactors. 
We note that most previous searches for the term $(V_2 + V_3)|_{VV}$ omitted the $V_2$ contribution. 
The $V_2$ contribution in $(V_2 + V_3)|_{VV}$ scales as $\propto M^2$ and hence vanishes in the limit $M \to 0$, meaning that it is sufficient to consider $V_3|_{VV}$ alone when placing limits on ${\mr{Re}(C_X)\mr{Re}(C_Y)}/{\Lambda^4}$. Alternatively, one may instead consider the $V_3|_{pp}$ term in the potential $V_{pp}$ [Eq.\,(\ref{pseudoscalar-pseudoscalar_potential})].

\begin{table}
\begin{threeparttable}
\centering
\caption{
Summary of limits on the coupling constant combinations $\mr{Re}(C_X)\mr{Re}(C_Y) / \Lambda^4$ and $\mr{Im}(C_X)\mr{Im}(C_Y) / \Lambda^4$ for a massless spin-1 boson. 
}
\renewcommand{\arraystretch}{1.8} 
\begin{tabular}{c|c|c} 

\hline
\hline
Fermion pairs & ${\mr{Re}(C_X)\mr{Re}(C_Y)}/{\Lambda^4}$ & Reference \\
& [${\mr{Im}(C_X)\mr{Im}(C_Y)}/{\Lambda^4}$] & \\
($X$-$Y$) & (eV$^{-4}$)& \\
\hline
$e$-$e$ & $2.2\times 10^{-52}$ (95\% C.L.) & \citet{terrano_short-range_2015}\\
$e$-$e^+$& $3.1 \times 10^{-41}$ (90\% C.L.) & \citet{fadeev_pseudovector_2022}\\
$e$-${\mu}^+$  & $4.0 \times 10^{-46}$ (90\% C.L.)& \citet{fadeev_pseudovector_2022}\\
\hline
$e$-$\overline{p}$& $4.0\times 10^{-42}$ (90\% C.L.)& \citet{ficek_constraints_2018}\\
$e$-$n$& $3.7\times 10^{-53}$ (95\% C.L.)& \citet{almasi_new_2020}\\
\hline
$p$-$p$  & $3.4\times 10^{-46}$ (90\% C.L.)& \citet{ramsey_tensor_1979}\\
$n$-$n$& $8.5\times 10^{-51}$ (1$\sigma$)& \citet{vasilakis_limits_2009}\\
\hline
\hline
\end{tabular}
\label{tabel_Massless-bounds1}
\end{threeparttable}
\end{table}

Therefore, we can translate constraints on $g_Vg_V$ into constraints on $\mr{Re}(C_X)\mr{Re}(C_Y)$ using the relation: 
\begin{equation}
\label{TT-translate}
   \frac{\mr{Re}(C_X)\mr{Re}(C_Y)}{\Lambda^4}=\frac{g_V^X g_V^Y}{16v_h^2m_Xm_Y} \, , 
\end{equation}
where $\Lambda$ is the ultraviolet energy-cutoff scale, and $v_h$ is the Higgs expectation value. 
We take the value $v_h = 246$\,GeV \cite{erler_electroweak_2004}. 
The energy scale $\Lambda$ is not universally defined and can vary depending on the specific theory. 
Experiments strictly constrain the combination ${\mr{Re}(C_X)\mr{Re}(C_Y)}/{\Lambda^4}$. 
Note that Eq.\,\eqref{TT-translate} is used to obtain bounds via correspondence (rather than being a strict relation that equates different sets of physical parameters), with the same being true for Eq.\,\eqref{PP-translate} below. 

Table \ref{tabel_Massless-bounds1} lists the existing best constraints on the tensor/tensor interaction parameters ${\mr{Re}(C_X)\mr{Re}(C_Y)}/{\Lambda^4}$ for various fermion pairs.

\subsubsection{\texorpdfstring{${\mathrm{Im}(C_X)\mathrm{Re}(C_Y)}/{\Lambda^4}$}{Im(CX)Re(CY)/Λ⁴}}

The potential $V_{\tilde{T}T}$ contains the terms $V_{9\pm10}$ and $V_{14q\pm15}$, which also appear in $V_{ps}$ [Eq.\,\eqref{scalar-pseudoscalar_potential}]. 
Note that the $V_{9\pm10}$ terms in the $V_{ps}$ potential give stronger bounds than the $V_{14q\pm15}$ terms, since the $V_{14q\pm15}$ terms in that case arise at higher order. 
However, the $V_{9\pm10}$ terms in the $V_{\tilde{T}T}$ potential will not generally give stronger bounds than the $V_{14q\pm15}$ terms, since the $V_{9\pm10}$ and $V_{14q\pm15}$ terms are formally of the same order in this case. 
Also, the structures of the $V_{9\pm10}$ terms in the $V_{ps}$ and $V_{\tilde{T}T}$ potentials in the massless limit are different: the $V_{9\pm10}|_{ps}$ term scales as $1/r^2$ in the massless limit, whereas $V_{9\pm10}|_{\tilde{T}T}$ scales as the spatial derivative of the $\delta(\v{r})$ function. 
This implies that macroscopic-scale experiments are practically insensitive to $V_{9\pm10}|_{\tilde{T}T}$.

Bounds
on $\mr{Im}(C_X)\mr{Re}(C_Y)$ 
could in principle be derived from constraints on
$V_{14q\pm15}|_{\tilde{T}T}$. 
Discussion of the necessity to consider $V_{14q\pm15}$ instead of $V_{15}$ can be found in 
in Sec.\,\ref{Subsec:Pairs-pots.}.
However, to date there are no experimental or observational constraints on $V_{14q\pm15}|_{\tilde{T}T}$ to our knowledge.
Furthermore,
the situation here is different compared to $(V_2+V_3)|_{VV}$ above, where the extra term added to $V_3$ vanishes in the massless limit. Therefore, we do not present limits for $\mr{Im}(C_X)\mr{Re}(C_Y)$ using $V_{15}|_{\tilde{T}T}$.
%
\subsubsection{\texorpdfstring{${\mathrm{Im}(C_X)\mathrm{Im}(C_Y)}/{\Lambda^4}$}{Im(CX)Im(CY)/Λ⁴}}

The potential $V_{\tilde{T}\tilde{T}}$ at leading order contains the term $V_3$, which is the same as the $V_3$ term in the potential $V_{pp}$ [Eq.\,(\ref{pseudoscalar-pseudoscalar_potential})] when taken in the massless limit $M \to 0$, except for different constant prefactors. 
Therefore, we can translate constraints on $g_p g_p$ into constraints on $\mr{Im}(C_X)\mr{Im}(C_Y)$ using the relation: 
\begin{equation}
\label{PP-translate}
   \frac{\mr{Im}(C_X)\mr{Im}(C_Y)}{\Lambda^4} = -\frac{g_p^X g_p^Y}{16 v_h^2 m_X m_Y} \, . 
\end{equation}

The bounds on $\mr{Im}(C_X)\mr{Im}(C_Y)/\Lambda^4$ are also presented in Tab.\,\ref{tabel_Massless-bounds1} since they are the same (numerically) as $\mr{Re}(C_X)\mr{Re}(C_Y)/\Lambda^4$, as both arise from a $V_3$ term. 


\subsubsection{\texorpdfstring{$g_Ag_A$ constraints for massless boson}{gAgA constraints for massless boson}}\label{gAgA_massless}

Additionally, we present a summary (Tab.\,\ref{tabel_Massless-bounds}) of constraints on axial-vector/axial-vector interactions for the special case when the spin-1 boson is massless -- in this case, the divergent $\propto 1/M^2$ $\mathcal{V}_3|_{AA}$ term in Eq.\,(\ref{pseudovector-pseudovector_potential}) is absent, because the massless spin-1 boson propagator lacks the momentum-dependent $k_\mu k_\nu / M^2$ term that is present in the massive spin-1 boson propagator. 
Most of the current best constraints on $g_A g_A$ for a massless spin-1 boson come from the leading-order $V_2$ term. See Figs.\,\ref{gaga-fig1} and \ref{gaga-fig2}, as well as Sec.\,\ref{subsec_g_Ag_A} for more details.

\begin{table}
\begin{threeparttable}
\centering
\caption{
Summary of limits on the coupling constant combination $g_A^X g_A^Y$ for a massless spin-1 boson. } 
\renewcommand{\arraystretch}{1.8} 
\begin{tabular}{c|c|c} 
\hline
\hline
Fermion & $g_A^X g_A^Y$ & Reference \\
pairs &&\\
\hline
$e$-$e$  & $5\times 10^{-47}$ (95\% C.L.)& \citet{hunter_using_2013}\\
$e$-$e^+$ & $2.1 \times 10^{-11}$ (1$\sigma$) & \citet{karshenboim_constraints_2010}\\
$e$-${\mu}^+$& $1.9 \times 10^{-15}$ (1$\sigma$) & \citet{karshenboim_constraints_2010}\\
\hline
$e$-$n$ & $5\times 10^{-47}$ (95\% C.L.)&\citet{hunter_using_2013}\\
$e$-$p$ & $1\times 10^{-46}$ (95\% C.L.) &\citet{hunter_using_2013}\\
$e$-$\overline{p}$ & $1.4 \times 10^{-11}$ (90\% C.L.)& \citet{ficek_constraints_2018}\\
\hline
$n$-$p$ & $2\times 10^{-10}$ (95\% C.L.) & \citet{jackson_kimball_constraints_2010}\\
$p$-$N$ &$8.2 \times 10^{-19}$ (95\% C.L.)&\citet{ledbetter_constraints_2013}\\
$p$-$N$  & $3 \times 10^{-28}$ (95\% C.L.) & \citet{clayburn_using_2023} \\
$n$-$n$ & $1.5 \times 10^{-40}$ (1$\sigma$) & \citet{vasilakis_limits_2009} \\
\hline
\hline
\end{tabular}
\label{tabel_Massless-bounds}
\end{threeparttable}
\end{table}


\subsection{\texorpdfstring{Pseudoscalar/Scalar interaction $g_p g_s$}{Pseudoscalar/Scalar interaction gpgs}}
\label{subsec_g_Pg_S}

A monopole-dipole force, sometimes called a mass-spin force, is a force between an unpolarized body and a spin-polarized one \cite{ohare_cornering_2020}. From the point of view of quantum field theory, the fundamental cause of the monopole-dipole force is a spin-0 field that has both a scalar interaction with some particles (characterized by the coupling constant $g_s$) and a pseudoscalar interaction with the same or other particles (characterized by the coupling constant $g_p$), see Eq.\,\eqref{Lagrangian_scalar}. 
The existence of a particle simultaneously possessing both scalar and pseudoscalar interactions necessitates CP violation; see, e.g., \citet{di_luzio_axion_2024} 
and references therein for some examples of such axion models.

The investigation of monopole-dipole interactions originated in the 1980s based on the theoretical considerations of \citet{moody_new_1984} who pointed out that axions could interact with nucleons and electrons via a scalar Yukawa interaction that violates $CP$. \citet{moody_new_1984} inspired a number of experiments, e.g., those of 
\citet{venema_search_1992}, \citet{youdin_limits_1996}, \citet{heckel_preferred-frame_2008}, \citet{bulatowicz_laboratory_2013}, \citet{afach_constraining_2015}, \citet{jackson_kimball_constraints_2017}, \citet{lee_improved_2018}, \citet{zhang_search_2023},
and the proposed targeted QCD axion search of \citet{arvanitaki_resonantly_2014}.
Recent review papers (\citealp{safronova_search_2018}; \citealp{ohare_cornering_2020}; \citealp{jackson_kimball_probing_2023})
provide presentations of constraints on monopole-dipole coupling constants $g_pg_s$ derived from laboratory experiments, as well as constraints derived from combinations of astrophysical observations and laboratory results \cite{raffelt_limits_2012}.

Since these previous reviews, several new experiments have been conducted. For instance, the QUAX team \citet{crescini_search_2022} has improved their pilot project \cite{crescini_improved_2017} with a larger-scale apparatus; experiments using single NV sensors \cite{rong_searching_2018} were enhanced by using ensembles \cite{liang_new_2022}, while \citet{stadnik_improved_2018} set new limits in the range $\lambda<10^{-4}$\,m based on results from atomic and molecular EDM experiments. 
Here we provide an up-to-date compilation of these constraints and examine the limits on the $g_pg_s$ coupling constants within the framework described in Sec.\,\ref{Subsec:Fformalism}.

This section is divided into four main categories, each representing different combinations of fermion species pairs: 
(1) electron-electron,
(2) electron-nucleon, (3) neutron-nucleon, and (4) proton-nucleon. 
Among them, couplings between electron or neutron spins and massive bodies has attracted the most attention. 
Experiments using torsion pendula \cite{heckel_preferred-frame_2008} and comagnetometers (\citealp{venema_search_1992}; \citealp{tullney_constraints_2013}; \citealp{jackson_kimball_constraints_2017}; \citealp{zhang_search_2023}),
among others, are discussed. 
The subsection on $g_p^pg_s^N$ examines the less studied proton-massive body interactions, which are derived from experiments with atomic vapor comagnetometers \cite{jackson_kimball_constraints_2017}. We note that some experimental constraints nominally reported for neutron couplings also imply constraints on proton couplings 
when the nuclear spin contents are decomposed into their underlying proton and neutron spin contents
\cite{jackson_kimball_nuclear_2015,stadnik_improved_2018}. 
The experiments used to search for monopole-dipole couplings span a wide range of interaction lengths, from $10^{-18}$\,m to $10^{14}$\,m.

\begin{figure*} [!htbp]
\begin{center}
\includegraphics[width=0.48\textwidth]{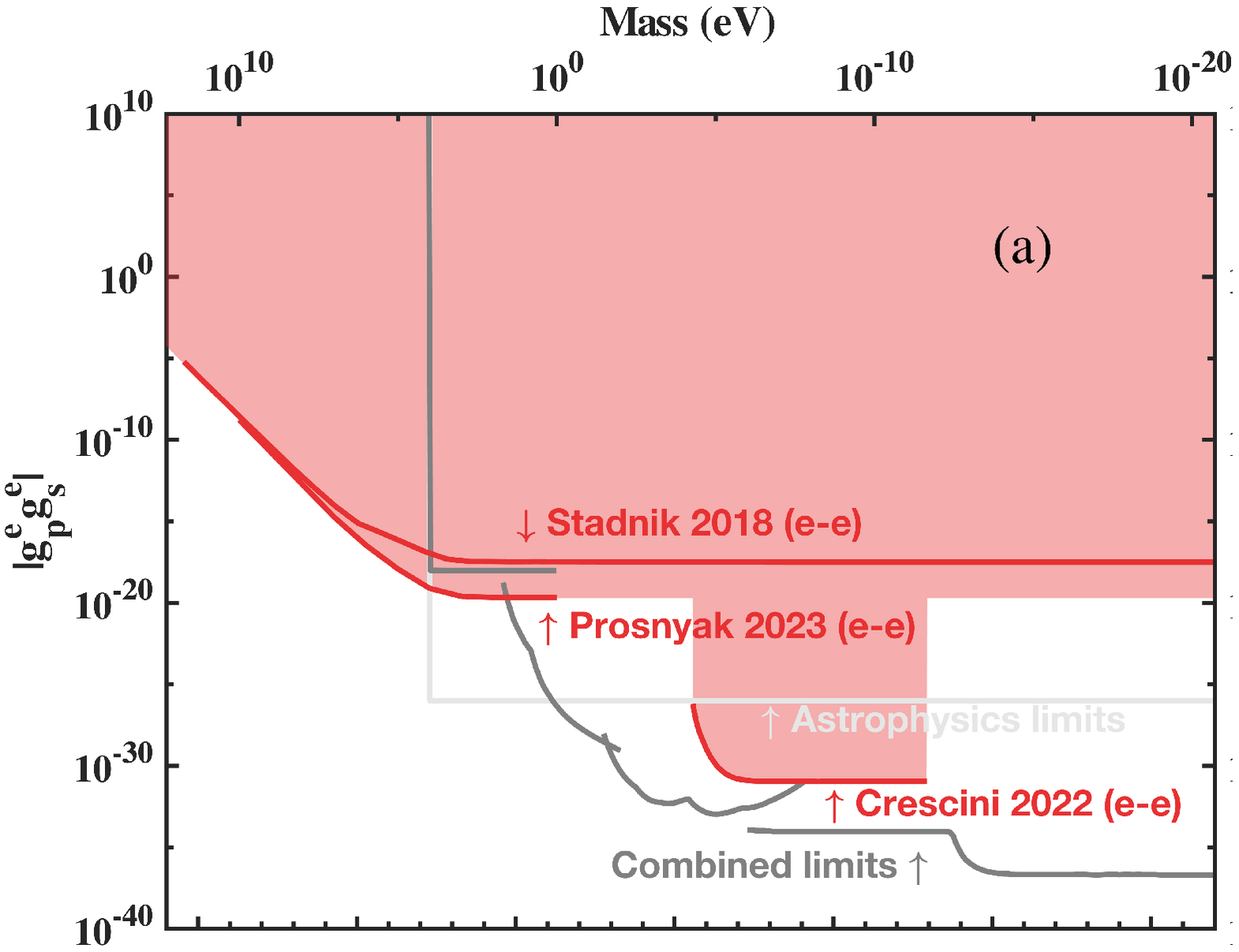}
\includegraphics[width=0.48\textwidth]{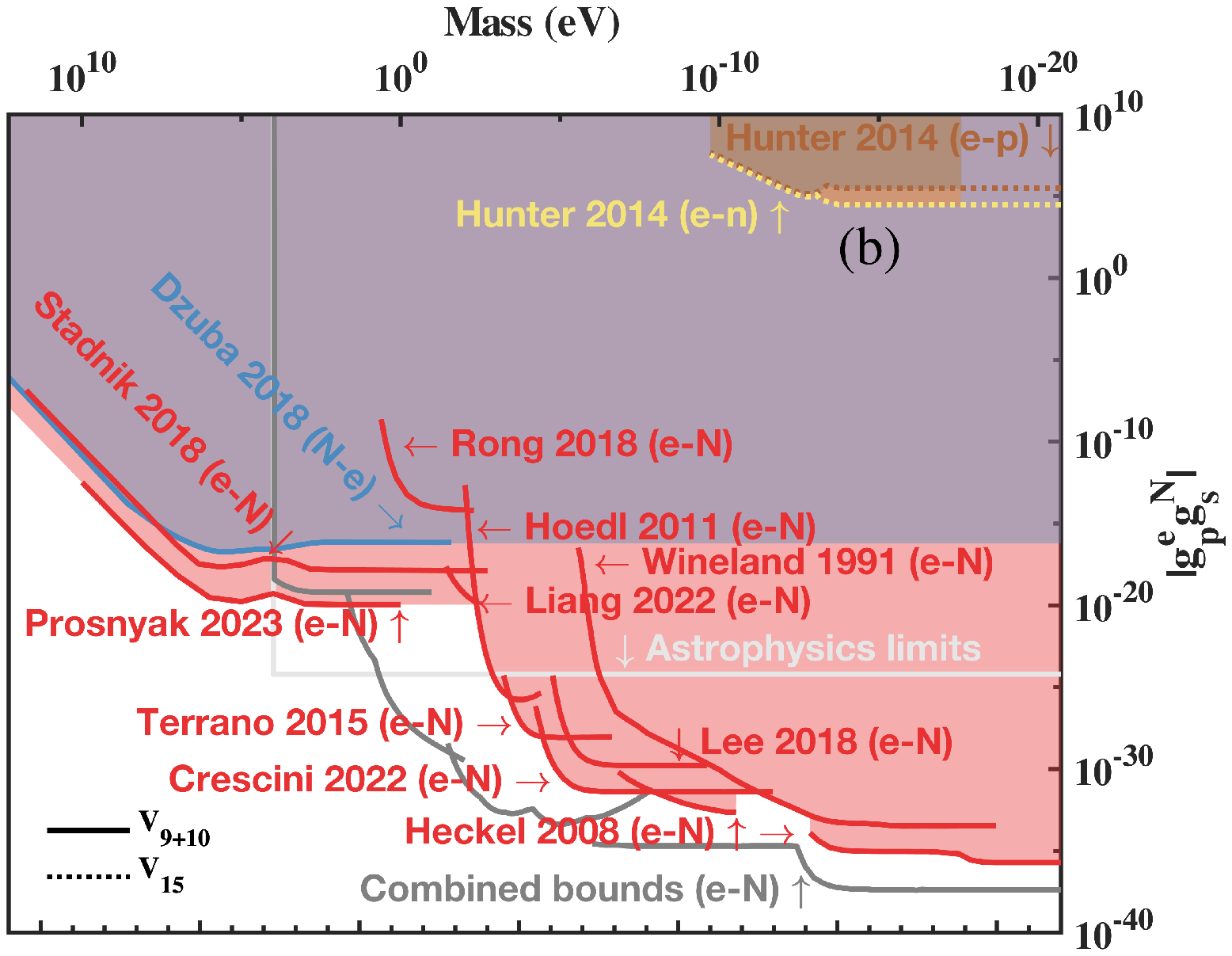}
\includegraphics[width=0.48\textwidth]{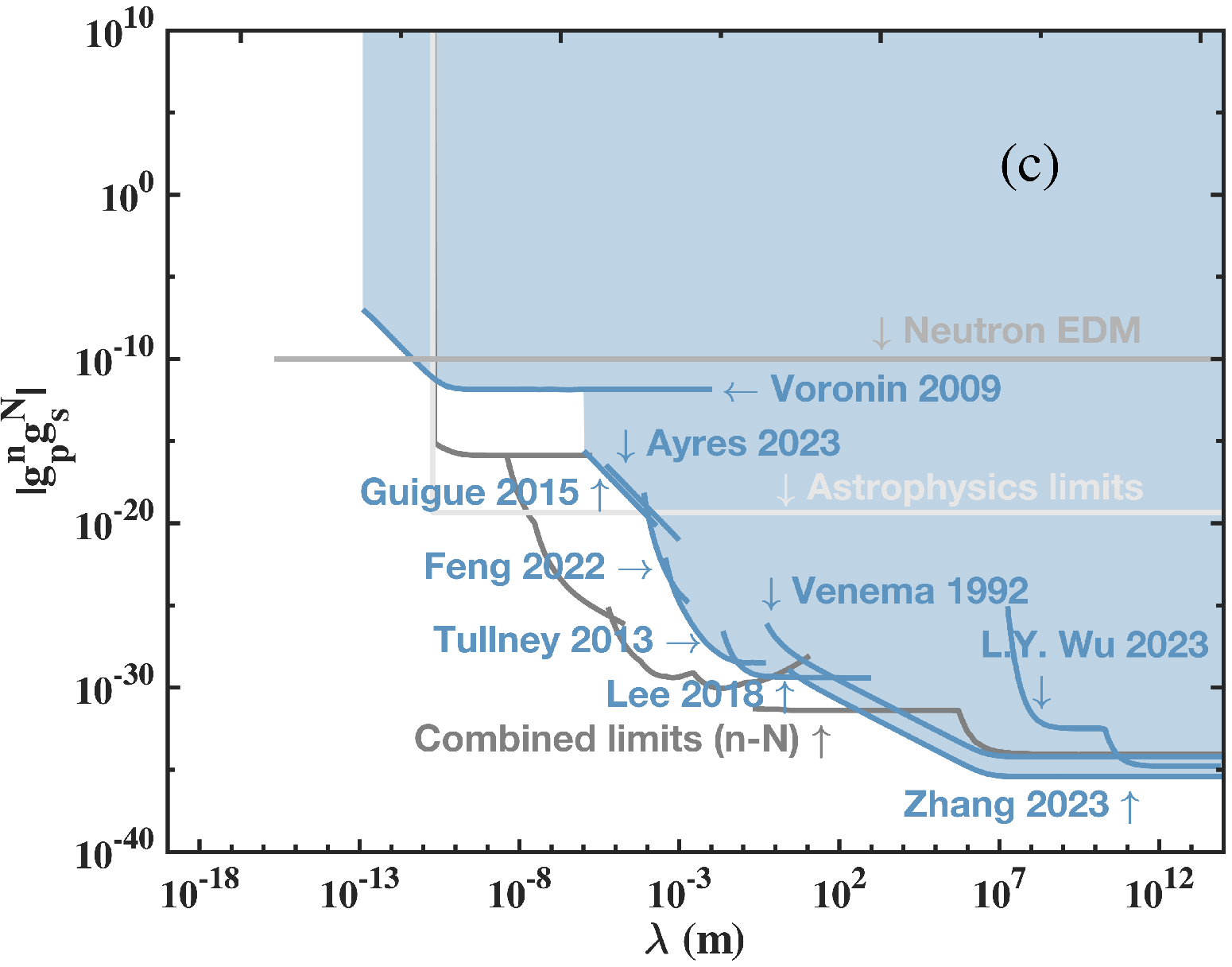}
\includegraphics[width=0.48\textwidth]{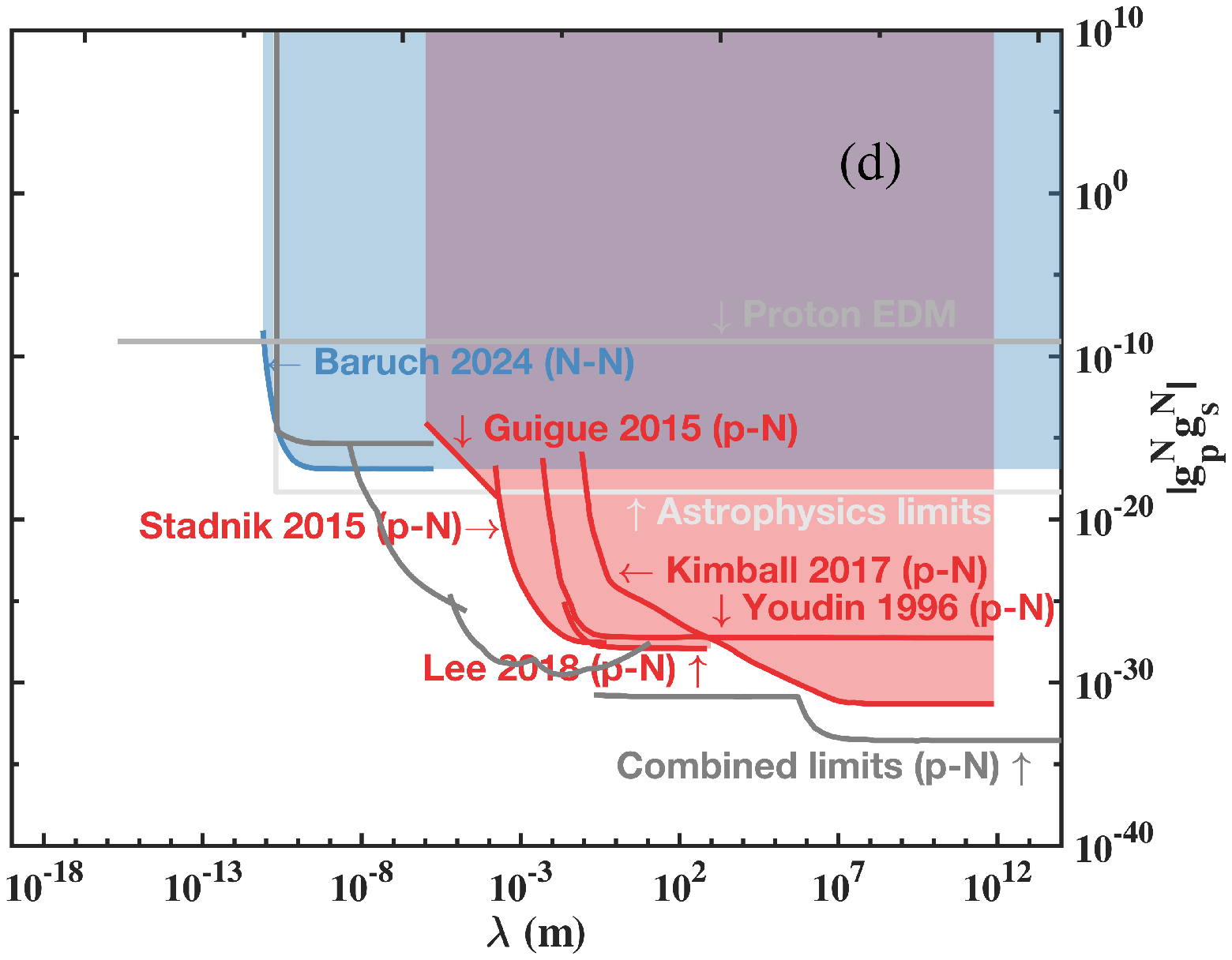}
\end{center}
\caption{Constraints, depicted by coloured regions, on the coupling constant product $g_p g_s$ as a function of the interaction range $\lambda$ shown on the bottom x-axis. 
The top x-axis represents the new spin-0 boson mass $M$. 
Here, the subscripts $p$ and $s$ denote the pseudoscalar and scalar couplings, respectively. 
Constraints shown for \textbf{(a)} $e$-$e$, \textbf{(b)} $e$-$N$, \textbf{(c)} $n$-$N$, \textbf{(d)} $p$-$N$ couplings. 
Results include $V_{9+10}$ (solid line) and $V_{15}$ (dotted line) terms. 
Combined bounds (solid, dark grey line) and astrophysical bounds (solid, light grey line) are elaborated in Appendices\,\ref{appendix_V1} and \ref{appendix_comp.}. 
}\label{gpgs-fig}
\end{figure*}

Figure\,\ref{gpgs-fig} shows the current laboratory constraints on $g_p^Xg_s^Y$ between the studied particle pairs (a) $e$-$e$, 
(b) $e$-$N$, $e$-$p$, and $e$-$n$, (c) $n$-$N$ and (d) $p$-$N$.
Experimental constraints on the exotic pseudoscalar/scalar combination of interaction parameters $g_pg_s$ can be found based on the exotic potential terms $V_{9+10}$ and $V_{14q+15}$: 

\begin{equation}\label{gPgS_V910}
\begin{aligned}
V_{9+10}&=-g_p^Xg_s^Y\frac{\hbar^2}{8\pi m_X}\boldsymbol{\sigma}_X\cdot \hat{\boldsymbol{r}}\,\, \left(\frac{1}{r^2}+\frac{1}{\lambda r} \right) \, e^{-{r}/{\lambda}} \, , 
\end{aligned}
\end{equation}

\begin{widetext}
{\small
\begin{equation}
\label{gPgS_V1415}
\begin{aligned}
&V_{14q+15}|_{ps}
=\frac{g^X_p g^Y_s}{8} \frac{\hbar^3}{c^2} \left[ \left\{ \left( \v{\sigma}_X \times \v\sigma_Y^{\,\prime}\right) \cdot \frac{\v{p}_Y}{m_Y} , \frac{1}{r^3} + \frac{1}{\lambda r^2} + \frac{4 \pi}{3} \delta(\v{r})  \right\} -\left\{ \left( \v{\sigma}_X \cdot \hat{\v{r}} \right) \v\sigma_Y^{\,\prime}\cdot \left( \frac{\v{p}_Y}{m_Y} \times \hat{\v{r}} \right) , \frac{3}{r^3} + \frac{3}{\lambda r^2} + \frac{1}{\lambda^2 r}  \right\}\right]\frac{e^{-r/\lambda}}{8 \pi m_X m_Y} \\ 
&\Rightarrow - {g^X_p g^Y_s} \frac{  \hbar^3  }{32\pi m_Y(m_X+m_Y)  c^2  } \left[ \left( \v{\sigma}_X \times \v\sigma_Y^{\,\prime}\right) \cdot \v{v}\, \left(\frac{1}{r^3} + \frac{1}{\lambda r^2} + \frac{4 \pi}{3} \delta(\v{r}) \right)  - \left( \v{\sigma}_X \cdot \hat{\v{r}} \right) \v\sigma_Y^{\,\prime}\cdot \left( \v{v} \times \hat{\v{r}} \right) \left(\frac{3}{r^3} + \frac{3}{\lambda r^2} + \frac{1}{\lambda^2 r} \right)  \right]e^{-r/\lambda} \, , 
\end{aligned}
\end{equation}
}
\end{widetext}

Since $V_{14q+15}|_{ps}$ has not been studied earlier, we present the traditional form of $V_{15}$ below in Eq.\,(\ref{gPgS_V15}) (see App.\,\ref{appendixA} for more details) and present the results from \citet{hunter_using_2014} based on this older $V_{15}$ form in Fig.\,\ref{gpgs-fig}\,(b). 
Note that $V_{15}$ vanishes in the case of identical fermions; therefore, we do not include the obtained $e$-$e$ constraints \cite{hunter_using_2014,ji_new_2018,xiao_exotic_2024}. For future experimental work, however, it is important to study the full term $V_{14q+15}|_{ps}$, see Sec.\,\ref{Subsec:Pairs-pots.} for further discussion.

\begin{widetext}
\begin{equation}
\label{gPgS_V15}
\begin{aligned}
V_{15}=
\left(\frac{g_s^Xg_p^Y}{m_X}-\frac{g_p^Xg_s^Y}{m_Y}\right) \frac{\hbar^3}{32\pi c^2} \frac{1}{m_X+m_Y}  \left[(\boldsymbol{\sigma}_X \cdot \hat{\boldsymbol{r}}) \, \v\sigma_Y^{\,\prime} \cdot (\hat{\boldsymbol{r}} \times \boldsymbol{v}) +(\v\sigma_Y^{\,\prime}\cdot\hat{\boldsymbol{r}}) \, \boldsymbol{\sigma}_X \cdot (\hat{\boldsymbol{r}} \times \boldsymbol{v}) \right]\,\left(3+3\frac{r}{\lambda}+\frac{r^2}{\lambda^2}\right) \frac{e^{-{r}/{\lambda}}}{r^3} \, . \\ 
\end{aligned}
\end{equation}
\end{widetext}

\subsubsection{\texorpdfstring{$e$-$e$}{e-e}}

We first turn our attention to the monopole-dipole lepton-lepton coupling for the fermion pair $e$-$e$, shown in Fig.\,\ref{gpgs-fig}\,(a). 

\citet{stadnik_improved_2018} present constraints on $g_p^e g_s^e$ based on the $V_{9+10}$ term by calculating axion-exchange-induced atomic EDMs using the relativistic Hartree-Fock Dirac method and then deriving limits from existing molecular EDM experiments. 
\citet{prosnyak_updated_2023} obtained constraints on $g_p^eg_s^e$ using the latest experimental result for HfF$^+$ cations \cite{roussy_improved_2023}. 
Their results are more stringent than the constraints from \citet{stadnik_improved_2018}, as well as their earlier result \cite{maison_axion-mediated_2021}, which investigated YbOH (not plotted for clarity).
\citet{crescini_search_2022} obtained constraints for $g_p^eg_s^e$ from the $V_{9+10}$ potential term in the range $3 \times 10^{-3} \,\textrm{m} <\lambda < 10^{5} \, \textrm{m}$
by scaling limits on the electron-nucleon interaction $g_p^eg_s^N$ by measuring the change of the magnetization of a paramagnetic crystal when the distance to a lead mass is modulated under the assumption that $A =2.5Z$ for the source mass. 
More details concerning the work of 
\citet{stadnik_improved_2018}; \citet{crescini_search_2022}; \citet{prosnyak_updated_2023}
are discussed later in this section.

\subsubsection{\texorpdfstring{$e$-$N$}{e-N}}
\label{gpgs_e-N}

In the case of $e$-$N$ couplings, as shown in Fig.\,\ref{gpgs-fig}\,(b), the strongest bounds in the interaction ranges $\lambda>10^{6}$\,m and $10\,\textrm{m} < \lambda < 10^{4}$\,m are set by the experiment of \citet{heckel_preferred-frame_2008} using a spin-polarized torsion pendulum (see Sec.\,\ref{METH1.SENS.MS}). 
Details about the experiment of \citet{heckel_preferred-frame_2008} can be found in Sec.\,\ref{subsec_g_Ag_V}. 
Another important experiment, covering the range $\lambda>4.9\times 10^{-4}$\,m, is that of \citet{terrano_short-range_2015} who also use a spin-polarized torsion-pendulum to search for spin-mass interactions. 
Details about the work of \citet{terrano_short-range_2015} are discussed in Sec.\,\ref{subsec_g_Pg_P}. 
In the range $4 \times 10^{-5} \, \textrm{m} < \lambda < 4.9\times 10^{-4}$\,m, \citet{hoedl_improved_2011} also conducted a spin-polarized torsion-pendulum experiment to search for exotic spin-dependent interactions. 
The experiment of \citet{hoedl_improved_2011} differed from those of \citet{heckel_preferred-frame_2008} and \citet{terrano_short-range_2015} in that the latter experiments used mu-metal shielding to minimize torques from stray magnetic fields, whereas \citet{hoedl_improved_2011} used a magnetically unshielded torsion pendulum in order to bring the source mass in closer proximity to the spins, thereby probing shorter-range forces than other experiments, including those in \cite{ni_search_1999,hammond_new_2007}.

In the range $3 \times 10^{5} \, \textrm{m} < \lambda<10^{6}$\,m, \citet{wineland_search_1991} 
placed the most stringent limit on the dipole-monopole interaction $V_{9+10}$ by examining the hyperfine resonances of trapped and cooled $^9$Be$^+$ ions and using Earth as the source body, see Sec.\,\ref{METH2.PM.TI} for further discussion.

As noted by \citet{hunter_using_2013}, Earth contains about $10^{42}$ spin-polarized electrons due to the geomagnetic field (also discussed in Sec.\,\ref{subsec_g_Ag_V}). 
For length scales $10^{3} \, \textrm{m} < \lambda < 10^{11}$\,m, \citet{hunter_using_2014} used a model of the spin-polarized geoelectrons to set constraints on $g_p^eg_s^p$ and $g_p^eg_s^n$ based on the exotic velocity-dependent potential term $V_{15}$. 
Recently, \citet{poddar_constraints_2023} pointed out that these electrons can interact with the unpolarized nucleons in the Sun, leading to a force caused by the monopole-dipole potential between the Sun and Earth. 
This force can affect Earth's motion. 
\citet{poddar_constraints_2023} used the well-measured perihelion precession of the Earth to set limits on the pseudoscalar-scalar electron-nucleon coupling at the level $g_p^e g_s^N \lesssim 2 \times 10^{-16}$ for the range $\lambda>10^{11}$\,m, and also derived stringent limits by considering Earth's perihelion precession in combination with limits from stellar cooling in the spirit of \citet{raffelt_limits_2012}.

\citet{crescini_search_2022}, as part of the QUAX experimental program (\citealp{ruoso_quax_2016}; \citealp{crescini_quax-gpgs_2017}),
conducted a search for spin-dependent interactions between lead masses (acting as the unpolarized ensemble of nuclei and electrons) and two paramagnetic Gd$_2$SiO$_5$ (GSO) crystals at $T=4.2$~K (acting as the spin-polarized ensemble). 
A set of 24 cylindrical lead masses (height = 6~cm, diameter = 5~cm) were arranged on a 1\,m diameter disc which was rotated above the pair of GSO crystals. 
The minimum separation between the GSO crystals and the lead masses was $\approx 3.5$\,cm. 
The GSO crystals were surrounded by two layers of mu-metal shielding and two additional layers of superconducting shielding. 
A monopole-dipole interaction between the lead masses and the spins in the GSO crystals would induce a change in the GSO magnetization \cite{landau_theoretical_2010} as the positions of the lead masses relative to the GSO crystals changed. The magnetization of the GSO crystals was measured with a state-of-the-art dc-SQUID (superconducting quantum interference device) magnetometer \cite{ni_scheme_1996}. 
For the interaction ranges $10^{-2}\,\textrm{m} < \lambda <  10$\,m and $10^{4}\,\textrm{m} < \lambda <  3\times10^{5}$\,m, \citet{crescini_search_2022} established the most stringent constraint on the $e$-$N$ monopole-dipole coupling, including improved sensitivity compared to their prior work \cite{crescini_improved_2017}.
Note that similar techniques were used in earlier searches by \citet{vorobyov_new_1988} and \citet{ni_search_1999}. \citet{crescini_search_2022} pointed out that their experimental method should be able to be improved by using an LC pickup coil \cite{vinante_dc_2001} or crystals with a higher magnetic susceptibility. If these improvements are successfully implemented in future experiments, the sensitivity of this method should be able to surpass the combined bounds.

In the range $\lambda < 4 \times 10^{-5}$\,m, \citet{stadnik_improved_2018} established limits on the pseudoscalar/scalar electron-nucleon interaction constant product $g_p^e g_s^N$ and electron-electron interaction constant product $g_p^eg_s^e$ mediated by ALPs. 
The $P,T$-violating potential $V_{9+10}$ can generate EDMs in atoms and molecules through the mixing of atomic states with opposite parity, see more in Sec.\,\ref{METH2.EDM}. 
\citet{stadnik_improved_2018} performed theoretical calculations of atomic EDMs using the relativistic Hartree-Fock-Dirac method, including electron core polarization corrections. 
Based on the experimental bounds on the EDMs of various atoms and molecules, including $^{133}$Cs, $^{205}$Tl, $^{129}$Xe, $^{199}$Hg, $^{171}$Yb$^{19}$F, $^{180}$Hf$^{19}$F$^+$, and $^{232}$Th$^{16}$O, \citet{stadnik_improved_2018} determined limits on $g_p^eg_s^N$ and $g_p^eg_s^e$ for a range of ALP masses. 
The most stringent limits on high-mass ALPs were derived from Hg and ThO, while the most stringent limits on low-mass ALPs were obtained from HfF$^+$. 

Building on the work of \citet{stadnik_improved_2018}, \citet{dzuba_new_2018} established further constraints on the interaction between an unpolarized electron and a polarized nucleon, denoted by $g_p^Ng_s^e$, which can induce EDMs of diamagnetic atoms. 
In contrast, the previous research \cite{stadnik_improved_2018} focused on the atomic EDMs of paramagnetic atoms. \citet{dzuba_new_2018} used the relativistic Hartree-Fock-Dirac method to calculate the atomic EDM of $^{199}$Hg, $^{129}$Xe, $^{211}$Rn, and $^{225}$Ra, and combined the results with experimental measurements of EDMs in these atoms to set constraints on the product of coupling constants $g_p^Ng_s^e$ for a wide range of ALP masses. 
The most stringent limits for any ALP mass $m_a \gtrsim 10^{-2}\,\textrm{eV}$ were derived from $^{199}$Hg. 
Compared to the constraints from macroscopic-scale experiments (\citealp{serebrov_new_2009}; \citealp{petukhov_polarized_2010}; \citealp{tullney_constraints_2013}; \citealp{bulatowicz_laboratory_2013}; \citealp{afach_constraining_2015}),
which reported constraints on the coupling parameters $g_p^ng_s^N$, \citet{dzuba_new_2018} provided significantly improved laboratory limits for $\lambda <2 \times 10^{-5}$\,m.

More recently, \citet{prosnyak_updated_2023} have used experimental data on the static $P,T$- violating molecular EDM of HfF$^+$ \citet{roussy_improved_2023} to constrain virtual ALP exchange between electrons and nuclei. 
This has resulted in a constraint on $g_p^e g_s^N$ that is more stringent than the one \citet{stadnik_improved_2018} derived from the ThO EDM experiment \citet{acme_collaboration_improved_2018}. 
Additionally, \citet{maison_electronic_2021} have calculated the sensitivity coefficients for a proposed YbOH experiment \cite{kozyryev_precision_2017}. 

In the range $10^{-7} \, \textrm{m} < \lambda < 2.3 \times 10^{-5}$\,m, \citet{rong_searching_2018} showed that a near-surface NV center in a diamond can be used as a quantum sensor to detect the monopole–dipole interaction $V_{9+10}$. 
In the range $6\times 10^{-6} \, \textrm{m} < \lambda <4.5 \times 10^{-5}$\,m, \citet{liang_new_2022} further improved the constraints using an ensemble of NV centers. 
Both of these sets of constraints have been surpassed by the EDM searches described above.

For other earlier constraints on $g_p^e g_s^N$ that are not presented in Fig.\,\ref{gpgs-fig}\,(b), such as earlier bounds from torsion-pendulum and comagnetometry experiments, such as \citet{youdin_limits_1996}; \citet{ni_search_1999}; \citet{hammond_new_2007}; \citet{crescini_improved_2017},
one can find a collection of references in the papers of \citet{stadnik_improved_2018}, \citet{safronova_search_2018}, and \citet{jackson_kimball_probing_2023}.

In addition, \citet{hunter_using_2014} employed a model of electron spins in Earth to explore long-range velocity-dependent interactions between spins related to the exchange of ultralight vector bosons and used existing laboratory experiments to set constraints on $V_{15}$. 
Here, using the theoretical framework discussed in earlier sections, we translate \citet{hunter_using_2014}'s constraints on $V_{15}$
to constraints on $g_p^eg_s^n$ and $g_p^eg_s^p$ and present them in Fig.\,\ref{gpgs-fig}\,(b), see Sec.\,\ref{g_Vg_A_e-n} for further discussion.

\subsubsection{\texorpdfstring{$n$-$N$}{n-N}}
\label{subsec_g_Pg_S_n-N} 

For monopole-dipole couplings between neutrons and nuclei, as shown in Fig.\,\ref{gpgs-fig}\,(c), the strongest bounds in the range $\lambda>10$\,m were established by \citet{zhang_search_2023}, who utilized a $^{129}$Xe-$^{131}$Xe-Rb clock-comparison-type atomic comagnetometer (Sec.\,\ref{METH1.SENS.AM}) to search for an exotic spin-dependent interaction between neutron spins and Earth. 
\citet{zhang_search_2023} measured the nuclear-spin precession frequency ratio of two Xe isotopes and its variation between field directions, indicating precession from non-magnetic effects like monopole-dipole coupling of the Xe spins to Earth. See more details in Sec.\,\ref{gpgs_spin-gravity}.
Their experimental limits surpassed those from experiments by \citet{wu_new_2023} using spin-polarized neutrons and the Sun and Moon as the monopole sources (with Earth's rotation providing modulation). 
The experiment of \citet{zhang_search_2023} also surpassed the prior work of \citet{venema_search_1992} using a $^{199}$Hg-$^{201}$Hg comagnetometer, which had long held the record for the best sensitivity to long-range monopole-dipole couplings. 

Earlier, the same team had used their $^{129}$Xe-$^{131}$Xe-Rb comagnetometer to search for an interaction between the Xe spins and the nuclei of a cm-scale, nonmagnetic BGO crystal source mass, enabling them to search for a monopole-dipole interaction in the range $10^{-4} \, \textrm{m} < \lambda < 5 \times 10^{-4}$\,m \cite{feng_search_2022}. 
Their experiment improved on an earlier experiment by \citet{bulatowicz_laboratory_2013} using a similar $^{129}$Xe-$^{131}$Xe-Rb comagnetometer. 
\citet{feng_search_2022} noted that further improvement of their sensitivity is possible by using multipass cavity technique \cite{hao_herriott-cavity-assisted_2021} to enhance the signal-to-noise performance of the comagnetometer. 
Future experiments could also probe shorter distances by employing microfabrication techniques \cite{kitching_chip-scale_2018}.

In the interaction range $10^{-1} \, \textrm{m} < \lambda < 10$\,m, \citet{lee_improved_2018} searched for monopole-dipole couplings of $^3$He (and K) spins to 
unpolarised source masses
and placed constraints on $g_p^ng_s^N$ (and $g_p^eg_s^N$). 
They utilized a K-$^3$He self-compensating SERF comagnetometer and movable Pb masses. 
\citet{lee_improved_2018} identified the scalar coupling to unpolarized fermions to be $g_s^N$ and assumed the coupling to be identical for both neutrons and protons and zero for electrons in the unpolarized mass [see the discussion after Eq.\,\eqref{ANZ}].

In the range around $5\times10^{-4} \, \textrm{m} < \lambda < 10^{-1}$\,m, \citet{tullney_constraints_2013} utilized 
a SQUID magnetometer
to detect the free precession of $^3$He and $^{129}$Xe nuclear spins and set constraints for $g_p^ng_s^N$ from $V_{9+10}$. See more details in Sec.\,\ref{Sec:EXP.METH.1}. 
\citet{tullney_constraints_2013} improved on the previous bounds (\citealp{chu_laboratory_2013}; \citealp{stadnik_nuclear_2015}; \citealp{youdin_limits_1996}) 
across most of the axion window.

In the range $10^{-6} \, \textrm{m} < \lambda < 10^{-4}$\,m, \citet{guigue_constraining_2015} searched for a monopole-dipole interaction between nucleons and hyperpolarized $^3$He based on the $V_{9+10}$ potential term. 
The nucleons in the walls of a $^3$He-filled glass cell served as the monopole source to generate a pseudomagnetic field. 
Helium spins evolving in the presence of the pseudomagnetic field would experience an additional longitudinal depolarization in addition to the usual depolarization mechanisms. 
Such an extra longitudinal depolarization was distinguished from the other standard contributions, such as atomic and wall collisions, by examining the longitudinal relaxation rate as a function of the applied magnetic field strength. 
The existence of a new boson would be revealed if a deviation in the behavior of the rate was observed and matched the calculated exotic-interaction-induced contribution to spin relaxation. 
\citet{guigue_constraining_2015} improved upon the constraints on $g_p^ng_s^N$ by a factor of 20 compared to the preliminary experiment \citet{petukhov_polarized_2010} performed at the same Institut Laue-Langevin (ILL). 
The long relaxation time of the polarized gas (up to several days) contributed to the increased sensitivity of the newer experiment. 
Note that \citet{fu_limits_2011} proposed and carried out a similar experiment to that of \citet{guigue_constraining_2015}, but with a moveable cm-scale polytetrafluoroethylene (PTFE) mass as the monopole source. 

\citet{voronin_neutron_2009} used neutron diffraction in a non-centrosymmetric crystal to search for a sub-micron range monopole-dipole interaction. 
When neutrons travel through the crystal over time $\tau$, the $V_{9+10}$ potential would lead to precession of neutron spins around the reciprocal lattice vector $\boldsymbol{g}$ by an angle $\phi={2 V_{9+10} \tau}/{\hbar}$. 
Based on the experimental result of \citet{fedorov_measurement_2009} from a search for the neutron EDM using the crystal-diffraction method, \citet{voronin_neutron_2009} constrained the coupling constant $g_p^ng_s^N$ over the range $10^{-13} \, \textrm{m} < \lambda < 10^{-6}$\,m. 

There are, in fact, a number of experiments that have probed monopole-dipole interactions using ultra-cold neutrons (UCN). 
These include Hg-UCN spin-precession-frequency comparison experiments \cite{ayres_search_2023,afach_constraining_2015}, UCN depolarization and precession experiments \cite{serebrov_new_2009}, and experiments employing the discrete UCN energy levels in Earth's gravitational potential \cite{jenke_gravity_2014,baesler_constraint_2007}. 
Even though the constraints these UCN experiments established have since been superseded by other measurements mentioned above, the use of UCN to probe exotic spin-dependent interactions is still a promising research direction.
For example, \citet{ayres_search_2023} aim to improve their current sensitivity by a factor of 64 
with a new apparatus \cite{ayres_design_2021} currently under construction at the Paul Scherrer Institute (PSI). 

\subsubsection{\texorpdfstring{$p$-$N$}{p-N}}
\label{subsec_g_Pg_S_p-N}

To complete our survey of experimental constraints on monopole-dipole couplings, we examine the less frequently studied case of $p$-$N$ couplings; limits on $g_p^p g_s^N$ are plotted in Fig.\,\ref{gpgs-fig}\,(d). 
Fewer experimental searches have reported constraints on $p$-$N$ monopole-dipole couplings, because most atomic vapor comagnetometers used to search for exotic spin-dependent interactions are based on polarized noble gases with valence neutrons and are hence primarily sensitive to $n$-$N$ couplings, while experiments using spin-polarized torsion pendula or solid-state systems are generally sensitive to $e$-$N$ monopole-dipole couplings rather than to nuclear couplings. 

\citet{jackson_kimball_constraints_2017} carried out a search specifically targeting a long-range monopole-dipole coupling of the proton spin to the mass of Earth, and established constraints on monopole-dipole interactions for $\lambda> 10^{-1}$\,m. 
\citet{jackson_kimball_constraints_2017} employed a $^{85}$Rb-$^{87}$Rb comagnetometer \cite{kimball_dual-isotope_2013}, 
a clock-comparison-type comagnetometer discussed in Sec.\,\ref{METH1.SENS.AM}. 
Notably, \citet{jackson_kimball_constraints_2017} improved on the long-range limits of \citet{youdin_limits_1996} for $p$-$N$ monopole-dipole couplings by about 4 orders of magnitude at the longest interaction ranges. 
Furthermore, their experiment provides the most stringent constraint on the proton GDM. 
The experimental sensitivity may be further improved by reducing systematic effects related to scattered light and magnetic-field gradients. 

In addition to the direct experimental study of \citet{jackson_kimball_constraints_2017} aimed at probing monopole-dipole interactions involving the proton spin, the results of other experiments
using polarised nuclear spins and
nominally probing monopole-dipole interactions involving the neutron spin can be re-scaled to constrain couplings to the proton spin. 
For example, \citet{youdin_limits_1996} originally interpreted the results of their experiment, which compared the spin-precession frequencies of $^{199}$Hg and $^{133}$Cs, to constrain only electron and neutron spin-mass couplings. 
However, the $^{133}$Cs nucleus has a valence proton, and \citet{jackson_kimball_nuclear_2015} noted that in such a case the single-particle Schmidt model, semi-empirical models, and large-scale nuclear shell-model calculations are all in reasonable agreement concerning the contribution of the valence proton spin to the nuclear spin of $^{133}$Cs.

Of particular interest is that the neutron and proton spin polarization of the $^3$He nucleus can be reliably determined as discussed by \citet{jackson_kimball_nuclear_2015}, since there are detailed calculations as well as direct experiments that are in good agreement with one another (\citealp{friar_neutron_1990}; \citealp{anthony_deep_1996}; \citealp{ethier_comparative_2013}).
Thus, as noted by \citet{safronova_search_2018}, the experiments of \citet{petukhov_polarized_2010} and \citet{chu_laboratory_2013}, which study spin-dependent interactions of $^3$He and nominally probed neutron couplings, also establish constraints for protons.

In this review, we apply this same logic to rescale the results of two other experiments utilizing $^3$He for the study of monopole-dipole interactions.
The fraction of neutron and proton polarization per $^3$He nucleus is calculated and measured to be $\eta_n\approx 0.87$ and 
$\eta_p\approx -0.027$, respectively \cite{friar_neutron_1990}. 
This generally means that constraints on neutron couplings from experiments using spin-polarized $^3$He also imply constraints on proton spin couplings that are a factor of $\approx 30$ times weaker than the neutron constraints, 
as long as $^3$He solely (or mainly) provides the sensitivity to the exotic spin-dependent coupling.\footnote{
In clock-comparison comagnetometers, this simple rescaling is generally not true since the monopole-dipole coupling to the proton spin of the other species must also be taken into account. Our rescalings are for those experiments using $^3$He spins as the sole or dominant species through which the coupling is measured, so there is no such ambiguity in the interpretation.
} 
\citet{lee_improved_2018} used a K-$^3$He self-compensating comagnetometer to constrain $n$-$N$ monopole-dipole couplings in the $10^{-1} \, \textrm{m} < \lambda < 10$\,m range. 
\citet{guigue_constraining_2015} studied relaxation of hyperpolarized $^3$He to constrain $n$-$N$ monopole-dipole couplings in the $10^{-6} \, \textrm{m} < \lambda < 10^{-4}$\,m range. 
The appropriately rescaled constraints on $p$-$N$ monopole-dipole couplings from both experiments are shown in Fig.\,\ref{gpgs-fig}\,(d). 

Besides the work mentioned above, there is still a debate about the reliability of rescaling the results from experiments using magnetometry with $^{129}$Xe, as in the case of the work of \citet{tullney_constraints_2013}, \citet{feng_search_2022}, and \citet{zhang_search_2023}. On the one hand, although the work of \citet{tullney_constraints_2013} also uses polarized $^3$He, because the technique of comagnetometry with $^{129}$Xe is employed and there is some 
uncertainty regarding the contribution of the proton spin to the $^{129}$Xe nuclear spin \cite{jackson_kimball_nuclear_2015}, one may prefer to avoid limiting the monopole-dipole interactions of the proton from this work. 
\citet{feng_search_2022} and \citet{zhang_search_2023} used a $^{129}$Xe-$^{131}$Xe-Rb comagnetometer and thus it is similarly difficult to interpret their results in terms of limits on proton couplings. 
In Fig.\,\ref{gpgs-fig}\,(d), we present constraints on the $p$-$N$ coupling by rescaling the $n$-$N$ results of \citet{tullney_constraints_2013} according to the nucleon spin content analysis of \citet{stadnik_nuclear_2015}. We note that there is an error in Fig.\,1 of \citet{stadnik_nuclear_2015}: in obtaining the $p$-$N$ bound there, the authors erroneously took the projection curve 9 from Fig.\,3 of \citet{tullney_constraints_2013} instead of the results curve 8. We have corrected this error in Fig.\,\ref{gpgs-fig}\,(d).

Since the monopole-dipole interaction violates both the $P$ and $T$ symmetries, experimental limits on $P,T$-violating EDMs also translate to limits on monopole-dipole interactions. 
\citet{mantry_distinguishing_2014} used EDM experiments to constrain on $P,T$-violating monopole-dipole interactions in two scenarios: one in which the monopole-dipole interaction is due specifically to the QCD axion and another in which it is due to a generic spin-0 boson with both scalar and pseudoscalar couplings. 
In the former case, because of the specific relationship of $g_sg_p$ to the $CP$-violating phase $\theta_{\textrm{eff}}$ and the Peccei-Quinn symmetry-breaking scale $f_a$ (see discussion in Sec.\,\ref{Sec:Intro-motiv-spin-0}), extremely strong limits on $g_sg_p$ for nucleons are obtained in the case of the QCD axion:\,at an interaction range of $\lambda \sim 0.2$\,m, \citet{mantry_distinguishing_2014} estimate the bound $g_s g_p \lesssim 10^{-41}$ based on the Hg EDM experiment of \citet{griffith_improved_2009},
which constrained $\theta_\textrm{eff} \lesssim 10^{-10}$.
On the other hand, for generic spin-0 bosons, as we consider in the present review, the implied constraints from EDM experiments
via intranuclear processes
are far weaker: 
at the level $g_s g_p \lesssim 10^{-9} - 10^{-11}$ for $\lambda \sim 0.2$\,m.
Thus for generic ALPs, the direct searches for monopole-dipole interactions described above offer the best sensitivity at longer interaction ranges. 
More recently, \citet{baruch_constraining_2024} studied the effects of a $CP$-violating nucleon-nucleon long-range interaction in diatomic molecules containing at least one nucleus with nonzero nuclear spin. Using data from molecular EDM experiments in HfF$^+$, ThO and YbF, \citet{baruch_constraining_2024} obtained the most stringent limits on $g_p^N g_s^N$ for generic ALPs in the interaction range $10^{-11} \, \textrm{m} < \lambda < 10^{-6} \, \textrm{m}$.
In addition, \citet{baruch_constraining_2024} also presented constraints on $g_s^p g_p^p$ obtained from the bound on the proton EDM [see Fig.\,\ref{gpgs-fig}\,(d)] and constraints on $g_s^n g_p^n$ obtained from the bound on the neutron EDM [see Fig.\,\ref{gpgs-fig}\,(c)]; see \citet{di_luzio_chiral_2024} for more details.

\subsubsection{Spin-gravity}\label{gpgs_spin-gravity}

As discussed in Sec.\,\ref{Sec:Intro-motiv-spin-gravity}, experimental searches for long-range monopole-dipole interactions 
(\citealp{wineland_search_1991}; \citealp{venema_search_1992}; \citealp{heckel_preferred-frame_2008}; \citealp{tarallo_test_2014}; \citealp{duan_test_2016}; \citealp{jackson_kimball_constraints_2017})
have often been inspired by early work of, for example, \citet{morgan_direct_1962}; \citet{kobzarev_gravitational_1963}; \citet{leitner_parity_1964}; \citet{dass_test_1976}; \citet{peres_test_1978}
suggesting a possible $P$- and $T$-violating spin-gravity interaction. 
At the outset, we stress the important points highlighted in Sec.\,\ref{Sec:Intro-motiv-spin-gravity}: in the event of an observation of a nonzero monopole-dipole interaction, it would be unclear without further experimentation if the new interaction was the result of a new aspect of gravity or the result of an entirely new interaction mediated by a new boson such as the axion. 
Tests of the equivalence principle for the new interaction, for example, may help distinguish between the two possibilities. A subtle issue is quantization: the theoretical framework presented in this review and by \citet{moody_new_1984}, \citet{dobrescu_spin-dependent_2006} and \citet{fadeev_revisiting_2019} for interpreting new spin-dependent interactions is based on quantum field theory. 
To date, there is no quantum theory of gravity, and the possibility remains open that gravity does not, in fact, have an underlying quantum nature. 
Therefore, a further clue as to whether a new monopole-dipole interaction is connected to gravity might be whether gravity 
can be experimentally proven to be quantum in nature.

Recently, \citet{zhang_search_2023} performed an experimental search for a spin-gravity coupling with unprecedented sensitivity, setting the most stringent limits to date for any particle. 
For a spin-gravity coupling of the form $V = \chi \, \v{\sigma} \cdot \v{g}(\v{r})$ with $\chi = 1$ [corresponding to an intrinsic strength equal to that of the usual gravitational interaction with the local gravitational field $\v{g}(\v{r})$], the precession frequency of a spin in Earth's gravitational field would be $\approx 10$\,nHz, see, for example, the discussion of \citet{peres_test_1978}. 
This frequency is over 10 billion times smaller than the typical nuclear Larmor frequency in Earth's magnetic field and is roughly a thousandth of our planet's once-per-day rotation rate.  
The approach used by \citet{zhang_search_2023} involved a spin-polarized gas consisting of two isotopes: $^{129}$Xe and $^{131}$Xe. 
In their experiment, the researchers measured the nuclear spin precession frequencies of both isotopes simultaneously in the presence of a bias magnetic field. 
To minimize the errors caused by gyroscopic effects, the direction of the bias magnetic field was carefully aligned to be parallel to Earth's rotation axis. 
By taking the ratio of the precession frequencies, the team was able to effectively eliminate magnetic-field-dependent effects. 
They repeated this measurement with the direction of the bias magnetic field reversed, and calculated the difference between the ratios obtained for the two field directions. 
This difference, to a first approximation, is directly proportional to the magnitude of precession resulting from nonmagnetic effects such as the torque induced by gravity on the spins. 

While the work of \citet{zhang_search_2023} surpassed all other existing constraints, including laboratory experiments and astrophysical observations, it has not yet reached sufficient sensitivity to probe a spin-gravity coupling with $\chi = \mathcal{O}(1)$, rather setting an upper bound about an order of magnitude away from the nominal strength of gravity. 
This benchmark sensitivity for a spin-gravity coupling appears to be within the reach of the next generation of experiments, 
as new ideas to use molecular comagnetometers \cite{wu_nuclear-spin_2018} and levitated ferromagnets \cite{fadeev_gravity_2021}, for example, are being pursued. 

\subsubsection{Future perspectives}
\label{gpgs_emerging}

As discussed in Sec.\,\ref{Sec:Intro-motiv-spin-0}, one of the leading motivations to search for exotic spin-dependent interactions is to test the hypothesis that the solution to the strong-$CP$ problem implies the existence of a QCD axion. 
The Axion Resonant InterAction Detection Experiment (ARIADNE) is an ongoing effort, originally proposed by \citet{arvanitaki_resonantly_2014}, to use NMR methods to search for a QCD-axion-mediated monopole-dipole interaction (in the form of the $V_{9+10}$ potential) or dipole-dipole interaction (in the form of the $V_{3}$ potential). 
The ARIADNE experiment utilizes an unpolarized source mass (or polarized spin source) along with a low-temperature, spin-polarized $^3$He gas. 
By exploiting the QCD-axion-mediated monopole-dipole (or dipole-dipole) interaction, it is possible to induce spin precession in the $^3$He gas. 
One method to achieve this is by rotating an unpolarized tungsten mass sprocket close to the $^3$He gas, which is placed within a magnetic field. 
When the sprocket's teeth move past the $^3$He sample at the Larmor frequency, the modulated monopole-dipole interaction has a frequency matching the NMR resonance frequency. 
In this case, the monopole-dipole interaction can tilt the $^3$He spins away from the leading field direction, and they will precess around the magnetic-field axis. 
The resulting precessing transverse magnetization can then be detected using a SQUID. 
The distance between the $^3$He gas and the source mass is at the $0.1 - 1$\,mm length scale, meaning that ARIADNE will investigate the QCD axion at the upper limit of the conventionally accepted axion mass range ($\sim 6$\,meV), a mass region that is difficult to probe with other techniques. 
The high spin density and long coherence time of $^3$He, along with the resonant enhancement of the NMR signal, can potentially offer sensitivity at the predicted scale of the QCD axion couplings \cite{ariadne_collaboration_characterization_2022}. 
The main distinction between the approach of the ARIADNE experiment and earlier ones based on precision magnetometry (\citealp{vasilakis_limits_2009}; \citealp{chu_laboratory_2013}; \citealp{tullney_constraints_2013}; \citealp{bulatowicz_laboratory_2013})
is the use of resonant enhancement.

There are a number of promising experimental proposals targeting monopole-dipole interactions. 
\citet{agrawal_searching_2023} analyzed the latest state-of-the-art spin precession experiments using cold atomic and molecular beams and trapped gases, and showed that if used to detect axion-mediated monopole-dipole forces between the spin sensor (electrons/nucleons in atoms and molecules) and the source (nucleons in localized sources or Earth), the expected sensitivity could cover a wide region of unexplored parameter space. 
\citet{chu_search_2015} proposed an experimental search for exotic spin-coupled interactions using a solid-state paramagnetic insulator, gadolinium gallium garnet (GGG), based on the concept used in a search for the electron EDM in GGG \cite{kim_new_2015}. 
The experiment proposed by \citet{chu_search_2015} could probe distances above the micron scale with sufficient sensitivity to surpass the limits of \citet{hoedl_improved_2011} for $g_p^eg_s^N$. 
\citet{chen_ultrasensitive_2021} proposed to study the electron–nucleon monopole–dipole interaction with a novel method: a hybrid spin-nanocantilever optomechanical system that can probe interactions at the $\gtrsim 10$\,nm scale. 
\citet{wu_spin-mechanical_2023} proposed a spin-mechanical quantum chip compatible with scalable on-chip detectors, which could be used for measurement of exotic monopole-dipole interactions at the $\gtrsim 0.1\,\mu\textrm{m}$ scale. 
\citet{wei_ultrasensitive_2023} proposed to use an optimized $^{21}$Ne-Rb-K comagnetometer to search for exotic spin-dependent interactions involving proton and neutron spins, and aim to search for long-range (Earth-scale) interactions in order to improve by orders of magnitude the only current limits on $g_p^pg_s^N$ at this length scale set by \citet{jackson_kimball_constraints_2017}. 
\citet{shortino_preliminary_2024} proposed to search for spin-dependent interactions by observing sidebands of nuclear spin-precession signals as a rotating source mass modulates the monopole-dipole interaction, with improved sensitivity to $g_p^ng_s^N$ expected at the $\sim 10^{-4}$\,m scale. 

Another area of research that can propel progress is further investigation of the nuclear spin content of various species of interest such as $^{129}$Xe, $^{131}$Xe, $^{199}$Hg, and $^{201}$Hg. 
More reliable understanding of the proton spin contribution to the nuclear spins of these species can maximize the utilization of the results of previous experiments and also guide the design of future studies. 

Finally, it is important to note the role of theoretical studies of EDMs. 
As seen in the work by \citet{dzuba_new_2018,baruch_constraining_2024}
once new experimental data become available from EDM searches, the theoretical calculation of proton and neutron spin contributions for nuclei can aid the interpretation of experimental data.

\subsection{\texorpdfstring{Pseudoscalar/Pseudoscalar interaction $g_p g_p$}{Pseudoscalar/Pseudoscalar interaction gpgp}}
\label{subsec_g_Pg_P}

The pseudoscalar potential $V_3|_{pp}$ is a dipole-dipole interaction that can be generated by the exchange of an axion or ALP between fermions: 
\begin{widetext}
\begin{equation}\label{gpgp_V3}
\begin{aligned}
V_3|_{pp} = -g_p^Xg_p^Y \frac{\hbar^3}{16\pi c}\frac{1}{m_Xm_Y}\left[\boldsymbol{\sigma}_X\cdot\v\sigma_Y^{\,\prime} \left(\frac{1}{r^3}+\frac{1}{\lambda r^2}+\frac{4\pi}{3}\delta(\v{r})\right)-(\boldsymbol{\sigma}_X\cdot \hat{\boldsymbol{r}})(\v\sigma_Y^{\,\prime}\cdot \hat{\boldsymbol{r}}) \left( \frac{3}{r^3} + \frac{3}{\lambda r^2} + \frac{1}{\lambda^2 r} \right)\right]e^{-{r}/{\lambda}} \, . 
\end{aligned}
\end{equation}
\end{widetext}

Note that $V_3|_{pp}$ has another equivalent form, which is useful for numerical atomic calculations, see \citet{fadeev_revisiting_2019} [around equation (13)] and \citet{fadeev_pseudovector_2022} [around equation (6)] for more discussion. 

Experimental searches for anomalous spin-spin interactions not mediated by the photon were first discussed by Ramsey \cite{ramsey_tensor_1979} and others, including \citet{ni_search_1994,pan_experimental_1992}. 
Such searches have been performed using a variety of systems, including atomic comagnetometers (\citealp{vasilakis_limits_2009}; \citealp{almasi_new_2020}),
torsion pendula \cite{terrano_short-range_2015}, and atomic spectroscopy (\citealp{ficek_constraints_2018}; \citealp{fadeev_pseudovector_2022}).

\begin{figure} [!htbp]
\begin{center}
\includegraphics[width=0.48\textwidth]{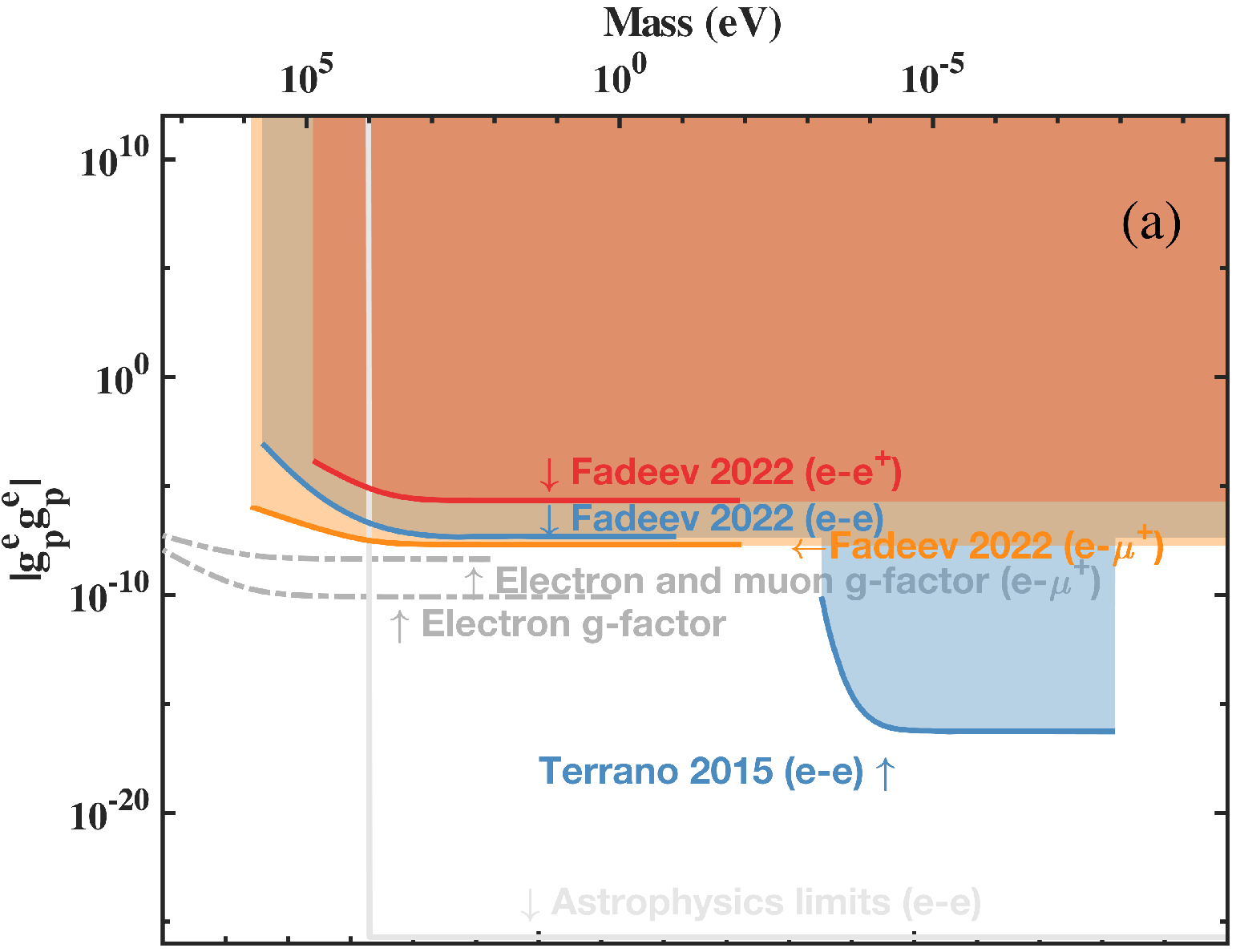}
\includegraphics[width=0.48\textwidth]{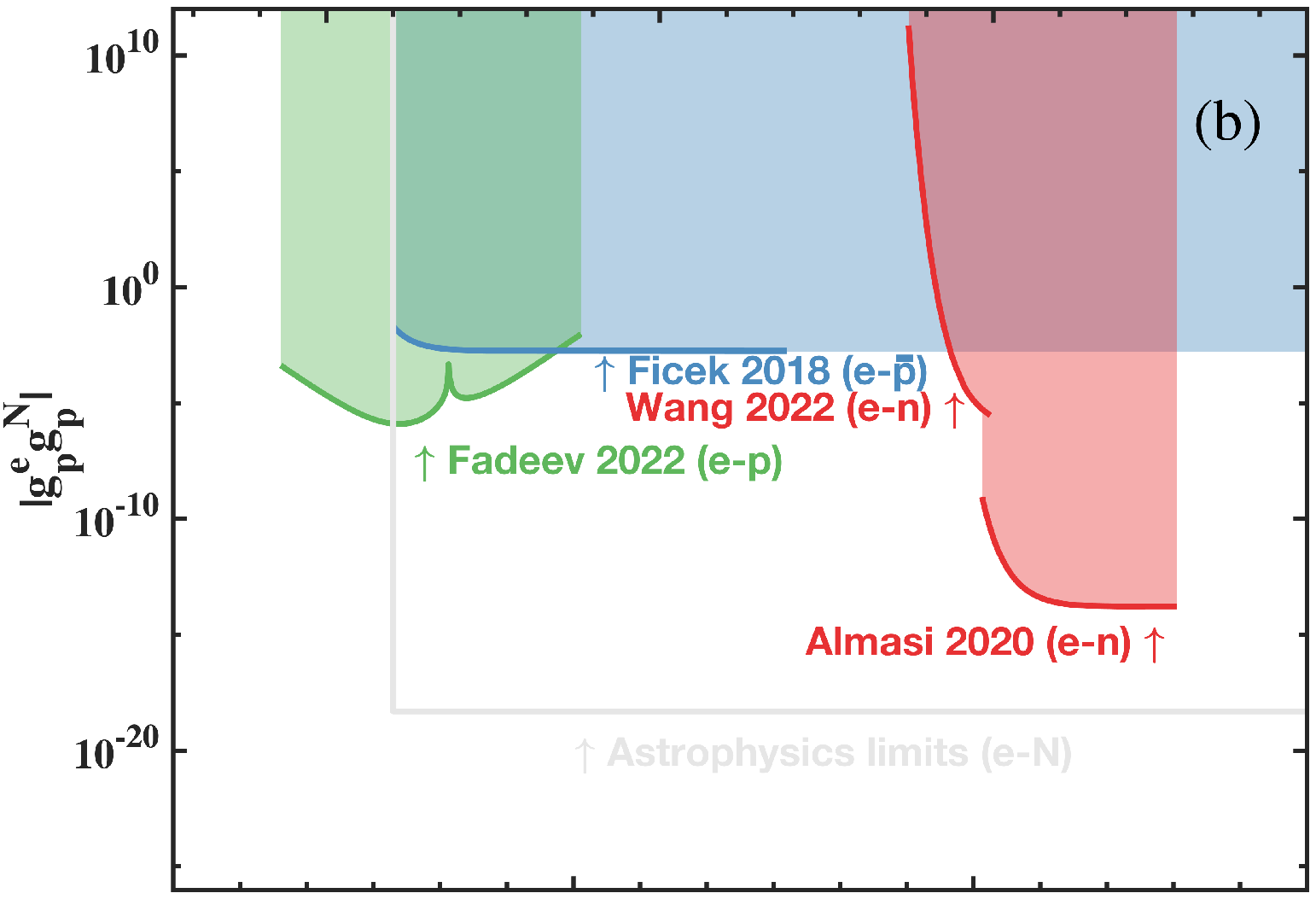}
\includegraphics[width=0.48\textwidth]{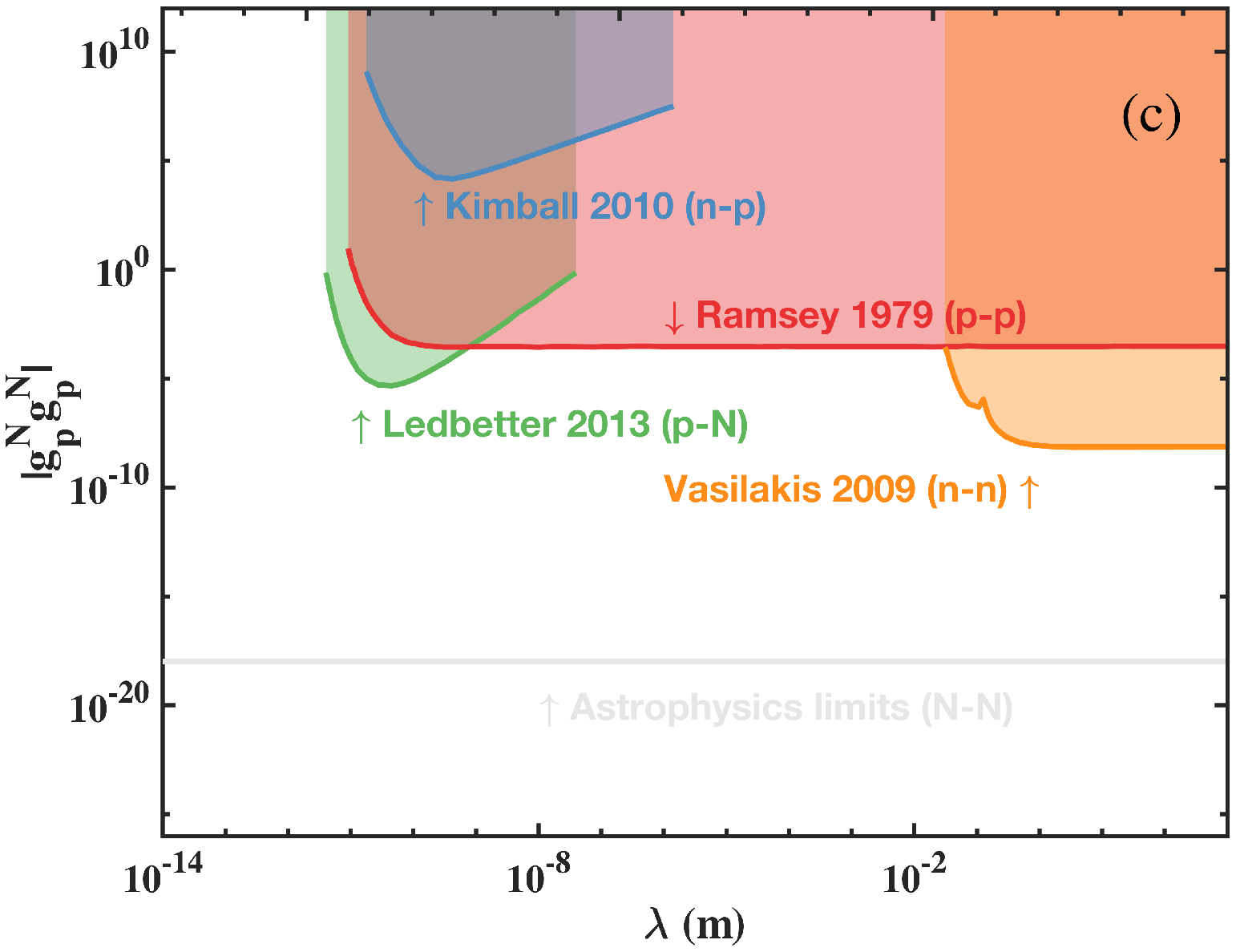}
\end{center}
\caption{Constraints, depicted by coloured regions, on the coupling constant product $g_p g_p$ as a function of the interaction range $\lambda$ shown on the bottom x-axis. 
The top x-axis represents the new spin-0 boson mass $M$. 
The subscript $p$ denotes the pseudoscalar coupling. 
Constraints shown for \textbf{(a)} $e$-$e$, $e$-$e^+$ and $e$-${\mu}^+$, \textbf{(b)} $e$-$n$, $e$-$p$ and $e$-$\overline{p}$, and \textbf{(c)} $n$-$n$, $n$-$p$, $p$-$p$ and $p$-$N$ couplings. 
Results are based on the $V_3$ term. 
Astrophysical bounds (light gray solid lines) and $g$-factor limits (gray dash-dotted lines) are elaborated in Appendix\,\ref{appendix_comp.}. } 
\label{gpgp-fig}
\end{figure}

Figure~\ref{gpgp-fig} shows the current laboratory constraints on the exotic pseudoscalar/pseudoscalar interaction $g_p^X g_p^Y$ between the studied particle pairs (a) $e$-$e$, $e$-$e^+$ and $e$-${\mu}^+$, (b) $e$-$n$, $e$-$p$ and $e$-$\overline{p}$, and (c) $n$-$n$, $n$-$p$, $p$-$p$ and $p$-$N$. 
Astrophysical bounds (gray solid lines) and $g$-factor limits (gray dash-dotted lines) are also presented for comparison, see App.\,\ref{appendix_comp.} for more details.

\subsubsection{\texorpdfstring{$e$-$e$, $e$-$e^+$, $e$-$\mu^+$}{e-e, e-e⁺, e-μ⁺}}

In Fig.\,\ref{gpgp-fig}\,(a), we present various limits on couplings between an electron and other charged leptons. 

In the range $\lambda > 3 \times 10^{-4}$\,m, \citet{terrano_short-range_2015} placed limits on the electron-electron coupling $g_p^e g_p^e$ by studying the pseudoscalar/pseudoscalar spin-spin interaction using a Eöt-Wash torsion-balance probe, described in detail in Sec.\,\ref{METH1.SENS.MS}. 
The results of \citet{terrano_short-range_2015} surpassed the earlier results reported in \citet{pan_experimental_1992}; \citet{ni_search_1994}; \citet{bergmann_spin_1998}.

In the range $\lambda > 4 \times 10^{-13}$\,m, \citet{fadeev_pseudovector_2022} set the best constraints on $g_p^eg_p^e$ and $g_p^eg_p^{e^+}$ using atomic spectroscopy methods. 
\citet{ficek_constraints_2017} obtained similar constraints on $g_p^e g_p^e$ as \citet{fadeev_pseudovector_2022}, while \citet{kotler_constraints_2015} set constraints on $g_p^e g_p^{e^+}$ similar to those in \citet{fadeev_pseudovector_2022}. 
For clarity, we have only presented the results from \citet{fadeev_pseudovector_2022} here. 
See Sec.\,\ref{subsec_g_Ag_A} for more details. 

In addition, \citet{fadeev_pseudovector_2022} 
set 
limits on the pseudoscalar/pseudoscalar coupling $g_p^e g_p^{{\mu}^+}$ between an electron and anti-muon.

\subsubsection{\texorpdfstring{$e$-$n$, $e$-$p$ and $e$-$\overline{p}$}{e-n, e-p and e-p̅}}

In Fig.\,\ref{gpgp-fig}\,(b), various limits on couplings between an electron and nucleons (or an anti-nucleon) are presented. 

In the interaction range $10^{-2} \, \textrm{m} < \lambda < 10$\,m, \citet{almasi_new_2020} utilized a rotatable SmCo$_5$ spin source and a Rb-$^{21}$Ne comagnetometer to establish the most stringent constraints on electron-neutron spin-dependent forces, specifically the pseudoscalar/pseudoscalar coupling $g_p^e g_p^n$. The constraints surpass the previous limits on $g_p^eg_p^n$ from \citet{wineland_search_1991}. 
More details about \citet{almasi_new_2020} can be found in Sec.\,\ref{subsec_g_Ag_A}. 

In the range $10^{-3} \, \textrm{m} < \lambda < 10^{-2}$\,m (within the classical QCD ``axion window''), \citet{wang_limits_2022} employed a spin-based amplifier (see Sec.\,\ref{METH1.SENS.AM.SMSA}) to search for axions and ALPs by measuring an exotic dipole-dipole interaction ($g_p^e g_p^n$) between polarized electron and neutron spins. 
In this experiment, the exotic dipole-dipole interaction is between polarized $^{87}$Rb electrons in one vapor cell and polarized $^{129}$Xe neutrons in another vapor cell. 
The interaction between the source electrons in $^{87}$Rb and the polarised neutrons in $^{129}$Xe produces a pseudomagnetic field that is measured by the latter. 
The spin-based amplifier in the second cell utilized a mixed ensemble of spin-polarized $^{87}$Rb and $^{129}$Xe to amplify the resonant signal generated by the pseudomagnetic fields, 
which are modulated by changes in the electron polarisation
\cite{jiang_search_2021}. 
The $^{87}$Rb spins acted as a magnetometer \cite{jiang_magnetic_2019,jiang_interference_2020} to detect the enhanced field. 
The use of a spin-based amplifier allowed for high sensitivity in the search for pseudomagnetic fields associated with the $V_{3}$ potential term.

\citet{ficek_constraints_2018} established stringent constraints on the exotic interaction between electrons and antiprotons 
via the $V_3$ potential using measurements and predictions of the spectrum of antiprotonic helium. 
Similar constraints have been obtained in \citet{fadeev_pseudovector_2022}. 
In addition, \citet{fadeev_pseudovector_2022} established stringent constraints on the exotic interaction between electrons and protons 
in the range $3.9 \times 10^{-13} \, \textrm{m} < \lambda < 1.3\times 10^{-8}$\,m using measurements and calculations of the spectrum of atomic hydrogen.
The details of these works are discussed in Sec.\,\ref{subsec_g_Ag_A}.

\subsubsection{\texorpdfstring{$N$-$N$}{N-N}}

In Fig.\,\ref{gpgp-fig}\,(c), various limits on couplings between nucleons are shown. 

For the interaction range $\lambda > 3.3 \times 10^{-2}$\,m, \citet{vasilakis_limits_2009} utilized a comagnetometer (see Sec.\,\ref{METH1.SENS.AM.AC}) based on spin-polarized K and ${^3}$He atoms to search for a pseudoscalar-boson-mediated neutron-neutron interaction which takes the form of the $V_3$ potential.
\citet{vasilakis_limits_2009} obtained constraints on $g_p^n g_p^n$ that improved on the earlier ${^3}$He-$^{129}$Xe maser experiment \cite{glenday_limits_2008} by a factor of about 500 and represent the currently most stringent constraints. 
In addition, \citet{vasilakis_limits_2009} also set constraints on couplings to light vector bosons through the potential terms $V_2$ and $V_{11}$, as well as placing constraints on unparticle couplings to neutrons and couplings to Goldstone bosons associated with the spontaneous breaking of the Lorentz symmetry. 
The results for $V_{2}$ and $V_{11}$ can be found in Secs.\,\ref{subsec_g_Ag_A} and \ref{subsec_g_Ag_V}, respectively. 
Later on, \citet{lee_laboratory_2023} reanalyzed data from \citet{vasilakis_limits_2009} to search for ultralight ALP dark matter.

Additionally, \citet{ramsey_tensor_1979}, \citet{jackson_kimball_constraints_2010} and \citet{ledbetter_constraints_2013} have established constraints on exotic interactions among the fermion pairs $p$-$p$, $p$-$N$ and $n$-$p$, respectively, based on the $V_3$ potential term. 
The details of those works can be found in Sec.\,\ref{subsec_g_Ag_A}.

\subsubsection{Future perspectives}

There are many intriguing possibilities for the field of probing exotic interactions mediated by the exchange of a new pseudoscalar boson in the future. 

In the force range $\lambda > 3 \times 10^{-5}$\,m, 
\citet{fadeev_ferromagnetic_2021,vinante_surpassing_2021} propose to search for spin precession induced by the pseudoscalar-mediated dipole-dipole interaction by modulating the distance between a polarized SmCo$_{5}$ electron spin source and a levitated ferromagnetic gyroscope, which contains electron spins. The precession of the ferromagnetic gyroscope can be measured with a SQUID. 

In the range $3\times 10^{-5} \, \textrm{m} < \lambda < 0.1$\,m,
\citet{arvanitaki_resonantly_2014} propose the use of NMR to obtain the best sensitivity to an exotic dipole-dipole interaction between an electron and nucleons. 
The sensitivity of their proposed method is expected to be at a level that is competitive with astrophysical bounds. 
See Sec.\,\ref{subsec_g_Pg_S} for more details. 

In the range $10^{-8} \, \textrm{m} < \lambda < 10^{-4}$\,m, \citet{guo_searching_2024} propose a diamond-based vector magnetometer to probe $g_p^e g_p^N$. 
See Sec.\,\ref{subsec_g_Ag_A} for further details.

\subsection{\texorpdfstring{Scalar/Scalar interaction $g_sg_s$}{Scalar/Scalar interaction gsgs}}
\label{subsec_g_sg_s}

Experimental constraints on the scalar/scalar interaction $g_s g_s$ arise from the ${V}_1$ and $V_{4+5}$ potential terms: 

\begin{equation}
\label{gsgs_V1}
V_1|_{ss} = - g^X_s g^Y_s \frac{\hbar c}{4\pi}\frac{e^{-{r}/{\lambda}}}{r} \, ,
\end{equation}

{\small
\begin{equation}
\label{gsgs_V45}
\begin{aligned}
&V_{4+5}|_{ss}=\frac{g_s^X g_s^Y}{4} \frac{\hbar^2}{c}\left\{ \boldsymbol{\sigma}_X \cdot \left( \frac{\boldsymbol{p}_X}{m_X^2} \times \hat{\boldsymbol{r}} \right), \left( \frac{1}{r^2} + \frac{M}{r} \right) \frac{e^{-Mr}}{8 \pi} \right\} \\
&\Rightarrow g_s^X g_s^Y \frac{\hbar^2}{16\pi c}\frac{m_Y}{m_X(m_X+m_Y)} \, \boldsymbol{\sigma}_X \cdot(\boldsymbol{v} \times \hat{\boldsymbol{r}}) \left(\frac{1}{r^2}+\frac{1}{\lambda r}\right) e^{-r/\lambda} \, , 
\end{aligned}
\end{equation}
}with the expression in the second line of Eq.\,\eqref{gsgs_V45} written in the two-body CM frame. 
Note that papers, such as \citet{heckel_preferred-frame_2008}; \citet{haddock_search_2018}; \citet{kim_experimental_2018}; \citet{wu_new_2023},
which studied the $V_{4+5}$ term but did not directly set constraints on $g_s^X g_s^Y$, can transfer their constraints to $g_s^X g_s^Y$ by comparing their equations with Eq.\,\eqref{gsgs_V45}. See Tab.\,\ref{tabel_f-gg} in App.\,\ref{appendixA} for more details.

The $V_{4+5}$ potential describes a new-boson-mediated spin- and velocity-dependent interaction between a spin and a moving source. 
While numerous earlier experiments searching for spin-dependent forces focused on static spin-dependent interactions \cite{safronova_search_2018}, recent investigations into the $V_{4+5}$ potential exhibit a noteworthy trend toward exploring the realm of spin- and velocity-dependent interactions. 
Existing experimental constraints on the $V_{4+5}$ term have been obtained using various experimental methods, including the torsion pendulum \cite{heckel_preferred-frame_2008}, atomic magnetometers \cite{kim_experimental_2018,wu_experimental_2022}, cantilevers \cite{ding_constraints_2020}, NV center-based quantum sensors \cite{chu_proposal_2022},
neutron spin rotation experiments
\cite{piegsa_limits_2012}, and atomic hyperfine spectroscopy \cite{ficek_constraints_2017}. 
These experiments cover a broad range of interaction length scales. 
Most of the experiments that study the $V_{4+5}$ term can also be used to study the spin- and velocity-dependent interaction $V_{12+13}$ without significant (or any) modifications to the experimental setup.

\begin{figure} [!htbp]
\begin{center}
\includegraphics[width=0.48\textwidth]{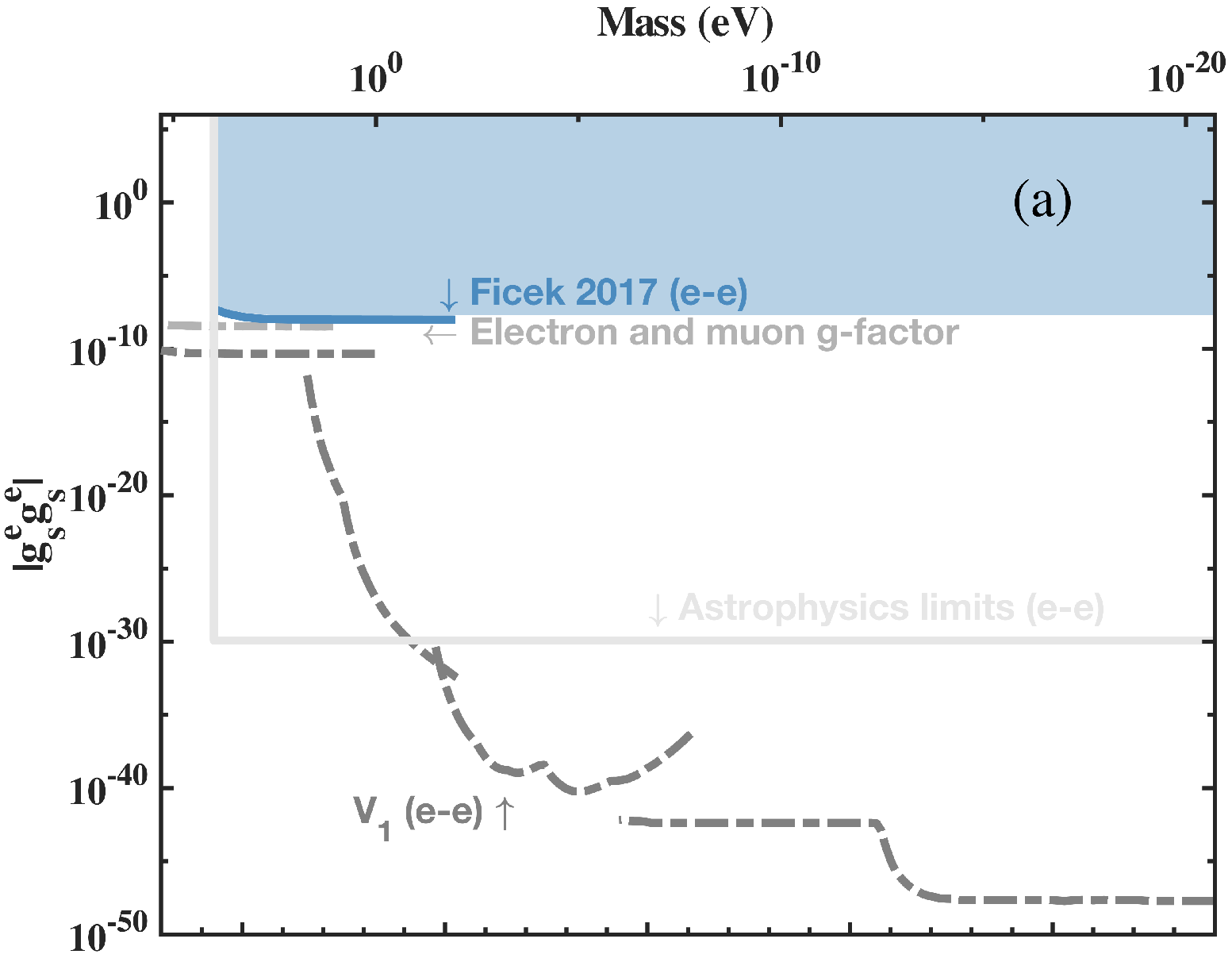}

\includegraphics[width=0.48\textwidth]{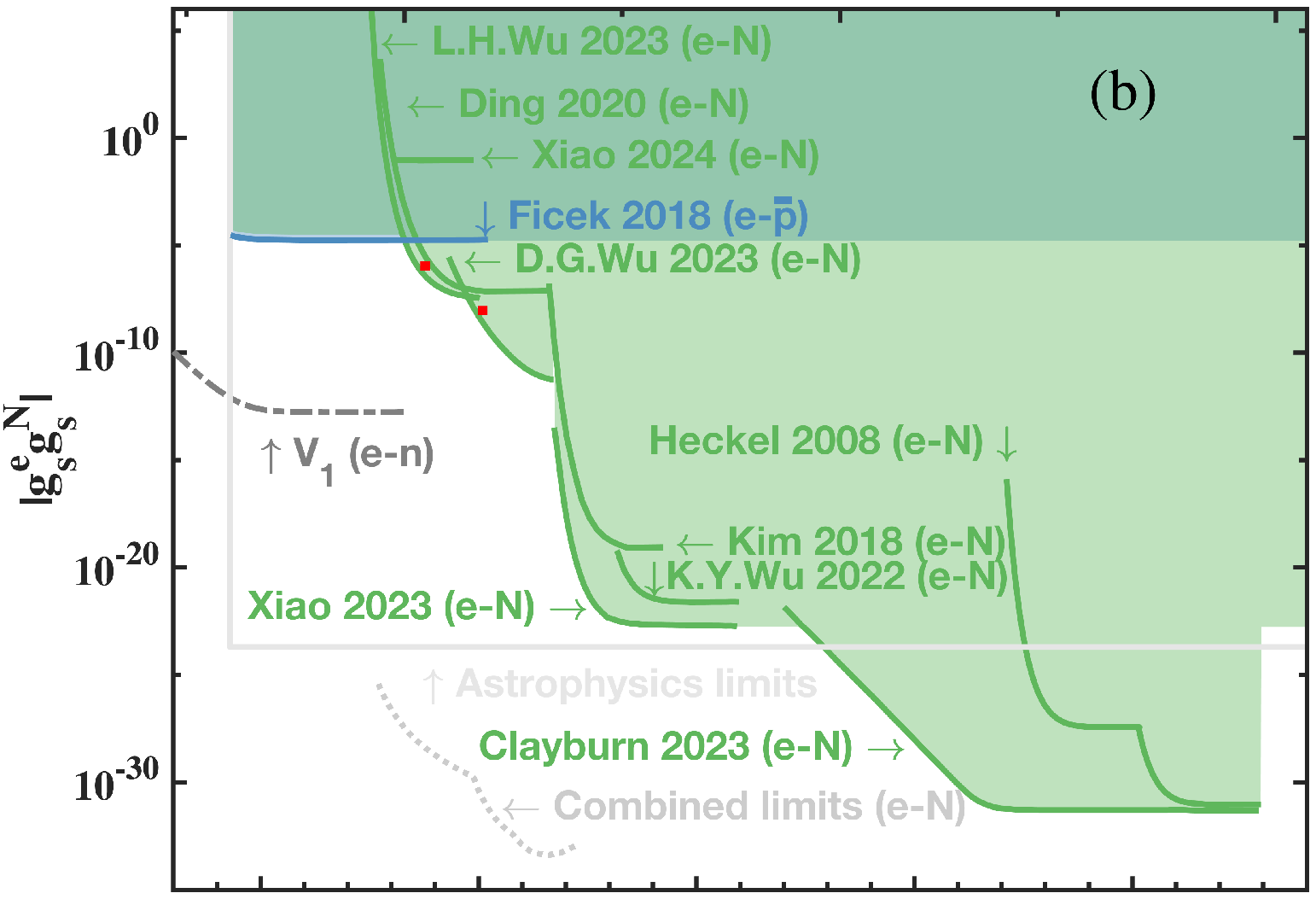}

\includegraphics[width=0.48\textwidth]{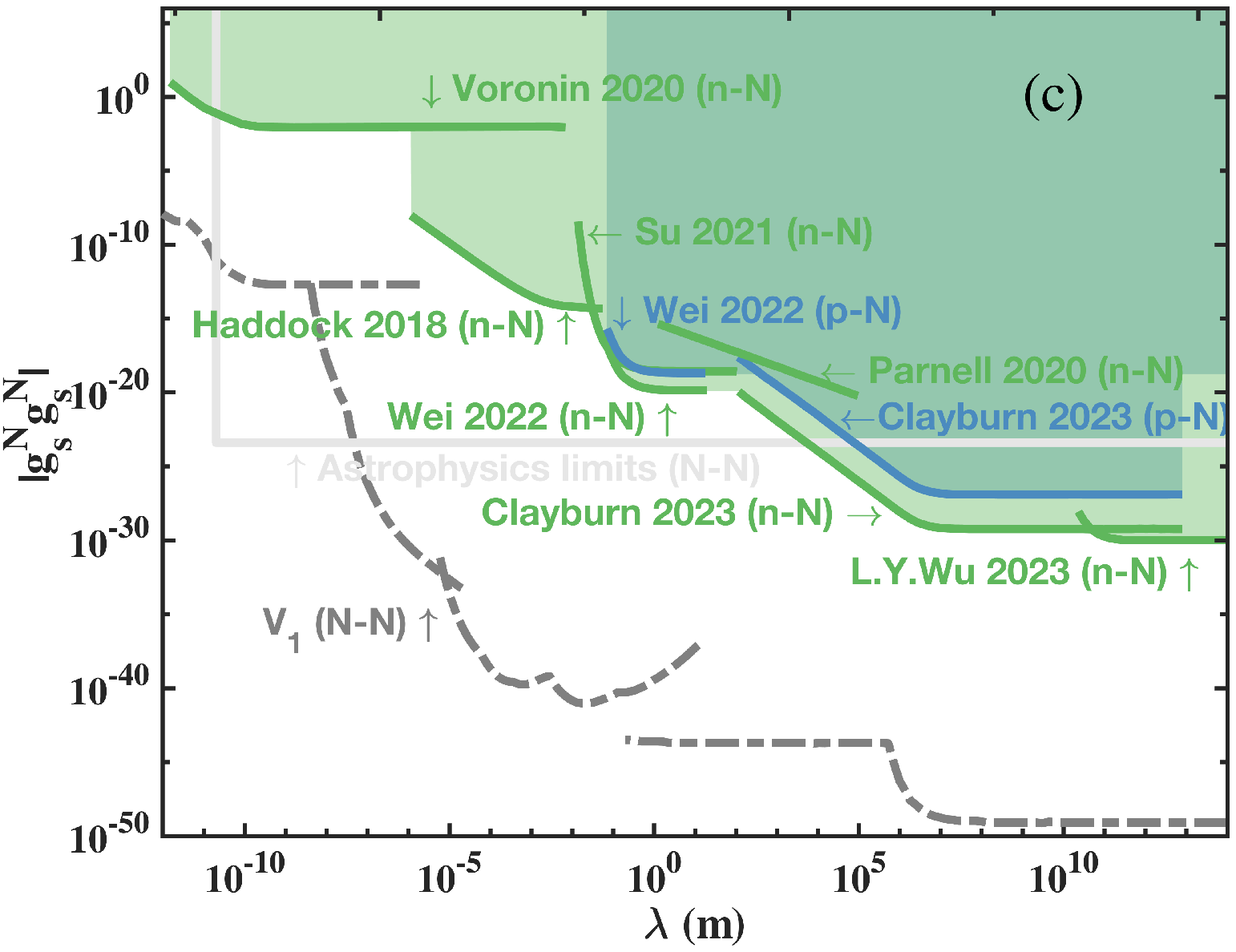}
\end{center}
\caption{
Constraints, depicted by coloured regions, on the coupling constant product $g_s g_s$ as a function of the interaction range $\lambda$ shown on the bottom x-axis. 
The top x-axis represents the new spin-0 boson mass $M$. 
The subscript $s$ denotes the scalar coupling. Constraints are shown for \textbf{(a)} $e$-$e$, \textbf{(b)} $e$-$N$ and $e$-$\overline{p}$, and \textbf{(c)} $n$-$N$ and $p$-$N$ couplings. 
Results are based on the $V_{4+5}$ term. 
The $V_1$ limits (dark grey dash-dotted lines), astrophysical limits (light grey solid lines) and combined astrophysical-$V_1$ limits (dotted grey line) are elaborated in Appendix~\ref{appendix_comp.}. 
In subfigure \textbf{(b)}, the two red dots represent the nonzero results from \citet{rong_observation_2020}. 
}
\label{gsgs-fig}
\end{figure}

Figure\,\ref{gsgs-fig} shows the current laboratory constraints on the exotic scalar/scalar interaction $g_s^X g_s^Y$ between the studied particle pairs (a) $e$-$e$, (b) $e$-$N$ and $e$-$\overline{p}$, and (c) $n$-$N$ and $p$-$N$. 
Constraints from spin-independent $V_1$ limits (dark grey dash-dotted lines), astrophysical limits (light grey solid lines) and combined astrophysical-$V_1$ limits (dotted grey line) are also shown for comparison, see Appendix~\ref{appendix_comp.} for more details.

\subsubsection{\texorpdfstring{$e$-$e$}{e-e}}
\citet{ficek_constraints_2017} established the only constraints so far on the spin-dependent scalar/scalar interaction $g_s g_s$ between electrons at the atomic scale through investigation of the electronic structure of $^4$He by studying the $V_{4+5}$ potential term. 
The details of this work are discussed in Sec.\,\ref{subsec_g_Ag_A}.

\subsubsection{\texorpdfstring{$e$-$N$}{e-N}}
\label{subsubsec:gsgs_eN}

In the range $10^{7} \, \textrm{m}<\lambda<10^{13}$\,m, \citet{heckel_preferred-frame_2008} were the first to study the exotic spin- and velocity-dependent interaction $V_{4+5}$ with their torsion-pendulum experiments (see Sec.\,\ref{METH1.SENS.MS}) and set stringent limits on the coupling $g_s^e g_s^N$. 
They studied interactions between polarized electrons in the pendulum and unpolarized matter in the Earth, Moon and Sun. 
Further details on the work of \citet{heckel_preferred-frame_2008} can be found in Sec.\,\ref{subsec_g_Ag_V}. 

In the range $10^{2} \, \textrm{m} < \lambda < 10^{13}$\,m, \citet{clayburn_using_2023} utilized the results from the E\"ot-Wash torsion-pendulum experiment \cite{heckel_preferred-frame_2008} and used their own model which describes Earth as a rotating, unpolarized source \cite{hunter_using_2013, hunter_using_2014} to set the most stringent constraints on $g_s^e g_s^N$. 
Further details can be found in Sec.\,\ref{subsec_g_Ag_V}. 

In the interaction range $5 \times 10^{-4} \, \textrm{m} < \lambda < 10$\,m, the most stringent constraints were set by
\citet{wu_experimental_2022,kim_experimental_2018} and \citet{xiao_femtotesla_2023}. 
All of these searches used atomic magnetometers (see Sec.\,\ref{METH1.SENS.AM}) and made a noteworthy contribution by placing constraints within the classical QCD ``axion window''. 

In the range $\sim 10^{-4} \, \textrm{m} < \lambda < 10^{-1}$\,m, \citet{kim_experimental_2018} used a SERF magnetometer (see Sec.\,\ref{METH1.SENS.AM}) to study the exotic spin- and velocity-dependent interaction $V_{4+5}$ and set limits on $g_s^e g_s^N$. 
They utilized an experimental approach based on their previous proposal \cite{chu_search_2016,karaulanov_spin-exchange_2016}. 
The $^{87}$Rb magnetometer detects the pseudomagnetic field arising from the exotic interaction between polarized electrons in the Rb atoms and unpolarized nucleons of a nearby solid-state BGO mass. 

In the range $10^{-2} \, \textrm{m} < \lambda < 10$\,m, \citet{wu_experimental_2022} employed an array of magnetometers and rotationally-modulated BGO source masses to search for exotic spin-dependent interactions and improved upon the result of \citet{kim_experimental_2018} by up to a factor of 140. 
Further details on the work of \citet{wu_experimental_2022} can be found in Sec.\,\ref{subsec_g_Ag_V}.

In the range $5\times10^{-4} \, \textrm{m} < \lambda < 10$\,m, \citet{xiao_femtotesla_2023} developed a diffusion optical pumping scheme to reduce the noise and systematic errors, 
such as light shifts and power-broadening effects,  in a Rb magnetometer and set the most stringent constraints on $V_{4+5}$ using a rotating non-magnetic BGO crystal as the source of unpolarised nucleons. 
It is likely that their method can be applied to further improve other kinds of magnetometers. 

For interaction ranges around $\lambda \sim 10^{-6}$\,m, \citet{ding_constraints_2020} searched for the exotic spin- and velocity-dependent interaction $V_{4+5}$ by measuring the force between a gold sphere and a microfabricated magnetic structure using a cantilever-based instrument (see Sec.\,\ref{METH1.SENS.MS.MMO}) and set limits on $g_s^e g_s^N$. 
This cantilever-based experiment, 
similar
to 
the experimental methods discussed by
\citet{leslie_prospects_2014}, \citet{chen_ultrasensitive_2021}, \citet{ren_search_2021}, and \citet{wang_proposal_2023},
directly measures the force 
between polarized electrons in a magnetic structure and unpolarized nucleons in a gold sphere,
an approach analogous to that used in torsion balance experiments such as those of \citet{heckel_new_2006}.
A periodic magnetic structure consisting of stripes with alternating antiparallel electron spin polarization was employed to suppress spurious background signals.
In their experiments, an exotic lateral force is modulated at harmonics of the driving frequency when the magnetic structure oscillates. 
\citet{ding_constraints_2020} chose to measure the sixth harmonic component, which due to the particular pattern of the magnetic structure and through the velocity-dependent exotic force is related solely to the out-of-phase component of the modulation. This feature enables discrimination of the exotic interaction from other known interactions, such as electrostatic, Casimir and magnetic forces, which in this setup do not contribute to the component of the signal at the sixth harmonic as they generate signals in phase with the modulation.

\citet{rong_observation_2020} reported a surprising observation of a nonzero coupling between a single NV center and a mechanically oscillating SiO$_2$ mass. 
The analysis suggested the presence of a new spin- and velocity-dependent force at two interaction lengths, namely $0.38\,\mu$m, and $8.07\,\mu$m, implying the existence of two new bosons with masses of about 0.5\,eV and 25\,meV, respectively, as denoted by the two red dots in Fig.\,\ref{gsgs-fig}\,(b). 
This nonzero result has now been ruled out \cite{wu_improved_2023, wu_spin-mechanical_2023} and the authors conclude that the earlier new physics signal may have come from instrumental artifacts. 

\citet{wu_improved_2023} focused on using an ensemble-NV-diamond magnetometer to search for exotic interactions between polarized electron spins and unpolarized nucleons at the micrometer scale. 
This study utilized diamond NV centers as both polarized electron sources and sensitive sensors, and used a vibrating lead sphere as the moving unpolarised nucleon source. 
Extending the measurements to an NV ensemble rather than a single NV center, \citet{rong_observation_2020} provided better magnetic detection sensitivity. Additionally, by averaging the magnetic field sensed by each NV center, the NV ensemble is insensitive to the effect of the diamagnetism of the material of the nucleon source \cite{liang_new_2022}.

\citet{wu_spin-mechanical_2023} demonstrated that their prototype chip, which integrates a mechanical resonator and a diamond with a single nitrogen vacancy at the microscale, sets constraints on the $V_{4+5}$ interaction that are a few orders of magnitude better compared to the sensitivity demonstrated in \citet{ding_constraints_2020}. 
Since the sensitivity improves with the number of sensors, integrating multiple sensors on a chip could significantly advance searches for exotic interactions when scaled up.

\subsubsection{\texorpdfstring{e-$\overline{p}$}{e-p̅}}
By comparing theoretical calculations and spectroscopic measurements of the hyperfine structure of antiprotonic helium, \citet{ficek_constraints_2018} set the only constraints on the $V_{4+5}$ potential term for the exotic semi-leptonic interaction between the fermion-antifermion pair e-$\overline{p}$. 
The details of this work are discussed in Sec.\,\ref{subsec_g_Ag_A}.

\subsubsection{\texorpdfstring{$N$-$N$}{N-N}}

At astronomical distances, \citet{wu_new_2023} analyzed existing data from laboratory measurements on Lorentz and $CPT$ violation to derive upper limits on the exotic spin-dependent interaction term $V_{4+5}$. 
Details of this work are discussed in Sec.\,\ref{subsec_g_Ag_V}. 

In the interaction range $10^{2} \, \textrm{m} < \lambda < 4.5 \times 10^{10}$\,m, \citet{clayburn_using_2023} set the most stringent constraints on $g_s^n g_s^N$ (as well as for $g_s^p g_s^N$ in the range $10^{2} \, \textrm{m} < \lambda < 10^{12}$\,m). 
Further details can be found in Sec.\,\ref{subsec_g_Ag_V}. 
In the range $1 \, \textrm{m} < \lambda < 10^{5}$\,m, \citet{parnell_search_2020} set constraints on the exotic scalar-scalar interaction $g_s^n g_s^N$. 
\citet{parnell_search_2020} used a spin-echo neutron spectrometer to detect possible exotic spin-dependent interactions of neutrons influenced by Earth's gravitational field. 
An earlier 
experiment 
by \citet*{colella_observation_1975}
demonstrated a gravitational phase shift in neutron interferometry. 
Experiments with a neutron spin-echo interferometer \cite{de_haan_measurement_2014} further improved upon these results and \citet{parnell_search_2020} used these data to obtain strong constraints on the $V_{4+5}$ and $V_{12+13}$ interactions. These constraints have been surpassed by those obtained by \citet{su_search_2021}; \citet{wei_constraints_2022}; \citet{clayburn_using_2023}.

In the range $10^{-2} \, \textrm{m} < \lambda < 10^{2}$\,m, \citet{su_search_2021} set the constraints on $g_s^n g_s^N$ based on the spin- and velocity-dependent interaction term $V_{4+5}$. 
The details are discussed in Sec.\,\ref{subsec_g_Ag_V}. 
For interaction ranges around $\lambda \sim 1$\,m, \citet{wei_constraints_2022} conducted an experiment using a high-mass-density tungsten ring as a source and a high-energy-resolution K-Rb-$^{21}$Ne self-compensating comagnetometer. This experiment set constraints on the $V_{4+5}$ $n$-$N$ interaction and improved on the previous result of \citet{su_search_2021} by about an order of magnitude. Taking into account the fraction of proton spin in the $^{21}$Ne nucleus, this experiment also placed constraints on the $p$-$N$ interaction. 

In the range $10^{-6} \, \textrm{m} < \lambda < 10^{-2}$\,\textrm{m}, \citet{haddock_search_2018} presented a search for spin-dependent interactions between the neutron (see more in Sec.\,\ref{METH1.SENS.PBS}) and unpolarized nucleons based on the spin- and velocity-dependent interaction $V_{4+5}$, which can be interpreted as a constraint on $g_s^n g_s^N$.
The central idea of \citet{haddock_search_2018} is that that when transversely polarized slow neutrons \cite{piegsa_limits_2012, yan_new_2013} pass near the surface of an unpolarized bulk material (copper), exotic interactions would cause the neutron polarization plane to tilt and develop a component along the neutron momentum. 
A Monte Carlo method was used to simulate the tilt induced by an exotic force. 
This experiment represents a two to four orders of magnitude improvement over the previous work \citet{piegsa_limits_2012}. 
The authors suggested that an additional two orders of magnitude improvement in sensitivity is possible by using higher intensity cold neutron beams, longer running times, and denser target materials, such as tungsten in place of copper. 
A similar experimental setup was proposed by \citet{yan_constraining_2019} to probe spin-dependent interactions of muons.

In the range $10^{-12} \, < \lambda < 10^{-6}$\,m, \citet{voronin_constraint_2020} set the best constraints on exotic $n$-$N$ interactions, 
using 
neutron diffraction data in a noncentrosymmetric
crystal (see Sec.\,\ref{subsec_g_Pg_S_n-N} for more details).

\subsubsection{Future perspectives}
The recent work by \citet{rong_observation_2020},  described in Sec.\,\ref{subsubsec:gsgs_eN}, on the search for exotic forces using diamond NV centers has inspired the implementation of new measurement schemes, for example, the NV-ensemble method \cite{wu_improved_2023} and spin-mechanical quantum chips \cite{wu_spin-mechanical_2023}. 
Further improvement on the result of \citet{wu_improved_2023,wu_spin-mechanical_2023} is possible (\citealp{leslie_prospects_2014}; \citealp{chu_search_2015}, \citeyear{chu_proposal_2022}).
\citet{chu_proposal_2022} proposed using multipulse quantum sensing protocols with NV spin ensembles to improve the sensitivity to the spin-dependent interaction $V_{4+5}$. 
The projected reach of the proposal of \citet{chu_proposal_2022} surpasses that of \citet{chu_search_2015} and \citet{leslie_prospects_2014}, which are proposals based on different techniques. 
The details of \citet{chu_proposal_2022} are introduced in Sec.\,\ref{subsec_g_Ag_V}.


\section{Conclusion and Outlook }
\label{Sec:Sum-Out}

\subsection*{A. Summary}
\label{Sec:Sum}

In this review, we discussed the motivation for searching for spin-dependent exotic interactions and surveyed the vast body of literature devoted to searches for such exotic interactions of various types and between various particles. 
This includes identifying relations between phenomena and experiments (such as atomic parity violation, searches for $P,T$-odd electric dipole moments, dark matter searches, and so on) that have not traditionally been directly associated with long-range exotic interaction searches. Historically, atomic PNC experiments and EDM searches have been interpreted as searches for (new) short-range forces (e.g., the usual $Z$ boson in the case of the SM or a heavy extra $Z'$ boson beyond the SM in the case of atomic PNC experiments, and a short-range/contact fermion-fermion interaction in the case of atomic EDMs). The possibility of long-range forces (on atomic scale and larger) in these types of phenomena has only been explored more recently. 

As the first step, we considered various ``catalogs'' of exotic interaction potentials and formulated the most complete and consistent set of potentials to date. 
We then described a broad suite of experiments of various types that can be used to probe these exotic interactions.

The strengths of exotic interactions are proportional to 
underlying
coupling constants, with various combinations of these generally appearing in multiple exotic potentials. 
Presenting the results of experiments and data analysis in terms of these coupling constants, as we have done here, allows for comparison of vastly different experiments, observations, and analyses in a systematic way.

Following this approach, we provided a global analysis and presented exclusion limits (where possible, in the form of exclusion plots) on combinations of coupling constants, identifying the most sensitive systems and experiments, thus providing guidance for future efforts in this rapidly developing field.

\subsection*{B. Future directions}
\label{Sec:Conclusion_future_directions}

After a comprehensive analysis of the current state of research, one may observe that the following topics deserve attention in future studies.

\paragraph*{a. Using a unified theoretical framework.}Working within a unified theoretical framework will allow all researchers to compare their results conveniently. 
Based on the unified theoretical framework presented in Sec.\,\ref{Sec:formalism}, we underscore the imperative for future experimental studies to comprehensively explore the nine different types of interactions 
that generate Eqs.\,\eqref{pseudovector-vector_potential} -- \eqref{pseudotensor-pseudotensor_potential}.
Potential terms such as $V_{2}+V_{3}$, $V_{3}$, $V_{4+5}$, $V_{9+10}$ and $V_{14q+15}$ 
can be generated by multiple different types of interactions.
For example, $V_{4+5}$ appears in axial-vector/axial-vector as well as vector/vector and scalar/scalar interactions. 
This unified theoretical framework overcomes the limitations of focusing solely on
parameters such as $f_{4+5}$, enabling the translation of results to specific interaction types described by coupling-constant combinations, such as $g_Ag_A$, $g_Vg_V$ and $g_sg_s$, thus enabling consistent use of the full body of experimental results and comparision to results from other subfields, such as astrophysics, see Sec.\,\ref{METH2.Astro.} and App.\,\ref{appendix_comp.}.

\paragraph*{b. Focus on understudied potentials.} Despite the intensive study of terms like $V_{4+5}$, $V_{9+10}$, and $V_{12+13}$, 
a crucial term $V_3$ for both pseudoscalar/pseudoscalar and axial-vector/axial-vector interactions, in our opinion has not garnered sufficient attention, see Sec.\,\ref{subsec_g_Pg_P}. 
This gap presents an opportunity to apply advanced quantum sensors, such as NV centers and cantilevers (see Sec.\,\ref{METH1.SENS.}), in new contexts. 
Notably, the combination $V_{2}+V_{3}$ appearing in the vector/vector interaction (as well as tensor/tensor interaction) is an unexplored area for experimental study, as explained in Sec.\,\ref{subsec_g_Vg_V}. 

Exotic interactions between certain fermion pairs warrant more attention. 
In particular, the realm of purely leptonic (such as $e$-$e$) interactions, with the exception of the axial-vector/axial-vector couplings (see Sec.\,\ref{subsec_g_Ag_A}), appears relatively unexplored. 
Investigations of other types of interactions, such as $g_A^e g_V^e$, for the $e$-$e$ fermion pair are limited, often constrained to one or two studies for each kind of $V_i$ term. 
Studies of interactions involving antimatter, for example, $e$-$\overline{p}$, are particularly scarce, with only a few studies to date \citet{fadeev_pseudovector_2022,ficek_constraints_2018}.

\paragraph*{c. Enhancing experimental techniques.} 
The field stands to benefit significantly from advances in the sensitivity and development of new protocols in source-sensor experiments (see Sec.\,\ref{Sec:EXP.METH.1}). 
Quantum NV-diamond sensors, for example, have enabled improvements in the sensitivity to exotic interactions and promise further improvements; see, e.g., \citet{jiao_experimental_2021}; \citet{chu_proposal_2022}.
Additionally, atom interferometry presents another promising avenue for exploration (\citealp{hamilton_atom-interferometry_2015}; \citealp{du_atom_2022}).
See \citet{ye_essay_2024} for more discussion of quantum sensing with atomic, molecular, and optical platforms for fundamental physics.
Levitated-magnet experiments \cite{ahrens_levitated_2024} with high sensitivity augment the tool box for exploring spin-dependent fifth forces.

\paragraph*{d. Attention to non-dedicated experiments.}
Besides dedicated source-sensor experiments, experiments that have not originally been designed to detect exotic spin-dependent interactions can also play an important role, especially in exploring exotic interactions on shorter length scales and between certain special fermion pairs, such as $e$-$\bar{p}$ and $e$-$\mu$. 
Constraints from such experiments are often derived by precisely comparing theory predictions and experimental results to search for a possible discrepancy which may originate from a spin-dependent exotic interaction. 
For example, EDM experiments and atomic parity-violation experiments are valuable methods worthy of further attention, see Sec.\,\ref{EXP.METH.2} for more details. 
Additionally, reinterpretation of results from studies such as those of \citet{jedi_collaboration_first_2023} could offer new insights into proton-nucleon and neutron-nucleon interactions, potentially surpassing or complementing existing constraints (\citealp{vasilakis_limits_2009}; \citealp{hunter_using_2013}).

\paragraph*{e. Theoretical framework expansion.}
The established theoretical framework has mainly been centered on effective Lagrangians involving spin-0 or spin-1 bosons. 
However, it is also interesting to consider possible extensions to bosons with higher spins. 
The study of exotic interactions mediated by massive spin-2 particles, ``heavy gravitons'', and the two-graviton exchange interaction \cite{holstein_spin_2008} may be of particular interest. 
Although these particles have not been pursued as extensively as their lower-spin counterparts, they represent a potential frontier in BSM physics.
Further theoretical development, including the incorporation of two-boson exchange (\citealp{fischbach_new_1999}; \citealp{adelberger_particle-physics_2007}),
is also of paramount importance. 

In models with operators that are quadratic in the boson field, besides the simultaneous exchange of a pair of bosons between two fermions in vacuum, the exchange of a single boson between two fermions becomes possible in the presence of a background of those bosons, with the strength and range of this in-medium force potentially greatly exceeding those of two-boson exchange processes in vacuum \citet{hees_violation_2018}. 
Since the bosonic background can come from the ``dark sector'', e.g., in the form of dark matter, these scenarios deserve attention in future work.

\addcontentsline{toc}{section}{Acknowledgements}
\section*{Acknowledgements}

L.~C. acknowledge helpful discussion with Younggeun Kim, Nathan Clayburn, Konstantin Gaul, and Man Jiao. W.~J. acknowledge helpful discussion with Dong Sheng and Yevgeny Kats.
L.~C. was partially supported by the OCPC-Helmholtz International Postdoctoral Exchange Fellowship Program (Grant No. ZD202116).
The work of Y.~V.~S.~was supported by the Australian Research Council under the Discovery Early Career Researcher Award No. DE210101593. 
The work of D.~F.~J.~K. was supported by the U.S. National Science Foundation under grant PHYS-2110388. The work of V. V. F.  was supported by the Australian Research Council Grants No. DP230101058 and DP200100150.
This research was supported in part by the DFG Project ID 390831469: EXC 2118 (PRISMA+ Cluster of Excellence), by the QuantERA project LEMAQUME (DFG Project No. 500314265) and by the COST Action within the project COSMIC WISPers (Grant No. CA21106), and the Munich Institute for Astro-, Particle and BioPhysics (MIAPbP), which is funded by the Deutsche Forschungsgemeinschaft (DFG, German Research Foundation) under Germany's Excellence Strategy – EXC-2094 – 390783311.

\newpage 

\section*{Appendix} 

\appendix 
\addcontentsline{toc}{section}{Appendix}

\section{\texorpdfstring{Commonly used format of $V_i$}{Commonly used format of Vi}}
\label{appendixA}

In Sec.\,\ref{Sec:formalism}, we present a complete form of the potentials, not only the leading-order terms but also the higher-order terms (e.g., $V_{4+5}$ and $V_8$) that have been studied experimentally. 
We use $V_i$ to label independent terms or combined terms to show the connection between the potentials in this review and those existing in the literature. 
To further clarify this point, here we present the potentials appearing in the main text [Eqs.\,\eqref{pseudovector-vector_potential} -- \eqref{pseudotensor-pseudotensor_potential}] as they appear in earlier references. 
For historical reasons, in contrast to the potentials in this review properly classified by the coupling-strength coefficients $g^Xg^Y$, the previous commonly used potentials are presented with the coefficients $f_i^{XY}$. 
We do the same here and then include a table (Tab.\,\ref{tabel_f-gg}) summarizing the relationship between the coefficients $f_i^{XY}$ and $g^Xg^Y$. 
In addition, we also compare the potentials in this review to the earliest ones in \citet{dobrescu_spin-dependent_2006} with an example of $V_{4+5}$, and show that we have the same result. 
In the end, we briefly point out the differences between the equations in the sections here and those in the literature.

\subsection{Potentials sorted by number}
\label{App:f^{XY}_i}

We start by presenting the set of potentials in the center-of-mass frame, following the format $V_i=f_i\sV_i$, where $\sV_i$ are the potentials from \citet{dobrescu_spin-dependent_2006} and are presented in Eqs.\,\eqref{DM_eqV1} -- \eqref{DM_eqV16}. 
For paired potentials, $\sV_{i\pm j}=\frac{1}{2}(\sV_{i}\pm\sV_{j})$. 
The first several potentials enumerated by the index $i = 1,...,8$ encompass all possible $P$-even (scalar) rotational invariants, which in the nonrelativistic limit are given by: 
\begin{equation}\label{eqV1}
V_1  = f^{XY}_1 \frac{\hbar c}{4\pi} \frac{1}{r}\,e^{-{r}/{\lambda}} \, ,
\end{equation}
\begin{equation}\label{eqV2}
V_2  = f^{XY}_2\frac{\hbar c}{4\pi}\boldsymbol{\sigma}_X\cdot\v\sigma_Y^{\,\prime}\frac{1}{r}\,e^{-{r}/{\lambda}} \, ,
\end{equation}
\begin{align}\label{eqV3}
V_3  = f^{XY}_3 \frac{\hbar^3}{4\pi m_X^2c}\left[\boldsymbol{\sigma}_X\cdot\v\sigma_Y^{\,\prime} \left(\frac{1}{r^3}+\frac{1}{\lambda r^2}+\frac{4\pi}{3} \delta(\boldsymbol{r}) \right)\nonumber \right. \\\left. -(\boldsymbol{\sigma}_X\cdot \hat{\boldsymbol{r}})(\v\sigma_Y^{\,\prime}\cdot \hat{\boldsymbol{r}})\left(\frac{1}{\lambda^2 r}+\frac{3}{\lambda r^2}+\frac{3}{r^3}\right)\right]e^{-{r}/{\lambda}} \, ,
\end{align}
\begin{equation}\label{eqV45}
V_{4+5}  = -f^{XY}_{4+5}\frac{\hbar^2}{16\pi m_X^2c} \left\{ \boldsymbol{\sigma}_X \cdot \left( \boldsymbol{p}_X \times \hat{\boldsymbol{r}} \right), \left(\frac{1}{r^2}+\frac{1}{\lambda r}\right) e^{-r/\lambda} \right\} \, , 
\end{equation}
\begin{equation}\label{eqV8}
V_{8}  = f^{XY}_8\frac{\hbar}{16\pi m_X^2 c}\left\{\boldsymbol{\sigma}_X\cdot \boldsymbol{p}_X,\left\{\v\sigma_Y^{\,\prime}\cdot \boldsymbol{p}_X, \frac{1}{r}e^{-{r}/{\lambda}}    \right\}\right\} \, . 
\end{equation}
The rest of the potentials enumerated by the index $i = 9,...,16$ encompass all possible $P$-odd (pseudoscalar) rotational invariants, given in the nonrelativistic limit by:\footnote{Note that unlike other coefficients $f_i$, the coefficient $f_{14}$ in the potential $V_{14}$
vanishes in the lowest order of momenta $f_i(0,0) \propto q^0$ considered by \citet{dobrescu_spin-dependent_2006}, and becomes nonzero only at order $q^2$
as noted in \cite{fadeev_neue_2018, fadeev_revisiting_2019}. 
This is because, while symmetries can determine the spin structure of a potential term like $V_{14}$, they do not uniquely predict the power-law dependence on the inter-particle separation (or equivalently, the momentum transfer).} 
\begin{equation}\label{eqV910}
V_{9+10}=-f^{XY}_{9+10}\frac{\hbar^2}{8\pi m_X}\boldsymbol{\sigma}_X\cdot \hat{\boldsymbol{r}}\,\, \left(\frac{1}{r^2}+\frac{1}{\lambda r}\right)\, e^{-{r}/{\lambda}} \, , 
\end{equation}
\begin{equation}\label{eqV11}
V_{11}=-f^{XY}_{11}\frac{\hbar^2}{4\pi m_X} \left(\boldsymbol{\sigma}_X\times\v\sigma_Y^{\,\prime} \right) \cdot \hat{\boldsymbol{r}}\left(\frac{1}{r^2}+\frac{1}{\lambda r}\right)e^{-{r}/{\lambda}} \, ,
\end{equation}
\begin{equation}\label{eqV1213}
V_{12+13}=f^{XY}_{12+13}\frac{\hbar}{16\pi m_X} \left\{\boldsymbol{\sigma}_X \cdot \boldsymbol{p}_X, \frac{1}{r}e^{-{r}/{\lambda}}    \right\} \, , 
\end{equation}
\begin{widetext}
\begin{equation}\label{eqV1415}
\begin{aligned}
V_{14q+15} 
= f_{14q+15}^{XY} \frac{ \hbar^3} {16 \pi m_X^3 c^2} \left[ \left\{ \left( \v{\sigma}_X \times \v\sigma_Y^{\,\prime}\right) \cdot \v{p}_X, \frac{1}{r^3} + \frac{1}{\lambda r^2} + \frac{4 \pi}{3} \delta(\v{r})  \right\} -\left\{ \left( \v{\sigma}_X \cdot \hat{\v{r}} \right) \v\sigma_Y^{\,\prime}\cdot \left(\v{p}_X \times \hat{\v{r}} \right) , \frac{3}{r^3} + \frac{3}{\lambda r^2} + \frac{1}{\lambda^2 r}  \right\}\right]e^{-{r}/{\lambda}} \, , \\
\end{aligned}
\end{equation}

\begin{equation}
\label{eqV1116}
\begin{aligned}
V_{11p+16}
= 
-f_{11p+16}^{XY} \frac{ \hbar^2}{32 \pi m_X^3 c^2} \left\{ \boldsymbol{\sigma}_X \cdot \boldsymbol{p}_X , \left\{ \v\sigma_Y^{\,\prime} \cdot \left( \v{p}_X \times \hat{\v{r}} \right) , \left( \frac{1}{r^2} + \frac{1}{\lambda r} \right)   e^{-{r}/{\lambda}} 
  \right\} \right\} \, , \\
\end{aligned}
\end{equation}
\begin{equation}\label{eqV15}
\begin{aligned}
V_{15} = f^{XY}_{15} \frac{\hbar^3}{16\pi m_X^3 c^2} \left\{(\boldsymbol{\sigma}_X \cdot \hat{\boldsymbol{r}})[\v\sigma_Y^{\,\prime} \cdot ( \hat{\boldsymbol{r}} \times \boldsymbol{p}_X)] + (\v\sigma_Y^{\,\prime}\cdot\hat{\boldsymbol{r}}) [\boldsymbol{\sigma}_X \cdot (\hat{\boldsymbol{r}} \times \boldsymbol{p}_X)], \left(\frac{3}{r^3}+\frac{3}{\lambda r^2}+\frac{1}{\lambda^2 r}\right) e^{-{r}/{\lambda}}    \right\} 
\, , 
\end{aligned}
\end{equation}
{\small
\begin{equation}\label{eqV16}
V_{16} = - f^{XY}_{16} \frac{\hbar^2}{32\pi m_X^3 c^2}\left[\left\{\boldsymbol{\sigma}_X\cdot\boldsymbol{p}_X, \left\{\v\sigma_Y^{\,\prime}\cdot \left( \boldsymbol{p}_X \times \hat{\boldsymbol{r}} \right), \left(\frac{1}{r^2}+\frac{1}{\lambda r}\right)\, e^{-{r}/{\lambda}}\right\}\right\}
+
\left\{\v\sigma_Y^{\,\prime}\cdot\boldsymbol{p}_X, \left\{\boldsymbol{\sigma}_X\cdot \left(\boldsymbol{p}_X \times \hat{\boldsymbol{r}}\right), \left(\frac{1}{r^2}+\frac{1}{\lambda r}\right)\,e^{-{r}/{\lambda}}\right\}\right\} 
\right] \, .
\end{equation}
}
\end{widetext}
Here, all the 
symbols or notations
are defined in the same way as in the main text, see the description below Eqs.\,\eqref{pseudovector-vector_potential} -- \eqref{pseudotensor-pseudotensor_potential}. 
We also present the terms $V_{15}$ and $V_{16}$ separately [sourced from \cite{dobrescu_spin-dependent_2006,fadeev_neue_2018}], in addition to $V_{14q+15}$ and $V_{11p+16}$. While we recommend the use of $V_{14q+15}$ and $V_{11p+16}$, the former are needed to present the results of some earlier experimental works.

\subsection{Potentials for macroscopic experiments}

Here we present the exotic potentials of Eqs.~\eqref{eqV1} -- \eqref{eqV16} with $\boldsymbol{p}_X = \boldsymbol{v} m_Xm_Y/(m_X+m_Y)$ in the 
two-body centre-of-mass reference frame, with $\boldsymbol{v} = \boldsymbol{v}_X - \boldsymbol{v}_Y$ being the relative velocity vector between the two particles, 
which we further assume to be macroscopic bodies.
Note that when the classical regime is relevant, momentum can be treated as a classical variable 
which commutes with the classical position vector $\v{r}$,
while in the quantum treatment (\citealp{ficek_constraints_2017}; \citealp{ficek_constraints_2018}),
the momentum is an operator and we substitute $\boldsymbol{p}_X=-i\hbar\boldsymbol{\nabla}_X$ and $\boldsymbol{p}_Y=-i\hbar\boldsymbol{\nabla}_Y$. 
In the case of macroscopic, non-overlapping systems, we can remove the contact term $4\pi \delta(\boldsymbol{r})/3$ contained in $V_3$ 
and $V_{14q+15}$, which cannot extend over spatially separated regions.

\begin{gather}
V_1  = f^{XY}_1 \frac{\hbar c}{4\pi} \frac{1}{r}\,e^{-{r}/{\lambda}} \label{eqV1f}\, ,\\
V_2  = f^{XY}_2\frac{\hbar c}{4\pi}\boldsymbol{\sigma}_X\cdot\v\sigma_Y^{\,\prime}\frac{1}{r}\,e^{-{r}/{\lambda}}\label{eqV2f} \, , \\
V_3  = f^{XY}_3 \frac{\hbar^3}{4\pi m_X^2c} \left[\boldsymbol{\sigma}_X\cdot\v\sigma_Y^{\,\prime} \left(\frac{1}{r^3}+\frac{1}{\lambda r^2}\right)\nonumber \right. \\\left.-(\boldsymbol{\sigma}_X\cdot \hat{\boldsymbol{r}})(\v\sigma_Y^{\,\prime}\cdot \hat{\boldsymbol{r}})\left(\frac{1}{\lambda^2 r}+\frac{3}{\lambda r^2}+\frac{3}{r^3}\right)  \right] e^{-{r}/{\lambda}}\label{eqV3f} \, , \\
V_{4+5} =\notag\\ -f^{XY}_{4+5} \frac{\hbar^2m_Y}{8\pi m_X(m_X+m_Y)c}\boldsymbol{\sigma}_X \cdot(\boldsymbol{v} \times \hat{\boldsymbol{r}}) \left(\frac{1}{r^2}+\frac{1}{\lambda r}\right) e^{-{r}/{\lambda}}\label{eqV45f} \, , \\
V_{8}  = f^{XY}_8\frac{\hbar}{4\pi c}\frac{m_Y^2}{(m_X+m_Y)^2}( \boldsymbol{\sigma}_X\cdot\boldsymbol{v})( \v\sigma_Y^{\,\prime}\cdot\boldsymbol{v})\frac{1}{r}e^{-{r}/{\lambda}} \label{eqV8f} \, , \\
V_{9+10}=-f^{XY}_{9+10}\frac{\hbar^2}{8\pi m_X}\boldsymbol{\sigma}_X\cdot \hat{\boldsymbol{r}}\,\, \left(\frac{1}{r^2}+\frac{1}{\lambda r}\right)\,\, e^{-{r}/{\lambda}} \label{eqV910f} \, , \\
V_{11}=-f^{XY}_{11}\frac{\hbar^2}{4\pi m_X} \left(\boldsymbol{\sigma}_X\times\v\sigma_Y^{\,\prime}\right) \cdot \hat{\boldsymbol{r}} \left(\frac{1}{r^2}+\frac{1}{\lambda r} \right)e^{-{r}/{\lambda}}\label{eqV11f} \, , \\
V_{12+13}=f^{XY}_{12+13}\frac{\hbar}{8\pi}\frac{m_Y}{m_X+m_Y} \boldsymbol{\sigma}_X\cdot\boldsymbol{v}\frac{1}{r}e^{-{r}/{\lambda}} \, , 
\label{eqV1213f}
\end{gather}
\begin{widetext}
\begin{gather}
V_{14q+15} = f_{14q+15}^{XY} \frac{ m_Y \hbar^3} {8 \pi m_X^2 (m_X+m_Y) c^2} \left[ \left( \v{\sigma}_X \times \v\sigma_Y^{\,\prime}\right) \cdot \v{v} \left( \frac{1}{r^3} + \frac{1}{\lambda r^2} \right)  - \left( \v{\sigma}_X \cdot \hat{\v{r}} \right) [\v\sigma_Y^{\,\prime}\cdot \left(\v{v} \times \hat{\v{r}} \right)] \left( \frac{3}{r^3} + \frac{3}{\lambda r^2} + \frac{1}{\lambda^2 r}  \right)\right]e^{-{r}/{\lambda}}\label{eqV1415f} \, , \\
V_{11p+16}|_{AV} = - f_{11p+16}^{XY} \frac{m_Y^2 \hbar^2}{8 \pi m_X (m_X+m_Y)^2 c^2}  (\boldsymbol{\sigma}_X \cdot \v{v}) [\v\sigma_Y^{\,\prime} \cdot \left( \v{v} \times \hat{\v{r}} \right)] \left( \frac{1}{r^2} + \frac{1}{\lambda r} \right)   e^{-{r}/{\lambda}}  \label{eqV1116f}  \, , \\
V_{15}=f^{XY}_{15}\frac{\hbar^3}{8\pi m_X^2 c^2}\frac{m_Y}{(m_X+m_Y)} \left\{ (\boldsymbol{\sigma}_X \cdot \hat{\boldsymbol{r}})[\v\sigma_Y^{\,\prime} \cdot (\hat{\boldsymbol{r}} \times \boldsymbol{v})] + (\v\sigma_Y^{\,\prime}\cdot\hat{\boldsymbol{r}}) [\boldsymbol{\sigma}_X \cdot (\hat{\boldsymbol{r}} \times \boldsymbol{v})] \right\} \left(\frac{3}{r^3} + \frac{3}{\lambda r^2} + \frac{1}{\lambda^2 r}\right) e^{-{r}/{\lambda}}\label{eqV15f} \, , \\
V_{16}=-f^{XY}_{16}\frac{\hbar^2}{8\pi m_X c^2}\frac{m_Y^2}{(m_X+m_Y)^2}\left\{ (\boldsymbol{\sigma}_X\cdot\boldsymbol{v}) [\v\sigma_Y^{\,\prime} \cdot (\boldsymbol{v}\times \hat{\boldsymbol{r}})] + (\v\sigma_Y^{\,\prime}\cdot\boldsymbol{v}) [\boldsymbol{\sigma}_X \cdot (\boldsymbol{v}\times \hat{\boldsymbol{r}})] \right\} \left(\frac{1}{r^2}+\frac{1}{\lambda r}\right)\,\, e^{-{r}/{\lambda}}\label{eqV16f} \, . 
\end{gather}
\end{widetext}

\subsection{\texorpdfstring{Relationship between $f_i^{XY}$ and $g^Xg^Y$}{Relationship between fiXY and gXgY}}
\label{App:appendix_f_gg_table}

The details of the relationship between the sets of formulae in Eqs.~\eqref{eqV1} -- \eqref{eqV16} and Eqs.~\eqref{eqV1f} -- \eqref{eqV16f} and other formats used in the literature, especially the relationship between $f_i^{XY}$ and coupling constants $g^Xg^Y$ are presented in Tab.\,\ref{tabel_f-gg}. 
In this review, we present the formulae in a way that the coefficients $f_i^{XY}$ or $g^Xg^Y$ are dimensionless. 
So neither of them shall be divided by $\hbar c$ again while plotting, which is different from how this is done in other sources, for example, \citet{safronova_search_2018}.

\renewcommand\arraystretch{3}
\begin{table}\normalsize
\centering
\caption{Relationship between the sets of coefficients $f_i^{XY}$ and $g^Xg^Y$. 
} 
\renewcommand{\arraystretch}{1.8} 
\begin{threeparttable}
\scalebox{0.85}{
\begin{tabular}{c|c|c} 
\hline
\hline
$f_i^{XY}$& Coupling constants & Fermion pairs studied\\
\hline
$f_2^{XY}$& $g_A^Xg_A^Y\times(-1)$& $e$-$e$ ($e^+$), $e$-${\mu}^+$, $e$-$n$ \\
&&$e$-$p$($\overline{p}$), $p$-$N$, $n$-$p$, $n$-$n$ \\
\hline
$f_3^{XY}$& $g_A^Xg_A^Y\times[-(\frac{c}{\hbar} m_X \lambda)^2]$ &$e$-$e$($e^+$), $e$-${\mu}^+$, $e$-$p$($\overline{p}$) \\
&$g_p^Xg_p^Y\times(-\frac{m_X}{4m_Y})$& $e$-$n$, $p$-$N(p)$, $n$-$p$, $n$-$n$ \\
&$g_V^Xg_V^Y \times\frac{m_X}{4m_Y}$&\\
\hline
$f_{4+5}^{XY}$&  $- g_A^Xg_A^Y\times( \frac{1}{2}\frac{m_X^2}{m_Y^2})$&$e$-$e$, $e$-$\overline{p}$, $e$-$N$, $p$-$N$, $n$-$N$ \\
&$- g_V^Xg_V^Y \times (\frac{1}{2}+\frac{m_X}{m_Y})$ &  \\
&$- g_s^eg_s^N\times(\frac{1}{2})$&\\
\hline
$f_{8}^{XY}$& $g_A^Xg_A^Y\times[-\frac{1}{2}(1+\frac{2m_X}{m_Y}+\frac{m_X^2}{m_Y^2})]$& $e$-$p(\overline{p})$, $n$-$p$, $e$-$e$, $e$-$n$ \\
\hline
$f_{9+10}^{XY}$&$g^X_p g_s^Y$&$e$-$e$, $e$-$N$, $n$-$N$, $p$-$N$, $N$-$e$ \\
\hline
$f_{11}^{XY}$&$g_V^X g_A^Y\times\frac{1}{2}+g_A^X g_V^Y\times\frac{m_X}{2m_Y}$ & $e$-$e$, $e$-$n$, $e$-$p$, $n$-$p$, $n$-$n$ \\
\hline
$f_{12+13}^{XY}$&$2 g_A^X g_V^Y\times (1+\frac{m_X}{m_Y})$ &$e$-$N$, $e$-$p$, $e$-$n$, $p$-$N$, $n$-$N$\\
\hline
$f_{14q+15}^{XY}$ & $-\frac{g^X_p g^Y_s}{4}\frac{m_X^2}{m_Y^2}$ & ---\\
\hline
$f_{11p+16}^{XY}$ & $\frac{g_A^X g_V^Y}{2} \frac{m_X(m_X+m_Y)}{m_Y^2}$  & ---\\
\hline
$f_{15}^{XY}$& $g_s^X g_p^Y \times \frac{m_X}{4m_Y} + g_p^X g_s^Y \times (-\frac{m_X^2}{4m_Y^2})$ &
$e$-$p$, $e$-$n$\\
\hline
$f_{16}^{XY}$& $(g_V^X g_A^Y - g_V^Y g_A^X) \times\frac{m_X}{4m_Y}(1+\frac{m_X}{m_Y})$ & $e$-$p$, $e$-$n$ \\
\hline
\hline
\end{tabular}
\label{tabel_f-gg}
}
\end{threeparttable}
\end{table}

\subsection{Differences among various forms of the potentials appearing in the literature and their consequences}

Several key references 
(\citealp{moody_new_1984}; \citealp{dobrescu_spin-dependent_2006}; \citealp{leslie_prospects_2014}; \citealp{daido_alp_2017}; \citealp{safronova_search_2018}; \citealp{fadeev_revisiting_2019})
have discussed the forms of various potentials. 

As two examples, compared to our Eqs.\,\eqref{eqV1} -- \eqref{eqV16}, \citet{dobrescu_spin-dependent_2006} present the same coefficients $f_i$. 
\citet{safronova_search_2018} also use the coefficients $f_i$, but absorb an extra factor of $1/2$ into their coefficients $f_{4+5}$, $f_{9+10}$, $f_{12+13}$ and $f_{15}$.
In addition, Eq.\,\eqref{eqV3} contains the contact term $4\pi\delta(\boldsymbol{r})/3$, 
which is missing in \cite{dobrescu_spin-dependent_2006} and \cite{safronova_search_2018},
and we present 
the new terms
$V_{14q+15}$ and $V_{11p+16}$. 

Using different formulae can lead to confusion when comparing the results of different experiments and a deviation in the constraints that one could set. 
We advocate the use of the set of potentials given by Eqs.\,\eqref{pseudovector-vector_potential} -- \eqref{pseudotensor-pseudotensor_potential} (or the equivalent ones in Sec.\,\ref{Sec:limits_main}) and presentation of the results in terms of the combinations of the nine coupling constants, such as $g_Ag_V$, $g_Ag_A$ and so on. 
This bypasses the differences in the definitions of the potentials discussed in this subsection.

\subsection{Equivalence of different forms}
\label{App:appendix_f_gg}

We take $V_{4+5}$ as an example to show the correspondence of the potentials from \citet{dobrescu_spin-dependent_2006} with the ones in this review. 
In the derivations below, we apply the approximation $m_e \ll m_N$ for the $e$-$N$ fermion pair.

\subsubsection*{a. $g_sg_s$}
${V}_{4+5}^\textrm{DM}$ from \citet{dobrescu_spin-dependent_2006} is given by $f_{4+5} {\sV}_{4+5}$. Starting from 
\begin{equation}
{\sV}_{4,5} =-\frac{1}{2 m_X\,r^2}
\left(\v{\sigma}_X \pm \v\sigma_Y^{\,\prime}\right) 
\cdot \left(\v v\times \hat{\v{r}} \right)  
\left(1 - r\frac{d}{dr}\right) y(r)\,,
\end{equation}
where $y(r) = \exp(-r/\lambda)/(4\pi)$,
we have
\begin{align}
&{\sV}_{4+5} =\frac{1}{2}({\sV}_{4}+{\sV}_{5}) \\
&=  -\frac{1}{ 8\pi m_X\,r^2}
\v{\sigma}_X 
\cdot \left(\v v\times \hat{\v{r}} \right)  
\left(\frac{1}{r^2}+\frac{1}{\lambda r}\right) e^{-\frac{r}{\lambda}} \, . 
\end{align}

For the exotic interaction involving the fermion pair $e$-$N$ for $g_sg_s$, taking the coefficients $f_{4,5}^{eN}(0,0)$ from Eq.\,(6.2) of \citet{dobrescu_spin-dependent_2006},
we have 
\begin{align}
f_{4+5}^{eN}&=f_{4}^{eN}(0,0)+f_{5}^{eN}(0,0)=-\frac{1}{2} g_s^eg_s^N\,.
\end{align}
Therefore
\begin{align}\label{V45-ss-Dob}
{V}_{4+5}|^\textrm{DM}_{ss} &=f_{4+5}^{eN} {\sV}_{4+5} \notag\\
&= \frac{g_s^e g_s^N}{16\pi m_e} 
\v{\sigma}_e 
\cdot \left(\v v\times \hat{\v{r}} \right)  
\left(\frac{1}{r^2}+\frac{1}{\lambda r}\right) e^{-\frac{r}{\lambda}}\,. 
\end{align}

On the other hand, 
Eq.\,\eqref{eqV45f} from this review gives $V_{4+5}$ as (using the relationship between $f_{4+5}^{XY}$ and $g_s^X g_s^Y$ presented in Tab.\,\ref{tabel_f-gg}, with $\hbar=c=1$): 
{\small
\begin{align}\label{V45-ss-Fad}
&V_{4+5}|_{ss}\notag\\
&= \left(\frac{-g_s^e g_s^N}{2}\right) \left[ \frac{-m_N}{8\pi m_e(m_e+m_N)} \right] \boldsymbol{\sigma}_e \cdot(\boldsymbol{v} \times \hat{\boldsymbol{r}}) \left( \frac{1}{r^2}+\frac{1}{\lambda r} \right) e^{-\frac{r}{\lambda}} \notag\\ 
&\approx \frac{g_s^eg_s^N}{16\pi m_e}\boldsymbol{\sigma}_e \cdot(\boldsymbol{v} \times \hat{\boldsymbol{r}}) \left(\frac{1}{r^2}+\frac{1}{\lambda r}\right) e^{-\frac{r}{\lambda}}\,. 
\end{align}
}
Therefore, Eqs.\,\eqref{V45-ss-Dob} and \eqref{V45-ss-Fad} correspond to the same potential.

\subsubsection*{b. $g_Ag_A$}
Similarly, for $g_Ag_A$, based on Eqs.\,(4.3)
and (5.28) of \citet{dobrescu_spin-dependent_2006}, we have 
\begin{align}\label{V45-AA-Dob}
&{V}_{4+5}|^\textrm{DM}_{AA} 
= \frac{g_A^e g_A^N}{16\pi} \frac{m_e}{m_N^2} 
\v{\sigma}_e 
\cdot \left(\v v\times \hat{\v{r}} \right)  
\left(\frac{1}{r^2}+\frac{1}{\lambda r}\right) e^{-\frac{r}{\lambda}}\,. 
\end{align}

Based on Eq.\,\eqref{eqV45f} and the relationship between $f_{4+5}^{XY}$ and $g_A^Xg_A^Y$ presented in Tab.\,\ref{tabel_f-gg}, we have
\begin{equation}
\begin{aligned}\label{V45-AA-Fad}
&V_{4+5}|_{AA}\\
&=  \left(\frac{- g_A^eg_A^N }{2 } \frac{m_e^2}{m_N^2}\right) \left[ \frac{-m_N}{8\pi m_e(m_e+m_N)} \right] \boldsymbol{\sigma}_e \cdot(\boldsymbol{v} \times \hat{\boldsymbol{r}})\\ &\left(\frac{1}{r^2}+\frac{1}{\lambda r} \right) e^{-\frac{r}{\lambda}} \\ 
&\approx \frac{g_A^e g_A^N}{16\pi } \frac{m_e}{m_N^2}\boldsymbol{\sigma}_e \cdot(\boldsymbol{v} \times \hat{\boldsymbol{r}}) \left(\frac{1}{r^2}+\frac{1}{\lambda r} \right) e^{-\frac{r}{\lambda}} \, . 
\end{aligned}
\end{equation}
Therefore, Eqs.\,\eqref{V45-AA-Dob} and \eqref{V45-AA-Fad} have the same form.

\subsubsection*{c. $g_Vg_V$}
In addition, for $g_Vg_V$, based on Eqs.\,(4.3)
and (5.28) of \citet{dobrescu_spin-dependent_2006}, we have 
\begin{align}\label{V45-VV-Dob}
{V}_{4+5}|^\textrm{DM}_{VV}
&= \frac{g_V^e g_V^N}{16\pi} \frac{2m_e+m_N}{ m_em_N}
\v{\sigma}_e 
\cdot \left(\v v\times \hat{\v{r}} \right)  
\left(\frac{1}{r^2}+\frac{1}{\lambda r}\right)  e^{-\frac{r}{\lambda}} \notag\\
&\approx \frac{g_V^e g_V^N}{16\pi m_e }\boldsymbol{\sigma}_e \cdot(\boldsymbol{v} \times \hat{\boldsymbol{r}}) \left(\frac{1}{r^2}+\frac{1}{\lambda r}\right) e^{-\frac{r}{\lambda}}\, . 
\end{align}

Based on Eq.\,\eqref{eqV45f} and the relationship between $f_{4+5}^{XY}$ and $g_V^Xg_V^Y$ presented in Tab.\,\ref{tabel_f-gg}, we have
{\small
\begin{align}\label{V45-VV-Fad}
&V_{4+5}|_{VV}\notag\\
&= \left[ -\frac{g_V^e g_V^N}{2} \left( \frac{2m_e + m_N}{m_N} \right) \right] \left[ \frac{-m_N}{8\pi m_e(m_e+m_N)} \right]  \boldsymbol{\sigma}_e \cdot(\boldsymbol{v} \times \hat{\boldsymbol{r}}) \notag\\
&\left(\frac{1}{r^2}+\frac{1}{\lambda r}\right) e^{-\frac{r}{\lambda}}\notag\\
&\approx \frac{g_V^e g_V^N}{16\pi m_e }\boldsymbol{\sigma}_e \cdot(\boldsymbol{v} \times \hat{\boldsymbol{r}}) \left(\frac{1}{r^2}+\frac{1}{\lambda r}\right) e^{-\frac{r}{\lambda}} \, . 
\end{align}
}

Therefore, Eqs.\,\eqref{V45-VV-Dob} and \eqref{V45-VV-Fad} correspond to the same potential.



\section{\texorpdfstring{Spin-independent $V_1$}{Spin-independent V1}}\label{appendix_V1}

The study of spin-independent interactions has a long history, with early research primarily focusing on testing the inverse-square law (ISL) \cite{safronova_search_2018}. 
Presently, a commonly employed approach involves considering a Yukawa-like deviation from the ISL according to \cite{talmadge_model-independent_1988}: 
\begin{equation}
\label{gg-alpha_definition}
    V=-G\frac{M_1 M_2}{r}(1+\alpha e^{-{r}/{\lambda}}) \, , 
\end{equation}
where $M_1$ and $M_2$ are the masses of the two objects. 
Note that alternative modifications to the ISL exist beyond the form represented in Eq.\,(\ref{gg-alpha_definition}). 
Further details on these modifications can be found in \cite{adelberger_tests_2003}. 

\begin{figure*} [!htbp]
\begin{center}
\includegraphics[width=0.98\textwidth]{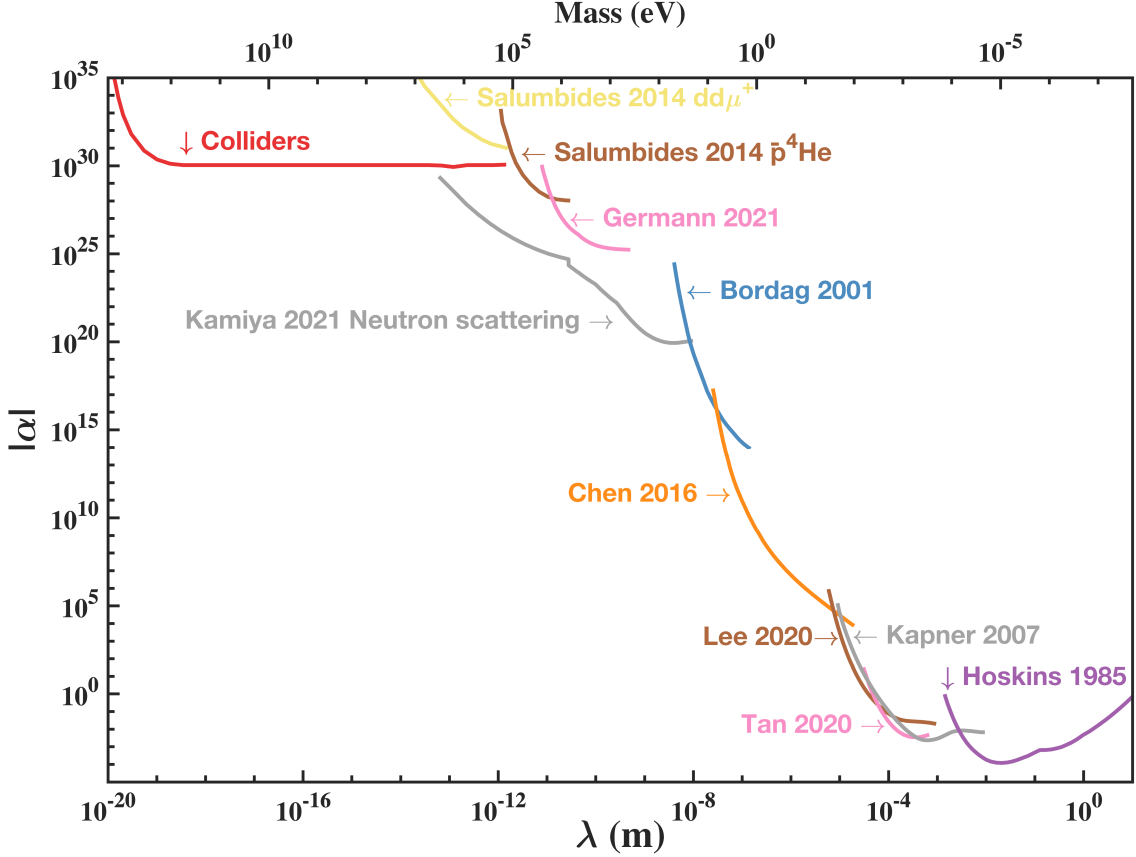}
\end{center}
\caption{
Constraints on the universal dimensionless coupling parameter $\alpha$, defined in Eq.\,\eqref{gg-alpha_definition}, as a function of the interaction range $\lambda$ (or alternatively, new boson mass) for the potential term $V_{1}$. 
}
\label{alpha-fig}
\end{figure*}

In this section, we provide an overview of the latest experimental research based on Eq.\,\eqref{gg-alpha_definition}, as depicted in Fig.\,\ref{alpha-fig} (details can be found in App.\,\ref{appendix_V1_Torsion} -- \ref{appendix_V1_exotic}). 
Additionally, in App.\,\ref{appendix_V1_alpha-to-gsgs} -- \ref{appendix_V1_PS}, we show how constraints on the coupling strength $\alpha$ can be explicitly related to the scalar/scalar coupling constant combination $g_sg_s$ and demonstrate the translation of constraints from precision spectroscopy 
to $g_sg_s$. 
In App.\,\ref{Appendix:WEP_tests}, we discuss bounds on $g_s g_s$ arising from tests of the WEP. 
Furthermore, in App.\,\ref{appendix_V1_gsgs-and-gvgv}, we address how $\alpha$ can also be translated to the vector/vector coupling constant combination $g_Vg_V$. 

\subsection{Torsion-pendulum experiments}\label{appendix_V1_Torsion}

An overview of the results is presented in Fig.\,\ref{alpha-fig}. 

In the interaction range $3 \times 10^{-3}\,\textrm{m} < \lambda < 10$\,m, there are results from Irvine, where 
\citet{spero_test_1980}; \citet{hoskins_experimental_1985} 
conducted torsion balance experiments. 

For interaction ranges around $\lambda \sim 10^{-4}$\,m, the Huazhong University of Science and Technology (HUST) group \cite{tan_improvement_2020} improved on their earlier research using a torsion balance \cite{tan_new_2016}. 

For the force range $10^{-5} \, \textrm{m} < \lambda <10^{-4}$\,m, the Washington group \cite{lee_new_2020} improved on their earlier Cavendish-type experiment results \cite{kapner_tests_2007} in a part of the range. 
\citet{kapner_tests_2007} outperformed the earlier work, such as Eöt-Wash 2004 \cite{hoyle_submillimeter_2004} and the results from Colorado \cite{long_upper_2003}, and partially on the Cavendish-type experimental result from Stanford \cite{chiaverini_new_2003,smullin_constraints_2005}.

\subsection{Casimir force}
For the interaction range $10^{-7} \, \textrm{m} < \lambda <10^{-5}$\,m, the Indiana University–Purdue University (IUPU) collaboration \cite{chen_stronger_2016} 
employed a technique that subtracts the otherwise dominant Casimir force, as demonstrated by IUPU \cite{decca_novel_2007,decca_tests_2007}, Yale \cite{sushkov_new_2011} and Stanford \cite{geraci_improved_2008} groups. 

In the force range around $\lambda \sim 10^{-8}$\,m, the Riverside group (\citealp{mohideen_precision_1998};
\citealp{roy_demonstration_1999}; 
\citealp{klimchitskaya_complete_1999}; \citealp{harris_precision_2000}; \citealp{bordag_new_2001}, \citeyear{bordag_advances_2009})
conducted a series of precision measurements of the Casimir force, providing the most stringent constraints on the interaction constants of Yukawa-type interactions using atomic force microscopy (AFM).

\subsection{Neutron scattering, spectroscopy of exotic atoms, collider experiments}\label{appendix_V1_exotic}



For the range $6 \times 10^{-14} \, \textrm{m} < \lambda < 8 \times 10^{-9}$\,m, neutron scattering (\citealp{pokotilovski_constraints_2006}; \citealp{kamiya_constraints_2015}; \citealp{haddock_search_2018-1}; \citealp{kamiya_experimental_2021})
established the most stringent constraints, involving scattering off xenon gas and $^{208}$Pb nuclei. 
Other neutron-scattering experiments include those of \citet{nesvizhevsky_neutron_2008,voronin_search_2023}. 
Within a similar range, research has been conducted based on antiprotonic helium atoms (for $\bar{p}$-$N$), $dd\mu^+$ \cite{salumbides_bounds_2014} (for $\mu^+$-$N$) and HD$^+$ \cite{germann_three-body_2021} (for $p$-$N$). 
Furthermore, \citet{lemos_submillimeter_2021} has provided prospective bounds (i.e., projections) for non-Newtonian gravity from spectroscopy by analyzing transitions between Rydberg states. 

Over the range $10^{-20} \, \textrm{m} < \lambda < 6 \times 10^{-14}$\,m, 
the most stringent constraints are derived from collider experiments \cite{murata_review_2015}. 

For additional earlier constraints on short-range modifications of gravity or additional interactions, one can also refer to the reviews by \citet{adelberger_tests_2003,mostepanenko_state_2020}, among others.

\subsection{\texorpdfstring{Conversion of bounds on $\alpha$ to $g_sg_s$}{Conversion of bounds on α to gsgs}}
\label{appendix_V1_alpha-to-gsgs}

The bounds on $|\alpha|$ are given with respect to the usual gravitational interaction strength, so in experiments that measure forces on test bodies, we can use the following relations [obtained by comparing Eq.\,(\ref{gg-alpha_definition}) and $V_1$ shown in Eq.\,(\ref{scalar-scalar_potential})] to obtain bounds on $\alpha$: 
\begin{equation}
\label{gsN-alpha}
    (g_s^N)^2 \approx 4 \pi |\alpha| \left(\frac{m_N}{M_\textrm{Pl}}\right)^2 \, , 
\end{equation}
where $M_\textrm{Pl}=\sqrt{\hbar c /G} \approx 1.2 \times 10^{19}\,\textrm{GeV}$ is the Planck mass, and 
\begin{equation}
\label{gse-alpha}
    (g_s^e)^2 \approx 4 \pi |\alpha| \left(\frac{m_N}{M_\textrm{Pl}}\right)^2  \frac{A_1 A_2}{Z_1  Z_2} \, , 
\end{equation}
where $A_{1,2}$ and $Z_{1,2}$ are the nucleon and electron numbers, respectively, per atom in bodies 1 and 2.
For typical moderately heavy elements, the average ratio of number of electrons to nucleons is $Z / A \approx 0.4$. 
Here $m_N$ denotes the average nucleon mass. 

Note that we generally cannot use the above procedure to place bounds on cross-combinations of parameters like $g_s^Ng_s^e$. 
The reason is that to extract a bound on $g_s^N$ from one of the vertices, we make an assumption like $|g_s^N| \gg |g_s^e|$, while to extract a bound on $g_s^e$ from the other vertex, we make an assumption $|g_s^e| \gg |g_s^N|$, and so it is generally not possible to simultaneously satisfy both of these conditions. 
If, for example, $g_s^N$ dominates the force contribution at one vertex (body), then $g_s^N$ will also generally dominate at the other vertex (body), and hence the (self-consistent) bound in this case would be on the product $g_s^N g_s^N$. 

For the constraints on $|\alpha|$ shown in Fig.\,\ref{alpha-fig}, at boson masses greater than $\sim 10^6$ eV, 
the most stringent constraints come from collider-based experiments. 
These collider-based bounds may not necessarily be simply translated to bounds on $(g_s^e)^2$ and $(g_s^N)^2$ in the way this is done for bounds from macroscopic tests of gravity. 
In high-energy collisions using protons at the Large Hadron Collider (LHC), it may not be appropriate to use nucleons as the degrees of freedom to describe the underlying interactions. 
Instead, one should describe interactions of exotic bosons in terms of their underlying interactions with quarks and gluons. 
In the case of electron interactions, one should consider other collider-based experiments that use accelerated electrons [e.g., the Large Electron-Positron (LEP) collider] -- these types of experiments operated at lower energies than the LHC (which reached peak energies of around 10\,TeV, whereas LEP operated at energies of about 100\,GeV), so the limits on electron couplings would be expected to start degrading at smaller boson masses. 
Therefore, since boson masses greater than $10^6$\,eV exceed the mass ranges in our $g_sg_s$ 
figures, we have not presented collider-based bounds in those figures.

\subsection{\texorpdfstring{Converting other constraints from precision spectroscopy to bounds on $g_sg_s$}{Converting other constraints from precision spectroscopy to bounds on gsgs}}
\label{appendix_V1_PS}

There are constraints from precision spectroscopy presented as bounds on spin-independent couplings of exotic bosons to electrons, protons, and neutrons. 
These couplings are occasionally denoted as $y_e$, $y_p$, and $y_n$, respectively, where within the theoretical framework utilized in this review, $y_X$ is equivalent to $g_s^X$.

\subsubsection*{$g_s^e g_s^e$ and $g_s^e g_s^{e^+}$:} 
Bounds on $g_s^e g_s^{e^+}$ from positronium spectroscopy were presented by \citet{adkins_precision_2022} and are not much weaker than bounds on $g_s^e g_s^e$ from the electron magnetic moment ($g$-factor) reported by \citet{delaunay_probing_2017}. 
There are other bounds on electron-electron interactions, see \citet{frugiuele_current_2019}; \citet{debierre_fifth-force_2020}.

\subsubsection*{$g_s^eg_s^n$:} 
\citet{delaunay_probing_2017} provided one of the most stringent constraints on $g_s^eg_s^n$ and there are other constraints from \citet{berengut_probing_2018}; \citet{debierre_fifth-force_2020}; \citet{solaro_improved_2020}; \citet{sailer_measurement_2022}; \citet{door_search_2024}.

\subsubsection*{$g_s^eg_s^{{\mu}^+}$:} 
The Mu-MASS Collaboration \cite{ohayon_precision_2022} and \citet{frugiuele_current_2019} both studied $1S-2S$ measurements in muonium. 
In our figures, for clarity, we have only included the results from \citet{ohayon_precision_2022}. 
These measurements strictly apply to the matter-antimatter combination $g_s^eg_s^{\mu^+}$. 

In addition, there are astrophysical bounds on $(g_s^{\mu})^2$ provided by \citet{caputo_muonic_2022}, which can be found in Tab.\,\ref{tabel_Astrophysical-bounds}. 
There are also ongoing and proposed experiments, such as muonium free-fall experiments in the context of the combination of parameters $g_s^eg_s^{{\mu}^+}$, as detailed by \citet{stadnik_searching_2023} and references therein. 
The parameters $\Lambda_e$ and $\Lambda_{\mu}$ in Fig.\,2 of \citet{stadnik_searching_2023} are related to the parameters $g_s^e$ and $g_s^{\mu}$ via the relationships $g_s^e = m_e/\Lambda_e$ and $g_s^{\mu} = m_{\mu}/\Lambda_{\mu}$, respectively. 

\begin{figure*} [!htbp]
\begin{center}
\includegraphics[width=0.98\textwidth]{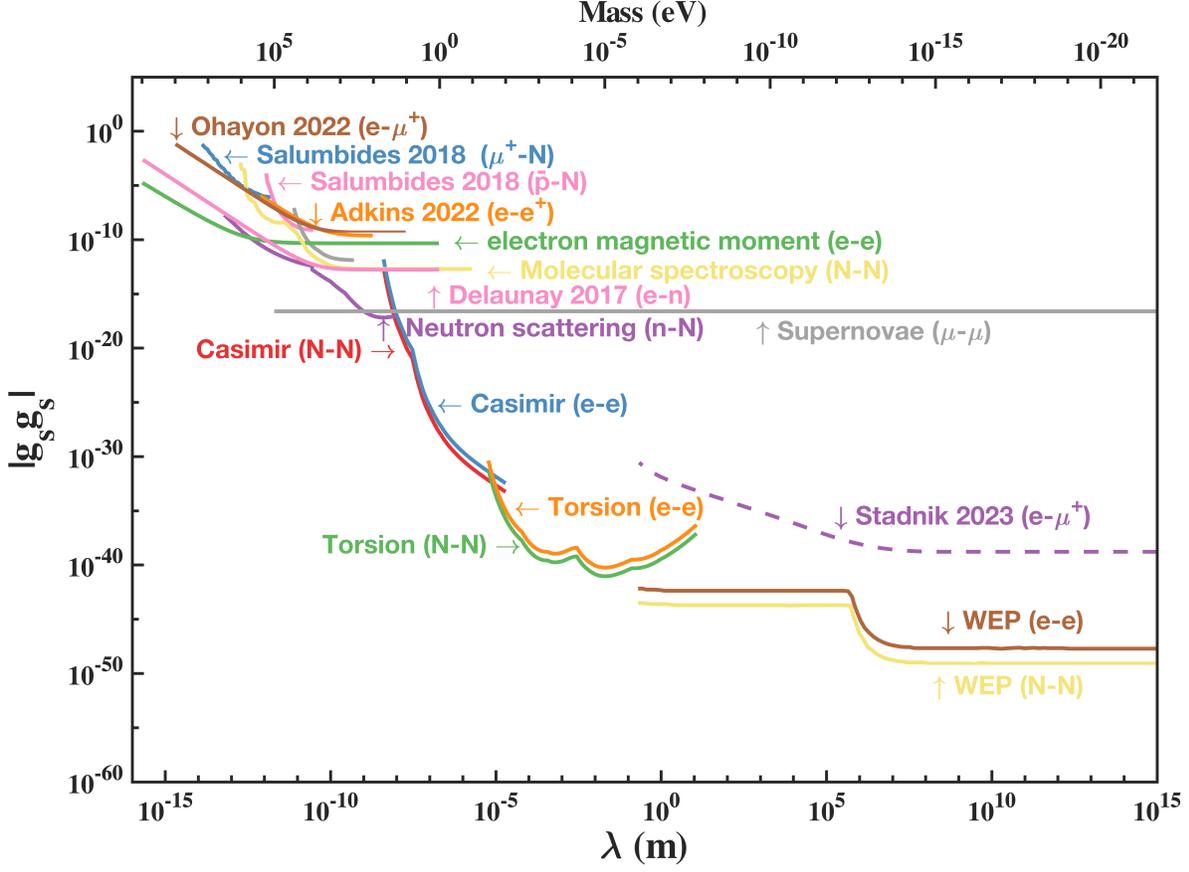}
\end{center}
\caption{
Constraints on $g_sg_s$ as a function of the interaction range $\lambda$ (or alternatively, the mass of the new scalar boson). 
}
\label{V1-fig}
\end{figure*}

We have provided a comprehensive overview of bounds on Yukawa-type interactions in Fig.\,\ref{V1-fig} for reference. 
The strongest bounds from this collection are featured in the main-text figures. 

\subsection{\texorpdfstring{Constraints on $g_sg_s$ from tests of the weak equivalence principle (WEP)}{Constraints on gsgs from tests of the weak equivalence principle (WEP)}}
\label{Appendix:WEP_tests}

The satellite-based MICROSCOPE experiment \cite{touboul_microscope_2017} measured differential accelerations between two pairs of test bodies made of different materials, which are sensitive to WEP-violating forces but insensitive to  universal gravitational-type forces. 
In particular, Eqs.\,\eqref{gsN-alpha} and \eqref{gse-alpha} are not applicable to WEP tests. 
Nonetheless, we can still place bounds from WEP tests on $g_s g_s$ in Fig.\,\ref{V1-fig} by translating the bounds, for example, from \citet{leefer_search_2016}; \citet{grote_novel_2019}.
It is noteworthy that the final MICROSCOPE bound of \citet{touboul_microscope_2022} can be re-scaled from their 2017 result in \citet{touboul_microscope_2017} by a factor of 4.6 improvement, though we do not present this improved bound here. 

\subsection{\texorpdfstring{Equivalence of constraints on $g_sg_s$ and $g_Vg_V$ from ${V}_1$}{Equivalence of constraints on gsgs and gVgV from V1}}
\label{appendix_V1_gsgs-and-gvgv}
The ${V}_1$ potential term in the scalar/scalar [see Eq.\,(\ref{gsgs_V1})] and vector/vector [see Eq.\,(\ref{gVgV_V1})] cases only differ by an overall sign, so there's no difference when presenting bounds on the modulus of coupling parameters $|g_s g_s|$ and $|g_V g_V|$. 
Thus, we can translate limits in the $\alpha-\lambda$ parameter space to limits in the $g^X_Vg^Y_V-\lambda$ parameter space in the same way.

\section{Bounds for the purpose of comparison}
\label{appendix_comp.}

In Appendix \ref{appendix_V1}, we present the constraints arising from the spin-independent potential term $V_1$, which can serve as a basis for comparison with constraints on $g_sg_s$ and $g_Vg_V$ from other spin-dependent potential terms. 
Additionally, in Sec.\,\ref{METH2.Astro.}, we discuss astrophysical limits, which are also valuable for comparison. 
In this section, we discuss combined astrophysical-laboratory bounds, which are also used for comparison, in addition to presenting individual constraints from the $V_1$ term and from astrophysics.

\subsubsection*{$g_Vg_A$}
For $g_V g_A$ (see Fig.\,\ref{gvga-fig}), we present combined astrophysical-laboratory bounds. 
There are two ways of constructing such bounds. 

One way is to take astrophysical bounds on $g_A^e$ and $g_A^N$ (see Tab.\,\ref{tabel_Astrophysical-bounds}). We also take laboratory bounds on $g_V$ from the $V_1$ potential term; 
specifically, we take constraints on $g_V^e$ from the electron magnetic moment, Casimir experiments, torsion-pendulum experiments and WEP experiments, and $g_V^N$ from the molecular spectroscopy, Casimir experiments, torsion-pendulum experiments and WEP experiments.
Then we multiply these bounds to obtain constraints on $g_A g_V$. 

Another way is to take the best laboratory bounds for $g_A$ (from the spin-dependent 
$V_2$ and $V_3$ terms) 
and astrophysical bounds from $g_V$ (see Tab.\,\ref{tabel_Astrophysical-bounds}). 
The resulting bounds on $g_A g_V$ diverge as $\propto 1/M$ in the limit as $M \to 0$. 
(In contrast, the scaling is $\propto 1/M^2$ for the $g_A g_A$ case, and is independent of $M$ for all of the other vertex combinations considered in our review). 
We do not present this result for clarity 
in Fig.\,\ref{gvga-fig}.

\subsubsection*{$g_Ag_A$}\label{appendix_comp.gAgA}

For $g_Ag_A$ (see Figs.\,\ref{gaga-fig1} and \ref{gaga-fig2}), we can present both purely astrophysical bounds and combined astrophysical-laboratory bounds. 

Regarding astrophysical bounds, Eq.\,(31) from \citet{dror_dark_2017} gives bounds on the electron and nucleon axial-vector couplings (see Tab.\,\ref{tabel_Astrophysical-bounds}). 
We plot $g_A^e g_A^e$, $g_A^e g_A^N$, and $g_A^N g_A^N$ for comparison. 
We use, for example, $g_A^eg_A^N \leq [(g_A^e)^2+(g_A^N)^2]/2$ obtained from the identity $(g_A^e-g_A^N)^2\geq0$. 

Regarding combined astrophysical-laboratory bounds, for $g_A g_A$, we can use the best available $g_A^e$ lab bounds (from the 
$V_2$ and $V_3$ terms)
and combine them with a bound on $g_A^e$ or $g_A^N$ from astrophysics. 
We do not present this result in the figure to avoid clutter. 

We also present analogous electron $g$-factor bounds for the purely axial-vector case in Fig.\,\ref{gaga-fig1} using Eq.\,(6) from \citet{karshenboim_constraints_2014}, which for our purposes reads: $g_A^eg_A^e=[g_V^e M/(\sqrt{2} m_e)]^2$ in the limit $M \ll m_e$. 
$g_V^e$ here are electron $g$-factor bounds based on a $V_1$-type term [taken from \citet{delaunay_probing_2017}]. 
Note that the resulting bound on $g_A^2$ diverges like $\propto 1/M^2$ in the limit of small $M$, which is due to the non-conservation of the axial-vector current.

\subsubsection*{$g_Vg_V$}

For $g_Vg_V$ (see Fig.\,\ref{gVgV-fig}), we present astrophysical bounds, as well as laboratory bounds on the spin-independent $V_1$ term. 
We do not present the (weaker) combined astrophysical-laboratory bounds to avoid cluttering the figure. 

In the case of astrophysical bounds, we plot constraints on $g_V^e g_V^e$, $g_V^e g_V^N$ and $g_V^N g_V^N$. 
Constraints on $g_V^e g_V^N$ can be calculated just like for $g_A^e g_A^N$ as discussed above. 

For laboratory bounds based on the spin-independent $V_1$ term, see App.\,\ref{appendix_V1}.

\subsubsection*{$g_pg_s$}
For $g_pg_s$ (see Fig.\,\ref{gpgs-fig}), we present astrophysical and combined astrophysical-laboratory bounds.

For the combined astrophysical-laboratory bounds, we take astrophysical bounds on the $g_p$ couplings from the literature (see Tab.\,\ref{tabel_Astrophysical-bounds}) and multiply them by laboratory bounds on $g_s$ couplings from $V_1$-type searches. 

For the purely astrophysical bounds, we can proceed as follows. 
Assuming that both $g_s$ and $g_p$ couplings are present, astrophysical energy-loss-type bounds constrain combinations of parameters like $a g_s^2 + b g_p^2$. 
The ratio of the dimensionless coefficients $a,b$ is not necessarily equal to 1. 
For the sake of simplicity, let us suppose that $a=b=1$; in this case, we know that $(g_s - g_p)^2 \ge 0 \Rightarrow g_s g_p \le (g_s^2 + g_p^2) / 2$, which allows us to constrain the combination $g_s g_p$. 

Regarding the combination of different types of bounds, if one combines bounds from two independent measurements taken in vastly different environments (e.g., if the bounds on $g_s$ come from lab-based measurements, whereas the bounds on $g_p$ are derived from measurements in hot or dense astrophysical environments), then the resulting bounds on the product of coupling constants (in this example, on $g_s g_p$) may not be robust if the properties of the boson involved change in different environments. 
See, for example, \citet{poddar_constraints_2023} for discussion of this point.

\subsubsection*{$g_pg_p$}

For $g_pg_p$ (see Fig.\,\ref{gpgp-fig}), we present astrophysical bounds (see Tab.\,\ref{tabel_Astrophysical-bounds}) and $g$-factor-based bounds.

For the $e$-$e$ and $e$-$\mu$ $g$-factor-based bounds, we take the limits presented in 
Figs.\,3 and 6
of \citet{frugiuele_current_2019}, respectively. 
[Note that the captions to Figs.\,3 and 5 in \citet{frugiuele_current_2019} are mixed up.]

\subsubsection*{$g_sg_s$}
For $g_s g_s$ (see Fig.\,\ref{gsgs-fig}), we present the best constraints from Fig.\,\ref{V1-fig} and separate purely astrophysical constraints. 
We also include constraints in Fig.\,\ref{gsgs-fig}\,(a) for $e$-$\mu^+$ which is the $g$-factor result from \citet{ohayon_precision_2022}.

In the literature [see, e.g., \citet{kim_experimental_2018}], there exist combined astrophysical-laboratory bounds. These incorporate the bounds on $g_s^e$ derived from stellar cooling rates \cite{grifols_constraints_1986,jain_evading_2006} and laboratory bounds on $g_s^N$ from recent searches for hypothetical (spin-independent) Yukawa interactions (\citealp{chen_stronger_2016}; \citealp{tan_improvement_2020}; \citealp{lee_new_2020}),
as discussed by \citet{raffelt_limits_2012} and \citet{leslie_prospects_2014}.

The laboratory bounds of interest typically constrain the combination of parameters $(a_1 g_s^N + b_1 g_s^e)(a_2 g_s^N + b_2 g_s^e)$, where $a_{1,2}$ and $b_{1,2}$ are dimensionless coefficients that are specific to the source and test masses used in the experiment; e.g., for fifth-force searches using torsion-pendulum experiments and Casimir-force measurements, as well as WEP tests, the typical ratio of coefficients is $a_{1,2}/b_{1,2} \sim 1$. 
On the other hand, astrophysical bounds constrain the combination of parameters $c (g_s^N)^2 + d (g_s^e)^2$, where the typical ratio of coefficients is $d/c \sim 10^6 - 10^7$. 
Depending on the relative sizes of the coupling constants $g_s^e$ and $g_s^N$, the resulting (dominant) combined astrophysical-laboratory bounds may be on the combination $(g_s^N)_\textrm{lab} \times (g_s^e)_\textrm{astro}$, $(g_s^e)_\textrm{lab} \times (g_s^e)_\textrm{astro}$ or $(g_s^N)_\textrm{lab} \times (g_s^N)_\textrm{astro}$.

The situation is different when one combines, for example, astrophysical bounds on $g_p$ with $V_1$-type laboratory bounds on $g_s$. 
In that case, those particular laboratory experiments tend to be practically insensitive to $g_p$, so one may assume $g_p \gg g_s$ to isolate the $g_p$ part of the astrophysical bounds, which are strictly on the combination of parameters $a g_s^2+ b g_p^2$, with the ratio $a/b \sim 1$.


\section{Units, symbols and abbreviations}\label{appendixF}

\subsection*{1. Units}
The International System of Units or \textit{Système International} (SI) is used in the main results section Sec.\,\ref{Sec:limits_main} for the convenience of experimentalists. 
In the International System of Units, the units used are meter (m) for length, kilogram (kg) for mass, and second (s) for time. 
The fundamental constants $\hbar$ and $c$ appear explicitly in these units. 

Natural units are predominantly employed in high-energy physics, particularly in particle physics and cosmology. 
The natural units are used in the theoretical framework section (Sec.\,\ref{Sec:formalism}), where $\hbar=c=1$. 

Atomic units are commonly used in atomic, molecular and optical physics. 
In these units, the elementary charge  $e$, electron mass $m_{\rm e}$, reduced Planck constant $\hbar$ and vacuum permittivity $\epsilon_0$ are set to $e=m_{\rm e}=\hbar = 4\pi\epsilon_0=1$, which implies that the speed of light in vacuum is $c= 137$. 

Each unit system helps to make particular calculations or theoretical developments more intuitive and/or less cluttered.

\subsection*{2. Symbols and abbreviations}
For ease of reference, the most common mathematical abbreviations and symbols are systematically listed in Tables \ref{Table:symbols} and \ref{Table:abbreviations}. 
These tables do not include terms that are limited to specific subtopics or that are only occasionally used; however, these terms are defined when they are first used. It should be noted that the chosen notations and abbreviations are consistent with those that are commonly used in the existing literature.

\begin{table}
\caption{Mathematical symbols used and their meanings. 
}
\renewcommand{\arraystretch}{1.2} 
\medskip \begin{tabular}{ll} \hline \hline
Symbol~~~~~~ & Meaning \\
\hline
\rule{0ex}{2.6ex} $c$ & speed of light in vacuum\\
\rule{0ex}{2.6ex} $G$ & Newtonian constant of gravitation\\
\rule{0ex}{2.6ex} $h$ & Planck constant\\
               &also reduced Planck constant $\hbar=h/(2\pi$) \\
\rule{0ex}{2.6ex} $g$  & local acceleration due to Earth's gravity \\ 
                       &or 
                       (Land\'{e}) $g$-factor of a particle\\
\rule{0ex}{2.6ex}  $m_{e}$ & electron mass \\
\rule{0ex}{2.6ex}  $m_{p}$ & proton mass \\
\rule{0ex}{2.6ex}  $m_{N}$ & average nucleon mass \\
\rule{0ex}{2.6ex}  $\sigma_{i}$ & Pauli matrices, $i=1,2,3$ \\
\rule{0ex}{2.6ex} $\gamma_{e,N}$ & gyromagnetic ratios of the electron or nucleus \\
\rule{0ex}{2.6ex} $\gamma _\mu$ & Dirac matrices, $\mu=0,1,2,3$ \\
\rule{0ex}{2.6ex} $\gamma _5$ & $\gamma_5 = i \gamma^0 \gamma^1 \gamma^2 \gamma^3$ \\ 
\rule{0ex}{2.6ex} $\sigma^{\mu \nu}$ & $\sigma^{\mu \nu}=\frac{i}{2}\left( \gamma^\mu \gamma^\nu - \gamma^\nu \gamma^\mu \right)$\\
\rule{0ex}{2.6ex} $\v{p}$ & linear momentum \\
\rule{0ex}{2.6ex} $\v{E}$ & electric field (vector)\\
\rule{0ex}{2.6ex} $\v{B}$ & magnetic field (vector)\\
\rule{0ex}{2.6ex} $C, P, T$ & charge conjugation, spatial inversion, and \\
\rule{0ex}{2.6ex} & time reversal discrete transformations\\
\rule{0ex}{2.6ex}  $\Lambda_{\mr{QCD}}$& quantum chromodynamics (QCD) energy scale\\
\rule{0ex}{2.6ex} $Q_\mr{W}$ & nuclear weak charge \\
\rule{0ex}{2.6ex} $\v{\sigma}$ & unit vector along spin\\
\hline \hline
\end{tabular}
\label{Table:symbols}
\end{table}

\begin{table}
\caption{Abbreviations and their meanings. 
}
\renewcommand{\arraystretch}{1.2} 
\medskip \begin{tabular}{ll} \hline \hline
Abbreviation~~ & Meaning \\
\hline
\rule{0ex}{2.6ex} ALPs & axionlike particles \\
\rule{0ex}{2.6ex} APV & atomic parity violation \\
\rule{0ex}{2.6ex} atm & atmosphere, a unit of measurement used\\
                      & to express pressure\\
\rule{0ex}{2.6ex} CPT & combined $CPT$ discrete transformation \\
\rule{0ex}{2.6ex} CPV & $CP$-violation \\
\rule{0ex}{2.6ex} CM & center-of-mass\\
\rule{0ex}{2.6ex} DFSZ & Dine-Fischler-Srednicki-Zhitnitsky (axion) \\
\rule{0ex}{2.6ex} DM & dark matter \\
\rule{0ex}{2.6ex} EDM & electric dipole moment \\
\rule{0ex}{2.6ex} ESR & electron spin resonance \\
\rule{0ex}{2.6ex} GDM & gravitational dipole moment \\
\rule{0ex}{2.6ex} ISL & inverse-square law \\
\rule{0ex}{2.6ex} KSVZ & Kim-Shifman-Vainshtein-Zakharov (axion) \\
\rule{0ex}{2.6ex} LHC & Large Hadron Collider \\
\rule{0ex}{2.6ex} NMR & nuclear magnetic resonance \\
\rule{0ex}{2.6ex} PNC & parity nonconservation\\
\rule{0ex}{2.6ex} PV & parity violation\\
\rule{0ex}{2.6ex} QCD & quantum chromodynamics \\
\rule{0ex}{2.6ex} QED & quantum electrodynamics \\
\rule{0ex}{2.6ex} SERF & spin-exchange relaxation-free \\
\rule{0ex}{2.6ex} SM & standard model \\
\rule{0ex}{2.6ex} SQUID & Superconducting Quantum\\
                        &Interference Device \\
\rule{0ex}{2.6ex} UBDM & ultralight bosonic dark matter\\
\rule{0ex}{2.6ex} UCN & ultra-cold neutrons\\
\rule{0ex}{2.6ex} WEP & weak equivalence principle\\
\rule{0ex}{2.6ex} WIMP & weakly-interacting massive particle \\
\hline \hline
\end{tabular}
\label{Table:abbreviations}
\end{table}

\newpage

\addcontentsline{toc}{section}{References}


\providecommand{\noopsort}[1]{}

\end{document}